\documentclass[12pt]{article}
\usepackage{amsmath,amssymb,amsfonts,amsthm,bbm}
\usepackage{epic,eepic,epsfig,longtable}
\usepackage{multirow,verbatim}
\usepackage{array}
\usepackage{graphicx}
\usepackage{mleftright}
\usepackage{xparse}
\usepackage{epsfig}
\usepackage{setspace}
\usepackage{CJK}
\usepackage{color}

\usepackage{lipsum}
\usepackage{mathtools}
\usepackage[authoryear,round]{natbib}
\usepackage{lineno,hyperref}
\usepackage{diagbox}

\usepackage{pdflscape}
\usepackage{stackengine}

\usepackage[normalem]{ulem}

\newcommand{\stkout}[1]{\ifmmode\text{\sout{\ensuremath{#1}}}\else\sout{#1}\fi}

\makeatletter
\def\singlespace{\def\baselinestretch{1}\@normalsize}

\numberwithin{equation}{section}

\renewcommand{\hat}{\widehat}

\renewcommand{\hat}{\widehat}

\newcommand{\bfm}[1]{\ensuremath{\mathbf{#1}}}

\def\bc{\bfm c}

\newcommand{\bfsym}[1]{\ensuremath{\boldsymbol{#1}}}

 \def\bmu{\bfsym {\mu}}

 \def\bvartheta{\bfsym \vartheta}
              \def\bSigma{\bfsym \Sigma}

 \def\bchi{\bfsym {\chi}}

\def\bvartheta{\bfsym{\vartheta}}	

\DeclareMathOperator{\argmax}{argmax}

\DeclarePairedDelimiter\norm{\lVert}{\rVert}

\def\today{\ifcase\month\or
  January\or February\or March\or April\or May\or June\or
  July\or August\or September\or October\or November\or December\fi
  \space\number\day, \number\year}

\newdimen\biblioindent    \biblioindent=30pt

 at 8truept

\newcommand{\beq}{\begin{equation}}
  \newcommand{\eeq}{\end{equation}}
\newcommand{\beqn}{\begin{eqnarray}}
  \newcommand{\eeqn}{\end{eqnarray}}
\newcommand{\beqnn}{\begin{eqnarray*}}
  \newcommand{\eeqnn}{\end{eqnarray*}}

\allowdisplaybreaks
\usepackage{verbatim}
\renewcommand{\baselinestretch}{1.66}
\def\tilde{\widetilde}

\def\[{\left [}  \def\]{\right ]} \def\({\left (}  \def\){\right )}
 \def\endpf{$\blacksquare$}
\def\hat{\widehat}

 \def \1 {\mathbf{1}}

\newtheorem{thm}{Theorem}

\newtheorem{lemma}{Lemma}

\theoremstyle{remark}
\newtheorem{remark}{Remark}
\theoremstyle{proposition}
\newtheorem{proposition}{Proposition}
\newtheorem{definition}{Definition}
\newtheorem{assumption}{Assumption}

\graphicspath{{figures/}{./images}}

\title{Dynamic Realized  Beta Models Using Robust Realized Integrated Beta Estimator}

\author{Donggyu Kim$^a$, Minseog Oh$^a$, Minjeong Song$^a$, and Yazhen Wang$^b$ \\
$^a$Korea Advanced Institute of Science and Technology (KAIST),\\
$^b$University of Wisconsin-Madison}

\begin{document}
\maketitle
\begin{spacing}{1.45}

\begin{abstract}
    This paper introduces a unified parametric modeling approach for time-varying market betas that can accommodate continuous-time diffusion and discrete-time series models based on a continuous-time series regression model to better capture the dynamic evolution of market betas.
    We call this the dynamic realized beta (DR Beta).
    We first develop a non-parametric realized integrated beta estimator using high-frequency financial data contaminated by microstructure noises, which is robust to the stylized features, such as the time-varying beta and  the dependence structure of microstructure noises, and construct the estimator's asymptotic properties. 
   Then,  with the robust realized integrated beta estimator, we propose a quasi-likelihood procedure for estimating the model parameters based on the combined high-frequency data and low-frequency dynamic structure. 
    We also establish asymptotic theorems for the proposed estimator and conduct a simulation study to check  the performance of finite samples of the estimator.
   The empirical study with the S\&P 500 index and the top 50 large trading volume stocks from the S\&P 500 illustrates that the proposed DR Beta model effectively accounts for dynamics in the market beta of individual stocks and better predicts future market betas.
\end{abstract}


\section{Introduction}\label{intro}
Market beta is a statistical measure of assets’ sensitivity to the overall market.
This measure plays a central role as the systemic risk measurement in financial applications such as asset pricing, risk management, and portfolio allocation \citep{fama2004capital,perold2004capital}.  
Thus, the characteristic of the market beta is a primary concern in empirical finance.
Especially, several empirical studies reported that  market betas vary over time \citep{bos1984empirical, breen1989economic, hansen1987role, keim1986predicting}.
To account for the  time-varying  property, low- and high-frequency finance modeling approaches have been independently adopted.
In the low-frequency financial modeling approach, we often employ discrete-time series regression models in either a non-parametric or parametric framework based on low-frequency data such as daily, weekly, and monthly return data.
For example, \citet{fama1973risk} used a rolling window regression approach with the ordinary least square (OLS) method, 
and \citet{black1992uk} employed the state-space model by using the Kalman filter method.
In addition, to account for market beta dynamics, several studies proposed auto-regressive time series models, such as generalized auto-regressive conditional heteroskedasticity (GARCH) model-type structures \citep{engle2016dynamic, gonzalez1996time, koutmos1994time, ng1991tests}.
In contrast,  \citet{bollerslev2016roughing} showed that incorporating high-frequency financial data offers more benefits while capturing beta dynamics.
Specifically, intraday data provide accurate estimations with sufficient data even within a short time period.
To exploit this property, several non-parametric market beta estimators based on high-frequency data  under  continuous-time series regression models have been developed.
For example, \citet{ barndorff2004econometric} employed the OLS method by calculating a ratio of the integrated covariance between assets and systematic factors to the integrated variation of systematic factors.
See also  \citet{andersen2006realized, li2017adaptive, reiss2015nonparametric}.
\citet{mykland2009inference} further computed the market beta as the aggregation of market betas estimated over local blocks. 
\citet{ait2020high} proposed an integrated beta approach, using spot market betas in the absence of market microstructure noise,
and \citet{andersen2020recalcitrant} investigated  intraday variation of spot market betas.
Recently, \citet{chen2018inference} introduced  the non-parametric inference for nonlinear volatility functionals of general multivariate It\^o semimartingales, in a high-frequency and noisy setting.
They do not allow any dependent structure for the microstructure noise.
However, several studies indicated that the microstructure noise has dependent structures \citep{jacod2019estimating,li2020robust, li2022remedi}.
 Thus, to measure the market beta accurately, we need to develop a robust realized beta estimation procedure.

Empirical studies have indicated that volatility has an auto-regressive time series structure \citep{bollerslev1986generalized, corsi2009simple,  engle1982autoregressive, engle2006multiple, hansen2012realized,  shephard2010realising}, and the market beta is calculated from the volatilities of the index and individual assets. 
Thus, we can conjecture that  market betas also have a  time series structure.
Some empirical studies supported this feature based on low-frequency stock return data \citep{engle2016dynamic, gonzalez1996time, koutmos1994time, ng1991tests}, and there are several studies that investigated the dynamics of the market beta based on the high-frequency financial data \citep{andersen2005framework, andersen2006realized,  hansen2014realized}. 
They found that incorporating high-frequency-based measures, such as  realized volatility and realized beta, helps account for the beta dynamics.
In these literatures, they often assumed that the market beta is constant over each low-frequency period, and they proposed low-frequency dynamic models using  the high-frequency-based measures without the mathematical connection with continuous diffusion models.
However, the high-frequency-based measures are developed, based on continuous diffusion processes, and we often observe that  the beta process is time-varying \citep{andersen2020recalcitrant, ait2020high}.
Thus, it is important to develop   a unified framework based on continuous diffusion models, which can account for the low-frequency time series dynamic structure based on the high-frequency financial data and time-varying beta.


%

In this paper, we introduce a novel parametric continuous-time regression model, which accommodates time-varying market betas, and we propose statistical inferences to better capture the financial features as follows.
First, we propose a spot market beta process, which is a continuous-time diffusion process and has the well-known autoregressive--moving-average (ARMA) form for the daily integrated beta. 
To capture the general dynamic structure, we propose ARMA$(p,q)$ structures.
Thus, our proposed model provides the rigorous mathematical background for unifying the general low-frequency time series dynamic models and the continuous diffusion processes.
Second, to make inferences for the proposed model, we first develop a non-parametric realized integrated beta ($RIB$) estimator for integrated betas with high-frequency data contaminated by microstructure noises, which is robust to the time-varying beta and the dependent structure of the microstructure noises.
For example, to handle the time-varying spot beta process and the dependent microstructure noise, we estimate spot volatilities using the robust pre-averaging method \citep{jacod2019estimating}.
Then,  we calculate the spot betas using spot volatility estimators and integrate them to obtain the realized integrated beta estimator. 
We show its asymptotic properties and obtain  the convergence rate $m^{-1/4}$, which is known as the optimal with the presence of  microstructure noises. 
To our best knowledge, the proposed $RIB$ is the first integrated beta estimator, which is robust to the financial features, such as the time-varying beta and the dependent micro-structure noises. 
Third, to estimate the model parameters, we suggest a quasi-maximum likelihood estimation procedure with the robust non-parametric $RIB$ estimator. 
For example, we use $RIB$ as the proxy for the corresponding conditional expected integrated beta and employ the well-known least square loss function.  
We establish asymptotic theorems for the proposed estimation procedure and discuss how to conduct hypothesis tests.

The rest of the paper is organized as follows.
In Section \ref{sec-2}, we introduce the dynamic realized beta (DR Beta) model and establish its properties.
In Section \ref{sec-3}, we propose statistical inference procedures for the integrated beta and the DR Beta model parameter, and we examine their asymptotic theorems. 
In Section \ref{sec-4}, we provide a simulation study to check the finite sample performance for the proposed estimators.
In Section \ref{sec-5}, we carry out an empirical study with the S\&P 500 index and 50 individual stocks to investigate the advantage of the proposed model. 
In Section \ref{sec-6}, we conclude.
All the proofs are contained in the \hyperref[Appendix]{Appendix}.

\section{Dynamic realized beta models} \label{sec-2}

In this section, we propose a diffusion process that enables forecasting and captures market dynamics.
We first fix some notations that we will use.
Let $\mathbb{R}_+=[0,\infty)$, and $\mathbb{N}$ be the set of all positive integers.
Let $A_{ij}$ denote the $(i,j)$th element of a matrix $A$, $A^\top$ denote its transpose matrix, and $\det(A)$ denote the determinant of $A$.
We use the superscripts $c$ and $d$ for the continuous and jump processes, respectively.

We consider  the following diffusion regression model:
\begin{equation}\label{Equation-2.1}
    dX_{2,t} = \beta_{t-}^c dX_{1,t}^c +  \beta^d_{t-} \Delta X^d_{1,t} +dV_t ,
\end{equation}
where $X_2$ is a dependent process, $X_1$ is a covariate process, and $V$ is a residual process. 
Further, $X^c_{1,t}$ denotes the continuous part of the covariate process, and $\Delta X^d_{1,t}$ is its jump at time $t$.  
Then, $\beta^c_t$ and $\beta^d_t$ are the time-varying factor loadings with respect to the continuous and jump parts of $X_1$, respectively.
We assume that $X_{1,t}$ and $V_t$ admit the following jump-diffusion models:
\begin{equation}\label{Equation-2.2}
    dX_{1,t}=\mu_{1,t}dt+\sigma_{t}dB_{t}+J_{1,t}d\Lambda_{1,t} \quad \text{and} \quad 
    dV_t=\mu_{2,t}dt+q _{t}dW_{t}+J_{2,t}d\Lambda_{2,t},
\end{equation}
where $\mu_{1,t}$ and $\mu_{2,t}$ are progressively measurable and locally bounded drifts, $\sigma_t$ and $q_t$ are adapted c\`adl\`ag processes, and  $B_t$ and $W_t$ are independent standard Brownian motions.
The stochastic processes $\mu_{1,t}$, $\mu_{2,t}$,  $\sigma_t$, and $q_t$ are defined on a filtered probability space $(\Omega, \mathcal{F} , \{\mathcal{F}_t, t \in [0,\infty) \}, P)$ with filtration $\mathcal{F}_t$ satisfying the general conditions.
Furthermore, $\sigma^2_t$ stays away from 0.
For the jump part, $\Lambda_{1,t}$ and $\Lambda_{2,t}$ are independent standard Poisson processes with bounded intensities $\lambda_{1,t}$ and $\lambda_{2,t}$, respectively, and $J_{1,t}$ and $J_{2,t}$ denote independent jump sizes, which are predictable and independent of Poisson and continuous diffusion processes.

In empirical finance, the regression-type model is widely used, and one of the most well-known discrete-time regression models is the capital asset pricing model (CAPM) \citep{lintner1965security, sharpe1964capital}.
By employing the individual asset price and market portfolio price processes as the dependent and covariate processes, respectively, we can consider the continuous-time series regression described in \eqref{Equation-2.1}, as the extension of the CAPM to the continuous-time diffusion process. 
Specifically, in the continuous-time CAPM in \eqref{Equation-2.1}, the market beta is time-varying, and the CAPM relationship holds at each time point. 
To measure the daily market beta, we use the following integrated beta ($I\beta$):
\begin{equation} \label{Equation-2.4}
    I\beta_i=\int_{i-1}^{i} \beta^c_t dt,    \quad i \in \mathbb{N}.
\end{equation} 
When the beta process is constant over time, the integrated beta returns to the usual market beta of the CAPM.  
That is, the continuous-time CAPM includes the traditional discrete-time CAPM. 

\begin{figure}[!ht] 
\centering
\includegraphics[width = 1\textwidth]{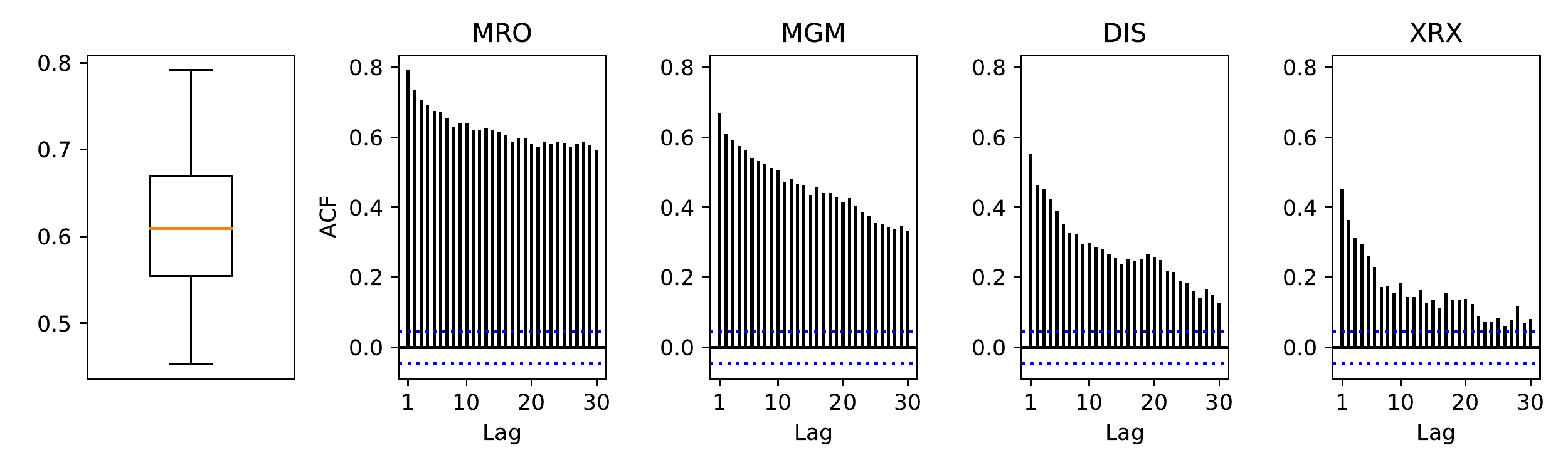}
\caption{The box plot (left) of the first-order auto-correlations of the daily realized betas for the top 50 large trading volume assets among the S\&P 500 from January 1, 2010, to December 31, 2016, and the ACF plots for the largest, 75th, 25th, and smallest first-order auto-correlation among the 50 assets excluding outliers.}
\label{Figure-1}
\end{figure} 
To check the low-frequency time series structure of high-frequency-based market betas, we draw auto-correlation function (ACF) plots for daily realized betas  for the top 50 large trading volume stocks  in Figure \ref{Figure-1}, where the specific form of the realized beta is defined in Section \ref{sec-3}.
Figure \ref{Figure-1} shows that the realized beta has a strong auto-regressive structure. 
To account for this beta dynamics, we consider the ARMA-type structure that is similar to the realized GARCH structure \citep{ hansen2012realized, song2020volatility} as follows.

\begin{definition}  \label{Definition-1}
For the proposed time-series regression model in \eqref{Equation-2.1}, a beta process $\beta_t^c(\theta)$,  $t \in \mathbb{R}_+$, follows the dynamic realized beta (DR Beta) model if it satisfies:
\begin{eqnarray} \label{EXTENDpq}
  \beta_{t}^{c}(\theta) &=& \beta_{[t]}^{c}(\theta) + \left( t - [t] \right) ^2 \left( \omega_1  + \sum_{i=1}^{p} \gamma_{i} \beta_{[t]+1-i}^{c} (\theta)  + \sum_{i=2}^{q} \alpha_{i} \int_{[t]-i+1}^{[t]-i+2} \beta_{s}^{c}(\theta) ds    \right)    \nonumber\\
  && - \left( t - [t] \right)    \(  \omega_ 2 +  \beta_{[t]}^{c}(\theta) \)  + \alpha_{1} \int_{[t]}^{t} \beta_s^c (\theta) ds  + \nu \left( [t] + 1 - t \right) \int_{[t]}^{t} dZ_t ,
\end{eqnarray}
where $[t]$ denotes the integer part of $t$ and $Z_t$ is a standard Brownian motion with $dZ_t dB_t=\rho dt$ and $dZ_t dW_t=0$  a.s.
We denote the model parameter by $\theta = \left( \omega_1, \omega_2, \gamma_1, \ldots, \gamma_{p} , \alpha_{1}, \ldots, \alpha_{q}, \nu \right) $.
\end{definition}

The spot beta process in the DR Beta model  is continuous at all times $t \in \mathbb{R}^+$
  and has a quadratic shape pattern within the intraday.
 For example, when considering the deterministic part of the spot beta process, by choosing appropriate parameters, high (low) initial betas form  downward (upward) convex shape with respect to time $t$.
This intraday structure can accommodate the intraday spot beta dynamics found in \citet{andersen2020recalcitrant}.
 We introduce $Z_t$ to account for the random fluctuations of the spot beta process.
 Moreover, the spot beta process can be considered as a generalized Ornstein-Uhlenbeck process.
 Specifically, for  $t \in (n-1, n]$, we have
\begin{eqnarray*} 
  d \beta_{t}^{c}(\theta) &=&  - \alpha_1  \Bigg \{  - \alpha_1   ^{-1} \Big (2  \left( t - [t] \right)  \left( \omega_1  + \sum_{i=1}^{p} \gamma_{i} \beta_{[t]+1-i}^{c} (\theta)  + \sum_{i=2}^{q} \alpha_{i} \int_{[t]-i+1}^{[t]-i+2} \beta_{s}^{c}(\theta) ds    \right)    \nonumber\\
  && -    \(  \omega_ 2 +  \beta_{[t]}^{c}(\theta) \)  - \nu  (Z_t - Z_{[t]})  \Big)   -  \beta_t^c (\theta)  \Bigg \}  dt   + \nu \left( [t] + 1 - t \right)  dZ_t.
\end{eqnarray*}
Finally, when the process is restricted to low-frequency time points, the spot beta adopts the following realized ARMA$(p,q)$ model-type structure: 
\begin{equation}\label{spot-int}
  \beta^c_n(\theta) = \omega + \sum_{i=1}^{p} \gamma_{i} \beta^c_{n-i} (\theta) + \sum_{j=1}^{q} \alpha_{j} \int_{n-j}^{n-j+1} \beta_{s}^{c} (\theta) ds \quad \text{ for any } n \in \mathbb{N} ,
\end{equation}
where $\omega= \omega_1- \omega_2$.
The spot beta process is along the lines of the GARCH-It\^o type processes  \citep{kim2016unified, song2020volatility}.
That is, the DR Beta process is developed to explain the low-frequency beta dynamics, which we find in the empirical study using the proposed robust non-parametric realized beta estimator, and fill the mathematical gap between the low-frequency discrete-time series and continuous-time series regression models.
Unlike the GARCH-It\^o type processes  \citep{kim2016unified, song2020volatility}, to capture more general dynamic structure, we propose the ARMA$(p,q)$ model-type structure.

\begin{remark}\label{remark-1}
In this paper, we separate the continuous and jump parts and mainly consider the continuous part. 
We also investigate market betas corresponding to the jump part in the empirical study, which is calculated based on the jump beta estimation method suggested by \citet{li2017robust}.
However, unlike the beta for the continuous part,  the beta for the jump part does not have significant time series structures (see Figure \ref{Figure-A}). 
Thus, we focus on developing the beta process for the continuous part. 
\end{remark}

The following proposition presents properties of the integrated betas, which will be employed for statistical inferences.
\begin{proposition}\label{Proposition-1}
  For $\sum_{i=1}^{p} \left|\gamma_{i} \right| < 1$, $\sum_{i=1}^{p \lor q} \left | \mathbf{1}_{\left\lbrace i \leq p \right\rbrace} \gamma_{i} + \mathbf{1}_{\left\lbrace i \leq p \lor q \right\rbrace} \alpha_{i}^{g} \right |   <1$, and $n \in \mathbb{N}$,  integrated betas for the DR Beta model in Definition \ref{Definition-1} have the following properties:
  \begin{enumerate}
    \item[(a)] We have
    \begin{equation}\label{HandD}
      I\beta_n(\theta)=\int ^{n}_{n-1}\beta_t^c(\theta)dt=h_n(\theta)+D_n\quad a.s.
      ,
    \end{equation}
    where
    \begin{align}
      & h_n(\theta) = \omega^g + \sum_{i=1}^{p}  \gamma_{i} h_{n-i}(\theta) + \sum_{i=1}^{p \lor q} \alpha_{i}^{g} I\beta_{n-i} , \label{GARCH-pq} \\
      & \omega^g = \left( \varrho_{1} - \varrho_{2} + 2 \varrho_{3} \right) \omega + (2 \varrho_{3} - \varrho_{2}) \left( 1 - \sum_{i=1}^{p} \gamma_{i} \right)  \omega_{2} , \nonumber \\
      & \alpha_{i}^{g} = \mathbf{1}_{\left\lbrace i \leq p \right\rbrace} 2 \varrho_{3} \gamma_{i} \alpha_{1} + \mathbf{1}_{\left\lbrace i \leq q \right\rbrace} (\varrho_{1} - \varrho_{2}) \alpha_{i} + \mathbf{1}_{\left\lbrace i \leq q-1 \right\rbrace} 2 \varrho_{3} \alpha_{i+1} , \nonumber \\
      & \varrho_1 = \alpha_{1}^{-1} \left( e^{\alpha_{1}} - 1 \right) , \quad \varrho_{2} = \alpha_{1}^{-2} \left( e^{\alpha_{1}} - 1 - \alpha_{1} \right) , \quad \varrho_{3} = \alpha_{1}^{-3} \left( e^{\alpha_{1}} - 1 - \alpha_{1} - \frac{\alpha_{1}^{2}}{2}  \right) , \nonumber
    \end{align}
    and
    \begin{equation*}
      D_n = \nu \int_{n-1}^{n} \left[ (n-t) \alpha_{1}^{-1} e^{\alpha_{1} (n-t)} - \alpha_{1}^{-2} e^{\alpha_{1}(n-t)} + \alpha_{1}^{-2} \right] dZ_t
    \end{equation*}
    is a martingale difference.

    \item[(b)] $\beta^c_n$ and $  I\beta_{n}$ have a finite moment for any given order, and we have 
    \begin{eqnarray*}
      &&\mathbb{E}[h_n(\theta)]=\frac{\omega^g}{1-\sum_{i=1}^{p}  \gamma_{i} -  \sum_{j=1}^{p \lor q} \alpha^g_{j}}, \\
      &&\mathbb{E}[\beta^c_n]=\frac{\omega \left( 1-\sum_{i=1}^{p}  \gamma_{i} -  \sum_{j=1}^{p \lor q} \alpha^g_{j} \right) +  \omega^g \sum_{i=1}^{p} \alpha_{i}}{\left(1-\sum_{i=1}^{p}  \gamma_{i} -  \sum_{j=1}^{p \lor q} \alpha^g_{j} \right)(1- \sum_{i=1}^{p} \gamma_{i})}, \quad \text{and} \\
      &&\mathbb{E}\left[ I\beta_{n}(\theta) | \mathcal{F}_{n-1} \right] = h_n(\theta) \quad \text{a.s.} 
    \end{eqnarray*}
  \end{enumerate}

\end{proposition}

Proposition \ref{Proposition-1}(a) indicates that the integrated betas $I\beta_n$'s can be decomposed into the conditional expectation $h_n(\theta)$ and the martingale difference $D_n$.
Further, $h_n(\theta)$ is adapted to the filtration $\mathcal{F}_{n-1}$. 
Furthermore, \eqref{HandD} and \eqref{GARCH-pq} imply that
\begin{equation*}
  I\beta_{n} = \omega^{g} + D_n + \sum_{i=1}^{p \lor q} (\mathbf{1}_{\left\lbrace i \leq p \right\rbrace} \gamma_{i} + \mathbf{1}_{\left\lbrace i \leq p \lor q \right\rbrace} \alpha_{i}^{g}) I\beta_{n-i} + \sum_{i=1}^{p} (-\gamma_{i}) D_{n-i}.
\end{equation*}
This means the daily integrated betas have the ARMA$(p \lor q, p)$ structure.
We impose the ARMA ($p,q$) structure on the spot beta process, but the integrated beta has the $p\vee q$ order instead of the  $q$ order. 
This is because, by the construction, each conditional expectation has the current innovation; thus, we have at least $p$ innovations as in Proposition \ref{Proposition-1}(a). 
In the empirical study, we find that this DR Beta structure helps to capture the beta dynamic  structure.
Details can be found in Section \ref{sec-5}.

We use the integrated beta relationship \eqref{HandD} to make statistical inferences and predict the future integrated beta. 
To investigate the low-frequency beta dynamics, we need only the following relationship:
\begin{equation*}  
I\beta_n(\theta)=\int ^{n}_{n-1}\beta_t^c(\theta)dt=h_n(\theta)+D_n\quad a.s.,
\end{equation*}
and  the specific form of the spot beta process $\beta_t^c(\theta)$  defined in Definition \ref{Definition-1} is not required. 
In fact, the intraday dynamics pattern, such as the convex pattern, is averaged out;  thus, its specific dynamic form of the intraday dynamics does not affect the low-frequency dynamics. 
That is, the DR Beta model is not the only solution and shows the existence of the diffusion process, satisfying \eqref{HandD}. 
For example, at each integer point, \eqref{spot-int} is satisfied, and between integer points, we can interpolate. 
In this paper, we adopt the quadratic interpolation. 
Alternatively, we can use linear interpolation or more higher interpolation.
On the other hand, we can also use a step function form for the spot beta over each low-frequency period--that is, it does not need to be continuous. 
In the following section, we discuss how to estimate model parameters under the integrated beta relationship in \eqref{HandD}.

\section{Statistical inferences} \label{sec-3}

In this section, we propose an estimation procedure for the DR Beta model and examine its statistical properties.
We first discuss the model set-up. 
For the proposed time series regression model in \eqref{Equation-2.1}, we assume that time-varying market betas follow a stochastic process defined on $(\Omega, \mathcal{F}, P)$ as follows:
\begin{equation} \label{Equation-3.beta}
	d \beta^c_t = \mu_{\beta,t}dt+\sigma_{\beta, t}dB_{\beta, t},
\end{equation} 
where $B_{\beta, t}$ is a standard Brownian motion with $dB_{\beta, t}dW_t=0$ and $dB_{\beta, t} dB_t=\rho_{\beta,t} dt$ a.s. 
For the high-frequency observations, one of the stylized features is that the transaction prices are polluted by the market microstructure noise due to the discreteness of the price, bid-ask spread bounce, and adverse selection effects, such as clearing costs \citep{ait2008high}.
To reflect this, we assume that the observed log prices have the additive microstructure noise as follows: 
\begin{equation} \label{Equation-3.1}
    Y_{1,i}^{m}=X_{1,t_i}+\epsilon_{1,i}^{m} \quad \text{and} \quad    Y_{2,i}^{m}=X_{2,t_i}+\epsilon_{2,i}^{m} ,  
\end{equation} 
where $\epsilon_{1,i}^{m}$ and $\epsilon_{2,i}^{m}$ are the noise.
Empirical studies reveal that the microstructure noise is dependent on the true price \citep{ait2011ultra,hansen2006realized,ubukata2009estimation} and has positive auto-correlation \citep{jacod2017statistical,li2021remedi}.
To capture this, we allow  microstructure noises to have dependence on the true latent price and diurnal features of the noise.
Before describing our assumption about microstructure noise, we state the $\rho$-mixing property of a stationary random vector $\bfsym\chi = (\bfsym\chi_i)_{i \in \mathbb{Z}} = ((\chi_{1,i},\chi_{2,i})^\top)_{i \in \mathbb{Z}}$.
\begin{definition}\label{rho-mixing}
  For a stationary process $\bfsym\chi$, let $\mathcal{G}_j = \sigma(\bfsym\chi_i : i \leq j)$ and $\mathcal{G}^j = \sigma(\bfsym\chi_i : i \geq j)$ be the pre- and post-$\sigma$-fields at time $j$.
  A stationary process $\bfsym\chi$ is $v$-polynomially $\rho$-mixing if for some $C > 0$, $\rho_k(\bfsym\chi) \leq C/k^v$ for all $k \geq 1$, where
  \begin{eqnarray*}
       \rho_k(\bfsym\chi)  & =& \sup \big\{ |\mathbb{E}(UV)| :  U \text{ and } V \text{ are random variables, measurable  with respect to}  \\
      && \mathcal{G}_0 \text{ and } \mathcal{G}^k, \text{ respectively} , \quad \mathbb{E}(U) = \mathbb{E}(V) = 0,    \mathbb{E}(U^2) \leq 1,  \mathbb{E}(V^2) \leq 1 \big\},
  \end{eqnarray*}
  $\mathcal{G}_j = \sigma(\bfsym\chi_i : i \leq j)$, and $\mathcal{G}^j = \sigma(\bfsym\chi_i : i \geq j)$.
\end{definition}

\begin{assumption}\label{assumption-noise}
  The noise $(\bfsym {\epsilon}_{i}^m)_{i \in \mathbb{Z}} = ((\epsilon_{1,i}^{m}, \epsilon_{2,i}^{m})^\top)_{i \in \mathbb{Z}}$ is realized as
  \begin{equation}\label{noise-form}
      \epsilon_{1,i}^{m} = \vartheta_{1,t_i} \chi_{1,i} \quad \text{and} \quad \epsilon_{2,i}^{m} = \vartheta_{2,t_i} \chi_{2,i} ,
  \end{equation}
  where $\vartheta_1$ and $\vartheta_2$ are locally bounded non-negative It\^{o} semimartingales.
  Furthermore, $(\bfsym\chi_i)_{i \in \mathbb{Z}}$ is a stationary process, independent of the $\sigma$-field $\mathcal{F}_{\infty} = \bigvee_{t>0} \mathcal{F}_{t}$ and $v$-polynomially $\rho$-mixing for some $v \geq 4 $.
  $\chi_{1,i}$ and $\chi_{2,i}$ are mean 0 and variance 1 with finite moments of all orders.
\end{assumption}
\begin{remark}
  Assumption \ref{assumption-noise} implies that there exists a constant $C$ such that $\left|r_{ab}(i)\right| \leq \frac{C}{\left( \left|i\right| +1 \right)^v }$ for all $i \in \mathbb{Z}$ and $a,b \in \left\lbrace 1,2 \right\rbrace$, where $r_{ab}(i) = \mathbb{E}[\chi_{a,0} \chi_{b,i}]$.
  Thus, $R_{ab} = \sum_{i\in \mathbb{Z}} r_{ab}(i)$ is well defined.
\end{remark}

\subsection{Non-parametric estimation: Realized integrated beta} \label{sec-3.1}

To evaluate the DR Beta model and its feasibility, the integrated beta is a key element.
Unfortunately, the integrated beta is unobservable, so we need its non-parametric estimator. 
When it comes to estimating the integrated beta based on the observed high-frequency financial data, there are a couple of obstacles. 
One is the microstructure noise, and the other  is the intraday dynamics of the spot beta process. 
In this section, we discuss how to overcome these issues for the general stochastic beta process in \eqref{Equation-3.beta}.

For simplicity, we temporarily assume that the distance between adjacent observations is equal to $\Delta_{m}={1}/{m}$, where $m$ is the number of high-frequency observations. 
We denote the high-frequency observed time points $t_l={l}/{m}$ for $l=1,  \ldots , m$. 
This equally spaced observation time assumption can be easily extended to irregular observation time points. 
We discuss this later.
To manage the intraday dynamics--that is, the time-varying spot beta process--we can use the following relationship:
$$
  \frac{d} {dt}[X^c_{2,t}, X^c_{1,t}] = \beta_t^c \frac{d} {dt} [X^c_{1,t}, X^c_{1,t}] ,
$$ 
where $X^c_{2,t}$ is the continuous part of the individual asset log price process, and   $[\cdot, \cdot]$ denotes the quadratic covariation.
Thus, we can obtain the spot beta by comparing the spot volatility of $X_1$ and the spot covolatility between $X_1$ and $X_2$  as follows:
\begin{equation}\label{eq-beta}
 \beta_t^c  = \frac{ \frac{d} {dt} [X^c_{2,t}, X^c_{1,t}]}{  \frac{d} {dt}[X^c_{1,t}, X^c_{1,t}]}.
\end{equation}
Then,  by integrating the spot beta process, we finally obtain the integrated beta. 
Therefore, as long as the spot volatility estimators perform well, we can estimate the integrated beta.

To estimate spot volatilities, we employ the estimation method  developed for estimating integrated volatility with microstructure noise \citep{ait2010high,barndorff2008designing,barndorff2011multivariate,christensen2010pre,fan2018robust,jacod2009microstructure,jacod2019estimating,xiu2010quasi, zhang2006efficient,zhang2005tale, zhang2016jump}.
In order to handle the auto-correlation structure  of the microstructure noise, we employ the pre-averaging method in \citet{jacod2019estimating} as follows.
We choose a sequence of integers, $k_m$, such that ${k_m}= C_k \Delta_m^{-1/2}$ for some positive constant $C_k$.
We select a weight function $g(\cdot)$ on $[0,1]$ satisfying that $g (\cdot)$
is continuous, piecewise continuously differentiable with a piecewise Lipschitz derivative ${g}'$ with $g(0)=g(1)=0$ and  $\int_{0}^{1}g^2(s)ds>0$. 
Let
\begin{eqnarray*}
    &&\phi_0(s)=\int ^1_s g(u)g(u-s)du,  \;  \psi_0=\phi_0(0), \cr
    &&\phi_1(s)=\int ^1_s g'(u)g'(u-s)du, \; \psi_1=\phi_1(0), \cr
    &&\Phi_{00}=\int^1_0 \phi_0^2(s)ds, \; \Phi_{01}=\int^1_0 \phi_0(s) \phi_1(s)ds, \; \Phi_{11}=\int^1_0 \phi_1^2(s)ds.
\end{eqnarray*}
We also choose a sequence of integers, $l_m$, such that $l_m=C_l \Delta_m ^{-\varsigma}$ for some positive constant $C_l$ and $\varsigma \in [\frac{1}{8}, \frac{1}{5} ]$.
Then, for $l =1, \ldots, m$, $d = 1, \ldots, l_m$, and any processes $P$ and $P'$, we define
\begin{eqnarray*}
  && g_{d}^{m} =  g \left( \frac{d}{k_m}  \right) , \quad P_{l}^{m} = P_{t_l }, \quad
  \tilde{P}_{l}^{m}=\sum_{j=1}^{k_m -1}g_{j}^{m} \left( P_{l+j}^{m} - P_{l+j-1}^{m} \right)  , \quad \bar{P}_{l}^{m}  = \frac{1}{l_m} \sum_{i=0}^{l_m-1} P_{l+i}^{m}, \\
  && \mathcal{E}_{PP',l}^{m,d}  = \left( P_{l}^{m} - \bar{P}_{l+2l_m}^{m}   \right) \left( P_{l+d}^{'m} - \bar{P}_{l+4l_m}^{'m} \right) .
\end{eqnarray*}
Then,  the spot covariance matrix $\bSigma$ of $(X_1^c, X_2^c)^\top$ at time $t_l$ is estimated with
\begin{equation*} \label{Equation-3.3}
    \hat{\bSigma}_{l}^{m}  = \hat{\bSigma}_{t_l} =
    \begin{pmatrix}
      v(Y_1,Y_1,u_{1,m},u_{1,m},u_{11,m},t_l) & v(Y_1,Y_2,u_{1,m},u_{2,m},u_{12,m},t_l) \\
      v(Y_2,Y_1,u_{2,m},u_{1,m},u_{21,m},t_l) & v(Y_2,Y_2,u_{2,m},u_{2,m},u_{22,m},t_l)
    \end{pmatrix},
  \end{equation*}
  where
  \begin{eqnarray*}
    && v(P,P',a, a',a'' t_l) = 
    \frac{1}{(b_m - 2 k_m) \Delta_m k_m \psi_0} 
    \Bigg\lbrace  \sum_{i=0}^{b_m - 2 k_m -1} \tilde{P}_{l+i}^{m}  \tilde{P}_{l+i}^{'m}  \mathbf{1}_{\{|\tilde{P}_{l+i}^{m}|\leq a, \, |\tilde{P}_{l+i}^{'m}|\leq a' \}} \\
    && \qquad \qquad \qquad \qquad \qquad \qquad \quad - \frac{1}{k_m}  \sum_{i=0}^{b_m - 6 l_m}  \hat{\mathcal{E}}_{PP',l+i}^{m} \mathbf{1}_{\{ |\hat{\mathcal{E}}_{PP',l+i}^{m}| \leq a''  \}}  \Bigg\rbrace,\\
    && \phi_{d}^{m} = k_m \sum_{i\in\mathbb{Z}} \left( g_{i+1}^{m} - g_{i}^{m} \right) \left( g_{i-d+1}^{m} - g_{i-d}^{m} \right)  , \quad \hat{\mathcal{E}}_{PP',i}^{m} = \sum_{d=-k_m'}^{k_m'} \phi_{d}^{m}  \mathcal{E}_{PP',i}^{m,d} ,
  \end{eqnarray*}
$b_m = C_b \Delta_m^{-\kappa}$, $k'_m = C_{k'} \Delta_m ^{-\tau}$ for some positive constant $C_b$ and $C_{k'}$, tuning parameters  
$\kappa \in ( \frac{5}{8}  , \frac{3}{4}  )$ and $\tau \in  (\frac{1}{4v-4}  , \frac{1}{8}   ]$, 
$u_{1,m}$, $u_{2,m}$, $u_{11,m}$, $u_{12,m}$, $u_{21,m}$, and $u_{22,m}$ are the thresholds chosen as $u_{1,m} = a_{1} (k_m \Delta_m )^{\varpi_1}$, $u_{2,m} = a_{2} (k_m \Delta_m )^{\varpi_1}$, $u_{11,m} = a_{11} \Delta_m^{-\varpi_2}$, $u_{12,m} = a_{12} \Delta_m^{-\varpi_2}$, $u_{21,m} = a_{21} \Delta_m^{-\varpi_2}$, and $u_{22,m} = a_{22} \Delta_m^{-\varpi_2}$ for some $a_{1}, a_{2}, a_{11}, a_{12}, a_{21}, a_{22}>0$, $\varpi_1 \in (\frac{5-4\kappa}{8}    , \frac{[v]+2\kappa -4}{2[v]-4})$, and $\varpi_2 \in (\tau, \frac{4\kappa - 1 - 2\varsigma}{8}  )$.
Under some mild conditions, we can show the consistency of the spot volatility estimator (see Theorem \ref{Theorem-1} and \citet{figueroa2020kernel}).
Then,  we can construct an realized integrated beta ($RIB$) estimator based on \eqref{eq-beta} and \eqref{Equation-3.3} as follows:
\begin{equation} \label{Equation-3.4}
RIB_{1}=\hat{I \beta}_{1}=b_m\Delta_m \sum ^{\lfloor \frac{1}{b_m\Delta_m} \rfloor -1 }_{i=0} 
\left[ \hat{\beta}_{i b_m }^{m}  - \hat{B}^{m}_{ib_m} \right], \quad \hat{\beta}^m_{i}=\frac{\hat{\bSigma}_{12,i}^{m}}{\hat{\bSigma}_{11,i}^{m,*}}, \quad  \hat{\bSigma}_{11,i}^{m,*} = \max( \hat{\bSigma}_{11,i}^{m}, \delta_m),
\end{equation}
where $\delta_m$ is a sequence of positive real numbers converging to zero and 
$\hat{B}^{m}_{ib_m}$ is a de-biasing term of the form
\begin{eqnarray}\label{estimators-for-debiasing}
    && \hat{B}^{m}_{i b_m } =   \frac{4}{\psi_0^{2} {C_k}^3 b_m\Delta_m^{1/2}} \left[ \left( \frac{{C_k}^2 \Phi_{01}}{\hat{\bSigma}_{11,ib_m}^{m,*}} + \frac{\Phi_{11} \hat{\bvartheta}_{11,ib_m}^{m}}{\left( \hat{\bSigma}_{11,ib_m}^{m,*} \right)^2 }   \right) \left( \hat{\bvartheta}_{11,ib_m}^{m} \frac{\hat{\bSigma}_{12,ib_m}^{m}}{\hat{\bSigma}_{11,ib_m}^{m,*}} - \hat{\bvartheta}_{12,ib_m}^{m}  \right)  \right]   ,\\
    && \hat{\bvartheta}^{m}_{11,i} = (b_m-6l_m)^{-1} \sum_{j=i}^{i + b_m - 6 l_m}  \dot{\mathcal{E}}_{Y_1 Y_1,j}^{m} \mathbf{1}_{\{ |\dot{\mathcal{E}}_{Y_1 Y_1,j}^{m}| \leq \dot{a}_{11} \Delta_m ^{\varpi_2} \}}, \quad \dot{\mathcal{E}}_{Y_1 Y_1,i}^{m} =  \sum_{d=-k_m'}^{k_m'} \mathcal{E}_{Y_1 Y_1,i}^{m,d}, \nonumber\\
    && \hat{\bvartheta}^{m}_{12,i} = (b_m-6l_m)^{-1} \sum_{j=i}^{i + b_m - 6 l_m}  \dot{\mathcal{E}}_{Y_1 Y_2,j}^{m} \mathbf{1}_{\{ |\dot{\mathcal{E}}_{Y_1 Y_2,j}^{m}| \leq \dot{a}_{12} \Delta_m ^{\varpi_2} \}}, \quad \dot{\mathcal{E}}_{Y_1 Y_2,i}^{m} =  \sum_{d=-k_m'}^{k_m'} \mathcal{E}_{Y_1 Y_2,i}^{m,d}, \nonumber\\
    && \hat{\bvartheta}^{m}_{22,i} = (b_m-6l_m)^{-1} \sum_{j=i}^{i + b_m - 6 l_m}  \dot{\mathcal{E}}_{Y_2 Y_2,j}^{m} \mathbf{1}_{\{ |\dot{\mathcal{E}}_{Y_2 Y_2,j}^{m}| \leq \dot{a}_{22} \Delta_m ^{\varpi_2} \}}, \quad \dot{\mathcal{E}}_{Y_2 Y_2,i}^{m} =  \sum_{d=-k_m'}^{k_m'} \mathcal{E}_{Y_2 Y_2,i}^{m,d} \nonumber
\end{eqnarray}
for some constant $\dot{a}_{11}, \dot{a}_{12}, \dot{a}_{22} > 0$.
We utilize $\hat{\bSigma}_{11,l}^{m,*}$ instead of $\hat{\bSigma}_{11,l}^{m}$ for estimating spot beta in order to prevent the denominator of $\hat{\beta}^m_{i}$ from being a non-positive value.

To investigate the asymptotic behavior of the $RIB$ estimator, we need the following technical conditions.
\begin{assumption}\label{assumption-formal}
~
\begin{enumerate}
\item[(a)]  The spot volatility matrix $\bSigma_{t}$ is locally bounded, and there exists $\delta > 0$ such that $\sigma_t > \delta$ a.s. for all $t \in \mathbb{R}_{+}$.


\item[(b)]  We have
\begin{equation*}
    d\sigma^2_{t}=\tilde{\mu}_{1,t}dt+\tilde{\sigma}_{t}d\tilde{B}_{t}+\tilde{J}_{1,t}(d{\tilde{\Lambda}}_{1,t}-\tilde{\lambda}_{1,t}dt) \;  \text{ and } \;  dq^2_{t}=\tilde{\mu}_{2,t}dt+\tilde{q}_{t}d\tilde{W}_{t}+\tilde{J}_{2,t}(d{\tilde{\Lambda}}_{2,t}-\tilde{\lambda}_{2,t}dt),
\end{equation*} 
where $\tilde{\mu}_{1,t}$ and  $\tilde{\mu}_{2,t}$ are progressively measurable and locally bounded drifts; $\tilde{\sigma}_{t}$ and   $\tilde{q}_{t}$ are adapted locally bounded processes.
 The standard Brownian motions $\tilde{B}_{t}$ and $\tilde{U}_{t}$ satisfy almost surely
\begin{eqnarray*}
    && d\tilde{B}_{t}dB_{\beta,t}=0, \quad d\tilde{B}_{t}dW_{t}=0, \quad d\tilde{B}_{t}d\tilde{W}_{t}=0, \quad d\tilde{B}_{t}dB_{t}=\tilde{\rho}_1dt, \\
    && d\tilde{W}_{t}dB_{\beta, t}=0, \quad d\tilde{W}_{t}dB_{t}=0, \quad d\tilde{W}_{t}d\tilde{B}_{t}=0, \quad d\tilde{W}_{t}dW_{t}=\tilde{\rho}_2dt,
\end{eqnarray*}
where $\tilde{\rho}_1$ and $\tilde{\rho}_2$ are bounded.
The stochastic processes $\tilde{\mu}_{1,t}, \tilde{\mu}_{2,t}, \tilde{\sigma}_{t}$, and $\tilde{q}_{t}$ are defined on $(\Omega,\mathcal{F}, P)$.
For the jump part, $\tilde{\Lambda}_{1,t}$ and $\tilde{\Lambda}_{2,t}$ are independent Poisson  processes with bounded intensities $\tilde{\lambda}_{1,t}$ and $\tilde{\lambda}_{2,t}$, respectively, and $\tilde{J}_{1,t}$ and $\tilde{J}_{2,t}$ denote independent jump sizes, which are predictable and independent of the Poisson processes and the continuous diffusion processes, $ \tilde{J}_{1,t} ^2 < \infty$  and $ \tilde{J}_{2,t} ^2 < \infty$ a.s.  

\item[(c)] 
The drift term of $  \bSigma_{t}$, $\tilde{\bmu}_t=\{\tilde{\bmu}_{pq,t}\}_{1\leq p,q\leq 2}$, the spot covariance of the continuous part of $\bSigma_{t}$, $\tilde{\bSigma}^c_{t}=\{\tilde{\bSigma}^c_{pq,dr,t}\}_{1\leq p,q,d,r \leq 2}$, and the spot covariance of the jump part of $\bSigma_{t}$, $\tilde{\bSigma}^d_{t}=\{\tilde{\bSigma}^d_{pq,dr,t}\}_{1\leq p,q,d,r \leq 2}$, are c\`adl\`ag.

\item[(d)]
If $P_{t}$ is one of the processes $\tilde{\sigma}_{t}$, $\tilde{q}_{t}$, $\tilde{\rho}_{1,t}$, or $\rho_{\beta,t}$, then it satsifies the following properties:
\begin{enumerate}
  \item [(P1)] $P_t$ is locally bounded;
  \item [(P2)] There exist $C > 0$, such that $\mathbb{E}\left[ (P_{t+s} - P_{t})^{2} | \mathcal{F}_{t} \right] \leq C s$ a.s.  for any $t,s \geq 0$.
\end{enumerate}
\end{enumerate}
\end{assumption} 
\begin{remark}
We restrict the intensities of the jumps to be bounded. 
However, we can relieve this condition as the jump activity can be infinity as long as the jump variation is finite. 
Under this condition, we can show the result in the similar way. 
\end{remark}


The following theorem establishes the convergence rate and asymptotic distributions for the proposed $RIB$ estimator.
\begin{thm} \label{Theorem-1}
    Under Assumptions \ref{assumption-noise} and \ref{assumption-formal}, we have
    \begin{equation*}
        m ^{1/4}(RIB_{1}-I \beta_1) \rightarrow \int_{0}^{1} \mathcal{R}_s d\tilde{Z}_s \quad \mathcal{F}_{\infty}\text{-stably as }  m \rightarrow \infty ,
    \end{equation*}
    where $\tilde{Z}$ is a standard Brownian motion independent of $\mathcal{F}$, $\mathcal{R}_s$ is the square root of
    \begin{equation*}
      \mathcal{R}_s = \frac{2}{\psi_0^2} \left( \Phi_{00}  \frac{{C_k} q_s^2}{\sigma_s^2} + \Phi_{01} \frac{A_{1,s}}{C_k} + \Phi_{11} \frac{A_{2,s}}{{C_k}^3}   \right) 
      ,
    \end{equation*}
    and
    \begin{eqnarray*}
      A_{1,s} &=&  \frac{\vartheta_{1,s}^2 R_{11} q_s^2 - 2 \beta_{s}^{c} \vartheta_{1,s}\vartheta_{2,s} R_{12} + \vartheta_{2,s}^2 R_{22}}{\sigma_s^2} + \vartheta_{1,s}^2 R_{11} (\beta_{s}^{c})^2 , \\
      A_{2,s} &=&  \frac{\vartheta_{1,s}^{2}}{\sigma_{s}^{4}} \left( 2(\beta_{s}^{c})^2 \vartheta_{1,s}^{2} R_{11}^{2} - 4 \beta_{s}^{c} \vartheta_{1,s} \vartheta_{2,s} R_{11} R_{12} + \vartheta_{2}^{2} (R_{11}R_{22} + R_{12}^2) \right) 
      .
    \end{eqnarray*}
\end{thm}

Theorem \ref{Theorem-1} shows that the convergence rate of the $RIB$ estimator is $m^{-1/4}$, which is known as the optimal rate with the presence of  microstructure noise and establishes its asymptotic normality.
To extend the estimator over all periods, we set $m_i$ to be the total number of high-frequency observations for the $i$th day and rewrite $m = \sum_{i=1}^n m_i /n$. 
We further let the $t_{i,j}$'s be the high-frequency observed time points for the $i$th day, such that $i-1=t_{i,0}<t_{i,1}< \cdots <t_{i,m_i}=i$, where   $\left| t_{i,j}-t_{i,j-1}\right| =O\left( m^{-1}\right)$ for all $i, j$. 
Then,  we can construct the $RIB$ estimator as follows:
\begin{equation}  \label{Equation-3.5}
RIB_{i}=\frac{b_{m_{i}}}{m_i} \sum_{l=0}^{\left [ m_i/b_{m_i} \right ]-1 } 
\hat{\beta}_{i,t_{i,lb_{m_i}}}\quad \text{and} \quad  \hat{\beta}_{i,t_{i,l}}=\frac{\hat{\bSigma}_{21}(t_{i,l})}{\max(\hat{\bSigma}_{22}(t_{i,l}),\delta_{m_i})}.
\end{equation}
Moreover, we can show Theorem \ref{Theorem-1} for each $RIB_i$ under the usual assumption in the asynchronous high-frequency data analysis (see Assumption \ref{Assumption-2}(d)). 
We utilize these well-performing $RIB$ estimators to build up a parametric estimation procedure in the following subsection.

To utilize the asymptotic distribution result, we need to construct a consisent asymptotic variance estimator.
In the following proposition, we propose the asymptotic variance estimator and show its consistency.

\begin{proposition}\label{prop-AsympVar}
  Under Assumptions \ref{assumption-noise} and \ref{assumption-formal}, $\hat{S}_m =  b_m \Delta_m  \sum_{i=0}^{\left[ \frac{1}{b_m\Delta_m} \right]  -1 } \hat{\mathcal{R}}^{2,m}_{i b_m} $ is a consistent asymptotic variance estimator of the $RIB$ estimator, where
  \begin{eqnarray*}
    \hat{\mathcal{R}}^{2,m}_{i} &=& \frac{2C_k}{\psi_{0}^{2}} \Biggl[ \Phi_{00}  \left( \frac{\hat{\bSigma}_{22,i}^{m}}{\hat{\bSigma}_{11,i}^{m,*}} - \frac{(\hat{\bSigma}_{12,i}^{m})^2}{(\hat{\bSigma}_{11,i}^{m,*})^2}  \right)     + \frac{\Phi_{01}}{C_k^{2}}  \left( \frac{\hat{\bvartheta}_{22,i}^{m}}{\hat{\bSigma}_{11,i}^{m,*}} -  \frac{2\hat{\bSigma}_{12,i}^{m} \hat{\bvartheta}_{12,i}^{m}  }{ ( \hat{\bSigma}_{11,i}^{m,*}  )^2  }  + \frac{\hat{\bSigma}_{22,i}^{m} \hat{\bvartheta}_{11,i}^{m} }{( \hat{\bSigma}_{11,i}^{m,*}  )^2 }   \right) \\
    &&  + \frac{\Phi_{11}}{C_k^3}  \left( \frac{2  (\hat{\bSigma}_{12,i}^{m} \hat{\bvartheta}_{11,i}^{m} )^2    }{(\hat{\bSigma}_{11,i}^{m,*} )^4} + \frac{\hat{\bvartheta}_{11,i}^{m} \hat{\bvartheta}_{12,i}^{m} }{(\hat{\bSigma}_{11,i}^{m,*} )^2} - 4 \frac{\hat{\bSigma}_{12,i}^{m} \hat{\bvartheta}_{11,i}^{m} \hat{\bvartheta}_{12,i}^{m}  }{(\hat{\bSigma}_{11,i}^{m,*} )^3}  + \frac{(\hat{\bvartheta}_{11,i}^{m} )^2}{(\hat{\bSigma}_{11,i}^{m,*} )^2}  \right)     \Biggl]
    ,
  \end{eqnarray*}
  where $\hat{\bvartheta}_{11,i}^{m}$, $\hat{\bvartheta}_{12,i}^{m}$, and $\hat{\bvartheta}_{12,i}^{m}$ are defined in \eqref{estimators-for-debiasing}.
  That is, we have $\hat{S}_m \xrightarrow[]{p} \int_{0}^{1} \mathcal{R}_{s}^2 ds$.
\end{proposition}


\subsection{Parametric estimation for the DR Beta model} \label{sec-3.2}

\subsubsection{Estimation procedure based on high-frequency data and a low-frequency structure}

According to the strong auto-regressive structure of the $RIB_i$'s in Figure \ref{Figure-1}, we now assume that time-varying market betas follow the DR Beta process defined in Definition 1.
Proposition \ref{Proposition-1}(a) indicates that the integrated beta over the $i$th period can be decomposed into the conditional expectation $h_i(\theta)$ and  the martingale difference $D_i$.
As the impact of the martingale difference term $D_i$ is negligible in the asymptotic sense, the $RIB_i$'s can be the observable proxies for the $h_i(\theta)$'s. 
Thus, to estimate the true parameters $\theta_0=(\omega^g_0, \gamma_{0,1}, \ldots, \gamma_{0,p}, \alpha^g_{0,1},\ldots, \alpha^g_{0,p\vee q})$, we consider the well-known  ordinary least squares estimation, which compares the proxies and conditional expectations, as follows:
\begin{equation}  \label{Equation-3.6}
    L_{n,m}(\theta)=-\frac{1}{n}\sum^n_{i=1} \left\{ RIB_i-h_i(\theta) \right\}^2.
\end{equation}
To harness the quasi-likelihood function above, we first need to evaluate the conditional expectation term $h_i(\theta)$.
Unfortunately, the true integrated betas are not observable.
Thus, we adopt their non-parametric estimators $RIB_{i}$ to evaluate $h_i(\theta)$ as follows:
\begin{eqnarray*}
  \hat{h}_n(\theta) = \omega_g + \sum_{i=1}^{p}  \gamma_{i} \hat{h}_{n-i}(\theta) + \sum_{j=1}^{p \lor q} \alpha_{j}^{g} RIB_{n-j}
  .
\end{eqnarray*}
Then,  we define the quasi-likelihood function as follows:
\begin{equation}  \label{Equation-3.7}
    \hat{L}_{n,m}(\theta)=-\frac{1}{n}\sum^n_{i=1} \left\{ RIB_i-\hat{h}_i(\theta) \right \}^2,
\end{equation}
and estimate the model parameters by maximizing  the quasi-likelihood function $\hat{L}_{n,m}(\theta)$ as follows:
 \begin{equation*}
    \hat{\theta} =\argmax_{\theta \in \Theta} \hat{L} _{n,m}(\theta),
\end{equation*}
 where  $\Theta$ is the parameter space of $\theta$.

\subsubsection{Asymptotic theory}

In this subsection, we establish asymptotic theorems for the proposed estimator $\hat{\theta}$.
We first define some notations. 
Define $\left\| A\right\| _{\max}=\smash{\displaystyle\max_{1 \leq i \leq k, 1 \leq j \leq k'}}  \left| A_{ij}\right|$ for a $k \times k'$ matrix $A$.
Let $C>0$ be generic constants whose values are free of $n$ and $m$ and may change from occurrence to occurrence.

To explore the asymptotic behaviors of $\hat{\theta}$, the following technical conditions are required.

\begin{assumption} \label{Assumption-2} 
~
\begin{enumerate}
\item [(a)] Let
\begin{eqnarray*}
   \Theta &=& \{ (\omega^g, \gamma_{1}, \ldots, \gamma_{p}, \alpha_{1}^{g}, \ldots, \alpha_{p \lor q}^{g}): \omega^g_l<\omega^g<\omega^g_u, \, \gamma_l< \gamma_{1} , \ldots, \gamma_{p}  <\gamma_u, \\
   && \alpha^g_l < \alpha_{1}^g, \ldots, \alpha_{p \lor q}^{g} <\alpha^g_u,  \sum_{i=1}^{p} \left|\gamma_{i} \right| < 1,  \sum_{i=1}^{p \lor q} \left | \mathbf{1}_{\left\lbrace i \leq p \right\rbrace} \gamma_{i} + \mathbf{1}_{\left\lbrace i \leq p \lor q \right\rbrace} \alpha_{i}^{g} \right |  <1 \},
\end{eqnarray*}
where $\omega^g_l, \omega^g_u, \alpha^g_l, \alpha^g_u, \gamma_l,\gamma_u$ are known constants such that $\alpha^g_l, \alpha^g_u, \gamma_l$ and $\gamma_u$ are in  $(-1,1)$.

\item [(b)] For all  $\theta \in \Theta$, $|\varphi_{\theta} (x) | = 0  \Rightarrow |x| > 1 $,  where
 $
\varphi _{\theta} (x)= 1-  \sum_{i=1}^{p \lor q}  (\mathbf{1}_{\left\lbrace i \leq p \right\rbrace} \gamma_{i} + \mathbf{1}_{\left\lbrace i \leq p \lor q \right\rbrace} \alpha_{i}^{g} ) x^i.
$

\item [(c)]    $\Upsilon_{ \theta_0} (x)$ does not have no common root with $\varphi _{\theta_0} (x)$, where 
 $
\Upsilon_{ \theta} (x)=  \sum_{i=1}^{p }  \gamma_i  x^i.
$

\item [(d)] There exist some fixed constants $C_1$, $C_2$ such that $C_1 m \leq m_i \leq C_2 m$, and \\
$\smash{\displaystyle\sup_{1\leq j \leq m_i}} \left| t_{i,j}-t_{i,j-1}\right| =O\left( m^{-1}\right)$ and $(n \log n )^2 m^{-1} \rightarrow 0$ as $m,n \rightarrow \infty$.

\item [(e)] We have $\smash{\displaystyle\sup_{i \leq n }} \left |RIB_i-I\beta_i \right|= O_p( m^{-1/4} \sqrt{\log n} )$.

\end{enumerate}
\end{assumption}

\begin{remark} 
Assumption \ref{Assumption-2}(a)--(c) are usually imposed when analyzing asymptotic properties of the ARMA-type models. 
For example, Assumption \ref{Assumption-2}(b) implies the stationarity of $I\beta_i$ and $h_i(\theta)$, and Assumption \ref{Assumption-2}(c) is required to identify the parameter space. 
 Finally, Assumption \ref{Assumption-2}(e) is  required to handle the estimation errors of the unobserved integrated betas. 
Under the locally bounded condition for the spot volatility and beta processes, similar to the proofs of \citet{fan2018robust, Kim2016SPCA} and Theorem \ref{Theorem-1}, we can show that the beta estimators have sub-Exponential tails.
Then, with the additional $\sqrt{ \log n}$ term, we can handle the maximum error of  $\left |RIB_i-I\beta_i \right|$ over the $n$ period. 
Alternatively, if we have $\smash{\displaystyle\sup_{i \leq n }} E \[ \left |RIB_i-I\beta_i \right|^2 \] \leq  C  m^{-1/2} $, we can establish the asymptotic results in Theorems \ref{Theorem-2} and \ref{Theorem-3}.
On the other hand, if we consider the estimated realized beta $RIB_i$'s as the observations, as in the realized GARCH model \citep{hansen2012realized}, we do not need this condition. 
\end{remark}

The following theorems provide the asymptotic results  including  the convergence rate and asymptotic normality for the proposed parameters $\hat{\theta}$. 

\begin{thm} \label{Theorem-2}
Under Assumptions \ref{Assumption-2} (except for $(n \log n )^2m^{-1} \rightarrow 0$), we have
\begin{equation*}
    \| \hat{\theta} -\theta_{ 0} \|_{max}=O_p( m^{-1/4} \sqrt{\log n }+n^{-1/2}).
\end{equation*}
\end{thm}

\begin{thm} \label{Theorem-3}
Under Assumptions \ref{Assumption-2}, we have, as $m,n\rightarrow \infty$, 
\begin{equation*}
    \sqrt{n}(\hat{\theta} -\theta_0) \overset{d}{\rightarrow} N(0,V),
\end{equation*} where
\begin{equation} \label{eq-V}
     V = 4 \alpha_{0,1}^{-4}\nu_0^2\int^1_0\{\alpha_{0,1} (1-t-\alpha_{0,1}^{-1})e^{\alpha_{0,1}(1-t)}+1\}^2tdt   \left ( \mathbb{E} \left[  \left. 
    {\dfrac{\partial h_1(\theta)}{\partial \theta} \dfrac{\partial h_1(\theta)}{\partial \theta^\top} }\right|_{\theta=\theta_0}  \right]\right )^{-1}.
\end{equation}
\end{thm}

\begin{remark}
    Theorem \ref{Theorem-2} shows that the quasi-maximum likelihood estimator $\hat{\theta}$ has the converge rate $m^{-1/4} \sqrt{\log n} +n^{-1/2}$. 
    The first term $m^{-1/4}$ comes from estimating the integrated beta, which is known as the optimal convergence rate  with the presence of  market microstructure noise.
    The second term $n^{-1/2}$ is the typical parametric convergence rate based on the low-frequency observations. 
    Theorem \ref{Theorem-3} establishes the asymptotic normality of $\hat{\theta}$.
\end{remark}

\subsubsection{Hypothesis tests}\label{sec-3.3}

In financial practices, we are interested in the validity of the model parameters $(\omega^g, \gamma_{1}, \ldots, \gamma_{p},$ $\alpha_{1}^{g}, \ldots, \alpha_{p \lor q}^{g})$ and make statistical inferences, such as hypothesis tests. 
To do this, we can harness   the asymptotic normality result in Theorem \ref{Theorem-3} as follows:
 \begin{equation*}
    T_n = \sqrt{n} \hat{V}^{-1/2}(\hat{\theta} -\theta_0)  \overset{d}{\rightarrow} N(0,\mathbf{I}),
\end{equation*}
where $\hat{V}$ is a consistent estimator of the asymptotic variance $V$ defined in \eqref{eq-V}, and $\mathbf{I}$ is a $3 \times 3$  identical matrix. 
Then,  with the test statistics $T_n$, we can conduct hypothesis tests based on the standard normal distribution.
To evaluate the statistics $T_n$, we use the following asymptotic variance estimator,
\begin{eqnarray} \label{Equation-3.8}
  \hat{V} &=& \frac{1}{n} \sum^{n}_{i=1} \left \{ RIB_i-\hat{h}_i (\hat{\theta}) \right \}^2 \left( \frac{1}{n} \sum_{j=1}^{n} \frac{\partial \hat{h}_{j}(\hat{\theta})}{\partial \theta} \frac{\partial \hat{h}_{j}(\hat{\theta})^{\top}}{\partial \theta} \right)^{-1} 
  .
\end{eqnarray}
Its consistency can be derived similarly to the proof of Theorem \ref{Theorem-3}.

\section{A simulation study} \label{sec-4}

We conducted simulations to check the finite sample performance of the proposed statistical inference procedures.  
For simplicity, we chose $p=q=1$ for the DR Beta model.
We generated the beta processes $\beta^c_{t_{i,j}} $ and  $\beta^d_{t_{i,j}}$ and the jump-diffusion processes $X_{1,t_{i,j}}$ and $X_{2,t_{i,j}}$ for $ t_{i,j}=i-1+j/m$, $i=1,2, \ldots, n, \, j=1,2, \ldots, m$ from the DR Beta model defined in Definition \ref{Definition-1}  as follows:
\begin{eqnarray*}
    && dX_{2,t}=\beta^c _{t} (\theta ) dX^c_{1,t}+\beta^d_t J_{1,t} d\Lambda_{1,t}+dV_t,  \cr  &&dX_{1,t}=\sigma_tdB_t+J_{1,t}d\Lambda_{1,t},  \quad   dV_t=q_tdW_t+J_{2,t}d\Lambda_{2,t}, \cr 
    && d \beta_{t}^{c}(\theta) =   \Bigg \{  2  \left( t - [t] \right)  \left( \omega_1  + \gamma_{1} \beta_{[t]}^{c} (\theta)      \right)  - \(  \omega_ 2 +  \beta_{[t]}^{c}(\theta) \)  - \nu  (Z_t - Z_{[t]}) + \alpha_1 \beta_t^c (\theta)  \Bigg \}  dt  \\
    && \qquad\qquad + \nu \left( [t] + 1 - t \right)  dZ_t \\
    &&dZ_t dB_t=\rho dt, \quad dZ_t dW_t= dB_t dW_t=0,
\end{eqnarray*}
where $(\omega_1, \omega_2, \gamma_1, \alpha_1, \nu, \rho)=(0.7, -0.5, 0.1, 0.37, 1.5, -0.6)$, $q_t = 0.012$, and $B_t$, $Z_t$, and $W_t$ are standard Brownian motions.
Then,  the model parameter is $\theta_0=(\omega^g_0, \gamma_0,\alpha^g_0)=(0.13, 0.25,$ $0.10)$. 
We generated the individual asset log price process $X_{1,t}$ based on the realized GARCH-It\^o model \citep{song2020volatility} as follows:
\begin{eqnarray*}
   d\sigma^2_t &= &\left \{2 \tilde{\gamma}(t-\lceil t-1\rceil)(\tilde{\omega}_1+\sigma^2_{\lceil  t-1\rceil})-(\tilde{\omega}_2+\sigma^2_{\lceil t-1\rceil})+\tilde{\alpha}\sigma^2_t -\tilde{\nu} \tilde{Z}_t^2 \right \} dt \cr
    &&  +\tilde{\beta}  J_{1,t}^2 d\Lambda_{1,t} +2\tilde{\nu}(\lceil t-1\rceil+1-t)\tilde{Z}_t d\tilde{B}_t,
\end{eqnarray*}
where a standard Brownian motion $\tilde{B}_t$  satisfies $d\tilde{B}_t dW_t=d\tilde{B}_t dU_t=0, d\tilde{B}_t dB_t=\tilde{\rho}dt$,       $(\tilde{\omega}_1,\tilde{\omega}_2,\tilde{\gamma},\tilde{\alpha}, \tilde{\beta},\tilde{\nu},\tilde{\rho})=(3.02\times 10^{-5}, 4.00\times 10^{-6}, 0.35, 0.4, 0.1,1 \times 10^{-5}, -0.424)$, and $\tilde{Z}_t=  \tilde{B}_t - \tilde{B}_{\lceil t-1\rceil}$.
The initial values for the simulation data were chosen to be $\beta^c_0 (\theta) = \mathbb{E}\left[ \beta^c_1 (\theta) \right] = 2.16$, $\sigma_0^2= \mathbb{E}\left[ \sigma_1^2 \right] = 3.12\times 10^{-5}$,  $X_{1,0}=16$, and $X_{2,0}=10$. 
 For the jump part, we took the intensities $\lambda_{1,t}=4$ and $\lambda_{2,t}=5$, and  the jump sizes $J_{1,t}$  and $J_{2,t}$ were generated as follows:
 \begin{eqnarray*}
 	J_{1,t}^2= \max(2 \times 10^{-5} + M_{1,t}, \quad 4 \times 10^{-5})  \quad  \text{and} \quad J^{2}_{2,t}   = \max(1 \times 10^{-5} + M_{2,t}, \quad 2 \times 10^{-5}) ,
 \end{eqnarray*}
 where $M_{1,t}$ and $M_{2,t}$ follow   $N(0,(2 \times 10^{-6})^2)$ and $N(0,(1 \times 10^{-6})^2)$, respectively.
For each $J_{1,t}$ and $J_{2,t}$, we further assigned a positive (negative) sign with probability $0.5$ to make a positive (negative) jump.
Finally, $\beta^d_t$ was chosen to be   $1.5$, and we generated Brownian motions using the Euler scheme.

The noisy high-frequency data $Y_{1,t_{i,j}}$ and  $Y_{2,t_{i,j}}$ were generated from the model \eqref{Equation-3.1}, where the true log price processes  $X_{1,{t_{i,j}}}$ and $X_{2,{t_{i,j}}}$ were generated from the DR Beta model above, and the microstructure noises $\epsilon_{1,t_{i,j}}$  and  $\epsilon_{2,t_{i,j}}$ follow \eqref{noise-form}, where $\vartheta_{1,t}$, $\vartheta_{2,t}$, and $\bchi_{i}$ follow Ornstein--Uhlenbeck-type processes with a U-shaped pattern and the AR$(1)$ process with Gaussian innovations as follows:
\begin{align*}
  & d\vartheta_{1,t} = 10(\mu_{\vartheta_1,t} - \vartheta_{1,t})dt + s_1 dB_t, \quad d\vartheta_{2,t} = 10(\mu_{\vartheta_2,t} - \vartheta_{2,t})dt + 0.6 s_2 dB_t + 0.8 s_2 dW_t, \\
  & \mu_{\vartheta_1,t} = s_1 (1 + 0.1\cos(2\pi t)), \quad  \mu_{\vartheta_2,t} = s_2 (1 + 0.1\cos(2\pi t)), \\
  & s_1 = 3.440 \times 10^{-4} ,\quad s_2 = 1.151 \times 10^{-3}, \\
  & \bchi_i = \begin{pmatrix} 0.8 & 0 \cr 0 & 0.8 \end{pmatrix} \bchi_{i-1} + e_{i}, \quad e_{i} \sim_{i.i.d.} N\left[\begin{pmatrix} 0 \cr 0 \end{pmatrix},\begin{pmatrix} 0.360 & 0.168 \cr 0.168 & 0.360 \end{pmatrix}\right]
  .
\end{align*}
We chose $n=125, \, 250, \, 500$ and $m=2340, \, 4680, \, 23400$.
For each combination of $n$ and $m$, we repeated the simulation process $1000$ times.

\begin{figure}[!ht] 
  \centering
  \includegraphics [width = 0.8\textwidth]{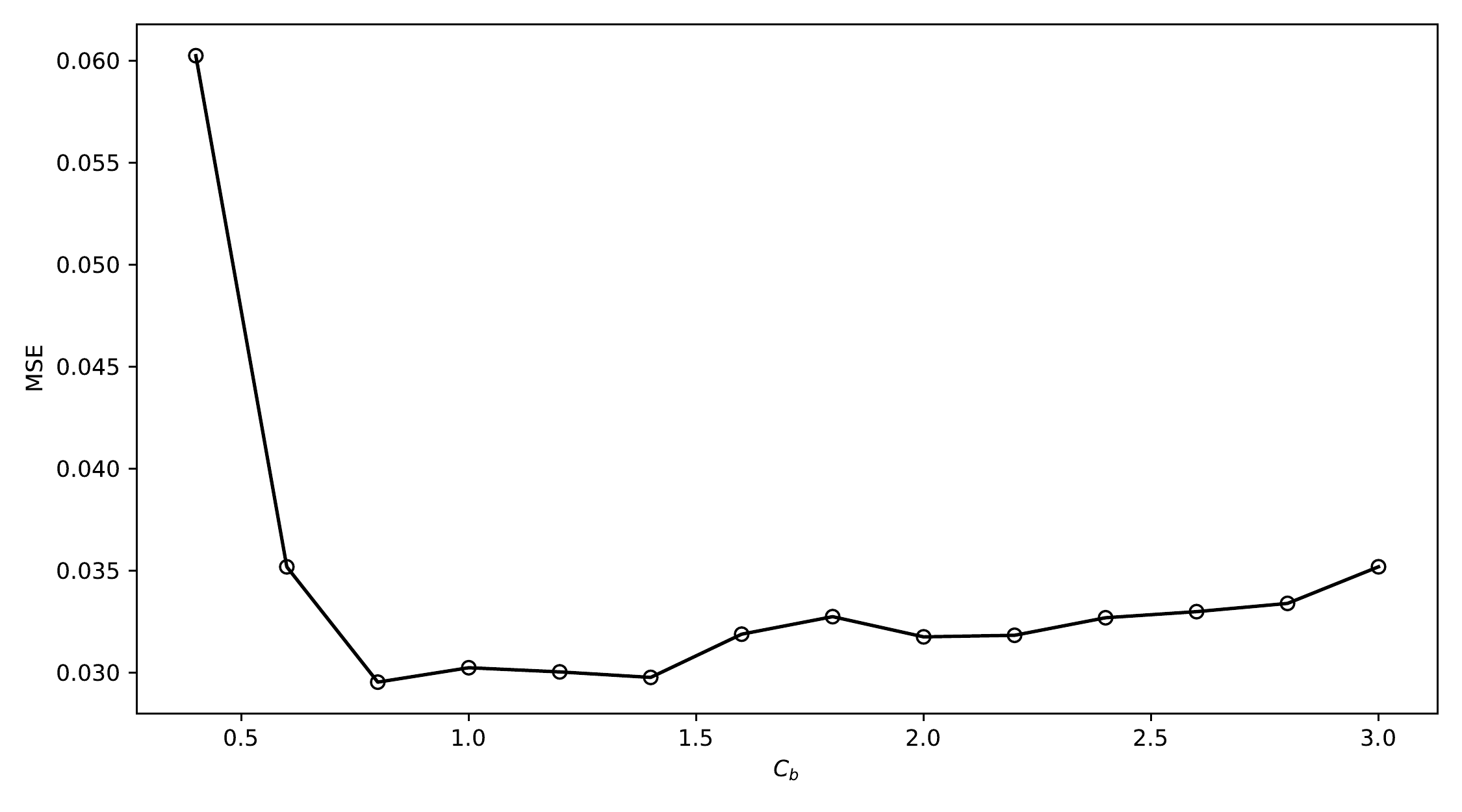}
  \caption{The MSEs of $RIB$ estimator with $m= 23400$ against varying $C_b$.}
  \label{Figure-sensitivity}
\end{figure}
For the $RIB$ estimator, we used the usual triangular weight function $g(x)=\{ x \wedge (1-x)\}$, and set $k_m = 0.8 [\Delta_m ^{-0.5}]$ and $\varpi_1=0.47$ as recommended by \citet{christensen2010pre} and \citet{ait2016increased}, respectively.
Then, we simply fixed $k_m' = [\Delta_m ^{-0.124}]$, $l_m = [\Delta_m ^{-0.15}]$, $\varpi_2=0.15$, and $\delta_m = 1.0 \times 10^{-5}$.
In addition, for the truncation, we chose $a_1$ and $a_2$ as four times the sample standard deviation of the pre-averaged prices $k_m^{-1/2} \tilde{Y}_{1,t_{d,k}}^{m}$ and $k_m^{-1/2} \tilde{Y}_{2,t_{d,k}}^{m}$, respectively, and $a_{11}$ and $a_{12}$ as $0.2$ times the sample standard deviation of $\hat{\mathcal{E}}_{Y_1 Y_1, i}$ and $\hat{\mathcal{E}}_{Y_1 Y_2, i}$, respectively.
We then needed to determine $C_b$.
To do this, we checked the effect of the choice of $C_b$ of the $RIB$ estimator.
Figure \ref{Figure-sensitivity} depicts the estimated mean squared errors (MSE) of the $RIB$ estimator with $m=23400$ against varying $C_b$ from $0.4$ to $3.0$, where the integrated beta $I\beta_i$ is calculated as the Riemann sum of the true beta values for each trading days.
From Figure \ref{Figure-sensitivity}, we find that for $C_b < 1$, the MSEs usually decrease as $C_b$ increases, and for $C_b \geq 1$, the MSEs usually increase as $C_b$ increases.
This may be because the window size for the spot betas should be large enough to estimate spot betas, but too large a window size hinders the capture of the intraday dynamics of the beta processes.
From this analysis, we set $C_b=1$.

We first checked the performance of the non-parametric integrated beta estimator, $RIB$,  proposed in Section \ref{sec-3.1}.
For comparison, we employed other integrated beta estimators  proposed by \citet{chen2018inference} and \citet{christensen2010pre}.
For example, \citet{chen2018inference} proposed the estimator for volatility functionals, and specifically, the integrated beta (CHEN) from \citet{chen2018inference} was estimated as follows:
\begin{equation*}
  \hat{I\beta}^{CHEN} = b_m \Delta_m  \sum_{i=0}^{[\frac{1}{b_m \Delta_m }] -1} \left[ \hat{\beta}_{ib_m}^{m,C} - \hat{B}_{ib_m}^{m,C} \right], \quad \hat{\beta}_{ib_m}^{m,C} = \frac{\hat{\bSigma}_{12,i}^{m,C}}{\hat{\bSigma}_{11,i}^{m,C,*}}, \quad \hat{\bSigma}_{11,i}^{m,C,*} = \max(\hat{\bSigma}_{11,i}^{m,C},\delta_m)
  ,
\end{equation*}
where
\begin{eqnarray}\label{def-chen}
  && \hat{\bSigma}_{i}^{m,C} = \frac{1}{(b_m - k_m) \Delta_m k_m \psi_0} \sum_{l=0}^{b_m - k_m + 1}  \left(  \tilde{\bold{Y}}_{i+l}^{m} \tilde{\bold{Y}}_{i+l}^{m\top} \mathbf{1}_{\left\lbrace \norm{\tilde{\bold{Y}}_{i+l}^{m}} \leq u_m \right\rbrace} - \hat{\bold{Y}}_{i+l}^{m} \right), \nonumber\\
  && \hat{\bold{Y}}_{i+l}^{m} = \frac{1}{2} \sum_{l=1}^{k_m} (g_{l}^{m} - g_{l-1}^{m})^2 (\bold{Y}_{i+l}^{m} - \bold{Y}_{i+l-1}^{m}) (\bold{Y}_{i+l}^{m} - \bold{Y}_{i+l-1}^{m})^{\top}, \nonumber\\
  && \hat{B}^{m, C}_{i b_m } =   \frac{4}{\psi_0^{2} {C_k}^3 b_m\Delta_m^{1/2}} \left[ \left( \frac{{C_k}^2 \Phi_{01}}{\hat{\bSigma}_{11,ib_m}^{m,*}} + \frac{\Phi_{11} \hat{\bvartheta}_{11,ib_m}^{m,C}}{\left( \hat{\bSigma}_{11,ib_m}^{m,C,*} \right)^2 }   \right) \left( \hat{\bvartheta}_{11,ib_m}^{m,C} \frac{\hat{\bSigma}_{12,ib_m}^{m,C}}{\hat{\bSigma}_{11,ib_m}^{m,C,*}} - \hat{\bvartheta}_{12,ib_m}^{m,C}  \right)  \right], \nonumber\\
  && \hat{\bvartheta}_{ib_m}^{m,C} = \frac{1}{2k_m}  \sum_{l=1}^{k_m} (\bold{Y}_{i+l}^{m} - \bold{Y}_{i+l-1}^{m}) (\bold{Y}_{i+l}^{m} - \bold{Y}_{i+l-1}^{m})^{\top}, \quad \bold{Y} = (Y_1, Y_2)^{\top} \nonumber
  ,
\end{eqnarray}
and $b_m$, $k_m$, and the truncation parameters are the same as that of the $RIB$ estimator.
  \begin{figure}[b] 
  \centering
  \includegraphics [height=0.5\textwidth,width = 0.5\textwidth]{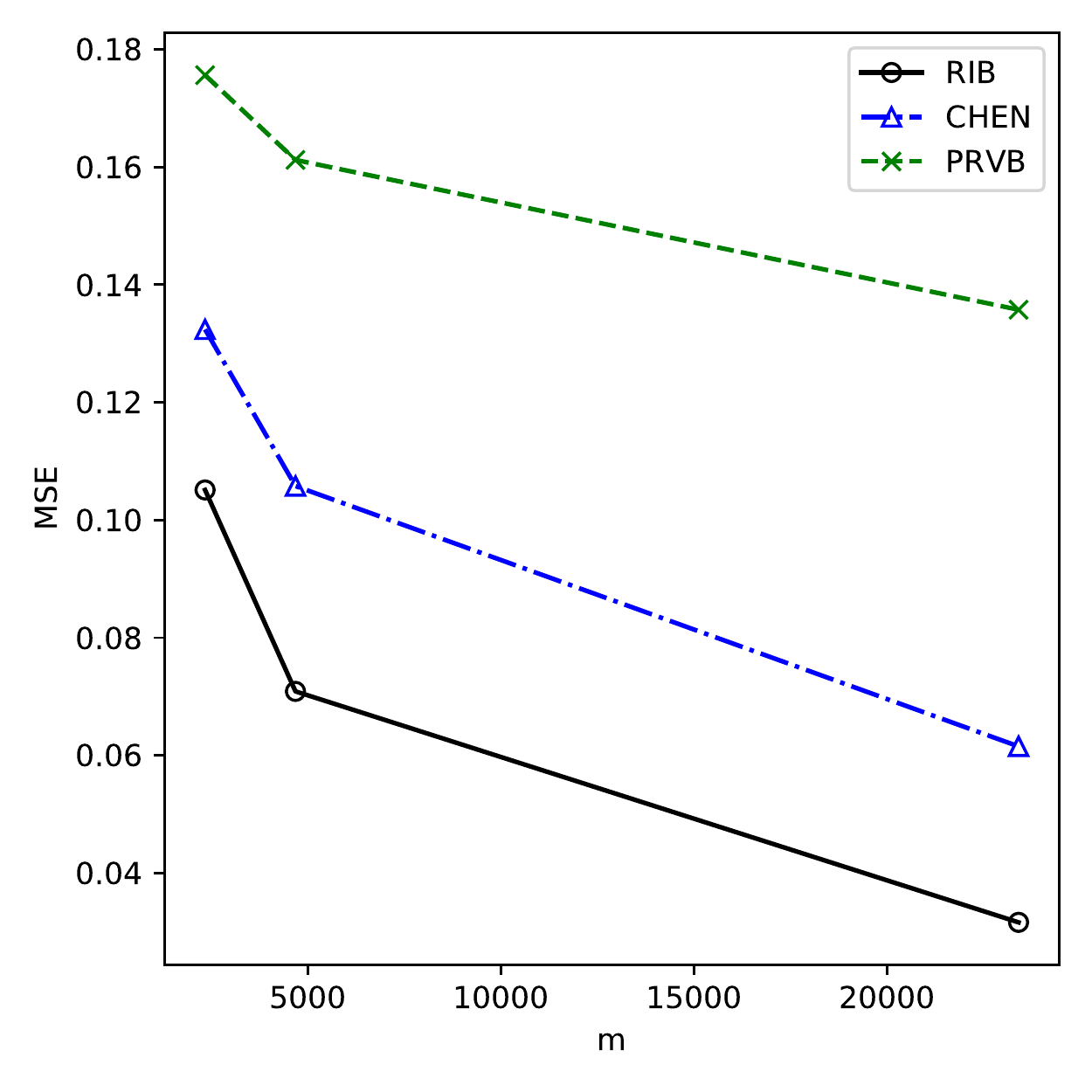}
  \caption{The  log MSEs  of  $RIB$, CHEN, and  PRVB  for  $m=2340, \, 4680, \, 23400$.}
  \label{Figure-nonparam}
\end{figure}
\citet{christensen2010pre} and \citet{jacod2009microstructure} proposed a pre-averaged realized covariance estimator that is robust to price jump and microsturcture noise. 
If we assume that the true beta value is constant during a trading day, the integrated beta can be estimated from the pre-averaged volatility matrix (PRVB) as follows:
\begin{eqnarray*}
  \hat{IB}^{PRVB} = \frac{\hat{\bSigma}_{12}^{m,P}}{\hat{\bSigma}_{11}^{m,P}}, \quad \hat{\bSigma}^{m,P} = \sum_{l=0}^{m - k_m + 1}  \left(  \tilde{\bold{Y}}_{l}^{m} \tilde{\bold{Y}}_{l}^{m\top} - \hat{\bold{Y}}_{l}^{m} \right)\mathbf{1}_{\left\lbrace \norm{\tilde{\bold{Y}}_{l}^{m}} \leq u_m \right\rbrace} ,
\end{eqnarray*}
where $k_m$ and the truncation parameters are the same as that of the $RIB$ estimator.
We note that PRVB assumes that the spot beta is constant over time. 
On the other hand, CHEN is designed for estimating time-varying beta, but it does not consider the dependent structure of the microstructure noise.
Figure  \ref{Figure-nonparam} shows the MSEs of the non-parametric integrated beta estimators,  $RIB$, CHEN, and PRVB, for   $m=2340, \, 4680, \, 23400$.
Figure \ref{Figure-nonparam} shows that the MSEs of $RIB$, CHEN, and PRVB decrease as the number of high-frequency observations increases, and $RIB$ exhibits better performance than CHEN and PRVB.
\begin{figure}[b] 
  \centering
  \includegraphics [width = 0.9\textwidth]{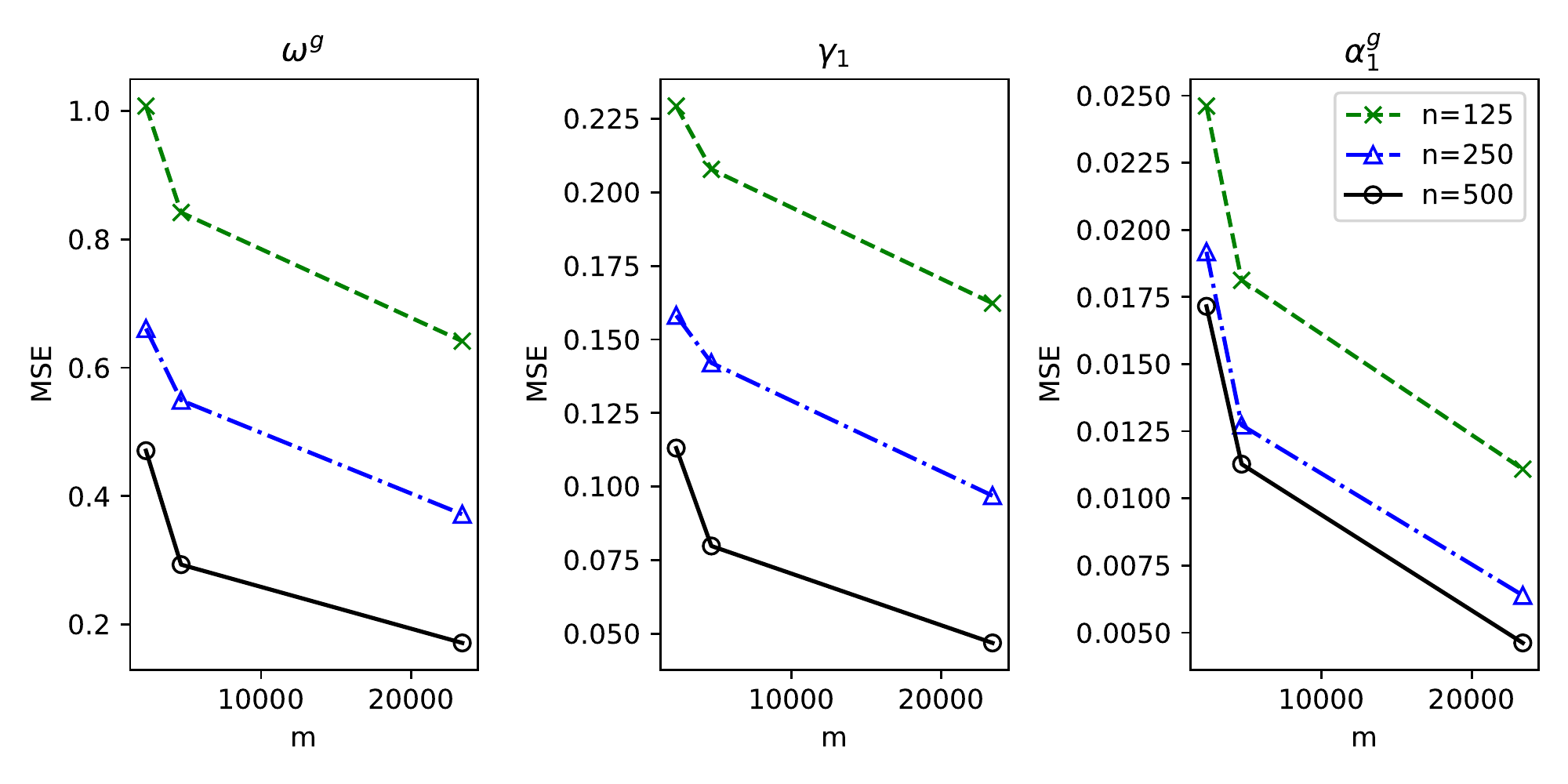}
  \caption{The MSEs of the least squared estimates with $m=2340, \,4680, \, 23400$ and $\, \, $ $n=100, \, 250, \, 500$.}
  \label{Figure-parMSE}
\end{figure}
This is because the proposed $RIB$ estimator can account for the auto-correlation structure of the microstructure noise and time-varying beta, while CHEN and PRVB fail to handle one of them. 
These results support the theoretical results derived in Section \ref{sec-3.1}.

Next, we checked the finite sample performances of the proposed DR Beta model.
We first estimated the model parameters using the proposed quasi-maximum likelihood estimation in Section \ref{sec-3.2}  for $n=100, \, 250, \, 500$ and $m=2340, \, 4680, \, 23400$. 
Figure \ref{Figure-parMSE} draws the MSEs of the least squared estimates $\hat{\theta}$'s for the model parameter $\theta_0$. 
From Figure \ref{Figure-parMSE}, we find that  the MSEs decrease as $n$ or $m$ increases.
These results match  the theoretical findings in Section \ref{sec-3.2}.

\begin{figure}[b] 
\centering
\includegraphics[width = 0.9\textwidth]{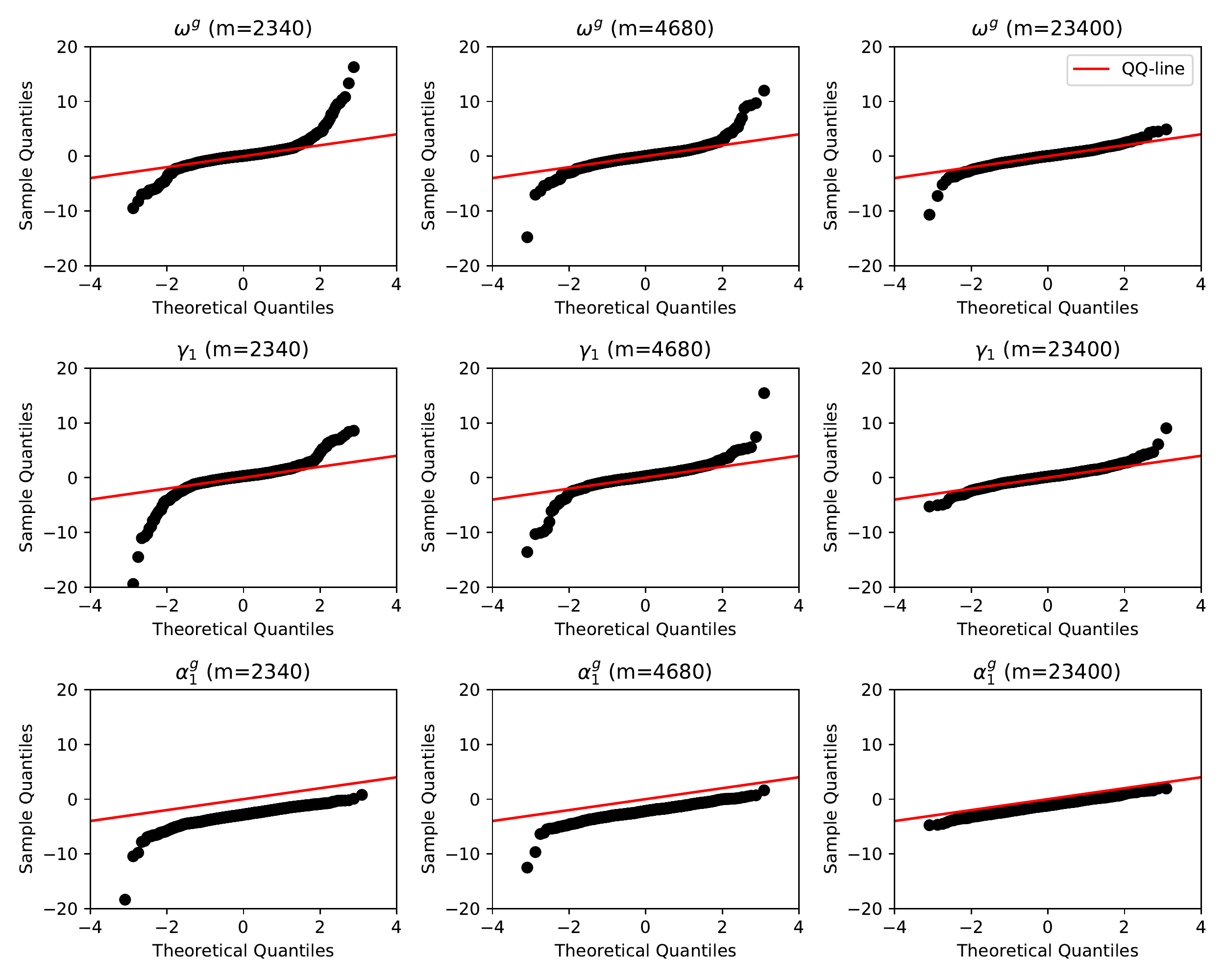}
 \caption{The standard normal (original) quantile-quantile plots of the $Z$-statistics estimates of $\omega^g$, $\gamma_1$, and $\alpha^g_1$ for $n=500$, $m=2340, 4680, 23400$.}
 \label{QQplot}
\end{figure}
 To check the asymptotic normality of the model parameters $(\omega^g, \gamma_1, \alpha^g_1)$, we calculated the $Z$-statistics  proposed in Section \ref{sec-3.3}.
Figure \ref{QQplot} shows standard normal quantile-quantile plots of the $Z$-statistics estimates of $\omega^g_1$, $\gamma_1$, and $\alpha^g_1$  for $n=500$ and $m=2340, 4680, 23400$. 
From Figure \ref{QQplot}, we find that the $Z$-statistics  close to the standard normal distribution as $m$ increases--that is,  the non-parametric integrated beta estimator $RIB$ closes to the true integrated beta $I\beta$.  
This result agrees with the theoretical findings in Section \ref{sec-3}.
Thus, based on the proposed $Z$-statistics, we can conduct hypothesis tests for the model parameters using the standard normal distribution.

\begin{figure}[!ht] 
\centering
\includegraphics[width = 0.9\textwidth]{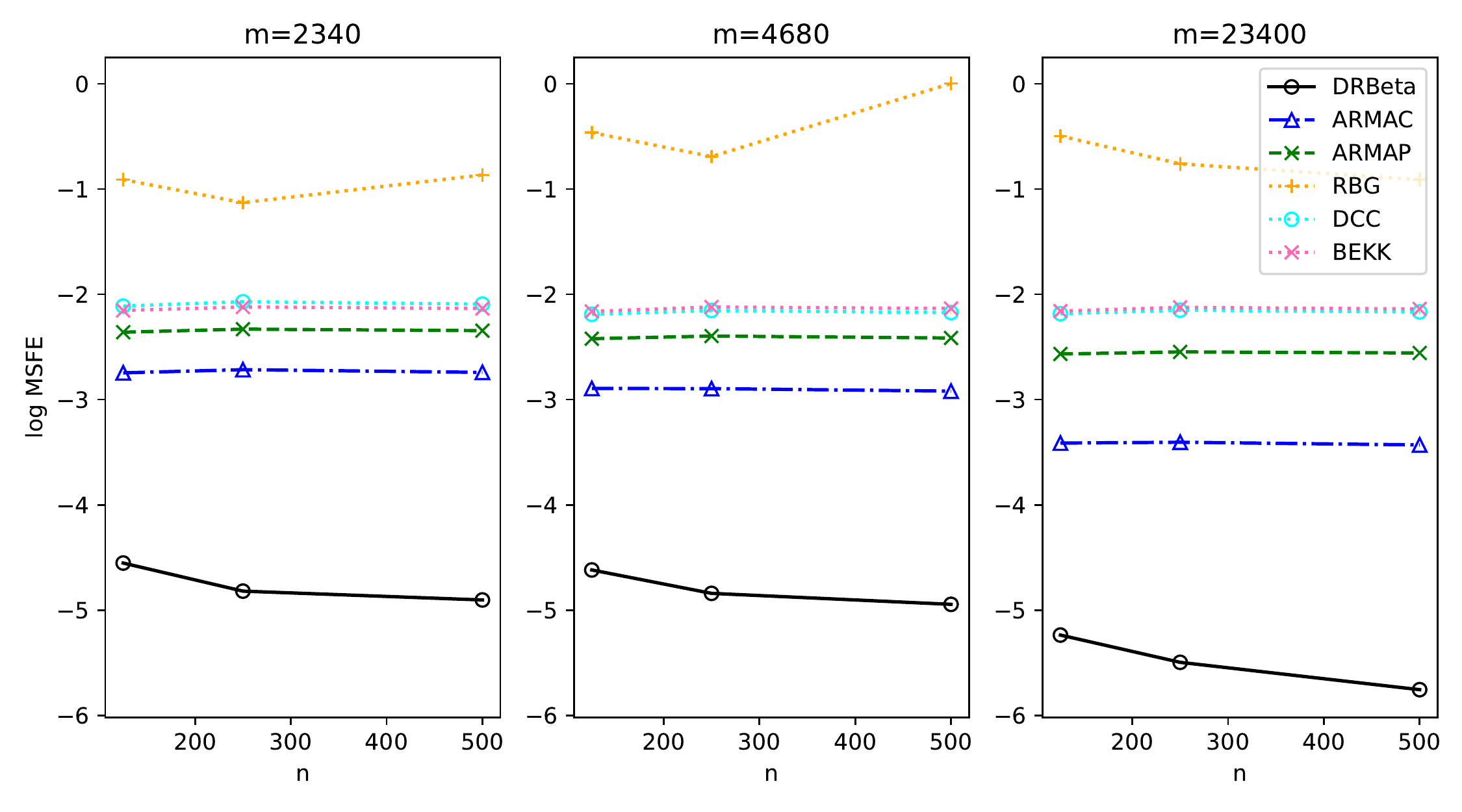} 
\caption{ The log MSFEs of DR Beta, ARMAC, ARMAP, RBG, DCC, and BEKK with $n=100, 250, 500$ and $m=2340, 4680, 23400$. }
\label{Figure-pred}
\end{figure}

One of the  major motivations of the DR Beta model is to predict future market betas by taking advantage of the  low-frequency auto-regressive dynamic structure.
Thus, we examined the out-of-sample performance of estimating the one-day-ahead conditional expected integrated beta $h_{n+1}(\theta_0)$ to check the predictability of the DR Beta model. 
We compared the DR Beta with three parametric models that employ the high-frequency data and two parametric models that employ the low-frequency data.
For the parametric model with high-frequency data,  we considered the ARMA$(1,1)$ models, which utilize CHEN (ARMAC) or PRVB (ARMAP) as daily realized beta, and Realized Beta GARCH (RBG) model \citep{hansen2014realized}, which is a multivariate GARCH model utilizing realized measures of volatility and correlation.
Details of the RBG model can be found in \citet{hansen2014realized}.
For the parametric models with low-frequency data, we used the dynamic conditional beta (DCB) model framework proposed by \citet{engle2016dynamic}.
Specifically, the beta can be established as follows:
\begin{eqnarray*}
    && \hat{I \beta}_i=\left(\mathbf{H}_{i,22}(\hat{\theta})\right)^{-1}\mathbf{H}_{i,12}(\hat{\theta}), \cr 
    &&\hat{\theta}=\argmax_{\theta} \left(-\frac{1}{2} \sum_{i=1}^{n}\log |\det(\mathbf{H}_{i}(\theta))| -\frac{1}{2} \sum_{i=1}^{n}(\mathbf{y}_{i}-\bar{\mathbf{y}})^\top \left(\mathbf{H}_{i}(\theta)\right)^{-1}(\mathbf{y}_{i}-\bar{\mathbf{y}}) \right ), 
\end{eqnarray*}
where $\mathbf{y}_{i} =(Q(i)-Q(i-1), \; P(i)-P(i-1))^{\top}$, $\bar{\mathbf{y}}=\frac{1}{n}\sum^{n}_{j=1}\mathbf{y}_j$, and $\mathbf{H}_i(\theta) $ is a conditional covariance matrix of $\mathbf{y}_{i}$.
As the conditional covariance matrix, we employed BEKK (1,1) and DCC(1,1)  suggested by \citet{engle1995multivariate} and \citet{bali2010intertemporal}, respectively.
We call the beta estimators with BEKK (1,1) and DCC(1,1)   BEKK and DCC, respectively. 
We measured the mean squared forecast errors (MSFEs) with the one-day-ahead sample period over 500 samples as follows:
\begin{equation*}
    \frac{1}{500}\sum_{i=1}^{500} \left (\hat{Beta}_{n+1, i} -h_{ n+1,i}(\theta_0) \right )^2,
\end{equation*}
where $\hat{Beta}_{n+1, i}$  is one of the future beta estimators above at the $i$th sample-path given the available information at time $n$.
 In Figure \ref{Figure-pred}, the log MSFEs of DR Beta, ARMAC, ARMAP, RBG, DCC, and BEKK are plotted for $n=100, \, 250, \, 500$ and $m=2340, \, 4680, \, 23400$.
 Figure \ref{Figure-pred} shows the log MSFEs of the DR Beta  decrease as $n$ or $m$ increases, but other estimators do not have any strong pattern. 
This may be because the other benchmarks cannot account for the beta dynamics well.
Meanwhile, the high-frequency-based ARMA models show better performance than other competitors.
One possible explanation is that they adopt the true model structure. 
However, since PRVB and CHEN cannot account for at least one of the time-varying beta and the auto-correlation structure of the microstructure noise, the DR Beta shows the best performance. 
From this result, we can conclude that the robust non-parametric beta estimator helps account for the market dynamics.

\section{Empirical analysis} \label{sec-5}

In this section, we apply the proposed DR Beta model to  real high-frequency trading data.
We obtained  high-frequency data for the top 50 large trading volume stocks among the S\&P  500  from the TAQ database in the Wharton Research Data Services (WRDS) system from January 1, 2010, to December 31, 2016,   $1762$ trading days in total. 
We used the S\&P 500 Composite index as the market portfolio, which was obtained   from the Chicago Board Options Exchange (CBOE).
We used 1-sec log-returns, which were subsampled by the previous tick \citep{zhang2011estimating} scheme.
High-frequency data were available between the open and close of the market, so that the number of high-frequency observations for a full trading day is $m = 23400$.

\begin{table}[!ht]
  \caption{ Estimated parameters from the DR Beta model. 
  The numbers in parentheses indicate the $p$-value, multiplied by 10, from the hypothesis tests.}\label{table-2}
  \centering
  \scalebox{0.9}{
  \stackengine{1pt}{
    \begin{tabular}{lrrrllrrr}
      \hline
      \multicolumn{1}{c}{Stock} & \multicolumn{1}{c}{$\omega^g$} & \multicolumn{1}{c}{$\gamma_1$} & \multicolumn{1}{c}{$\alpha_1^g$} & \multicolumn{1}{c}{} & \multicolumn{1}{c}{Stock} & \multicolumn{1}{c}{$\omega^g$} & \multicolumn{1}{c}{$\gamma_1$} & \multicolumn{1}{c}{$\alpha_1^g$} \\ \cline{1-4} \cline{6-9} 
      AAPL                    & 0.187 (0.0)                         & 0.463 (0.0)                         & 0.354 (0.0)                           &                      & INTC                    & 0.139 (0.0)                         & 0.478 (0.0)                         & 0.374 (0.0)                           \\
      AIG                    & 0.078 (0.0)                         & 0.578 (0.0)                         & 0.341 (0.0)                           &                      & JPM                    & 0.142 (0.0)                         & 0.503 (0.0)                         & 0.361 (0.0)                           \\
      AMAT                    & 0.241 (0.0)                         & 0.424 (0.0)                         & 0.358 (0.0)                           &                      & KEY                    & 0.133 (0.0)                         & 0.505 (0.0)                         & 0.379 (0.0)                           \\
      ATVI                    & 0.085 (0.0)                         & 0.624 (0.0)                         & 0.288 (0.0)                           &                      & KO                    & 0.051 (0.0)                         & 0.533 (0.0)                         & 0.376 (0.0)                           \\
      BAC                    & 0.101 (0.0)                         & 0.567 (0.0)                         & 0.348 (0.0)                           &                      & MGM                    & 0.059 (0.0)                         & 0.661 (0.0)                         & 0.294 (0.0)                           \\
      BMY                    & 0.101 (0.0)                         & 0.520 (0.0)                         & 0.365 (0.0)                           &                      & MRK                    & 0.093 (0.0)                         & 0.504 (0.0)                         & 0.376 (0.0)                           \\
      BSX                    & 0.155 (0.0)                         & 0.469 (0.0)                         & 0.378 (0.0)                           &                      & MS                    & 0.125 (0.0)                         & 0.570 (0.0)                         & 0.336 (0.0)                           \\
      CSCO                    & 0.130 (0.0)                         & 0.512 (0.0)                         & 0.348 (0.0)                           &                      & MSFT                    & 0.115 (0.0)                         & 0.535 (0.0)                         & 0.348 (0.0)                           \\
      CSX                    & 0.103 (0.0)                         & 0.516 (0.0)                         & 0.380 (0.0)                           &                      & MU                    & 0.122 (0.0)                         & 0.629 (0.0)                         & 0.292 (0.0)                           \\
      DAL                    & 0.132 (0.0)                         & 0.537 (0.0)                         & 0.345 (0.0)                           &                      & NEM                    & 0.027 (0.0)                         & 0.530 (0.0)                         & 0.397 (0.0)                           \\
      DIS                    & 0.105 (0.0)                         & 0.533 (0.0)                         & 0.346 (0.0)                           &                      & NFLX                    & 0.065 (0.0)                         & 0.658 (0.0)                         & 0.297 (0.0)                           \\
      DOW                    & 0.069 (0.0)                         & 0.637 (0.0)                         & 0.294 (0.0)                           &                      & NVDA                    & 0.213 (0.0)                         & 0.392 (0.0)                         & 0.417 (0.0)                           \\
      EBAY                    & 0.115 (0.0)                         & 0.513 (0.0)                         & 0.366 (0.0)                           &                      & ORCL                    & 0.126 (0.0)                         & 0.494 (0.0)                         & 0.365 (0.0)                           \\
      F                    & 0.114 (0.0)                         & 0.542 (0.0)                         & 0.338 (0.0)                           &                      & PFE                    & 0.083 (0.0)                         & 0.555 (0.0)                         & 0.340 (0.0)                           \\
      FCX                    & 0.036 (0.0)                         & 0.632 (0.0)                         & 0.343 (0.0)                           &                      & PG                    & 0.041 (0.0)                         & 0.552 (0.0)                         & 0.374 (0.0)                           \\
      FITB                    & 0.073 (0.0)                         & 0.592 (0.0)                         & 0.342 (0.0)                           &                      & QCOM                    & 0.107 (0.0)                         & 0.490 (0.0)                         & 0.386 (0.0)                           \\
      GE                    & 0.139 (0.0)                         & 0.469 (0.0)                         & 0.361 (0.0)                           &                      & RF                    & 0.091 (0.0)                         & 0.594 (0.0)                         & 0.332 (0.0)                           \\
      GILD                    & 0.100 (0.0)                         & 0.528 (0.0)                         & 0.386 (0.0)                           &                      & SCHW                    & 0.139 (0.0)                         & 0.525 (0.0)                         & 0.370 (0.0)                           \\
      GLW                    & 0.168 (0.0)                         & 0.498 (0.0)                         & 0.340 (0.0)                           &                      & T                    & 0.059 (0.0)                         & 0.594 (0.0)                         & 0.299 (0.0)                           \\
      HAL                    & 0.072 (0.0)                         & 0.618 (0.0)                         & 0.323 (0.0)                           &                      & WFC                    & 0.114 (0.0)                         & 0.534 (0.0)                         & 0.339 (0.0)                           \\
      HBAN                    & 0.085 (0.0)                         & 0.576 (0.0)                         & 0.344 (0.0)                           &                      & WMT                    & 0.049 (0.0)                         & 0.565 (0.0)                         & 0.344 (0.0)                           \\
      HPQ                    & 0.205 (0.0)                         & 0.416 (0.0)                         & 0.400 (0.0)                           &                      & XOM                    & 0.067 (0.0)                         & 0.541 (0.0)                         & 0.379 (0.0)                           \\
      HST                    & 0.097 (0.0)                         & 0.551 (0.0)                         & 0.346 (0.0)                           &                      & XRX                    & 0.187 (0.0)                         & 0.476 (0.0)                         & 0.341 (0.0)                           \\ \hline
      \end{tabular}      
  }{
  \begin{tabular}{lrrrrr}
  \hline
  \multicolumn{1}{c}{Stock} & \multicolumn{1}{c}{$\omega^g$} & \multicolumn{1}{c}{$\gamma_1$} & \multicolumn{1}{c}{$\gamma_2$} & \multicolumn{1}{c}{$\alpha_1^g$} & \multicolumn{1}{c}{$\alpha_2^g$} \\ \hline
  AMD                    & 0.008 (2.0)                         & 0.702 (0.0)                         & 0.150 (3.8)                         & 0.294 (0.0)                          & -0.153 (0.0)                         \\
MRO                    & 0.022 (0.6)                         & 0.564 (0.0)                         & 0.167 (1.8)                         & 0.429 (0.0)                          & -0.174 (0.1)                         \\
VZ                    & 0.028 (0.3)                         & 0.610 (0.1)                         & 0.132 (3.8)                         & 0.366 (0.0)                          & -0.156 (0.3)                         \\
WMB                    & 0.067 (0.0)                         & -0.332 (0.0)                         & 0.604 (0.0)                         & 0.319 (0.0)                          & 0.353 (0.0)                         \\ \hline
  \end{tabular}
  }{U}{l}{F}{F}{S}
  }
\end{table}

 \begin{figure}[b] \label{Figure-A}
\centering
\includegraphics[width = 1\textwidth]{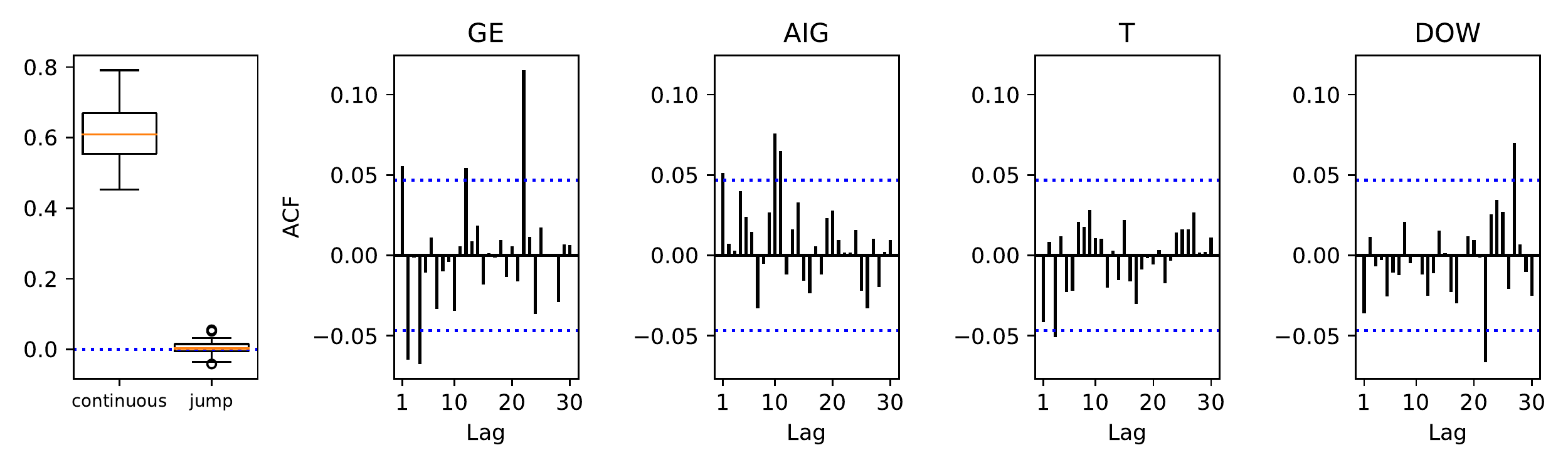} \label{Figure-A}
\caption{The box plot (left) of the first-order auto-correlations of $RIB$ (continuous) and  jump beta (jump) from January 1, 2010, to December 31, 2016, and the ACF plots for the top four first-order auto-correlation stocks.}
\end{figure} 
To examine the goodness of the fit, we  conducted in-sample validation.
We draw auto-correlation plots for $RIB$ and jump beta in Figures \ref{Figure-1} and \ref{Figure-A}, where the jump beta is estimated by the method suggested by \citet{li2017robust}.
As we discussed in Section \ref{sec-2}, the integrated beta for the continuous part has a strong auto-correlation structure, but the beta for the jumps does not.
Thus, it is reasonable to focus on modeling the beta for the continuous part. 
To conduct the validation of the DR Beta model, we first selected the $(p,q) \in \left\lbrace (p,q): 0 \leq p, q \leq 3 \right\rbrace$ of the DR Beta for each stock by BIC, and we estimated the parameters of the DR Beta using the sample over the last 1000 trading days.
Then, we conducted the hypothesis tests proposed in Section \ref{sec-3.3}.
Table \ref{table-2} reports the parameter estimates of the DR Beta model with the selected $(p,q)$ for each stock and their $p$-values.
The BIC values were minimized when $(p,q)=(1,1)$, except for four stocks whose BIC values were minimized when $(p,q)=(2,2)$.
Except for five coefficients, all of the $p$-values are lower than 0.05,  and all stocks have at least  three signficant coefficients.
Thus, we can conclude that the proposed DR Beta model is statistically valid and  may capture the auto-regressive structure.

  To check the out-of-sample performance, we measure the mean absolute prediction error (MAPE)  as follows: 
 \begin{equation*}
 \frac{1}{n-500}\sum^n_{i=501}\left|\hat{Beta}_{i}-RIB_i \right|,
\end{equation*}
where $\hat{Beta}_{i}$ is one of the beta estimators using DR Beta, ARMAC, ARMAP, RBG, DCC, and BEKK  defined in Section \ref{sec-4}, and for each stock, we used the selected $(p,q)$ order for the DR Beta, ARMAC, and ARMAP models. 
The in-sample period is 500 days, and we estimated the models using the rolling window scheme. 
Table \ref{table-3} reports the MAPEs for  DR Beta, ARMAC, ARMAP, RBG, DCC, and BEKK for 50 stocks.
From Table \ref{table-3}, we find that the ARMA-type models utilizing realized betas show better performance than the other models that use only low-frequency information or RBG.
When comparing the ARMA-type models using realized betas, MAPEs for the proposed DR Beta or ARMAC have the smallest values for every stock.
It may be because the proposed DR Beta and ARMAC model can account for the time-varying beta by incorporating the high-frequency data.
These results indicate that accomodating  the time-varying beta feature helps account for the beta dynamics, and the DR Beta holds advantages in predicting future beta by utilizing the auto-regressive structure with high-frequency data.

\begin{landscape}
  \begin{table}[!ht]
    \caption{The MAPEs for the DR Beta, ARMAC, ARMAP, RBG, DCC, and BEKK for each stock.
    }\label{table-3}
  \centering
  \scalebox{1}{
    \setlength\tabcolsep{3.5pt}
    \begin{tabular}{lrrrrrrllrrrrrr}
      \hline
    \multicolumn{1}{c}{Stock} & \multicolumn{1}{c}{DRBeta} & \multicolumn{1}{c}{ARMAC} & \multicolumn{1}{c}{ARMAP} & \multicolumn{1}{c}{RBG} & \multicolumn{1}{c}{DCC} & \multicolumn{1}{c}{BEKK} & \multicolumn{1}{c}{} & \multicolumn{1}{c}{Stock} & \multicolumn{1}{c}{DRBeta} & \multicolumn{1}{c}{ARMAC} & \multicolumn{1}{c}{ARMAP} & \multicolumn{1}{c}{RBG} & \multicolumn{1}{c}{DCC} & \multicolumn{1}{c}{BEKK} \\ \cline{1-7} \cline{9-15} 
    AAPL                      & 0.257                     & 0.259                    & 0.300                    & 0.826                  & 0.429                  & 0.489                   &                      & JPM                       & 0.179                     & 0.181                    & 0.238                    & 0.573                  & 0.323                  & 0.339                   \\
    AIG                       & 0.188                     & 0.190                    & 0.243                    & 0.656                  & 0.316                  & 0.311                   &                      & KEY                       & 0.232                     & 0.232                    & 0.292                    & 0.721                  & 0.376                  & 0.383                   \\
    AMAT                      & 0.226                     & 0.227                    & 0.277                    & 0.694                  & 0.334                  & 0.334                   &                      & KO                        & 0.124                     & 0.124                    & 0.141                    & 0.470                  & 0.208                  & 0.222                   \\
    AMD                       & 0.382                     & 0.382                    & 0.467                    & 1.704                  & 0.585                  & 0.683                   &                      & MGM                       & 0.281                     & 0.283                    & 0.363                    & 0.918                  & 0.498                  & 0.486                   \\
    ATVI                      & 0.215                     & 0.216                    & 0.258                    & 0.913                  & 0.352                  & 0.389                   &                      & MRK                       & 0.147                     & 0.147                    & 0.171                    & 0.541                  & 0.260                  & 0.282                   \\
    BAC                       & 0.253                     & 0.254                    & 0.322                    & 0.802                  & 0.407                  & 0.447                   &                      & MRO                       & 0.308                     & 0.312                    & 0.390                    & 1.157                  & 0.491                  & 0.703                   \\
    BMY                       & 0.165                     & 0.167                    & 0.202                    & 0.606                  & 0.308                  & 0.337                   &                      & MS                        & 0.241                     & 0.242                    & 0.316                    & 0.725                  & 0.400                  & 0.457                   \\
    BSX                       & 0.224                     & 0.227                    & 0.276                    & 0.873                  & 0.353                  & 0.321                   &                      & MSFT                      & 0.177                     & 0.177                    & 0.212                    & 0.517                  & 0.325                  & 0.337                   \\
    CSCO                      & 0.174                     & 0.173                    & 0.205                    & 0.468                  & 0.303                  & 0.337                   &                      & MU                        & 0.343                     & 0.343                    & 0.410                    & 1.370                  & 0.502                  & 0.557                   \\
    CSX                       & 0.185                     & 0.185                    & 0.236                    & 0.610                  & 0.327                  & 0.341                   &                      & NEM                       & 0.261                     & 0.262                    & 0.278                    & 1.417                  & 0.457                  & 0.480                   \\
    DAL                       & 0.268                     & 0.268                    & 0.325                    & 1.170                  & 0.398                  & 0.430                   &                      & NFLX                      & 0.318                     & 0.322                    & 0.377                    & 1.424                  & 0.677                  & 0.789                   \\
    DIS                       & 0.148                     & 0.149                    & 0.196                    & 0.455                  & 0.237                  & 0.238                   &                      & NVDA                      & 0.238                     & 0.241                    & 0.292                    & 0.917                  & 0.399                  & 0.407                   \\
    DOW                       & 0.196                     & 0.198                    & 0.252                    & 0.605                  & 0.285                  & 0.294                   &                      & ORCL                      & 0.168                     & 0.169                    & 0.209                    & 0.469                  & 0.284                  & 0.285                   \\
    EBAY                      & 0.193                     & 0.194                    & 0.231                    & 2.653                  & 0.338                  & 0.349                   &                      & PFE                       & 0.149                     & 0.149                    & 0.175                    & 0.682                  & 0.278                  & 0.310                   \\
    F                         & 0.214                     & 0.215                    & 0.266                    & 0.666                  & 0.330                  & 0.350                   &                      & PG                        & 0.122                     & 0.122                    & 0.136                    & 0.453                  & 0.205                  & 0.233                   \\
    FCX                       & 0.328                     & 0.326                    & 0.394                    & 1.358                  & 0.503                  & 0.661                   &                      & QCOM                      & 0.167                     & 0.168                    & 0.200                    & 0.516                  & 0.308                  & 0.324                   \\
    FITB                      & 0.205                     & 0.205                    & 0.264                    & 0.577                  & 0.333                  & 0.356                   &                      & RF                        & 0.265                     & 0.266                    & 0.334                    & 1.002                  & 0.427                  & 0.432                   \\
    GE                        & 0.158                     & 0.159                    & 0.195                    & 0.489                  & 0.259                  & 0.268                   &                      & SCHW                      & 0.230                     & 0.230                    & 0.299                    & 0.660                  & 0.380                  & 0.385                   \\
    GILD                      & 0.219                     & 0.219                    & 0.278                    & 1.094                  & 0.356                  & 0.440                   &                      & T                         & 0.126                     & 0.126                    & 0.144                    & 0.447                  & 0.199                  & 0.209                   \\
    GLW                       & 0.207                     & 0.209                    & 0.258                    & 0.627                  & 0.339                  & 0.340                   &                      & VZ                        & 0.131                     & 0.131                    & 0.148                    & 0.449                  & 0.235                  & 0.258                   \\
    HAL                       & 0.245                     & 0.249                    & 0.318                    & 0.980                  & 0.437                  & 0.475                   &                      & WFC                       & 0.163                     & 0.163                    & 0.206                    & 0.484                  & 0.263                  & 0.263                   \\
    HBAN                      & 0.222                     & 0.223                    & 0.292                    & 0.838                  & 0.327                  & 0.349                   &                      & WMB                       & 0.260                     & 0.261                    & 0.316                    & 1.376                  & 0.426                  & 0.525                   \\
    HPQ                       & 0.234                     & 0.235                    & 0.285                    & 0.894                  & 0.393                  & 0.460                   &                      & WMT                       & 0.117                     & 0.117                    & 0.132                    & 0.545                  & 0.214                  & 0.234                   \\
    HST                       & 0.199                     & 0.199                    & 0.247                    & 1.055                  & 0.294                  & 0.303                   &                      & XOM                       & 0.145                     & 0.146                    & 0.181                    & 0.401                  & 0.271                  & 0.300                   \\
    INTC                      & 0.171                     & 0.171                    & 0.211                    & 0.562                  & 0.327                  & 0.344                   &                      & XRX                       & 0.246                     & 0.246                    & 0.293                    & 0.808                  & 0.370                  & 0.321                   \\
    \hline
    \end{tabular}
  }
  \end{table}
\end{landscape}

Finally, we evaluated how well the proposed methodologies capture the auto-regressive structure.
Adopting the idea of the Durbin-Watson test, we took into account regression residuals between the non-parametric and out-of-sample predicted values using DR Beta, ARMAC, ARMAP, RBG, DCC, and BEKK. 
Specifically, for each model, we fitted the following linear regression model:
\begin{equation*}
    RIB_i = a+b \times  Beta _{i} + e_i , 
\end{equation*}
where the $Beta_i$'s are the predicted integrated betas using one of the DR Beta, ARMAC, ARMAP, RBG, DCC, and BEKK.
Then, we calculated the regression residuals for each model and checked their auto-correlations. 
Figure \ref{Figure-7} shows the ACF plots for $RIB$ and the models' regression residuals for six stocks, which have the smallest, $20$th, $40$th, $60$th, $80$th, and the largest first-order auto-correlations among the 50 assets.
Figure \ref{Figure-6} depicts the box plot of the first-order auto-correlations of the regression residuals for each model, and Table \ref{table-4} reports  their numerical values. 
From Figures  \ref{Figure-7} and \ref{Figure-6} and Table \ref{table-4}, we find that the proposed DR Beta and ARMAC models have  much smaller auto-correlations for most of the assets, but the other models still yield significantly non-zero auto-correlations for most of the assets.
This may be because the other competitors  do not accomodate the time-varying beta feature.  
When comparing the DR Beta and ARMAC models, the DR Beta model usually has smaller auto-correlation than the ARMAC model.
One of the possible explanations is that the CHEN estimator, which is used in the ARMAC model as the non-parametric beta estimator, cannot handle the auto-correlation structure of the microstructure noise; thus, some auto-correlation may remain in the regression residuals.
From these numerical results, we can conjecture  that  incorporating the stylized features, such as the time-varying beta and the auto-correlation structure of the micro-structure noise, helps account for the beta dynamics. 
Thus, the proposed DR Beta model can explain the beta dynamics well by incoporating the proposed robust realized integrated beta estimator.

 \begin{figure}[h] 
\centering
\includegraphics[width =.98\textwidth]{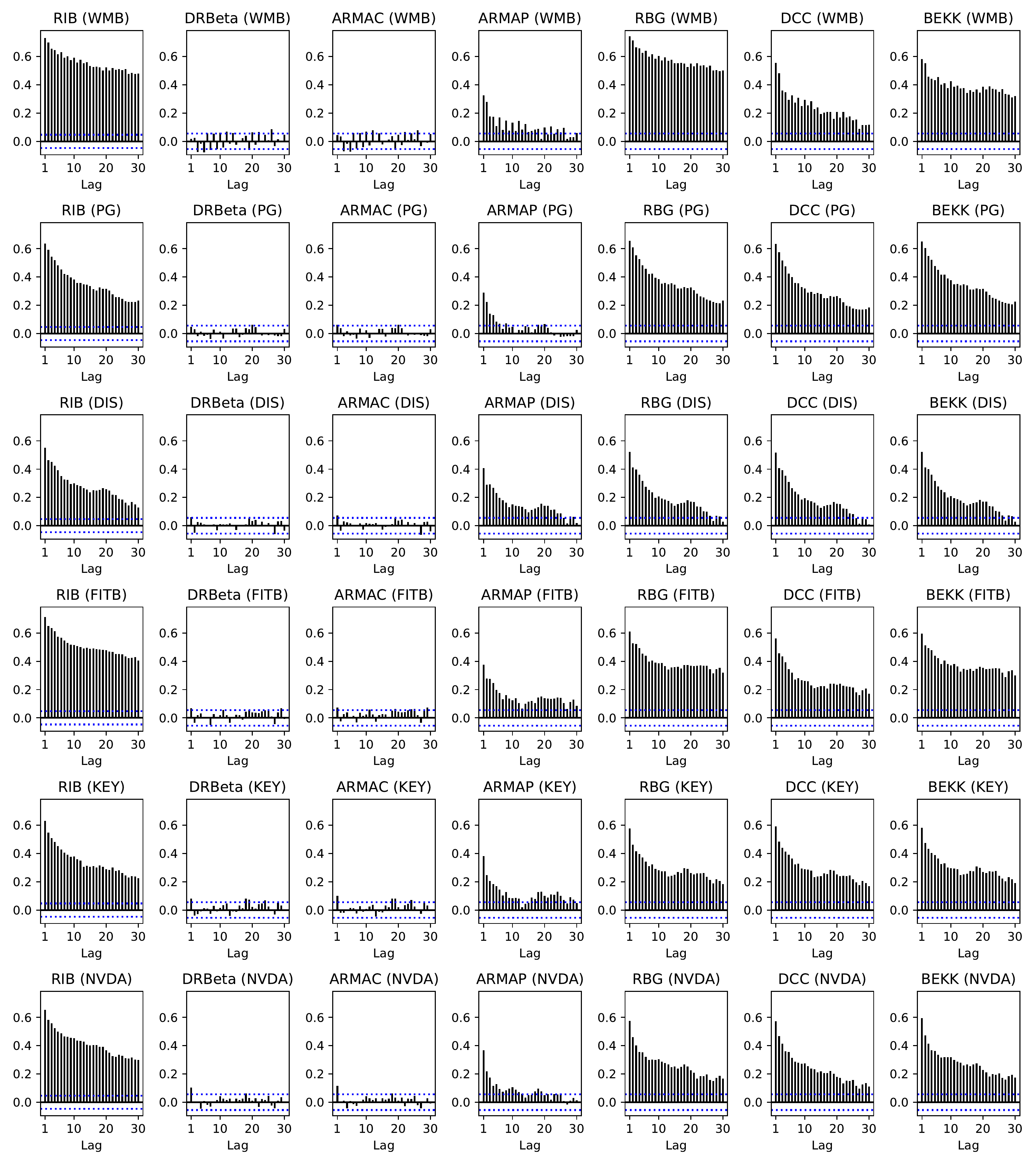} \label{Figure-7}
\caption{ACF plots for the non-parametric integrated beta, $RIB$, and the regression residuals between $RIB$ and the predicted integrated beta from DR Beta, ARMAC, ARMAP, RBG, DCC, and BEKK for six stocks, which have the smallest, $20$th, $40$th, $60$th, $80$th, and largest first-order auto-correlations among the 50 assets in order. } \label{Figure-7}
\end{figure} 

\begin{landscape}
  \begin{table}[!ht]
    \caption{The first-order auto-correlations for the regression residuals between the non-parametric integrated beta estimates, $RIB$, and the predicted integrated beta from DR Beta, ARMAC, ARMAP, RBG, DCC, and BEKK. 
  }\label{table-4}
  \centering
  \scalebox{1}{
    \setlength\tabcolsep{3.5pt}
    \begin{tabular}{lrrrrrrllrrrrrr}
      \hline
    \multicolumn{1}{c}{Stock} & \multicolumn{1}{c}{DRBeta} & \multicolumn{1}{c}{ARMAC} & \multicolumn{1}{c}{ARMAP} & \multicolumn{1}{c}{RBG} & \multicolumn{1}{c}{DCC} & \multicolumn{1}{c}{BEKK} & \multicolumn{1}{c}{} & \multicolumn{1}{c}{Stock} & \multicolumn{1}{c}{DRBeta} & \multicolumn{1}{c}{ARMAC} & \multicolumn{1}{c}{ARMAP} & \multicolumn{1}{c}{RBG} & \multicolumn{1}{c}{DCC} & \multicolumn{1}{c}{BEKK} \\ \cline{1-7} \cline{9-15} 
    AAPL                      & 0.059                     & 0.068                    & 0.271                    & 0.579                  & 0.552                  & 0.586                   &                      & JPM                       & 0.042                     & 0.055                    & 0.399                    & 0.577                  & 0.569                  & 0.587                   \\
    AIG                       & 0.038                     & 0.054                    & 0.348                    & 0.651                  & 0.622                  & 0.620                   &                      & KEY                       & 0.080                     & 0.101                    & 0.383                    & 0.576                  & 0.593                  & 0.582                   \\
    AMAT                      & 0.064                     & 0.079                    & 0.353                    & 0.443                  & 0.443                  & 0.442                   &                      & KO                        & 0.061                     & 0.070                    & 0.313                    & 0.635                  & 0.634                  & 0.636                   \\
    AMD                       & 0.085                     & 0.091                    & 0.223                    & 0.620                  & 0.567                  & 0.619                   &                      & MGM                       & 0.094                     & 0.108                    & 0.390                    & 0.638                  & 0.645                  & 0.645                   \\
    ATVI                      & 0.073                     & 0.092                    & 0.366                    & 0.503                  & 0.504                  & 0.504                   &                      & MRK                       & 0.034                     & 0.055                    & 0.364                    & 0.604                  & 0.570                  & 0.548                   \\
    BAC                       & 0.064                     & 0.072                    & 0.340                    & 0.614                  & 0.570                  & 0.677                   &                      & MRO                       & 0.084                     & 0.099                    & 0.371                    & 0.621                  & 0.647                  & 0.789                   \\
    BMY                       & 0.047                     & 0.073                    & 0.368                    & 0.622                  & 0.626                  & 0.621                   &                      & MS                        & 0.062                     & 0.081                    & 0.370                    & 0.584                  & 0.582                  & 0.662                   \\
    BSX                       & 0.074                     & 0.084                    & 0.338                    & 0.530                  & 0.516                  & 0.522                   &                      & MSFT                      & 0.050                     & 0.069                    & 0.390                    & 0.516                  & 0.516                  & 0.491                   \\
    CSCO                      & 0.055                     & 0.063                    & 0.352                    & 0.504                  & 0.502                  & 0.488                   &                      & MU                        & 0.080                     & 0.087                    & 0.317                    & 0.498                  & 0.488                  & 0.511                   \\
    CSX                       & 0.084                     & 0.094                    & 0.389                    & 0.556                  & 0.581                  & 0.581                   &                      & NEM                       & 0.030                     & 0.042                    & 0.171                    & 0.727                  & 0.720                  & 0.733                   \\
    DAL                       & 0.072                     & 0.083                    & 0.352                    & 0.507                  & 0.507                  & 0.508                   &                      & NFLX                      & 0.055                     & 0.078                    & 0.356                    & 0.595                  & 0.606                  & 0.590                   \\
    DIS                       & 0.061                     & 0.072                    & 0.408                    & 0.521                  & 0.518                  & 0.522                   &                      & NVDA                      & 0.105                     & 0.115                    & 0.368                    & 0.580                  & 0.574                  & 0.599                   \\
    DOW                       & 0.023                     & 0.043                    & 0.319                    & 0.572                  & 0.533                  & 0.556                   &                      & ORCL                      & 0.070                     & 0.081                    & 0.378                    & 0.563                  & 0.561                  & 0.555                   \\
    EBAY                      & 0.098                     & 0.124                    & 0.352                    & 0.571                  & 0.569                  & 0.570                   &                      & PFE                       & 0.050                     & 0.051                    & 0.315                    & 0.543                  & 0.539                  & 0.541                   \\
    F                         & 0.053                     & 0.069                    & 0.330                    & 0.541                  & 0.539                  & 0.537                   &                      & PG                        & 0.047                     & 0.063                    & 0.289                    & 0.657                  & 0.635                  & 0.651                   \\
    FCX                       & 0.048                     & 0.059                    & 0.330                    & 0.658                  & 0.612                  & 0.639                   &                      & QCOM                      & 0.051                     & 0.069                    & 0.384                    & 0.587                  & 0.598                  & 0.609                   \\
    FITB                      & 0.068                     & 0.073                    & 0.377                    & 0.609                  & 0.562                  & 0.597                   &                      & RF                        & 0.069                     & 0.077                    & 0.335                    & 0.587                  & 0.577                  & 0.589                   \\
    GE                        & 0.081                     & 0.096                    & 0.345                    & 0.532                  & 0.528                  & 0.527                   &                      & SCHW                      & 0.037                     & 0.058                    & 0.411                    & 0.537                  & 0.552                  & 0.551                   \\
    GILD                      & 0.087                     & 0.100                    & 0.429                    & 0.682                  & 0.643                  & 0.677                   &                      & T                         & 0.071                     & 0.095                    & 0.313                    & 0.514                  & 0.512                  & 0.512                   \\
    GLW                       & 0.035                     & 0.049                    & 0.357                    & 0.449                  & 0.427                  & 0.445                   &                      & VZ                        & 0.094                     & 0.107                    & 0.322                    & 0.525                  & 0.525                  & 0.525                   \\
    HAL                       & 0.072                     & 0.089                    & 0.396                    & 0.639                  & 0.644                  & 0.650                   &                      & WFC                       & 0.064                     & 0.070                    & 0.353                    & 0.572                  & 0.557                  & 0.574                   \\
    HBAN                      & 0.055                     & 0.078                    & 0.356                    & 0.560                  & 0.559                  & 0.546                   &                      & WMB                       & 0.017                     & 0.043                    & 0.324                    & 0.744                  & 0.554                  & 0.582                   \\
    HPQ                       & 0.084                     & 0.091                    & 0.385                    & 0.510                  & 0.504                  & 0.506                   &                      & WMT                       & 0.063                     & 0.061                    & 0.308                    & 0.575                  & 0.574                  & 0.573                   \\
    HST                       & 0.066                     & 0.081                    & 0.347                    & 0.589                  & 0.566                  & 0.590                   &                      & XOM                       & 0.038                     & 0.056                    & 0.421                    & 0.620                  & 0.629                  & 0.628                   \\
    INTC                      & 0.063                     & 0.082                    & 0.395                    & 0.543                  & 0.539                  & 0.533                   &                      & XRX                       & 0.069                     & 0.090                    & 0.327                    & 0.432                  & 0.424                  & 0.427                   \\
    \hline
    \end{tabular}
  }
  \end{table}
\end{landscape}

\begin{figure}[!ht] 
\centering
\includegraphics[width = 1\textwidth]{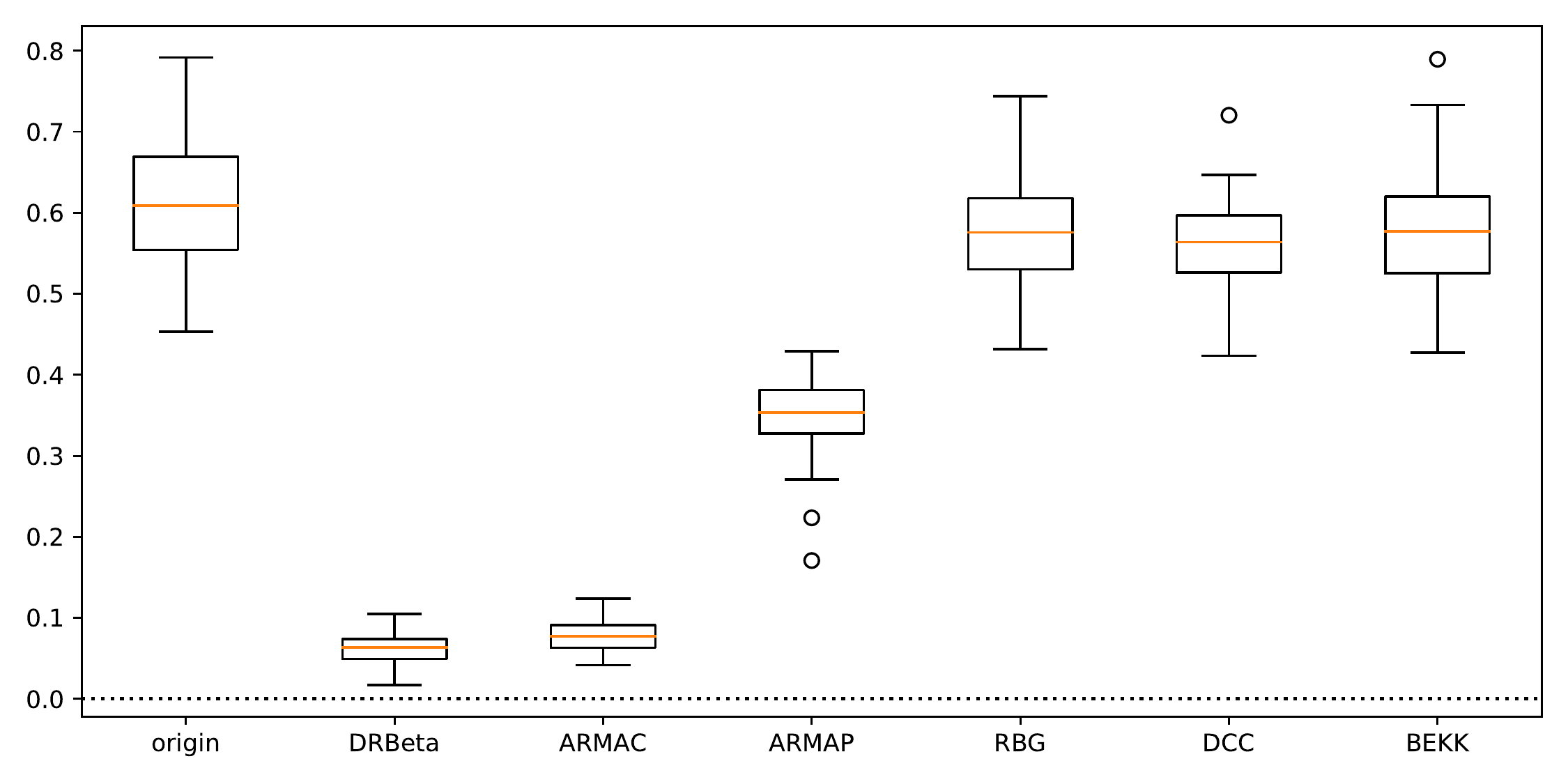} \label{Figure-6}
\caption{The box plots of the first-order auto-correlations for the regression residuals between the non-parametric integrated beta and the predicted integrated beta from DR Beta, ARMAC, ARMAP, RBG, DCC, and BEKK. 
The origin is  the first-order auto-correlations of the non-parametric integrated beta. } \label{Figure-6}
\end{figure}

\section{Conclusion}\label{sec-6}
This paper proposes a novel  time-varying beta model based on the continuous-time series regression  framework. 
To investigate  market betas based on the high-frequency financial data, we first develop a robust non-parametric integrated beta estimation procedure, $RIB$, which can handle the dependent microstructure noise and time-varying beta.
Then,  we establish its asymptotic properties. 
With this robust non-parametric $RIB$ estimator, we find the time-series structure of the integrated betas. 
To account for this beta dynamics, we develop a novel parametric continuous-time diffusion process, the DR Beta model. 
To estimate the model parameters, we propose a quasi-likelihood estimation procedure and establish its asymptotic theorems.
From the empirical study, we find that the proposed DR Beta model using the robust realized integrated beta estimator accounts effectively for the beta dynamics.

We employ the usual quasi-likelihood estimation procedure, which is already well-studied based on the high-frequency financial data \citep{li2018asymptotic}.
It would be an interesting future study to develop efficient or robust estimation procedures.
Furthermore, when making statistical inferences, we only use the information from the low-frequency dynamics. 
It would be very interesting to develop a unified model which can account for both the intraday dynamics and the low-frequency dynamics and to investigate its efficient estimation procedure.
We leave this for a future study.

\bibliography{references}

\begin{thebibliography}{}

\bibitem[A{\"\i}t-Sahalia et~al., 2010]{ait2010high}
A{\"\i}t-Sahalia, Y., Fan, J., and Xiu, D. (2010).
\newblock High-frequency covariance estimates with noisy and asynchronous
  financial data.
\newblock {\em Journal of the American Statistical Association},
  105(492):1504--1517.

\bibitem[A{\"\i}t-Sahalia et~al., 2020]{ait2020high}
A{\"\i}t-Sahalia, Y., Kalnina, I., and Xiu, D. (2020).
\newblock High-frequency factor models and regressions.
\newblock {\em Journal of Econometrics}.

\bibitem[A{\"\i}t-Sahalia et~al., 2011]{ait2011ultra}
A{\"\i}t-Sahalia, Y., Mykland, P.~A., and Zhang, L. (2011).
\newblock Ultra high frequency volatility estimation with dependent
  microstructure noise.
\newblock {\em Journal of Econometrics}, 160(1):160--175.

\bibitem[A{\"\i}t-Sahalia and Xiu, 2016]{ait2016increased}
A{\"\i}t-Sahalia, Y. and Xiu, D. (2016).
\newblock Increased correlation among asset classes: Are volatility or jumps to
  blame, or both?
\newblock {\em Journal of Econometrics}, 194(2):205--219.

\bibitem[Ait-Sahalia and Yu, 2008]{ait2008high}
Ait-Sahalia, Y. and Yu, J. (2008).
\newblock High frequency market microstructure noise estimates and liquidity
  measures.
\newblock Technical report, National Bureau of Economic Research.

\bibitem[Andersen et~al., 2006]{andersen2006realized}
Andersen, T.~G., Bollerslev, T., Diebold, F.~X., and Wu, G. (2006).
\newblock {\em Realized beta: Persistence and predictability}.
\newblock Emerald Group Publishing Limited.

\bibitem[Andersen et~al., 2005]{andersen2005framework}
Andersen, T.~G., Bollerslev, T., Diebold, F.~X., and Wu, J. (2005).
\newblock A framework for exploring the macroeconomic determinants of
  systematic risk.
\newblock {\em American Economic Review}, 95(2):398--404.

\bibitem[Andersen et~al., 2020]{andersen2020recalcitrant}
Andersen, T.~G., Thyrsgaard, M., and Todorov, V. (2020).
\newblock Recalcitrant betas: Intraday variation in the cross-sectional
  dispersion of systematic risk and expected returns.

\bibitem[Andrews, 1992]{andrews1992generic}
Andrews, D.~W. (1992).
\newblock Generic uniform convergence.
\newblock {\em Econometric theory}, pages 241--257.

\bibitem[Bali and Engle, 2010]{bali2010intertemporal}
Bali, T.~G. and Engle, R.~F. (2010).
\newblock The intertemporal capital asset pricing model with dynamic
  conditional correlations.
\newblock {\em Journal of Monetary Economics}, 57(4):377--390.

\bibitem[Barndorff-Nielsen et~al., 2008]{barndorff2008designing}
Barndorff-Nielsen, O.~E., Hansen, P.~R., Lunde, A., and Shephard, N. (2008).
\newblock Designing realized kernels to measure the ex post variation of equity
  prices in the presence of noise.
\newblock {\em Econometrica}, 76(6):1481--1536.

\bibitem[Barndorff-Nielsen et~al., 2011]{barndorff2011multivariate}
Barndorff-Nielsen, O.~E., Hansen, P.~R., Lunde, A., and Shephard, N. (2011).
\newblock Multivariate realised kernels: consistent positive semi-definite
  estimators of the covariation of equity prices with noise and non-synchronous
  trading.
\newblock {\em Journal of Econometrics}, 162(2):149--169.

\bibitem[Barndorff-Nielsen and Shephard, 2004]{barndorff2004econometric}
Barndorff-Nielsen, O.~E. and Shephard, N. (2004).
\newblock Econometric analysis of realized covariation: High frequency based
  covariance, regression, and correlation in financial economics.
\newblock {\em Econometrica}, 72(3):885--925.

\bibitem[Black et~al., 1992]{black1992uk}
Black, A., Fraser, P., and Power, D. (1992).
\newblock Uk unit trust performance 1980--1989: A passive time-varying
  approach.
\newblock {\em Journal of Banking \& Finance}, 16(5):1015--1033.

\bibitem[Bollerslev, 1986]{bollerslev1986generalized}
Bollerslev, T. (1986).
\newblock Generalized autoregressive conditional heteroskedasticity.
\newblock {\em Journal of econometrics}, 31(3):307--327.

\bibitem[Bollerslev et~al., 2016]{bollerslev2016roughing}
Bollerslev, T., Li, S.~Z., and Todorov, V. (2016).
\newblock Roughing up beta: Continuous versus discontinuous betas and the cross
  section of expected stock returns.
\newblock {\em Journal of Financial Economics}, 120(3):464--490.

\bibitem[Bos and Newbold, 1984]{bos1984empirical}
Bos, T. and Newbold, P. (1984).
\newblock An empirical investigation of the possibility of stochastic
  systematic risk in the market model.
\newblock {\em Journal of Business}, pages 35--41.

\bibitem[Breen et~al., 1989]{breen1989economic}
Breen, W., Glosten, L.~R., and Jagannathan, R. (1989).
\newblock Economic significance of predictable variations in stock index
  returns.
\newblock {\em The Journal of finance}, 44(5):1177--1189.

\bibitem[Chen, 2018]{chen2018inference}
Chen, R.~Y. (2018).
\newblock Inference for volatility functionals of multivariate it\^{o}
  semimartingales observed with jump and noise.
\newblock {\em arXiv preprint arXiv:1810.04725}.

\bibitem[Christensen et~al., 2010]{christensen2010pre}
Christensen, K., Kinnebrock, S., and Podolskij, M. (2010).
\newblock Pre-averaging estimators of the ex-post covariance matrix in noisy
  diffusion models with non-synchronous data.
\newblock {\em Journal of Econometrics}, 159(1):116--133.

\bibitem[Corsi, 2009]{corsi2009simple}
Corsi, F. (2009).
\newblock A simple approximate long-memory model of realized volatility.
\newblock {\em Journal of Financial Econometrics}, 7(2):174--196.

\bibitem[Engle, 1982]{engle1982autoregressive}
Engle, R.~F. (1982).
\newblock Autoregressive conditional heteroscedasticity with estimates of the
  variance of united kingdom inflation.
\newblock {\em Econometrica: Journal of the Econometric Society}, pages
  987--1007.

\bibitem[Engle, 2016]{engle2016dynamic}
Engle, R.~F. (2016).
\newblock Dynamic conditional beta.
\newblock {\em Journal of Financial Econometrics}, 14(4):643--667.

\bibitem[Engle and Gallo, 2006]{engle2006multiple}
Engle, R.~F. and Gallo, G.~M. (2006).
\newblock A multiple indicators model for volatility using intra-daily data.
\newblock {\em Journal of Econometrics}, 131(1):3--27.

\bibitem[Engle and Kroner, 1995]{engle1995multivariate}
Engle, R.~F. and Kroner, K.~F. (1995).
\newblock Multivariate simultaneous generalized arch.
\newblock {\em Econometric theory}, pages 122--150.

\bibitem[Fama and French, 2004]{fama2004capital}
Fama, E.~F. and French, K.~R. (2004).
\newblock The capital asset pricing model: Theory and evidence.
\newblock {\em Journal of economic perspectives}, 18(3):25--46.

\bibitem[Fama and MacBeth, 1973]{fama1973risk}
Fama, E.~F. and MacBeth, J.~D. (1973).
\newblock Risk, return, and equilibrium: Empirical tests.
\newblock {\em Journal of political economy}, 81(3):607--636.

\bibitem[Fan and Kim, 2018]{fan2018robust}
Fan, J. and Kim, D. (2018).
\newblock Robust high-dimensional volatility matrix estimation for
  high-frequency factor model.
\newblock {\em Journal of the American Statistical Association},
  113(523):1268--1283.

\bibitem[Figueroa-L{\'o}pez and Wu, 2020]{figueroa2020kernel}
Figueroa-L{\'o}pez, J.~E. and Wu, B. (2020).
\newblock Kernel estimation of spot volatility with microstructure noise using
  pre-averaging.
\newblock {\em arXiv preprint arXiv:2004.01865}.

\bibitem[Francq et~al., 2013]{francq2013garch}
Francq, C., Wintenberger, O., and Zakoian, J.-M. (2013).
\newblock Garch models without positivity constraints: Exponential or log
  garch?
\newblock {\em Journal of Econometrics}, 177(1):34--46.

\bibitem[Gonz{\'a}lez-Rivera, 1996]{gonzalez1996time}
Gonz{\'a}lez-Rivera, G. (1996).
\newblock Time-varying risk the case of the american computer industry.
\newblock {\em Journal of Empirical Finance}, 2(4):333--342.

\bibitem[Hall and Heyde, 2014]{hall2014martingale}
Hall, P. and Heyde, C.~C. (2014).
\newblock {\em Martingale limit theory and its application}.
\newblock Academic press.

\bibitem[Hansen and Richard, 1987]{hansen1987role}
Hansen, L.~P. and Richard, S.~F. (1987).
\newblock The role of conditioning information in deducing testable
  restrictions implied by dynamic asset pricing models.
\newblock {\em Econometrica: Journal of the Econometric Society}, pages
  587--613.

\bibitem[Hansen et~al., 2012]{hansen2012realized}
Hansen, P.~R., Huang, Z., and Shek, H.~H. (2012).
\newblock Realized garch: a joint model for returns and realized measures of
  volatility.
\newblock {\em Journal of Applied Econometrics}, 27(6):877--906.

\bibitem[Hansen and Lunde, 2006]{hansen2006realized}
Hansen, P.~R. and Lunde, A. (2006).
\newblock Realized variance and market microstructure noise.
\newblock {\em Journal of Business \& Economic Statistics}, 24(2):127--161.

\bibitem[Hansen et~al., 2014]{hansen2014realized}
Hansen, P.~R., Lunde, A., and Voev, V. (2014).
\newblock Realized beta garch: A multivariate garch model with realized
  measures of volatility.
\newblock {\em Journal of Applied Econometrics}, 29(5):774--799.

\bibitem[Jacod et~al., 2009]{jacod2009microstructure}
Jacod, J., Li, Y., Mykland, P.~A., Podolskij, M., and Vetter, M. (2009).
\newblock Microstructure noise in the continuous case: the pre-averaging
  approach.
\newblock {\em Stochastic processes and their applications}, 119(7):2249--2276.

\bibitem[Jacod et~al., 2017]{jacod2017statistical}
Jacod, J., Li, Y., and Zheng, X. (2017).
\newblock Statistical properties of microstructure noise.
\newblock {\em Econometrica}, 85(4):1133--1174.

\bibitem[Jacod et~al., 2019]{jacod2019estimating}
Jacod, J., Li, Y., and Zheng, X. (2019).
\newblock Estimating the integrated volatility with tick observations.
\newblock {\em Journal of Econometrics}, 208(1):80--100.

\bibitem[Jacod and Protter, 2012]{jacod2012discretization}
Jacod, J. and Protter, P. (2012).
\newblock {\em Discretization of Processes}.
\newblock Springer.

\bibitem[Keim and Stambaugh, 1986]{keim1986predicting}
Keim, D.~B. and Stambaugh, R.~F. (1986).
\newblock Predicting returns in the stock and bond markets.
\newblock {\em Journal of financial Economics}, 17(2):357--390.

\bibitem[Kim and Wang, 2016a]{Kim2016SPCA}
Kim, D. and Wang, Y. (2016a).
\newblock Sparse pca based on high-dimensional it\^o processes with measurement
  errors.
\newblock {\em Journal of Multivariate Analysis}, 152:172--18.

\bibitem[Kim and Wang, 2016b]{kim2016unified}
Kim, D. and Wang, Y. (2016b).
\newblock Unified discrete-time and continuous-time models and statistical
  inferences for merged low-frequency and high-frequency financial data.
\newblock {\em Journal of Econometrics}, 194(2):220--230.

\bibitem[Koutmos et~al., 1994]{koutmos1994time}
Koutmos, G., Lee, U., and Theodossiu, P. (1994).
\newblock Time-varying betas and volatility persistence in international stock
  markets.
\newblock {\em Journal of Economics and Business}, 46(2):101--112.

\bibitem[Li and Patton, 2018]{li2018asymptotic}
Li, J. and Patton, A.~J. (2018).
\newblock Asymptotic inference about predictive accuracy using high frequency
  data.
\newblock {\em Journal of Econometrics}, 203(2):223--240.

\bibitem[Li et~al., 2017a]{li2017adaptive}
Li, J., Todorov, V., and Tauchen, G. (2017a).
\newblock Adaptive estimation of continuous-time regression models using
  high-frequency data.
\newblock {\em Journal of Econometrics}, 200(1):36--47.

\bibitem[Li et~al., 2017b]{li2017robust}
Li, J., Todorov, V., and Tauchen, G. (2017b).
\newblock Robust jump regressions.
\newblock {\em Journal of the American Statistical Association},
  112(517):332--341.

\bibitem[Li and Linton, 2021]{li2021remedi}
Li, Z.~M. and Linton, O. (2021).
\newblock A remedi for microstructure noise.
\newblock {\em Econometrica}, forthcoming.

\bibitem[Li and Linton, 2022]{li2022remedi}
Li, Z.~M. and Linton, O. (2022).
\newblock A remedi for microstructure noise.
\newblock {\em Econometrica}, 90(1):367--389.

\bibitem[Li and Linton, 2020]{li2020robust}
Li, Z.~M. and Linton, O.~B. (2020).
\newblock Robust estimation of integrated volatility.
\newblock {\em Available at SSRN 3702143}.

\bibitem[Lintner, 1965]{lintner1965security}
Lintner, J. (1965).
\newblock Security prices, risk, and maximal gains from diversification.
\newblock {\em The journal of finance}, 20(4):587--615.

\bibitem[Mykland and Zhang, 2009]{mykland2009inference}
Mykland, P.~A. and Zhang, L. (2009).
\newblock Inference for continuous semimartingales observed at high frequency.
\newblock {\em Econometrica}, 77(5):1403--1445.

\bibitem[Ng, 1991]{ng1991tests}
Ng, L. (1991).
\newblock Tests of the capm with time-varying covariances: A multivariate garch
  approach.
\newblock {\em The Journal of Finance}, 46(4):1507--1521.

\bibitem[Perold, 2004]{perold2004capital}
Perold, A.~F. (2004).
\newblock The capital asset pricing model.
\newblock {\em Journal of economic perspectives}, 18(3):3--24.

\bibitem[Rei{\ss} et~al., 2015]{reiss2015nonparametric}
Rei{\ss}, M., Todorov, V., and Tauchen, G. (2015).
\newblock Nonparametric test for a constant beta between it{\^o}
  semi-martingales based on high-frequency data.
\newblock {\em Stochastic Processes and their Applications}, 125(8):2955--2988.

\bibitem[Shao, 1995]{shao1995maximal}
Shao, Q.-M. (1995).
\newblock Maximal inequalities for partial sums of $\rho$-mixing sequences.
\newblock {\em The Annals of Probability}, pages 948--965.

\bibitem[Sharpe, 1964]{sharpe1964capital}
Sharpe, W.~F. (1964).
\newblock Capital asset prices: A theory of market equilibrium under conditions
  of risk.
\newblock {\em The journal of finance}, 19(3):425--442.

\bibitem[Shephard and Sheppard, 2010]{shephard2010realising}
Shephard, N. and Sheppard, K. (2010).
\newblock Realising the future: forecasting with high-frequency-based
  volatility (heavy) models.
\newblock {\em Journal of Applied Econometrics}, 25(2):197--231.

\bibitem[Song et~al., 2020]{song2020volatility}
Song, X., Kim, D., Yuan, H., Cui, X., Lu, Z., Zhou, Y., and Wang, Y. (2020).
\newblock Volatility analysis with realized garch-it{\^o} models.
\newblock {\em Journal of Econometrics}.

\bibitem[Ubukata and Oya, 2009]{ubukata2009estimation}
Ubukata, M. and Oya, K. (2009).
\newblock Estimation and testing for dependence in market microstructure noise.
\newblock {\em Journal of Financial Econometrics}, 7(2):106--151.

\bibitem[Xiu, 2010]{xiu2010quasi}
Xiu, D. (2010).
\newblock Quasi-maximum likelihood estimation of volatility with high frequency
  data.
\newblock {\em Journal of Econometrics}, 159(1):235--250.

\bibitem[Zhang, 2006]{zhang2006efficient}
Zhang, L. (2006).
\newblock Efficient estimation of stochastic volatility using noisy
  observations: A multi-scale approach.
\newblock {\em Bernoulli}, 12(6):1019--1043.

\bibitem[Zhang, 2011]{zhang2011estimating}
Zhang, L. (2011).
\newblock Estimating covariation: Epps effect, microstructure noise.
\newblock {\em Journal of Econometrics}, 160(1):33--47.

\bibitem[Zhang et~al., 2005]{zhang2005tale}
Zhang, L., Mykland, P.~A., and A{\"\i}t-Sahalia, Y. (2005).
\newblock A tale of two time scales: Determining integrated volatility with
  noisy high-frequency data.
\newblock {\em Journal of the American Statistical Association},
  100(472):1394--1411.

\bibitem[Zhang et~al., 2016]{zhang2016jump}
Zhang, X., Kim, D., and Wang, Y. (2016).
\newblock Jump variation estimation with noisy high frequency financial data
  via wavelets.
\newblock {\em Econometrics}, 4(3):34.

\end{thebibliography}

\clearpage

\appendix

\section*{Appendix.} \label{Appendix}
\addcontentsline{toc}{section}{Appendix.}\stepcounter{section}  
Let $C$ be a generic constant whose values are free of  $n$, and $m$.
We denote the matrix differentiation $\partial_{jk}f(A) = \partial f(A)/ \partial A_{jk}$ for any $2 \times 2$ matrix $A$ and generic differentiable function $f$ defined on the $2 \times 2$ matrix space.
In addition, we define $\mathbf{1}_{\{statement\}}$ as follows:
\begin{equation*}
    \mathbf{1}_{\{statement\}}=\begin{cases}1,\quad \text{if the statement is true} \\ 0, \quad \text{otherwise.}\end{cases}
\end{equation*}
We use generic random variables $\varPsi_{par}^{m,w}$, depending on $m$ and parameters ``$par$'', nonnegative, $\mathcal{G}$-measurable, and satisfying $\mathbb{E}\left[ \left( \varPsi_{par}^{m,w} \right)^w  \right] \leq 1$.
Similarly, we use generic generic random variables $\varPsi_{par}^{m}$, depending on $m$ and parameters ``$par$'', nonnegative, $\mathcal{G}$-measurable, but satisfying $\mathbb{E}\left[ \left( \varPsi_{par}^{m} \right)^w  \right] \leq C_w$ for any $w > 0$.
We also use $O_u(x)$ for a random quantity smaller than $Cx$ for some constant $C$.

\subsection{Proof of Proposition \ref{Proposition-1}}
\textbf{Proof of Proposition \ref{Proposition-1}.}
For $k,n \in \mathbb{N}$, let 
\begin{equation*}
    R(k) \equiv \int ^n_{n-1} \frac{(n-t)^k}{k!}\beta^c_t(\theta)dt.
\end{equation*}
By It\^{o}'s lemma, we have almost surely
\begin{eqnarray*}
     R(k) &=& \frac{\beta_{n-1}^{c}(\theta)}{(k+1)!} + \frac{\omega_{1} + \sum_{i=1}^{p} \gamma_{i} \beta_{n-i}^{c} (\theta) + \sum_{j=2}^{q} \alpha_j \int_{n-j}^{n+1-j} \beta_{t}^{c}(\theta) dt  }{(k+3)! / 2} - \frac{\omega_{2} + \beta_{n-1}^{c}(\theta)}{(k+2)!}  \\
     &&   + \nu \int_{n-1}^{n} \left( \frac{(n-t)^{k+2}}{(k+1)!} - \frac{(n-t)^{k+2}}{(k+2)!}   \right) dZ_t + \alpha_1 R(k+1).
\end{eqnarray*}
Using the iterative relationship and the fact that $\alpha_{1}^{k} R(k) \xrightarrow[]{a.s.} 0$ for any $\alpha_{1} \in \mathbb{R}$, we have
\begin{equation*}
     \int ^n_{n-1} \beta^{c}_{t}(\theta)dt = R(0) = h_n(\theta) + D_n \quad \text{a.s.},
\end{equation*}
where
\begin{eqnarray*}
  h_n(\theta) &=&   \varrho_{1} \beta^{c}_{n-1}(\theta) - \varrho_{2} \left( \omega_{2} + \beta_{n-1}^{c}(\theta) \right)  \\
  && + 2 \varrho_3 \left( \omega_{1} - \beta^{c}_{n-1}(\theta) + \sum_{i=1}^{p} \gamma_{i} \beta^{c}_{n-i}(\theta) + \sum_{j=2}^{q} \alpha_j \int_{n-j}^{n+1-j} \beta_{t}^{c}(\theta) dt \right).
\end{eqnarray*}
By \eqref{EXTENDpq}, we have almost surely
\begin{eqnarray*}
  h_n(\theta) &=& \omega^{(1)} + \sum_{i=1}^{p}  \gamma_{i}^{(1)} \beta^{c}_{n-1-i}(\theta) + \sum_{j=1}^{q} \alpha_{j}^{(1)} \int_{n-j-1}^{n-j} \beta_{t}^{c}(\theta) dt \\
  &=& \omega^{(N)}  + \sum_{i=1}^{p}  \gamma_{i}^{(N)} \beta^{c}_{n-1-i}(\theta) + \sum_{j=1}^{q+N-1} \alpha_{j}^{(N)} \int_{n-j-1}^{n-j} \beta_{t}^{c}(\theta) dt
\end{eqnarray*}
for any integer $N \geq 2$, where $\omega^{(N)}$, $\gamma_{i}^{(N)}$ and $\alpha_{i}^{(N)}$ are recursively defined as follows:
\begin{eqnarray}
  && \gamma_{1}^{(-1)} = 2 \varrho_{3} , \quad \gamma_{1}^{(0)} = \varrho_{1} - \varrho_{2} + 2 \varrho_{3} \gamma_{1} , \nonumber\\
  && \gamma_{i}^{(1)} =  (\varrho_{1} - \varrho_{2} + 2 \varrho_{3} \gamma_{1})  \gamma_{i} + 2 \varrho_{3} \gamma_{i+1}, \quad \gamma_{k} = 0 \quad \text{for} \quad k \geq p+1, \label{gamma-(1)}\\
  && \gamma_{i}^{(N)} = \gamma_{1}^{(N-1)} \gamma_{i} + \gamma_{i+1}^{(N-1)} = \sum_{k=0}^{N \land (p-i)} \gamma_{1}^{(N-1-k)} \gamma_{i+k}, \label{gamma-(N)} \\
  && \omega^{(1)} = \left( \varrho_{1} - \varrho_{2} + 2 \varrho_{3} ( 1+ \gamma_{1})  \right) \omega_{1} - (\varrho_{1} + 2\varrho_{3} \gamma_{1}) \omega_{2}, \quad \omega^{(N)} = \omega^{(1)} + \omega \sum_{k=1}^{N-1} \gamma_1^{(k)}, \nonumber\\
  && \alpha_{j}^{(1)} = (\varrho_{1} - \varrho_{2} + 2 \varrho_{3} \gamma_{1})  \alpha_{j} + 2 \varrho_{3} \alpha_{j+1}, \quad \alpha_{k} = 0 \quad \text{for} \quad k \notin [1,q], \nonumber\\
  && \alpha_{j}^{(N)} = \alpha_{j}^{(N-1)} + \gamma_{1}^{(N-1)} \alpha_{j-N+1} = \sum_{k=0}^{N \land j} \gamma_{1}^{(k-1)} \alpha_{j-k+1} \label{alpha-(N)}.
\end{eqnarray}
Let $\bar{\gamma} = \sum_{i=1}^{p} \left|\gamma_{i}\right| < 1$.
Using the mathematical induction method and \eqref{gamma-(N)} with $i=1$, we can show that there exists $C > 0$ such that
\begin{equation}\label{gamma-1}
  \left|\gamma_{1}^{(N)}\right| \leq C \left( \bar{\gamma} \right)^{\frac{N}{2p} } \quad  \text{for any} \quad N \in \mathbb{N}
  ,
\end{equation}
which implies that there exists $C > 0$ such that
\begin{equation}\label{gamma-i}
  \left|\gamma_{i}^{(N)}\right| \leq C \left( \bar{\gamma} \right)^{\frac{N}{2p} } \quad  \text{for any} \quad i, N \in \mathbb{N}
  .
\end{equation}
Thus, we have
\begin{equation*}
  h_n(\theta) = \omega^{(\infty)} + \sum_{j=1}^{\infty} \alpha_{j}^{(\infty)} \int_{n-j-1}^{n-j} \beta_{t}^{c}(\theta) dt \quad \text{a.s.}
  ,
\end{equation*}
where $\omega^{(\infty)} = \omega^{(1)} + \omega \sum_{k=1}^{\infty} \gamma_{1}^{(k)} $ and $\alpha_{j}^{(\infty)} = \sum_{s=1}^{(j+1) \land q} \gamma_{1}^{(j-s)} \alpha_{s} $.
Simple algebra shows that
\begin{equation*}
  h_n(\theta) - \sum_{i=1}^{p} \gamma_{i} h_{n-i}(\theta) = \bar{\omega}  + \sum_{j=1}^{\infty} \bar{\alpha}_{j} \int_{n-j-1}^{n-j} \beta_{t}^{c}(\theta) dt
  ,
\end{equation*}
where
\begin{equation*}
  \bar{\omega} = \omega^{(\infty)} \left( 1 - \sum_{i=1}^{p} \gamma_{i} \right) 
\end{equation*}
and
\begin{eqnarray*}
  \bar{\alpha}_{j} &=&  \alpha_{j}^{(\infty)} - \sum_{k=1}^{(j-1) \land p} \gamma_{k} \alpha_{j-k}^{(\infty)} \\
  &=& \sum_{s=1}^{(j+1) \land q} \gamma_{1}^{(j-s)} \alpha_{s} - \sum_{k=1}^{(j-1) \land p} \gamma_{k} \sum_{s=1}^{(j-k+1) \land q}  \gamma_{1}^{(j-k-s)}\alpha_{s}
  .
\end{eqnarray*}
For $j \geq (p \lor q) + 1$, we have
\begin{eqnarray*}
  \bar{\alpha}_{j} &=&   \sum_{s=1}^{q} \gamma_{1}^{(j-s)} \alpha_{s} - \sum_{k=1}^{p} \gamma_{k} \sum_{s=1}^{(j-k+1) \land q}  \gamma_{1}^{(j-k-s)}\alpha_{s} \\
  &=& \sum_{s=1}^{q} \left(  \gamma_{1}^{(j-s)} - \sum_{k=1}^{(j-q+1) \land p} \gamma_{k} \gamma_{1}^{(j-s-k)}   \right)  \alpha_{s}  -   \sum_{k=(j-q+1) \land p + 1}^{p} \sum_{s=1}^{(j-k+1) \land q} \gamma_{k} \gamma_{1}^{(j-s-k)} \alpha_{s} \\
  &=& \mathbf{1}_{\left\lbrace j \leq p+q-2 \right\rbrace} \left[ \sum_{s=1}^{q-1} \sum_{k=j-q+2}^{p \land (j-s+1)} \gamma_{k} \gamma_{1}^{(j-s-k)}     \alpha_{s}  -   \sum_{k=j-q+2}^{p} \sum_{s=1}^{j-k+1} \gamma_{k} \gamma_{1}^{(j-s-k)} \alpha_{s} \right]  \\
  &=&  0 
  ,
\end{eqnarray*}
where the third equality is due to \eqref{gamma-(N)}.
Furthermore, after some tedious algebra, we have
\begin{eqnarray*}
  \bar{\alpha}_{j} &=& \mathbf{1}_{\left\lbrace j \leq p \right\rbrace} \gamma_{1}^{(-1)} \gamma_{j} \alpha_{1} + \mathbf{1}_{\left\lbrace j \leq q \right\rbrace} (\varrho_{1} - \varrho_{2}) \alpha_{j} + \mathbf{1}_{\left\lbrace j \leq q-1 \right\rbrace} \gamma_{1}^{(-1)} \alpha_{j+1} \\
  &=& \mathbf{1}_{\left\lbrace j \leq p \right\rbrace} 2 \varrho_{3} \gamma_{j} \alpha_{1} + \mathbf{1}_{\left\lbrace j \leq q \right\rbrace} (\varrho_{1} - \varrho_{2}) \alpha_{j} + \mathbf{1}_{\left\lbrace j \leq q-1 \right\rbrace} 2 \varrho_{3} \alpha_{j+1} \\
  &=& \alpha_{j}^{g}. 
\end{eqnarray*}

On the other hand, by \eqref{gamma-(N)}, we have for any integer $N \geq 2$
\begin{equation*}
  \sum_{i=1}^{p} \gamma_{i}^{(N)} = \gamma_{1}^{(N-1)}\sum_{i=1}^{p} \gamma_{i} + \sum_{i=1}^{p} \gamma_{i}^{(N-1)} - \gamma_{1}^{(N-1)}
  ,
\end{equation*}
which implies that
\begin{equation*}
  \sum_{i=1}^{p} \gamma_{i}^{(1)} - \sum_{i=1}^{p} \gamma_{i}^{(N)} = (1 - \sum_{i=1}^{p} \gamma_{i}) \sum_{k=1}^{N-1} \gamma_{1}^{(k)}
  .
\end{equation*}
By \eqref{gamma-i} and \eqref{gamma-(1)}, we have
\begin{eqnarray*}
  \left(1 - \sum_{i=1}^{p} \gamma_{i}\right) \sum_{k=1}^{\infty} \gamma_{1}^{(k)} &=&  \sum_{i=1}^{p} \gamma_{i}^{(1)} \\
  &=& \left( \varrho_{1} - \varrho_{2} + 2 \varrho_{3} ( \gamma_{1} + 1) \right) \sum_{i=1}^{p} \gamma_{i} - 2 \varrho_{3}\gamma_{1}
  .
\end{eqnarray*}
Thus, we have
\begin{eqnarray*}
  \bar{\omega} &=&  \left( \omega^{(1)} + \omega \sum_{k=1}^{\infty} \gamma_{1}^{(k)}  \right) \left( 1 - \sum_{i=1}^{p} \gamma_{i} \right)  \\
  &=&  \left( \varrho_{1} - \varrho_{2} + 2 \varrho_{3} ( 1+ \gamma_{1})  \right) \omega  \left( 1 - \sum_{i=1}^{p} \gamma_{i} \right) + (2 \varrho_{3} - \varrho_{2}) \omega_{2}  \left( 1 - \sum_{i=1}^{p} \gamma_{i} \right) \\
  && + \omega \sum_{k=1}^{\infty} \gamma_{1}^{(k)}   \left( 1 - \sum_{i=1}^{p} \gamma_{i} \right)  \\
  &=&  \left( \varrho_{1} - \varrho_{2} + 2 \varrho_{3} ( 1+ \gamma_{1})  \right) \omega  \left( 1 - \sum_{i=1}^{p} \gamma_{i} \right) + (2 \varrho_{3} - \varrho_{2}) \omega_{2}  \left( 1 - \sum_{i=1}^{p} \gamma_{i} \right) \\
  && + \omega \left( \left( \varrho_{1} - \varrho_{2} + 2 \varrho_{3} ( \gamma_{1} + 1) \right) \sum_{i=1}^{p} \gamma_{i} - 2 \varrho_{3}\gamma_{1} \right)   \\
  &=& (\varrho_{1} - \varrho_{2} + 2 \varrho_{3}) \omega + (2\varrho_{3} - \varrho_{2}) \left( 1 - \sum_{i=1}^{p} \gamma_{i} \right)  \omega_2 \\
  &=& \omega^g 
  ,
\end{eqnarray*}
and the proof of Proposition \ref{Proposition-1}(a) is complete.

Similar to the proof of  Proposition 2.3 \citep{francq2013garch}, Proposition \ref{Proposition-1}(b) can be showed with the result of Proposition \ref{Proposition-1}(a).
\endpf

\subsection{Proof of Theorem \ref{Theorem-1}}\label{pf-thm1}

Note that the spot covariance matrix $\bSigma_t$ of $(X_1^c, X_2^c)^\top$ can be written as 
\begin{equation*}
    \bSigma_t
    =\begin{pmatrix} \sigma^2_t & \beta^c_t \sigma^2_t \\ \beta^c_t \sigma^2_t & (\beta^c_t)^2 \sigma^2_t+q_t^2 \end{pmatrix},
\end{equation*}
for all $t \in \mathbb{R}^+$.
Moreover, similar to \eqref{Equation-3.3}, we can construct an estimator for $\bSigma(t)$ as follows:
\begin{eqnarray*}
    \hat{\bSigma}^c(t)
= \frac{1}{(b_m - 2k_m)\Delta _m  k_m \psi_0} \begin{pmatrix} v(Y_1^c,Y_1^c,\infty,\infty,\infty,t) & v(Y_1^c,Y_2^c,\infty,\infty,\infty,t)  \\
  v(Y_2^c,Y_1^c,\infty,\infty,\infty,t) & v(Y_2^c,Y_2^c,\infty,\infty,\infty,t)  \end{pmatrix} ,
\end{eqnarray*}
where continuous processes $Y_1^c$ and $Y_2^c$ satisfy $Y_{1,t}^{c,m}=X^c_{1,t}+\epsilon_{1,i}^{m}$ and  $Y_{2,t}^{c,m}=X^c_{2,t}+\epsilon_{2,i}^{m}$.
Note that $Y_{1,t}=Y^c_{1,t}+X^d_{1,t}$, $Y_{2,t}=Y^c_{2,t}+X^d_{2,t}$. 

Define $\Xi (v, \mu) : \mathbb{R}^{2\times 2} \times \mathbb{R}^{2\times 2} \rightarrow \mathbb{R}^{2\times 2}$ such that for any $x,y \in \left\lbrace 1,2 \right\rbrace$
\begin{eqnarray*}
  \Xi (v, \mu)_{x,y} &=&   \frac{2}{\psi_0^2 {C_k}^3} \big[ {C_k}^4 \Phi_{00} \left( v_{11} v_{xy} + v_{1x} v_{1y} \right)  + {C_k}^2 \Phi_{01} \left( v_{11} \mu_{xy} + v_{1x} \mu_{1y} + v_{1y} \mu_{1x} + v_{xy} \mu_{11}  \right) \\
  &&  + \Phi_{11} \left( \mu_{11} \mu_{xy} + \mu_{1x} \mu_{1y} \right)   \big]
  ,
\end{eqnarray*}
and for any matrix $A^{m} \in \mathbb{R}^{2\times 2}$
\begin{equation*}
  A^{m,*} =
  \begin{pmatrix}
    \max (A_{11}^{m}, \delta_m) & A_{12}^{m} \\
    A_{21}^{m} & A_{22}^{m}
  \end{pmatrix}.
\end{equation*}
Then, we obtain
\begin{equation*}
  \hat{B}^{m}_{ib_m} = \frac{1}{2 b_m \Delta_m ^{1/2}} \sum_{x,y=1}^{2} \Xi (\hat{\bSigma}_{ib_m}^{m,*} , \hat{\bvartheta}_{i b_m }^{m} )_{x,y} \partial^2_{1x,1y}f(\hat{\bSigma}_{ib_m}^{m,*})  ,
\end{equation*}
where $f(\bc) = (c_{11})^{-1} c_{12}$.
Furthermore, let
\begin{eqnarray*}
  && \hat{B}^{c,m}_{ib_m} = \frac{1}{2 b_m \Delta_m ^{1/2}} \sum_{x,y=1}^{2}  \Xi (\hat{\bSigma}_{ib_m}^{c,m,*} , \hat{\bvartheta}_{i b_m }^{c,m} )_{x,y} \partial^2_{1x,1y}f(\hat{\bSigma}_{ib_m}^{c,m,*}) ,\\
  && {B}^{c,m}_{ib_m} = \frac{1}{2 b_m \Delta_m ^{1/2}} \sum_{x,y=1}^{2}  \Xi ({\bSigma}_{ib_m}^{m,*} , {\bvartheta}_{i b_m }^{m} )_{x,y} \partial^2_{1x,1y}f({\bSigma}_{ib_m}^{m,*}) ,
\end{eqnarray*}
$\hat{\bvartheta}^{m}_{xy,i} = (b_m-6l_m)^{-1} \sum_{j=i}^{i + b_m - 6 l_m}  \dot{\mathcal{E}}_{Y_x^c Y_y^c,j}^{m}$, and $\bvartheta_{xy,i}^{m}  = \vartheta_{x,i}^{m} \vartheta_{y,i}^{m} R_{x y}  $ for any $x, y \in \left\lbrace 1,2 \right\rbrace$.

Lemma 4.4.9 in \citet{jacod2012discretization} indicates that if the asymptotic result such as convergence in probability or stable convergence in law is satisfied under the boundedness condition, it is also satisfied under the locally boundedness condition.
Thus, without loss of generality, we assume that the drift and spot volatility processes are bounded in the following proofs. 


\subsubsection{Properties of spot volatility: Continuous part}

We first show some properties of spot volatility estimator $\hat{\bSigma}^{c,m}_{xy,i}$.
We introduce some notations to follow the ``big blocks and small blocks"-technique \citep{jacod2009microstructure}.
For $p,i \in \mathbb{N}$ and $x,y \in \left\lbrace 1,2 \right\rbrace$, we define
\begin{eqnarray*}\label{block-notation}
  &&C_{xy,t} =  \int _{0}^{t} \bSigma_{xy,s} ds, \qquad \breve{C}_{xy,i}^{m} = \sum_{l=1}^{k_m-1} (g _{l}^{m})^2 \left( C_{xy,i+l}^{m} - C_{xy,i+l-1}^{m} \right) , \\
  &&\tilde{\Gamma}^{m}_{xy,i} = \vartheta^{m}_{x,i} \vartheta^{m}_{y,i} \sum_{l_1,l_2 = 0}^{k_m-1} r_{xy}(l_1,l_2) h_{l_1}^m h_{l_2}^m ,\qquad 
  r_{xy}(l_1,l_2) = \mathbb{E}\left[ \chi_{x,l_1} \chi_{y,l_2} \right]. \\
  &&\zeta_{xy,i}^{m} = \tilde{Y}_{x,i}^{c} \tilde{Y}_{y,i}^{c} - \breve{C}_{xy,i}^{m} - \tilde{\Gamma}^{m}_{xy,i}, \qquad
\zeta(p)_{xy,i}^{m} = \sum_{l=i}^{i+pk_m-1} \zeta _{xy,l}^{m}. 
\end{eqnarray*}
The estimation error of spot volatility, $\hat{\bSigma}_{xy,i}^{c,m} - \bSigma_{xy,i}^{m}$ can be decomposed as follows:
\begin{equation}\label{SpotErrorDecomp}
  e_{xy,i}^{m} = \hat{\bSigma}_{xy,i}^{c,m} - \bSigma_{xy,i}^{m} = M(p)_{xy,i}^{m} + M'(p)_{xy,i}^{m} + \bar{M}(p)_{xy,i}^{m} + \bar{M}'(p)_{xy,i}^{m} + \xi_{xy,i}^{m,1} + \xi_{xy,i}^{m,2},
\end{equation}
where
\begin{align*}\label{details}
  &L(m,p) = \left[ \frac{b_m - 2 k_m}{(p+2)k_m} \right] , \quad \mathcal{K}_{i}^{m} = \mathcal{F}_{i}^{m} \otimes \mathcal{G}_{i-k_m}, \\
  & \mathcal{H}(p)_{j}^{m,i} = \mathcal{K}_{i+j(p+2)k_m}^{m} ,\quad \mathcal{H}'(p)_{j}^{m,i} = \mathcal{K}_{i+j(p+2)k_m+p}^{m},  \\
  &\eta(p)_{xy,j}^{m,i} =  \frac{1}{(b_m - 2 k_m)\Delta_m k_m \psi_0} \zeta(p)_{xy,i+(j-1)(p+2)k_m}^{m}, \quad \bar{\eta}(p)_{xy,j}^{m,i} = \mathbb{E}\left[ \eta(p)_{xy,j}^{m,i} | \mathcal{H}(p)_{j}^{m,i} \right], \\
  &\eta'(p)_{xy,j}^{m,i} =  \frac{1}{(b_m - 2 k_m)\Delta_m k_m \psi_0} \zeta(2)_{xy,i+(j-1)(p+2)k_m}^{m}, \quad \bar{\eta}'(p)_{xy,j}^{m,i} = \mathbb{E}\left[ \eta'(p)_{xy,j}^{m,i} | \mathcal{H}'(p)_{j}^{m,i} \right], \\
  &\hat{\eta}(p)_{xy,j}^{m,i} = \eta(p)_{xy,j}^{m,i} - \bar{\eta}(p)_{xy,j}^{m,i}, \quad M(p)_{xy,i}^{m} =   \sum_{j=0}^{L(m,p)-1} \hat{\eta}(p)_{xy,j}^{m,i} , \quad \bar{M}(p)_{xy,i}^{m} =   \sum_{l=0}^{L(m,p)-1} \bar{\eta}(p)_{xy,j}^{m,i}, \\
  &\hat{\eta}'(p)_{xy,j}^{m,i} = \eta'(p)_{xy,j}^{m,i} - \bar{\eta}'(p)_{xy,j}^{m,i}, \quad M'(p)_{xy,i}^{m} =   \sum_{j=0}^{L(m,p)-1} \hat{\eta}'(p)_{xy,j}^{m,i} , \\
  & \bar{M}'(p)_{xy,i}^{m} =   \sum_{l=0}^{L(m,p)-1} \bar{\eta}'(p)_{xy,j}^{m,i}, \quad \xi_{xy,i}^{m,1} = \frac{1}{(b_m - 2 k_m) \Delta_m k_m \psi_0} \sum_{l=0}^{b_m- 2 k_m - 1}  \breve{C}_{xy,i}^{m} -\bSigma_{xy,i}^{m}, \\
  &\xi_{xy,i}^{m,2} = \frac{1}{(b_m - 2 k_m) \Delta_m k_m \psi_0}  \sum_{l=1}^{b_m - 2 k_m} \tilde{\Gamma}_{xy,i}^{m} - \frac{1}{(b_m - 2 k_m) \Delta_m k_m^2 \psi_0}  \sum_{d=-k_m'}^{k_m'} \phi_{d}^{m}  U_{m,i}^{Y_x^{c} Y_y^{c}} (|d|), \\
  & U_{m,i}^{Y_x^{c} Y_y^{c}} (|d|) = \sum_{l=0}^{b_m - 6 l_m} \mathcal{E}_{Y_x^c Y_y^c,i+l}^{m,d}
  .
\end{align*}

\begin{lemma}\label{smooth-process}
  If $P_{1,t},\ldots,P_{k,t}$, $k \in \mathbb{N}$ are stochastic processes defined on a filtered probability space $(\Omega, \mathcal{F}, \left\lbrace \mathcal{F}_{t}, t \in [0,\infty) \right\rbrace, P)$, satisfying the properties $(P1)$ and $(P2)$ defined in Assumption \ref{assumption-formal}(b),
   then $Q_{t} = \prod_{i=1}^{k} P_{i,t}$ also satisfy (P1) and (P2).
\end{lemma}  
\textbf{Proof of Lemma \ref{smooth-process}.}
$(P1)$ is trivial.
If $k=2$, $(P2)$ is satisfied for $Q_{t}$, since
\begin{eqnarray*}
  \mathbb{E}\left[ \left( Q_{t+s} - Q_{t} \right)^2  | \mathcal{F}_{t} \right] &\leq& C \mathbb{E}\left[P_{1,t+s}^{2} \left( P_{2,t+s} - P_{2,t}   \right)^2  | \mathcal{F}_{t} \right] + C \mathbb{E}\left[P_{2,t}^{2} \left( P_{1,t+s} - P_{1,t}   \right)^2  | \mathcal{F}_{t} \right] \\
  &\leq& C \left(  \mathbb{E}\left[ \left( P_{2,t+s} - P_{2,t}   \right)^2 | \mathcal{F}_{t} \right] +   \mathbb{E}\left[ \left( P_{1,t+s} - P_{1,t}   \right)^2 | \mathcal{F}_{t} \right] \right) \\
  &\leq& Cs \text{ a.s.}
  ,
\end{eqnarray*}
where the first inequality is due to Jensen's inequality, and the second and third inequalities are due to the fact that the process $P_1$ and $P_2$ satisfy $(P1)$ and $(P2)$, respectively.
We can prove for the cases $k>2$, by using mathematical induction method.
$\blacksquare$

\begin{lemma}\label{gmeasurable-bound}
  Let $\xi_{i}^{m} $ be random variables, measurable with respect to $\mathcal{G}^{i}$.
  We have
  \begin{enumerate}
    \item [(a)] $\left| \mathbb{E}\left[ \xi_{i}^{m} | \mathcal{K}_{i}^{m}  \right] \right|  \leq \varPsi_{i}^{m} \text{ a.s.}$ for any $w \in \mathbb{N}$, if $\mathbb{E}\left[ \left|\xi_{i}^{m}\right| ^{w} \right] \leq C_{w}$;
    \item [(b)] $\left|\mathbb{E}\left[ \xi_{i}^{m} | \mathcal{K}_{i}^{m}  \right]\right|^{w}  \leq C \varPsi_{i,1}^{m} \Delta_m ^{7v/8} \text{ a.s.}$ for any $w \geq 3$, if $\xi _{i}^{m} $ is centered with finite moment of all orders;
    \item [(c)] if $k,i', j \in \mathbb{N}$, $i \leq i' < i + j$, and $\xi_{i}^{m}$ and $\xi_{i'+j}^{m}$ are measurable with respect to $\mathcal{G}_{i'}$ and $\mathcal{G}^{i'+j}$, respectively, then we have 
    \begin{eqnarray*}
      \mathbb{E}\left[\xi_{i}^{m} \xi_{i'+j}^{m} | \mathcal{G}_{i-k}  \right] &\leq& \left| \mathbb{E}\left[\xi_{i}^{m} \right] \mathbb{E}\left[ \xi_{i'+j}^{m}   \right] \right|  +  C j^{-v} \mathbb{E}\left[ \left( \xi_{i'+j}^{m} \right)^2  \right]^{1/2} \mathbb{E}\left[ \left( \xi_{i}^{m} \right)^2  \right]^{1/2}   \\
      && + C \varPsi_{i,2}^{m} k^{-v} \mathbb{E}\left[ \left( \xi_{i'+j}^{m} \right)^4  \right]^{1/4} \mathbb{E}\left[ \left( \xi_{i}^{m} \right)^4  \right]^{1/4}\text{ a.s.}
    \end{eqnarray*}
    
  \end{enumerate}
\end{lemma}
\textbf{Proof of Lemma \ref{gmeasurable-bound}.}
Consider (a).
By (A.3) of \citet{jacod2017statistical}, we have 
\begin{eqnarray}\label{rho-mixing2}
  \mathbb{E}\left[ \left|\mathbb{E}\left[ \xi_{i}^{m} | \mathcal{K}_{i}^{m} \right]\right|^{2}   \right] &\leq& \frac{C}{k_m^{v}} \mathbb{E}\left[ \mathbb{E}\left[ \xi_{i}^{m}  | \mathcal{K}_{i}^{m}  \right]^{2} \right]^{1/2} \mathbb{E}\left[ \mathbb{E}\left[ \xi_{i}^{m} \right]^{2} \right]^{1/2} + \left|\mathbb{E}\left[ \mathbb{E}\left[ \xi_{i}^{m} | \mathcal{K}_{i}^{m}  \right] \right] \mathbb{E}\left[ \xi_{i}^{m} \right] \right| \nonumber\\
  &\leq& \frac{C}{k_m^{v}} \mathbb{E}\left[  \left(   \xi_{i}^{m} \right)^2  \right]  + \mathbb{E}\left[ \xi_{i}^{m} \right]^2 \nonumber \\
  &\leq& C_2 \text{ a.s.}
\end{eqnarray}
Similarly, we have  
\begin{eqnarray}\label{rho-mixing3}
  \mathbb{E}\left[ \left| \mathbb{E}\left[ \xi_{i}^{m} | \mathcal{G}_{i} \right]\right|^3  \right] &\leq& \frac{C}{k_m^v} \mathbb{E}\left[ \mathbb{E}\left[ \xi_{i}^{m} | \mathcal{G}_{i} \right]^4 \right]^{1/2} \mathbb{E}\left[ \left( \xi_{i}^{m} \right)^2 \right]^{1/2} + \left| \mathbb{E}\left[ \mathbb{E}\left[ \xi_{i}^{m} | \mathcal{G}_{i} \right]^2  \right] \mathbb{E}\left[ \xi_{i}^{m} \right] \right| \nonumber\\
  &\leq& \frac{C}{k_m^v} \mathbb{E}\left[ \left( \xi_{i}^{m} \right) ^4 \right]^{1/2} \mathbb{E}\left[ \left( \xi_{i}^{m} \right)^2 \right]^{1/2}  + \left| \mathbb{E}\left[ \left( \xi_{i}^{m} \right) ^2  \right] \mathbb{E}\left[ \xi_{i}^{m} \right] \right| \nonumber\\
  &\leq& C_3 \text{ a.s.}
\end{eqnarray}
and  
\begin{eqnarray}\label{rho-mixing4}
  \left|\mathbb{E}\left[ \mathbb{E}\left[ \xi_{i}^{m} | \mathcal{G}_{i} \right]^4 \right]\right| &\leq& \frac{C}{k_m^v} \mathbb{E}\left[ \mathbb{E}\left[ \xi_{i}^{m} | \mathcal{G}_{i} \right]^6 \right]^{1/2} \mathbb{E}\left[ \left( \xi_{i}^{m} \right)^2 \right]^{1/2} + \left| \mathbb{E}\left[ \mathbb{E}\left[ \xi_{i}^{m} | \mathcal{G}_{i} \right]^3  \right] \mathbb{E}\left[ \xi_{i}^{m} \right] \right| \nonumber \\
  &\leq& \frac{C}{k_m^v} \mathbb{E}\left[ \mathbb{E}\left[ \xi_{i}^{m} | \mathcal{G}_{i} \right]^6 \right]^{1/2} \mathbb{E}\left[ \left( \xi_{i}^{m} \right)^2 \right]^{1/2} + \frac{C}{k_m^v} \mathbb{E}\left[ \left( \xi_{i}^{m} \right) ^4 \right]^{1/2} \mathbb{E}\left[ \left( \xi_{i}^{m} \right)^2 \right]^{1/2} \left| \mathbb{E}\left[ \xi_{i}^{m} \right] \right| \nonumber\\
  && +  \mathbb{E}\left[ \left( \xi_{i}^{m} \right) ^2  \right] \left| \mathbb{E}\left[ \xi_{i}^{m} \right] \right|^2 \nonumber \\
  &\leq& \frac{C}{k_m^v} \mathbb{E}\left[ \left( \xi_{i}^{m} \right) ^6 \right]^{1/2} \mathbb{E}\left[ \left( \xi_{i}^{m} \right)^2 \right]^{1/2} + \frac{C}{k_m^v} \mathbb{E}\left[ \left( \xi_{i}^{m} \right) ^4 \right]^{1/2} \mathbb{E}\left[ \left( \xi_{i}^{m} \right)^2 \right]^{1/2} \left| \mathbb{E}\left[ \xi_{i}^{m} \right] \right| \nonumber\\
  && +  \mathbb{E}\left[ \left( \xi_{i}^{m} \right) ^2  \right] \left| \mathbb{E}\left[ \xi_{i}^{m} \right] \right|^2 \nonumber \\
  &\leq& C_4 \text{ a.s.}
\end{eqnarray}
Using the iterative relationship, we can that 
$\mathbb{E}\left[ \mathbb{E}\left[ \xi_{i}^{m} | \mathcal{K}_{i}^{m}  \right]^{w} \right] \leq C_w \text{ a.s.}$ for any $w \in \mathbb{N}$
which implies Lemma \ref{gmeasurable-bound}(a).

Consider (b).
We have
\begin{equation}\label{rho-mixing6}
  \left|\mathbb{E}\left[ \mathbb{E}\left[ \xi_{i}^{m} | \mathcal{K}_{i}^{m} \right]^6 \right]\right| \leq C \Delta_m ^{v/2} \mathbb{E}\left[ \mathbb{E}\left[ \xi_{i}^{m} | \mathcal{K}_{i}^{m} \right]^{10} \right]^{1/2} \mathbb{E}\left[ \left( \xi_{i}^{m} \right)^2  \right]^{1/2} \text{ a.s.}
\end{equation}
Using \eqref{rho-mixing3}, \eqref{rho-mixing4}, and \eqref{rho-mixing6}, we have
\begin{eqnarray*}
  \mathbb{E}\left[ \left| \mathbb{E}\left[ \xi_{i}^{m} | \mathcal{K}_{i}^{m} \right]\right|^3  \right] &\leq& C \Delta_m ^{v/2} \mathbb{E}\left[ \mathbb{E}\left[ \xi_{i}^{m} | \mathcal{K}_{i}^{m} \right]^4 \right]^{1/2} \mathbb{E}\left[ \left( \xi_{i}^{m} \right)^2  \right]^{1/2} \\
  &\leq& C \Delta_m ^{3v/4} \mathbb{E}\left[ \mathbb{E}\left[ \xi_{i}^{m} | \mathcal{K}_{i}^{m} \right]^{6} \right]^{1/4} \mathbb{E}\left[ \left( \xi_{i}^{m} \right)^2  \right]^{3/4} \\
  &\leq& C \Delta_m ^{7v/8} \mathbb{E}\left[ \mathbb{E}\left[ \xi_{i}^{m} | \mathcal{K}_{i}^{m} \right]^{10} \right]^{1/8} \mathbb{E}\left[ \left( \xi_{i}^{m} \right)^2  \right]^{7/8} \\
  &\leq& C \Delta_m ^{7v/8}\text{ a.s.},
\end{eqnarray*}
where the fourth inequality is due to Jensen's inequality and the fact that $\xi_{i}^{m}$ has finite moment of all orders.
Thus, we have $\left|\mathbb{E}\left[ \xi_{i}^{m} | \mathcal{K}_{i}^{m} \right]\right|^{3}  \leq C \varPsi_{i,1}^{m} \Delta_m ^{7v/8} \text{ a.s.}$
Similarly, we can show that for any integer $w \geq 3$,
\begin{equation*}
  \left|\mathbb{E}\left[ \xi_{i}^{m} | \mathcal{K}_{i}^{m} \right]\right|^{w}  \leq C \varPsi_{i,1}^{m} \Delta_m ^{7v/8} \text{ a.s.}
\end{equation*}

Consider (c).
Since $\xi_{i}^{m} \xi_{i'+j}^{m}$ is $\mathcal{G}^{i}$-measurable, we have
\begin{equation*}
  \left| \mathbb{E}\left[ \xi_{i}^{m} \xi_{i'+j}^{m} \right] \right| \leq C j^{-v} \mathbb{E}\left[ \left( \xi_{i}^{m} \right)^2  \right]^{1/2} \mathbb{E}\left[ \left( \xi_{i'+j}^{m} \right)^2  \right]^{1/2} + \left|\mathbb{E}\left[ \xi_{i}^{m} \right] \mathbb{E}\left[ \xi_{i'+j}^{m} \right] \right| 
  .
\end{equation*}
By (A.4) of \citet{jacod2019estimating}, we have
\begin{eqnarray*}
  \mathbb{E}\left[ \left( \mathbb{E}\left[ \xi_{i}^{m} \xi_{i'+j}^{m} | \mathcal{G}_{i-k} \right] - \mathbb{E}\left[ \xi_{i}^{m} \xi_{i'+j}^{m} \right] \right)^2  \right] &=& \mathbb{E}\left[ \mathbb{E}\left[ \xi_{i}^{m} \xi_{i'+j}^{m} - \mathbb{E}\left[ \xi_{i}^{m} \xi_{i'+j}^{m} \right] | \mathcal{G}_{i-k} \right]^{2} \right] \\
  &\leq& C k^{-2v} \mathbb{E}\left[ \left( \xi_{i}^{m} \xi_{i'+j}^{m} \right)^2  \right] \\
  &\leq& C  k^{-2v} \mathbb{E}\left[ \left( \xi_{i}^{m} \right)^{4} \right]^{1/2} \mathbb{E}\left[ \left( \xi_{i'+j}^{m} \right)^4  \right]^{1/2} \text{ a.s.}
  ,
\end{eqnarray*}
where the second inequality is due to H\"older's inequality.
Thus, we have
\begin{eqnarray*}
  \mathbb{E}\left[\xi_{i}^{m} \xi_{i'+j}^{m} | \mathcal{G}_{i-k}  \right] &\leq& \left| \mathbb{E}\left[ \xi_{i}^{m} \xi_{i'+j}^{m} \right] \right| + \left|\mathbb{E}\left[ \xi_{i}^{m} \xi_{i'+j}^{m} | \mathcal{G}_{i-k} \right] - \mathbb{E}\left[ \xi_{i}^{m} \xi_{i'+j}^{m} \right]\right|  \\
  &\leq&  C j^{-v} \mathbb{E}\left[ \left( \xi_{i'+j}^{m} \right)^2  \right]^{1/2} \mathbb{E}\left[ \left( \xi_{i}^{m} \right)^2  \right]^{1/2} + C \varPsi_{i,2}^{m} k^{-v} \mathbb{E}\left[ \left( \xi_{i'+j}^{m} \right)^4  \right]^{1/4} \mathbb{E}\left[ \left( \xi_{i}^{m} \right)^4  \right]^{1/4} \\
  && + \left| \mathbb{E}\left[\xi_{i}^{m} \right] \mathbb{E}\left[ \xi_{i'+j}^{m}   \right] \right| 
\text{ a.s.}
\end{eqnarray*}
$\blacksquare$

\begin{lemma}\label{chi-property}
  Let $\mathcal{X}_{xy,i,d} = \chi_{x,i} \chi_{y,i+d} - r_{xy}(|d|)$ for any $x,y \in \left\lbrace 1,2 \right\rbrace$.
  Under Assumption \ref{assumption-noise}, for any $i,j,d,w \in \mathbb{N}$, we have
  \begin{enumerate}
    \item [(a)] $ \mathbb{E}[ |\bar{\chi}_{x,i}^m |^{w}  ] \leq C l_m^{-w/2} $, $ \mathbb{E}[ |\bar{\chi}_{x,i}^m |^{w} | \mathcal{K}_{i}^{m}  ] \leq C \varPsi_{i}^{m,2} l_m ^{-w/2}   $, and $\mathbb{E}\left[ \chi_{x,i} \bar{\chi}_{y,i+j}^m | \mathcal{G}_{i-k_m} \right]  \leq C j^{-v} l_m^{-1/2} + C \varPsi_{i}^{m,2} k_m^{-v} l_m ^{-1/2} $ \text{ a.s.};
    \item [(b)] $|\mathbb{E}\[\left( \mathcal{X}_{xy,i,d} \mathcal{X}_{xy,j,d} \right)^w  \]| \leq C_w$;
    \item [(c)] $|\mathbb{E}\[\mathcal{X}_{xy,i,d} \mathcal{X}_{xy,j,d} | \mathcal{G}_{i-k_m} \]| \leq C \left( (j-i-d)^{-v} + \varPsi_{i}^{m,2}  k_m^{-v} \right)  \text{ a.s.} $ if $j-i > d$;
  \end{enumerate}
\end{lemma}

\textbf{Proof of Lemma \ref{chi-property}.}
Consider (a).
By Theorem 1.1 in \citet{shao1995maximal}, the first part of Lemma \ref{chi-property}(a) holds.
For the second part of Lemma \ref{chi-property}(a), using (A.4) in \citet{jacod2019estimating}, we can show that
\begin{eqnarray}\label{unconditional-to-conditional}
  \mathbb{E}\left[ \left( \mathbb{E}\left[ |\bar{\chi}_{x,i}^m |^{w} | \mathcal{K}_{i}^{m}  \right] - \mathbb{E}\left[ |\bar{\chi}_{x,i}^m |^{w} \right] \right)^2  \right] &=&  \mathbb{E}\left[ \mathbb{E}\left[ |\bar{\chi}_{x,i}^m |^{w} -  \mathbb{E}\left[ |\bar{\chi}_{x,i}^m |^{w}   \right] | \mathcal{K}_{i}^{m}  \right]^2  \right] \nonumber\\
  &\leq& C k_m^{-2v} \mathbb{E}\left[ |\bar{\chi}_{x,i}^m |^{2w} \right]
  \text{ a.s.}
\end{eqnarray}
By the first part of Lemma \ref{chi-property}(a) and \eqref{unconditional-to-conditional}, the second part of Lemma \ref{chi-property}(a) holds.
For the third part of the Lemma \ref{chi-property}(a), using Lemma \ref{gmeasurable-bound}(c) with the finiteness of all moments of $\chi_i$, we have
\begin{eqnarray*}
  \mathbb{E}[\chi_{x,i} \bar{\chi}_{y,i+j}^m | \mathcal{G}_{i-k_m} ] &\leq&  C j^{-v} + C \varPsi_{i}^{m,2} k_m^{-v}\text{ a.s.}
\end{eqnarray*}

Consider (b) and (c).
By the finiteness of all moments of $\chi_{i}$, Lemma \ref{chi-property}(b) holds.
Using Lemma \ref{gmeasurable-bound}(c), we can show Lemma \ref{chi-property}(c).
$\blacksquare$

\begin{lemma}\label{negligible-xi}
  Under Assumption 1, we have for any $x,y \in \left\lbrace 1,2 \right\rbrace$
  \begin{enumerate}
    \item[(a)] $\left| \mathbb{E} \left[ \xi_{xy,i}^{m,1} | \mathcal{F}_{i}^{m} \right] \right| \leq C  b_m \Delta_m$ and $\left| \mathbb{E}\[ | \xi_{11,i}^{m,1} |^q \] | \mathcal{F}_{i}^{m} \right| \leq C_q (b_m \Delta_m)^{(q/2) \wedge 1} \text{ a.s.}$ for any $q \in \mathbb{N}$;
    \item[(b)] $\left| \mathbb{E}\[ \xi_{xy,i}^{m,2} | \mathcal{K}_{i}^{m} \]  \right| \leq C  (k_m'^{-(v-1)} + \varPsi_{i}^{m,2}  k'_m l_m ^{-(v+\frac{1}{2} )})$, $E[ \left| \xi_{xy,i}^{m,2} \right|^{2} | \mathcal{K}_{i}^{m} ] \leq C (k_m'^{-2(v-1)} + \varPsi_{i}^{m,2} b_m ^{-1} k_m'^{3} )$, and 
    $ E[ \left| \xi_{xy,i}^{m,2} \right|^{w}  ]  \leq C k_m'^w b_m^{-w/2} l_m^{w/2} \text{ a.s.}$ for any $w\in\{3,4\}$.
  \end{enumerate}
\end{lemma}

\textbf{Proof of Lemma \ref{negligible-xi}.}
Lemma \ref{negligible-xi}(a) is trivial consequence of (B.8) and (B.9) in \citet{chen2018inference}, so we only need to prove (b).
We consider the case $x=y=1$.
Simple algebra shows that
$\tilde{\Gamma}^{m}_{11,i} = \frac{\left( \vartheta_{1,i}^{m} \right)^2 }{k_m} \sum_{m\in\mathbb{Z}} \phi_{d}^{m} r_{11}(m)$.
Then, we can write
\begin{equation*}
  \xi_{11,i}^{m,2} = \frac{1}{(b_m- 2 k_m)\Delta_m k_m^2 \psi_0}  \left(  V_{11,i}^{m} + V_{11,i}^{'m}    \right),
\end{equation*}
where $V_{11,i}^{m} = \mathcal{U}_{11,i}^{m} \sum_{|d|>k_m'} \phi _{d}^{m} r_{11}(d)$, $V_{11,i}^{'m} = \sum_{d=-k_m'}^{k_m'} \phi _{d}^{m} \left\{ r_{11}(|d|) \mathcal{U} _{11,i}^{m} - U _{m,i}^{Y_{1}^{c} Y_{1}^{c}} (|d|) \right\}$, and $\mathcal{U} _{11,i}^{m} = \sum_{l=0}^{b_m - 6l_m} (\vartheta _{1,i+l}^{n} )^2$.
Using $\rho$-mixing property and the facts that $|\mathcal{U} _{11,i}^{m}| \leq C b_m $ and $|\phi _{d}^{m}| \leq C$, we can show that
\begin{equation}\label{V11-bound}
  |V_{11,i}^{m}| \leq C b_m (k_m') ^{-(v-1)} \quad \text{a.s.}
\end{equation}
Let
\begin{eqnarray}\label{micro-autocorrelation-theta}
  T (j,l) _{i}^{m,1} &=& X _{1,i+j}^{c,m} - \bar{X} _{1,i+l}^{c,m} + \chi_{1,i+j} (\vartheta_{1,i+j}^{m} - \vartheta_{1,i}^{m}) - \frac{1}{l_m} \sum_{s=0}^{l_m -1} \chi_{1,i+l+s} (\vartheta_{1,i+l+s}^{m} - \vartheta_{1,i}^{m}) \quad \text{and} \nonumber \\
  T (j,l) _{i}^{m,2} &=& \vartheta _{1,i}^{m} (\chi_{1,i+j} - \bar{\chi} _{1,i+l}^{m}).
\end{eqnarray}
By the finiteness of all moments of $\chi_{i}$ and the Assumption 1, we have for any $q \geq 2$ and $l > j$,
\begin{equation}\label{theta-moment}
\mathbb{E}\left[ | T (j,l) _{i}^{m,1} |^q | \mathcal{K}_{i}^{m} \right] \leq C_q \varPsi_{i}^{m,2} (l-j+ l_m )^{q/2} \Delta_m^{q/2}, \quad \mathbb{E}\left[ | T (j,l) _{i}^{m,2} |^q  | \mathcal{K}_{i}^{m} \right] \leq C_q \varPsi_{i}^{m,2} \text{ a.s.}
\end{equation}
On the other hand, we have
\begin{align}
  U _{m,i}^{Y_{1}^{c}Y_{1}^{c}} (|d|) - r_{11} (|d| ) \mathcal{U} _{11,i}^{m} =& \sum_{l=0}^{b_m -6 l_m} (Y _{1,i+l}^{m} - \bar{Y} _{1,i+l+2 l_m}^{m}) (Y _{1,i+l+d}^{m} - \bar{Y} _{1,i+l+4 l_m}^{m})- r_{11}(|d|) \mathcal{U} _{11,i}^{m} \notag\\
  =& \sum_{l=0}^{b_m -6 l_m } \left( T (0,2 l_m ) _{i+l}^{m,1} + T (0, 2 l_m ) _{i+l}^{m,2} \right)  \notag \\
  & \times \left( T (d,4 l_m ) _{i+l}^{m,1} + T (d, 4 l_m ) _{i+l}^{m,2} \right) - r_{11}(|d|) \mathcal{U} _{11,i}^{m} \notag\\
  =& \mathcal{V}_{11,i}^{m,1}(d) + \mathcal{V}_{11,i}^{m,2}(d) + \mathcal{V}_{11,i}^{m,3}(d) + \mathcal{V}_{11,i}^{m,4}(d), \notag
\end{align}
where
\begin{eqnarray*}\label{mathcalVs}
  \mathcal{V}_{11,i}^{m,1}(d) &=& \sum_{l=0}^{b_m -6 l_m} T (0,2 l_m ) _{i+l}^{m,1} T (d,4 l_m ) _{i+l}^{m,1}, \\
  \mathcal{V}_{11,i}^{m,2}(d) &=&  \sum_{l=0}^{b_m -6 l_m} T (0,2 l_m ) _{i+l}^{m,1} T (d,4 l_m ) _{i+l}^{m,2}, \\
  \mathcal{V}_{11,i}^{m,3}(d) &=& \sum_{l=0}^{b_m -6 l_m} T (0,2 l_m ) _{i+l}^{m,2} T (d,4 l_m ) _{i+l}^{m,1}, \\ 
  \mathcal{V}_{11,i}^{m,4}(d) &=&  \sum_{l=0}^{b_m -6 l_m} T (0,2 l_m ) _{i+l}^{m,2} T (d,4 l_m ) _{i+l}^{m,2}   - r_{11}(|d|) \mathcal{U} _{11,i}^{m}.
\end{eqnarray*}
By \eqref{theta-moment}, we have for any $w \in \mathbb{N}$
\begin{eqnarray}\label{mathV1-w}
  &&\mathbb{E}\left[ \left( \mathcal{V}_{11,i}^{m,1}(d) \right)^w  | \mathcal{K}_{i}^{m} \right] \cr
   &&\leq C_w b_m^{w-1}  \sum_{l=0}^{b_m -6 l_m} \mathbb{E}\left[ \left( T (0,2 l_m ) _{i+l}^{m,1} T (d,4 l_m ) _{i+l}^{m,1} \right)^w  | \mathcal{K}_{i}^{m}  \right]  \nonumber\\
  &&\leq C_w b_m^{w-1}  \sum_{l=0}^{b_m -6 l_m} \mathbb{E}\left[ \left( T (0,2 l_m ) _{i+l}^{m,1}\right)^{2w}  | \mathcal{K}_{i}^{m}  \right]^{1/2} \mathbb{E}\left[ \left( T (d,4 l_m ) _{i+l}^{m,1} \right)^{2w}  | \mathcal{K}_{i}^{m}  \right]^{1/2}\nonumber\\
  &&\leq C_w \varPsi_{i,2}^{m} b_m ^{w} l_m^{w} \Delta_m^{w} \text{ a.s.}
  \end{eqnarray}
where the first and second inequality is due to Jensen's inequality and H\"older's inequality.
Thus, by H\"older's inequality, we have 
\begin{equation}\label{mathV1E}
  \left| \mathbb{E}\left[ \mathcal{V}_{11,i}^{m,1}(d) | \mathcal{K}_{i}^{m} \right] \right| \leq C \varPsi_{i}^{m,2} b_m l_m \Delta_m \text{ a.s.}
\end{equation}
Let $\mathcal{T}(d)_{i,l}^{m,12} = T (0,2 l_m ) _{i+l}^{m,1} T (d,4 l_m ) _{i+l}^{m,2}$ and $\mathcal{T}(d)_{i,l}^{m,21} = T (0,2 l_m ) _{i+l}^{m,2} T (d,4 l_m ) _{i+l}^{m,1}$.
By H\"older's inequality and \eqref{theta-moment}, we have almost surely
\begin{equation}\label{mathT-w}
  \mathbb{E}\left[ \left|\mathcal{T}(d)_{i,l}^{m,12}\right|^{w}   | \mathcal{K}_{i}^{m}  \right] \leq C_w \varPsi_{i}^{m,2} l_m^{w/2} \Delta_m ^{w/2} \quad \text{and} \quad \mathbb{E}\left[ \left|\mathcal{T}(d)_{i,l}^{m,21}\right|^{w}   | \mathcal{K}_{i}^{m}  \right] \leq C_w \varPsi_{i}^{m,2}  l_m^{w/2} \Delta_m ^{w/2}  \text{ a.s.}
\end{equation}
Since the process $\chi$ is independent of the $\sigma$-field $\mathcal{F}_{\infty}$, we obtain
\begin{align*}
  & \quad \left|\mathbb{E}\left[ \mathcal{T}(d)_{i,l}^{m,21}  | \mathcal{K}_{i}^{m}  \right]\right|  \\
  & \leq
  \left|\mathbb{E}\left[ \mathbb{E}\left[  X _{1,i+l+d}^{c,m} - \bar{X} _{1,i+l+4 l_m}^{c,m}  | \mathcal{K}_{i+l+d}^{m}  \right]  \vartheta _{1,i+l}^{m} | \mathcal{K}_{i}^{m}  \right] E\left[   (\chi_{1,i+l} - \bar{\chi} _{1,i+l+2 l_m}^{m}) | \mathcal{K}_{i}^{m}  \right]\right|  \\
  &\quad + \left|\mathbb{E}\left[ \mathbb{E}\left[ \vartheta_{1,i+l+d}^{m} - \vartheta_{1,i+l}^{m} | \mathcal{K}_{i+l}^{m}  \right] \vartheta _{1,i}^{m} | \mathcal{K}_{i}^{m}  \right] E\left[ (\chi_{1,i+j} - \bar{\chi} _{1,i+l}^{m}) \chi_{1,i+l+d} | \mathcal{K}_{i}^{m} \right]\right|  \\
  &\quad + \frac{1}{l_m} \sum_{s=0}^{l_m -1} \left| \mathbb{E}\left[ \mathbb{E}\left[ \vartheta_{1,i+l+4 l_m+s}^{m} - \vartheta_{1,i+l}^{m} | \mathcal{K}_{i+l}^{m}  \right] \vartheta _{1,i}^{m} | \mathcal{K}_{i}^{m}  \right] E\left[  \chi_{1,i+l+4 l_m+s}^{w} (\chi_{1,i+j} - \bar{\chi} _{1,i+l}^{m}) | \mathcal{K}_{i}^{m}  \right]\right|  \\
  &\leq C l_m \Delta_m \left( \left| E\left[   (\chi_{1,i+l} - \bar{\chi} _{1,i+l+2 l_m}^{m}) | \mathcal{K}_{i}^{m}  \right]\right| + \left| E\left[ (\chi_{1,i+j} - \bar{\chi} _{1,i+l}^{m}) \chi_{1,i+l+d} | \mathcal{K}_{i}^{m} \right]\right| \right) \\
  &\quad + C \Delta_m  \sum_{s=0}^{l_m -1} \left|  E\left[  \chi_{1,i+l+4 l_m+s}^{w} (\chi_{1,i+j} - \bar{\chi} _{1,i+l}^{m}) | \mathcal{K}_{i}^{m}  \right]\right|  \\
  & \leq C \varPsi_{i}^{m}  l_m \Delta_m \text{ a.s.} ,
\end{align*}
where the third inequality is due to Lemma \ref{gmeasurable-bound}(a).
Similarly, we can show that
\begin{equation}\label{mathcalT-CE}
  \left|\mathbb{E}\left[ \mathcal{T}(d)_{i,l}^{m,12} | \mathcal{K}_{i}^{m} \right]\right|  \leq C \varPsi_{i}^{m}   l_m \Delta_m   \text{ a.s.}
  \end{equation}
Thus, we have
\begin{equation}\label{mathV23E}
  \left| \mathbb{E}\left[ \mathcal{V}_{11,i}^{m,2}(d) | \mathcal{K}_{i}^{m} \right] \right| \leq C \varPsi_{i}^{m} b_m l_m \Delta_m \quad \text{and} \quad \left| \mathbb{E}\left[ \mathcal{V}_{11,i}^{m,3}(d) | \mathcal{K}_{i}^{m} \right] \right| \leq C \varPsi_{i}^{m} b_m l_m \Delta_m \text{ a.s.}
\end{equation}
Let $\mathcal{V}_{11,i}^{m,2}(d) = \hat{\mathcal{V}}_{11,i}^{m,2}(d) + \bar{\mathcal{V}}_{11,i}^{m,2}(d) + \tilde{\mathcal{V}}_{11,i}^{m,2}(d) $, where
\begin{eqnarray*}
  && \hat{\mathcal{V}}_{11,i}^{m,2}(d) = \sum_{l=0}^{5 k_m - 1} \sum_{j=0}^{\lfloor \frac{(b_m -6 k_m)}{5 k_m}  \rfloor  -1} \bar{\mathcal{T}}(d)_{i,5 k_mj + l}^{m,12},  \\
  && \bar{\mathcal{V}}_{11,i}^{m,2}(d) = \sum_{l=0}^{5 k_m - 1} \sum_{j=0}^{\lfloor \frac{(b_m -6 k_m)}{5 k_m} \rfloor  -1}  \mathbb{E}\left[ \mathcal{T}(d)_{i,5 k_mj + l}^{m,12} | \mathcal{K}_{i+5 k_mj + l}^{m} \right], \\
  && \tilde{\mathcal{V}}_{11,i}^{m,2}(d) = \sum_{l=\lfloor \frac{(b_m -6 k_m)}{5 k_m}  \rfloor 5k_m}^{b_m - 6 k_m}   \mathcal{T}(d)_{i,5 k_mj + l}^{m,12},  \\
  && \bar{\mathcal{T}}(d)_{i,5 k_mj + l}^{m,12} =  \mathcal{T}(d)_{i,5 k_mj + l}^{m,12} - \mathbb{E}\left[ \mathcal{T}(d)_{i,5 k_mj + l}^{m,12} | \mathcal{K}_{i+5 k_mj + l}^{m} \right]
  .
\end{eqnarray*}
For $\hat{\mathcal{V}}_{11,i}^{m,2}(d)$, we have
\begin{eqnarray*}
  \mathbb{E}\left[ \left( \hat{\mathcal{V}}_{11,i}^{m,2}(d) \right)^2 | \mathcal{K}_{i}^{m}  \right] &\leq&  5 l_m  \sum_{l=0}^{5 l_m - 1} \mathbb{E}\left[ \left( \sum_{j=0}^{\lfloor \frac{(b_m -6 l_m)}{5 l_m}  \rfloor} \bar{\mathcal{T}}(d)_{i,5 l_mj + l}^{m,12} \right)^2 | \mathcal{K}_{i}^{m} \right] \\
  &=& 5 l_m \sum_{l=0}^{5 l_m - 1} \sum_{j=0}^{\lfloor \frac{(b_m -6 l_m)}{5 l_m}  \rfloor} \mathbb{E}\left[  \left(  \bar{\mathcal{T}}(d)_{i,5 l_mj + l}^{m,12} \right)^2 | \mathcal{K}_{i}^{m} \right] \\
  && + 10 l_m \sum_{l=0}^{5 l_m - 1}  \sum_{j > j'}^{\lfloor \frac{(b_m -6 l_m)}{5 l_m}  \rfloor} \mathbb{E}\left[  \bar{\mathcal{T}}(d)_{i,5 l_mj' + l}^{m,12} \bar{\mathcal{T}}(d)_{i,5 l_mj + l}^{m,12}| \mathcal{K}_{i}^{m} \right] \\
  &=& 5 l_m \sum_{l=0}^{5 l_m - 1} \sum_{j=0}^{\lfloor \frac{(b_m -6 l_m)}{5 l_m}  \rfloor} \mathbb{E}\left[  \left(  \bar{\mathcal{T}}(d)_{i,5 l_mj + l}^{m,12} \right)^2 | \mathcal{K}_{i}^{m} \right] \\
  &\leq& C \varPsi_{i}^{m,2}   b_m l_m^2 \Delta_m  \text{ a.s.},
\end{eqnarray*}
where the first and second inequality is due to Jensen's inequality and \eqref{mathT-w}, respectively, and the second inequality is due to the orthogonality property of the martingale.
Furthermore, we have for any integer $w > 2$
\begin{eqnarray*}
  \mathbb{E}\left[ \left| \hat{\mathcal{V}}_{11,i}^{m,2}(d) \right|^w  \right] &\leq& \left( 5 l_m \right)^{w-1}   \sum_{l=0}^{5 l_m - 1} \mathbb{E}\left[ \left| \sum_{j=0}^{\lfloor \frac{(b_m -6 l_m)}{5 l_m}  \rfloor} \bar{\mathcal{T}}(d)_{i,5 l_mj + l}^{m,12} \right|^w  \right] \\
  &\leq& C_w \left( 5 l_m \right)^{w-1}   \sum_{l=0}^{5 l_m - 1} \mathbb{E}\left[ \left| \sum_{j=0}^{\lfloor \frac{(b_m -6 l_m)}{5 l_m}  \rfloor} \left( \bar{\mathcal{T}}(d)_{i,5 l_mj + l}^{m,12} \right)^2  \right|^{w/2}  \right] \\
  &\leq& C_w \left( 5 l_m \right)^{w-1} \frac{b_m ^{w/2 -1}}{\left( 5 l_m \right)^{w/2 -1} }  \sum_{l=0}^{5 l_m - 1}   \sum_{j=0}^{\lfloor \frac{(b_m -6 l_m)}{5 l_m}  \rfloor} \mathbb{E}\left[ \left| \bar{\mathcal{T}}(d)_{i,5 l_mj + l}^{m,12} \right|^w    \right] \\
  &\leq& C_w (b_m \Delta_m )^{w/2} l_m^{w},
\end{eqnarray*}
where the first, second, third, and fourth inequalities are due to Jensen's inequality, Burkholder-Davis-Gundy inequality, Jensen's inequality, and \eqref{mathT-w}, respectively.
In case of $\bar{\mathcal{V}}_{11,i}^{m,2}(d)$, we have
\begin{eqnarray*}
  \mathbb{E}\left[ \left| \bar{\mathcal{V}}_{11,i}^{m,2}(d) \right|^w | \mathcal{K}_{i}^{m}  \right] &\leq& C b_m ^{w-1} \sum_{l=0}^{b_m - 6 l_m} \mathbb{E}\left[ \left|\mathbb{E}\left[ \mathcal{T}(d)_{i,5 k_mj + l}^{m,12} | \mathcal{K}_{i+5 k_mj + l}^{m} \right]\right| ^{w} | \mathcal{K}_{i}^{m}  \right] \\
  &\leq& C_w \varPsi_{i}^{m}  b_m ^{w} \left(  l_m \Delta_m \right)^w \text{ a.s.},
\end{eqnarray*}
where the first and second inequalities are due to Jensen's inequality and \eqref{mathcalT-CE}, respectively.
For $\tilde{\mathcal{V}}_{11,i}^{m,2}(d)$, we have
\begin{eqnarray*}
  \mathbb{E}\left[ \left( \tilde{\mathcal{V}}_{11,i}^{m,2}(d) \right)^w | \mathcal{K}_{i}^{m}  \right] &\leq&  (5 l_m)^{w-1} \sum_{l=\lfloor \frac{(b_m -6 l_m)}{5 l_m}  \rfloor 5l_m}^{b_m - 6 l_m} \mathbb{E}\left[ \left( \mathcal{T}(d)_{i,5 l_mj + l}^{m,12} \right)^w | \mathcal{K}_{i}^{m}  \right] \\
  &\leq& C_w \varPsi_{i}^{m,2}  l_m^{3w/2} \Delta_m ^{w/2}
  \text{ a.s.},
\end{eqnarray*}
where the first and second inequalities are due to Jensen's inequality and H\"older's inequality with \eqref{theta-moment}, respectively.
Therefore, we have
\begin{equation*}
  \mathbb{E}\left[ \left( \mathcal{V}_{11,i}^{m,2}(d) \right)^2  | \mathcal{K}_{i}^{m}  \right] \leq C \varPsi_{i}^{m,2} b_m l_m^2 \Delta_m  \text{ a.s.}
\end{equation*}
and
\begin{eqnarray*}
  \mathbb{E}\left[ \left( \mathcal{V}_{11,i}^{m,2}(d) \right)^w  \right] &\leq& C_w \left( \mathbb{E}\left[ \left( \hat{\mathcal{V}}_{11,i}^{m,2}(d) \right)^w  \right] + \mathbb{E}\left[ \left( \bar{\mathcal{V}}_{11,i}^{m,2}(d) \right)^w  \right] + \mathbb{E}\left[ \left( \tilde{\mathcal{V}}_{11,i}^{m,2}(d) \right)^w  \right] \right)  \\
  &\leq& C_w (b_m \Delta_m ) ^{w/2} l_m^{w}
  ,
\end{eqnarray*}
for any $w > 2$.
Similarly, we can show that for any $z \in \left\lbrace 2,3 \right\rbrace$
\begin{equation}\label{mathV23-2w}
  \mathbb{E}\left[ \left( \mathcal{V}_{11,i}^{m,z}(d) \right)^2  | \mathcal{K}_{i}^{m}  \right] \leq C \varPsi_{i}^{m,2} b_m l_m^2 \Delta_m   \text{ a.s.} \quad \text{and} \quad \mathbb{E}\left[ \left( \mathcal{V}_{11,i}^{m,z}(d) \right)^w  \right] \leq C_w (b_m \Delta_m ) ^{w/2} l_m^{w} .
\end{equation}
In case of $\mathcal{V}_{11,i}^{m,4}(d)$, we have
\begin{eqnarray*}
  \mathcal{V}_{11,i}^{m,4}(d) &=&  \sum_{l=0}^{b_m -6 l_m}  (\vartheta _{1,i+l}^{m})^2 \left\{ (\chi_{1,i+l} - \bar{\chi} _{1,i+l+2 l_m }^{m}) (\chi_{1,i+l+d} - \bar{\chi} _{1,i+l+ 4 l_m }^{m})   - r_{11}(|d|) \right\} \\
  &=& \mathcal{S} _{11,i}^{m,1}(d) + \mathcal{S} _{11,i}^{m,2} + \mathcal{S} _{11,i}^{m,3} + \mathcal{S} _{11,i}^{m,4},
\end{eqnarray*}
where
\begin{align*}\label{S-sub}
  & \mathcal{S} _{11,i}^{m,1}(d) = \sum_{l=0}^{b_m -6 l_m}  (\vartheta _{1,i+l}^{m})^2 (\chi_{1,i+l} \chi_{1,i+l+d} - r_{11}(|d|)), \qquad
  \mathcal{S} _{11,i}^{m,2} = \sum_{l=0}^{b_m -6 l_m}  (\vartheta _{1,i+l}^{m})^2 \chi_{1,i+l} \bar{\chi} _{1,i+l+ 4 l_m }^{m}, \\
  &\mathcal{S} _{11,i}^{m,3} = \sum_{l=0}^{b_m -6 l_m}  (\vartheta _{1,i+l}^{m})^2 \chi_{1,i+l} \bar{\chi} _{1,i+l+ 2 l_m }^{m}, \quad \text{and} \quad
  \mathcal{S} _{11,i}^{m,4} = \sum_{l=0}^{b_m -6 l_m}  (\vartheta _{1,i+l}^{m})^2 \bar{\chi} _{1,i+l+ 2 l_m }^{m} \bar{\chi} _{1,i+l+ 4 l_m }^{m}.
\end{align*}
Since the process $\chi$ is independent of the $\sigma$-field of the process $\vartheta$ and $\vartheta$ and all finite moments of $\chi$ are bounded, we have
\begin{eqnarray*}
  \left|\mathbb{E}\left[ \mathcal{S} _{11,i}^{m,1}(d)  | \mathcal{K}_{i}^{m}  \right]\right|  &\leq& C \sum_{l=0}^{b_m -6 l_m} \mathbb{E}\left[  \chi_{1,i+l} \chi_{1,i+l+d} - r_{11}(|d|) | \mathcal{K}_{i}^{m}  \right] \\
  &\leq& C \varPsi_{i}^{m,2}  b_m \Delta_m ^{v/2} \text{ a.s.}
  ,
\end{eqnarray*}
where the last inequality is due to (A.4) of \citet{jacod2019estimating}.
Furthermore, by Lemma \ref{chi-property}(a), we have
\begin{eqnarray*}
  \left|\mathbb{E}\left[ \mathcal{S} _{11,i}^{m,2}(d)  | \mathcal{K}_{i}^{m}  \right]\right|  &\leq&  C \sum_{l=0}^{b_m -6 l_m}  \mathbb{E}\left[ \chi_{1,i+l} \bar{\chi} _{1,i+l+ 4 l_m }^{m} | \mathcal{K}_{i}^{m}  \right] \\
  &&\leq C \varPsi_{i}^{m,2} b_m  l_m ^{-(v+\frac{1}{2} )} \text{ a.s.}
\end{eqnarray*}
Similarly, we can show that
\begin{equation*}
  \left| \mathbb{E}\left[ \mathcal{S} _{11,i}^{m,z}(d) | \mathcal{K}_{i}^{m} \right]  \right| \leq C \varPsi_{i}^{m,2} b_m l_m ^{-(v+\frac{1}{2} )}  \text{ a.s.} \text{ for any }   z \in \{1,2,3,4\}.
\end{equation*}
Thus, we have
\begin{eqnarray}\label{mathV4E}
  \left| \mathbb{E}\left[ \mathcal{V}_{11,i}^{m,4}(d) | \mathcal{K}_{i}^{m} \right] \right| &\leq& \sum _{z=1}^{4} \left| \mathbb{E}\left[ \mathcal{S} _{11,i}^{m,z}  | \mathcal{K}_{i}^{m} \right] \right| \nonumber\\
  &\leq& C \varPsi_{i}^{m,2} b_m l_m ^{-(v+\frac{1}{2} )} \text{ a.s.}
  \end{eqnarray}
By \eqref{V11-bound}, \eqref{mathV1E}, \eqref{mathV23E}, \eqref{mathV4E}, we have
\begin{eqnarray*}
  \left| \mathbb{E}\left[ \xi_{11,i}^{m,2} | \mathcal{K}_{i}^{m} \right] \right| &\leq&  \frac{\left|E  \left[  V_{11,i}^{m} | \mathcal{K}_{i}^{m} \right] \right| + \left| E  \left[ V_{11,i}^{'m} | \mathcal{K}_{i}^{m} \right] \right|}{(b_m- 2 k_m)\Delta_m k_m^2 \psi_0}  , \\
  &\leq& C k_m'^{-(v-1)} + \frac{C \sum_{d=-k'_m}^{k'_m} \sum _{z=1}^{4} \left| \mathbb{E}\left[ \mathcal{V}_{11,i}^{m,z}(d) | \mathcal{K}_{i}^{m} \right] \right|}{(b_m- 2 k_m)\Delta_m k_m^2 \psi_0}\\
  &\leq& C (k_m'^{-(v-1)} + \varPsi_{i}^{m,2} k'_m l_m ^{-(v+\frac{1}{2} )} ) \text{ a.s.}
\end{eqnarray*}
By Lemmas \ref{negligible-xi}(a) and \ref{chi-property}(c) and the boundedness of $\vartheta_{1}$, we have
\begin{eqnarray*}\label{S11}
  \mathbb{E}\left[ \left( \mathcal{S} _{11,i}^{m,1} (d) \right)^2 | \mathcal{K}_{i}^{m}   \right]  &=&  \sum_{l,l' = 0}^{b_m -6 l_m} \mathbb{E}\left[ (\vartheta _{1,i+l}^{m})^2 (\vartheta _{1,i+l'}^{m})^2 | \mathcal{K}_{i}^{m} \right] E\left[ \mathcal{X}_{11,i+l,d} \mathcal{X}_{11,i+l',d}  | \mathcal{K}_{i}^{m} \right] \\
  &\leq& C \varPsi_{i}^{m,2} b_m d\text{ a.s.}
\end{eqnarray*}
Furthermore, we have
\begin{eqnarray}\label{S1-4}
  &&\mathbb{E}\left[ \left( \mathcal{S} _{11,i}^{m,1} (d) \right)^4  \right] \cr
  &&=  \mathbb{E}\[ \left( \sum_{l=0}^{b_m -6 l_m}  (\vartheta _{1,i+l}^{m})^2 \mathcal{X}_{11,i+l,d}  \right)^{4} \]  \nonumber\\
  &&\leq C \sum_{i_1,i_2,i_3,i_4=0}^{b_m-6l_m} \left|\mathbb{E}\[ \mathcal{X}_{11,i+i_1,d} \mathcal{X}_{11,i+i_2,d} \mathcal{X}_{11,i+i_3,d} \mathcal{X}_{11,i+i_4,d}  \] \right|\nonumber\\
  &&= C \sum_{j=0}^{b_m-6l_m} \mathbb{E}\[ \mathcal{X}_{11,i+j,d}^4 \] \nonumber\\
  &&\quad  + C \sum_{(j_1,j_2,j_3) \in \mathcal{J}} \sum_{i=0}^{\substack{b_m-6l_m \nonumber\\ -j_1-j_2-j_3}} \left| \mathbb{E}\left[ \mathcal{X}_{11,i,d} \mathcal{X}_{11,i+j_1,d} \mathcal{X}_{11,i+j_1+j_2,d} \mathcal{X}_{11,i+j_1+j_2+j_3,d} \right] \right|\nonumber\\
  &&\leq C b_m   + C \sum_{(j_1,j_2,j_3) \in \mathcal{J}} \sum_{i=0}^{\substack{b_m-6l_m \nonumber\\ -j_1-j_2-j_3}} (j_1\lor j_2\lor j_3\lor (d+1) - d)^{-v} \nonumber\\
  &&\quad  +  C \sum_{(j_1,j_2,j_3) \in \mathcal{J}} \sum_{i=0}^{\substack{b_m-6l_m \nonumber\\ -j_1-j_2-j_3}} \mathbf{1}\{ j_2 \geq j_1 \lor j_3, j_2 \geq d+1\}\nonumber\\
  && \qquad \qquad \qquad  \times \left| \mathbb{E}\left[ \mathcal{X}_{11,i,d} \mathcal{X}_{11,i+j_1,d} \right] \right| \left| \mathbb{E}\left[ \mathcal{X}_{11,i+j_1+j_2,d} \mathcal{X}_{11,i+j_1+j_2+j_3,d} \right] \right| \nonumber\\
  &&\leq Cb_md^3 + Cb_m^2 d^2  \nonumber\\
  &&\quad + Cb_m \sum_{j_2=d+1}^{b_m - 6 l_m} \sum_{j_1,j_3 = 0}^{j_2}\left| \mathbb{E}\left[ \mathcal{X}_{11,i,d} \mathcal{X}_{11,i+j_1,d} \right] \right| \left| \mathbb{E}\left[ \mathcal{X}_{11,i+j_1+j_2,d} \mathcal{X}_{11,i+j_1+j_2+j_3,d} \right] \right| \nonumber\\
  &&\leq Cb_m^2 d^2 + Cb_m \sum_{j_2=d+1}^{b_m - 6 l_m} \sum_{j_1,j_3 = 0}^{j_2}\left| \mathbb{E}\left[ \mathcal{X}_{11,i,d} \mathcal{X}_{11,i+j_1,d} \right] \right| \left| \mathbb{E}\left[ \mathcal{X}_{11,i+j_1+j_2,d} \mathcal{X}_{11,i+j_1+j_2+j_3,d} \right] \right| \nonumber\\
  &&= Cb_m^2 d^2 + Cb_m \Biggl( \sum_{j_2=d+1}^{b_m - 6 l_m} \sum_{j_1 = 0}^{d} \sum_{j_3 = 0}^{d} \left| \mathbb{E}\left[ \mathcal{X}_{11,0,d} \mathcal{X}_{11,j_1,d} \right] \right| \left| \mathbb{E}\left[ \mathcal{X}_{11,0,d} \mathcal{X}_{11,j_3,d} \right] \right|   \nonumber\\
  &&\quad \quad + \sum_{j_2=d+1}^{b_m - 6 l_m} \sum_{j_1 = d+1}^{j_2} \sum_{j_3 = 0}^{d} \left| \mathbb{E}\left[ \mathcal{X}_{11,0,d} \mathcal{X}_{11,j_1,d} \right] \right| \left| \mathbb{E}\left[ \mathcal{X}_{11,0,d} \mathcal{X}_{11,j_3,d} \right] \right| \nonumber\\
  &&\quad \quad + \sum_{j_2=d+1}^{b_m - 6 l_m} \sum_{j_1 = 0}^{d} \sum_{j_3 = d+1}^{j_2} \left| \mathbb{E}\left[ \mathcal{X}_{11,0,d} \mathcal{X}_{11,j_1,d} \right] \right|\left| \mathbb{E}\left[ \mathcal{X}_{11,0,d} \mathcal{X}_{11,j_3,d} \right] \right| \nonumber\\
  &&\quad \quad + \sum_{j_2=d+1}^{b_m - 6 l_m} \sum_{j_1 = d+1}^{j_2} \sum_{j_3 = d+1}^{j_2} \left| \mathbb{E}\left[ \mathcal{X}_{11,0,d} \mathcal{X}_{11,j_1,d} \right] \right| \left| \mathbb{E}\left[ \mathcal{X}_{11,0,d} \mathcal{X}_{11,j_3,d} \right] \right| \Biggl) \nonumber\\
  &&\leq C b_m ^{2} d^2 
  ,
\end{eqnarray}
where $\mathcal{J} = \left\{ (j_1, j_2 , j_3) \in \mathbb{Z}_{\geq 0}^{3} : 0 < j_1 + j_2 + j_3 \leq b_m - 6l_m \right\}$, the third equality is due to stationarity of $\chi_{1}$, and the first, second, and fifth inequalities are due to the boundedness of $\vartheta_{1}$, (A.3) in \citet{jacod2019estimating}, and Lemma \ref{chi-property}(c), respectively.
By Theorem 1.1 in \citet{shao1995maximal} and the finiteness of all moments of $\chi_{i}$, we have $\mathbb{E}\left[ \left( \bar{\chi}_{i}^{m} \right)^w  \right] \leq C l_m^{-w/2}$ for any $w \in \mathbb{N}$.
Furthermore, using Lemma \ref{gmeasurable-bound}(c), we can show that
\begin{eqnarray*}
  &&\mathbb{E}\left[ \left( \mathcal{S} _{11,i}^{m,2} \right)^2 | \mathcal{K}_{i}^{m} \right]\cr
   &&=  \sum_{l,l'=0}^{b_m -6 l_m} \mathbb{E}\left[  (\vartheta _{1,i+l}^{m})^2 (\vartheta _{1,i+l'}^{m})^2 | \mathcal{K}_{i}^{m} \right] \mathbb{E}\left[ \chi_{1,i+l} \bar{\chi} _{1,i+l+ 4 l_m }^{m} \chi_{1,i+l'} \bar{\chi} _{1,i+l'+ 4 l_m }^{m}  | \mathcal{K}_{i}^{m} \right]\\
  &&\leq C\sum_{l=0}^{b_m -6 l_m}\sum_{l' =0 \lor (l - 6l_m)}^{(l+6 l_m) \land (b_m - 6l_m)} \left|\mathbb{E}\left[ \chi_{1,i+l} \bar{\chi} _{1,i+l+ 4 l_m }^{m} \chi_{1,i+l'} \bar{\chi} _{1,i+l'+ 4 l_m }^{m} | \mathcal{K}_{i}^{m} \right]\right|  \\
  &&\quad + C \sum_{l=0}^{b_m -6 l_m} \sum_{l'=0}^{0 \lor (l - 6 l_m) - 1} \left|\mathbb{E}\left[ \chi_{1,i+l} \bar{\chi} _{1,i+l+ 4 l_m }^{m} \chi_{1,i+l'} \bar{\chi} _{1,i+l'+ 4 l_m }^{m} | \mathcal{K}_{i}^{m} \right]\right| \\
  && \quad+ C \sum_{l=0}^{b_m -6 l_m} \sum_{l'=(l+6 l_m) \land (b_m - 6l_m) + 1}^{b_m - 6 l_m}  \left|\mathbb{E}\left[ \chi_{1,i+l} \bar{\chi} _{1,i+l+ 4 l_m }^{m} \chi_{1,i+l'} \bar{\chi} _{1,i+l'+ 4 l_m }^{m} | \mathcal{K}_{i}^{m} \right]\right|  \\
  &&\leq C\sum_{l=0}^{b_m -6 l_m}\sum_{l' =0 \lor (l - 6l_m)}^{(l+6 l_m) \land (b_m - 6l_m)} \left|\mathbb{E}\left[ \chi_{1,i+l} \bar{\chi} _{1,i+l+ 4 l_m }^{m} \chi_{1,i+l'} \bar{\chi} _{1,i+l'+ 4 l_m }^{m} | \mathcal{K}_{i}^{m} \right]\right|  \\
  &&\quad + C \sum_{l=0}^{b_m -6 l_m} \sum_{l'=0}^{0 \lor (l - 6 l_m) - 1} \Big( \left|\mathbb{E}\left[ \chi_{1,i+l} \bar{\chi} _{1,i+l+ 4 l_m }^{m}   \right] \mathbb{E}\left[ \chi_{1,i+l'} \bar{\chi} _{1,i+l'+ 4 l_m }^{m} \right]\right| \\
  &&\qquad + C l_m^{-v} \mathbb{E}\left[ ( \chi_{1,i+l} \bar{\chi} _{1,i+l+ 4 l_m }^{m} )^2  \right]^{1/2} \mathbb{E}\left[ ( \chi_{1,i+l'} \bar{\chi} _{1,i+l'+ 4 l_m }^{m} )^2  \right]^{1/2} \\
  &&\qquad + C \varPsi_{i}^{m,2}  l_m^{-v} \mathbb{E}\left[ ( \chi_{1,i+l} \bar{\chi} _{1,i+l+ 4 l_m }^{m} )^4  \right]^{1/4} \mathbb{E}\left[ ( \chi_{1,i+l'} \bar{\chi} _{1,i+l'+ 4 l_m }^{m} )^4  \right]^{1/4} \Big) \\
  &&\quad + C \sum_{l=0}^{b_m -6 l_m} \sum_{l'=(l+6 l_m) \land (b_m - 6l_m) + 1}^{b_m - 6 l_m}  \Big( \left|\mathbb{E}\left[ \chi_{1,i+l} \bar{\chi} _{1,i+l+ 4 l_m }^{m}   \right] \mathbb{E}\left[ \chi_{1,i+l'} \bar{\chi} _{1,i+l'+ 4 l_m }^{m}  \right]\right| \\
  &&\qquad + C l_m^{-v} \mathbb{E}\left[ ( \chi_{1,i+l} \bar{\chi} _{1,i+l+ 4 l_m }^{m} )^2  \right]^{1/2} \mathbb{E}\left[ ( \chi_{1,i+l'} \bar{\chi} _{1,i+l'+ 4 l_m }^{m} )^2  \right]^{1/2} \\
  &&\qquad + C \varPsi_{i}^{m,2}  l_m^{-v} \mathbb{E}\left[ ( \chi_{1,i+l} \bar{\chi} _{1,i+l+ 4 l_m }^{m} )^4  \right]^{1/4} \mathbb{E}\left[ ( \chi_{1,i+l'} \bar{\chi} _{1,i+l'+ 4 l_m }^{m} )^4  \right]^{1/4} \Big) \\
  &&\qquad \times \mathbb{E}\left[ \left( \chi_{1,i+l'} \right)^4  | \mathcal{K}_{i}^{m} \right]^{1/4} \mathbb{E}\left[ \left( \bar{\chi} _{1,i+l'+ 4 l_m }^{m} \right)^4  | \mathcal{K}_{i}^{m} \right]^{1/4} \\
  &&\leq C \varPsi_{i}^{m,2}  b_m \text{ a.s.}
  ,
\end{eqnarray*}
where the first and second inequalities are due to the boundedness of $\vartheta_{1}$ and Lemma \ref{gmeasurable-bound}(c), respectively, and the third inequality is due to H\"older's inequality with Lemma \ref{chi-property}(a).
Similarly, we can show that $\mathbb{E}[ ( \mathcal{S} _{11,i}^{m,z} )^2  | \mathcal{K}_{i}^{m}  ] \leq C \varPsi_{i}^{m,2}  b_m$ for any $z \in \{2,3,4\}$.
This implies that
\begin{equation}\label{mathV4-2}
  \mathbb{E}\left[ \left( \mathcal{V}_{11,i}^{m,4}(d) \right)^2 | \mathcal{K}_{i}^{m}   \right] \leq C \varPsi_{i}^{m,2} b_m d \text{ a.s.}
\end{equation}
Furthermore, similar to \eqref{S1-4}, we can show that
\begin{equation}\label{mathS123-w}
  \mathbb{E}\left[ \left( \mathcal{S}_{11,r}^{m,z}(d) \right)^w  \right]  \leq C b_m^{w/2}  l_m^{w/2} \quad \text{for any } w \in \{3,4\} \text{ and } z \in \{2,3,4\}.
\end{equation}
\eqref{S1-4} and \eqref{mathS123-w} imply that
\begin{equation}\label{mathV4-w}
  \norm{\mathcal{V}_{11,i}^{m,4}(d)}_{L_w} \leq C b_m^{1/2} l_m ^{1/2}
\end{equation}
for any $w \in \{3,4\}$.
By \eqref{mathV1-w}, \eqref{mathV23-2w}, and \eqref{mathV4-2}, we have
\begin{eqnarray*}
  \mathbb{E}\left[ \left( V_{11}^{'m} \right)^2 | \mathcal{K}_{i}^{m}   \right] &\leq& k'_m  \sum_{d=-k_m'}^{k_m'} \left( \phi _{d}^{m} \right)^2   \mathbb{E}\left[ \left( r_{11}(|d|) \mathcal{U} _{11,i}^{m} - U _{m,i}^{Y_{1}^{c} Y_{1}^{c}} (|d|) \right)^2  | \mathcal{K}_{i}^{m}  \right] \\
  &\leq& C k'_m \sum_{d=-k_m'}^{k_m'} \sum_{z=1}^{4} \mathbb{E}\left[ \left( \mathcal{V}_{11,i}^{m,z}(d) \right)^2 | \mathcal{K}_{i}^{m}   \right] \\
  &\leq& C \varPsi_{i}^{m,2}  b_m  k_m ^{'3} \text{ a.s.}
  ,
\end{eqnarray*}
where the first and second inequalities are due to Jensen's inequality and the fact that $|\phi _{d}^{m} | \leq C$ for any $d \in \mathbb{Z}$.
Similarly, using \eqref{mathV1-w}, \eqref{mathV23-2w}, and \eqref{mathV4-w} we can show that
\begin{equation*}
  \norm{V_{11}^{'m}}_{L_w} \leq C k_m' b_m^{1/2} l_m^{1/2} \text{ for any } w \in \{3,4\}
  .
\end{equation*}
Thus, we have
\begin{eqnarray*}
  E[ \left| \xi_{11,i}^{m,2} \right|^{w} ] &\leq& \left(\frac{\norm{V_{11,i}^{m}}_{L_w} + \norm{V_{11,i}^{'m}}_{L_w}}{(b_m- 2 k_m)\Delta_m k_m^2 \psi_0} \right)^w \\
  &\leq& C (k_m'^{-w(v-1)} + k_m'^w b_m^{-w/2} l_m^{w/2})\\
  &\leq& C k_m'^w b_m^{-w/2} l_m^{w/2} \text{ for any } w \in \{3,4\} \quad \text{ and }\\
  E[ \left| \xi_{11,i}^{m,2} \right|^{2} | \mathcal{K}_{i}^{m} ] &\leq& C (k_m'^{-2(v-1)} + \varPsi_{i}^{m,2}  b_m ^{-1} k_m'^3) \text{ a.s.}
\end{eqnarray*}
Similarly, we can show the statement for the other cases of $x$ and $y$.
$\blacksquare$

\begin{lemma}\label{lemma:zeta}
  Under Assumption 1, we have for any $x,y \in \left\lbrace 1, 2 \right\rbrace$ and $p \in \mathbb{N} \cap [2,b_m /k_m -2]$,
  \begin{eqnarray}\label{zeta-moment}
    && \left|\mathbb{E}\left[ \zeta(p)_{1x,i}^{m} | \mathcal{K}_{i}^{m} \right]\right| \leq C_w p \left( \Delta_m ^{1/2} + \varPsi_{i}^{m,w} \Delta_m ^{7v/8w - 1/2} \right) \quad \text{for any} \quad w \geq 2, \nonumber\\
    && \mathbb{E}\left[ \left| \zeta(p)_{1x,i}^{m} \right|^w | \mathcal{K}_{i}^{m}  \right] \leq C \varPsi_{i}^{m,2}  p^w   \quad \text{for any} \quad w\in\{1,2,3\}, \nonumber\\
    && \mathbb{E}\left[ \left| \zeta(p)_{1x,i}^{m} \right|^4 | \mathcal{K}_{i}^{m} \right] \leq C \varPsi_{i,2}^{m}  p^w  \Delta_{m}^{v/2 - 2}, \nonumber\\
    && \left| \mathbb{E}\left[  \zeta(p)_{1x,i}^{m} \zeta(p)_{1y,i}^{m}    - \varXi(p)_{x,y,i}^{m} \Big| \mathcal{K}_{i}^{m}  \right] \right| \leq C p^2 \varPsi_{i}^{m,2} \Delta_m^{1/4} \text{ a.s.}
    ,
  \end{eqnarray}
  where $\varXi(p)_{x,y,i}^{m} = 2 (p\Phi_{00}-\bar{\Phi}_{00})  (\bSigma_{11,i}^{m} \bSigma_{xy,i}^{m} + \bSigma_{1x,i}^{m} \bSigma_{1y,i}^{m} )      k_m^4 \Delta_m^2  + 2 (p\Phi_{01}-\bar{\Phi}_{01}) (\bSigma_{11,i}^{m} \bvartheta_{xy,i}^{m} + \bSigma_{1x,i}^{c,m} \bvartheta_{1y,i}^{m} + \bSigma_{1y,i}^{m} \bvartheta_{1x,i}^{m} + \bSigma_{xy,i}^{m} \bvartheta_{11,i}^{m}  ) k_m^2 \Delta_m  + 2 \left( p \Phi_{11} - \bar{\Phi}_{11} \right) (\bvartheta_{11,i}^{m} \bvartheta_{xy,i}^{m} + \bvartheta_{1x,i}^{m} \bvartheta_{1y,i}^{m})$.
  Furthermore, if $p_1, p_2 \in \mathbb{N} \cap [2,b_m /k_m -2]$ and $i_1 + (p_1 +2) k_m \leq i_2$, we have
  \begin{eqnarray}\label{zeta-p1p2}
    && \left|\mathbb{E}\left[ \zeta(p_1)_{1x,i_1}^{m} \zeta(p_2)_{1y,i_2}^{m} | \mathcal{K}_{i_1}^{m} \right]\right|  \leq C \varPsi_{i}^{m,1} p_{1}p_{2} \Delta_m ^{1/2} , \nonumber\\
    && \mathbb{E}\left[ \left( \zeta(p_1)_{1x,i_1}^{m} \right)^2  \left( \zeta(p_2)_{1y,i_2}^{m} \right)^2  | \mathcal{K}_{i_1}^{m} \right] \leq C \varPsi_{i}^{m,1}  p_{1}^{2}p_{2}^{2} \Delta_m ^{v/4 - 1}   \text{ a.s.}  \end{eqnarray}
\end{lemma}
\textbf{Proof of Lemma \ref{lemma:zeta}.}
Consider \eqref{zeta-moment}.
Let $\hat{X}_{1x,i}^{c,n} = \tilde{X}_{1,i}^{c,n} \tilde{X}_{x,i}^{c,n}  - \breve{C}_{1x,i}^{m}$, $\hat{\epsilon}_{1x,i}^{m} = \tilde{\epsilon}_{1,i}^{m} \tilde{\epsilon}_{x,i}^{m} - \tilde{\Gamma}^{m}_{1x,i}$, and $\hat{X^{m}\epsilon}^{m}_{xy,i} = \tilde{X}_{x,i}^{c,m} \tilde{\epsilon}_{y,i}^{m} $.
Then, we have $ \zeta(p)_{1x,i}^{m} = \sum_{l=i}^{i+pk_m-1}  \zeta_{1x,l}^{m} = \sum_{l=i}^{i+pk_m-1} \hat{X}_{1x,l}^{c,m} + \hat{\epsilon}_{1x,l}^{m} + \hat{X^{c}\epsilon}^{m}_{1x,l} + \hat{X^{c}\epsilon}^{m}_{x1,l} $.
Similar to the proof of (A.29) of \citet{jacod2019estimating}, using Lemma \ref{gmeasurable-bound}(b), we can show that
\begin{equation}\label{decomped-zeta-bound}
  \left|\mathbb{E}\left[ \hat{\epsilon}_{1x,i}^{m} | \mathcal{K}_{i}^{m} \right]\right| \leq C \left(  \varPsi_{i}^{m,w} \Delta_m ^{7v/8w} +  \Delta_m  \right) \quad \text{and} \quad \left| \mathbb{E}\left[ \hat{X^{c}\epsilon}^{m}_{xy,l}  | \mathcal{K}_{i}^{m} \right] \right| \leq C \varPsi_{i}^{m,w} \Delta_m ^{7v/8w + 1/4} \text{ a.s.}
\end{equation}
for any integer $w \geq 3$.
Then, using Lemma A.2 of \citet{jacod2019estimating} and \eqref{decomped-zeta-bound}, we have for any integer $w \geq 3$,
\begin{eqnarray*}
  \left|\mathbb{E}\left[ \zeta(p)_{1x,i}^{m} | \mathcal{K}_{i}^{m} \right]\right| &\leq& \left|\mathbb{E}\left[ \sum_{l=i}^{i+pk_m-1} \hat{X}_{1x,l}^{c,m} + \hat{\epsilon}_{1x,l}^{m} + \hat{X^{c}\epsilon}^{m}_{1x,l} + \hat{X^{c}\epsilon}^{m}_{x1,l} | \mathcal{K}_{i}^{m} \right]\right| \\
  &\leq& C p \left( \Delta_m ^{1/2} + \varPsi_{i}^{m,w} \Delta_m ^{7v/8w - 1/2} \right) 
  \text{ a.s.}
\end{eqnarray*}
The second and third part of \eqref{zeta-moment} are trivial consequence of Lemma A.5 of \citet{jacod2019estimating}.
For the fourth part \eqref{zeta-moment}, we have
\begin{eqnarray*}
  && \left| \mathbb{E}\left[ \left(  \zeta(p)_{11,i}^{m} \right)^2  - \varXi\left(  p \right)_{1,1,i}^{m} | \mathcal{K}_{i}^{m}  \right] \right| \\
  && \leq \left| \mathbb{E}\left[ \left(  \zeta(p)_{11,i}^{m} \right)^2 - 4 (\bSigma_{11,i}^{c,m})^2 \rho(p,1)_{i}^{m} - 4 \bSigma_{11,i}^{c,m} (\vartheta _{1,i}^{m} )^2 \rho(p,3)_{i}^{m} - (\vartheta _{1,i}^{m}) ^4 \rho(p,2)_{i}^{m}  | \mathcal{K}_{i}^{m}   \right] \right| \\
  &&\quad + \left| \mathbb{E}\left[ 4 \left(\bSigma_{11,i}^{c,m}\right)^2 \rho(p,1)_{i}^{m}  - 4 \left( \bSigma_{11,i}^c \right)^{2}   k_m^4 \Delta_m^2\left(p\Phi_{00}-\bar{\Phi}_{00}\right) \right] \right| \\
  &&\quad + \left| \mathbb{E}\left[ 4 \bSigma_{11,i}^{c,m} \left(\vartheta _{1,i}^{m} \right)^2 \rho(p,3)_{i}^{m}  -  8 \bSigma_{11,i}^c  \left(  \vartheta_{11,i}^{m} \right)^2 R k_m^2 \Delta_m \left(p\Phi_{01}-\bar{\Phi}_{01}\right) \right] \right| \\
  &&\quad + \left| \mathbb{E}\left[ (\vartheta _{1,i}^{m}) ^4 \rho(p,2)_{i}^{m} -  4 \left( \vartheta_{11,i}^{m} \right)^{4} R^2 \left( p \Phi_{11} - \bar{\Phi}_{11} \right)  \right] \right| \\
  && \leq C p^2 \varPsi_{i,p}^{m,2} \Delta_m^{1/4}
  \text{ a.s.}
\end{eqnarray*}
where the second inequality is due to Lemmas A.8--10 of \citet{jacod2019estimating} and the locally bounded $\bSigma _{11,i}^{c,m}$ and $\vartheta _{1,i}^{m}$.
Similarly, we can show the statement for the other cases of $x$ and $y$.

Consider \eqref{zeta-p1p2}.
Now, we only consider the case $x=y=1$, since we can similarly show the other cases.
Simple algebra shows that
\begin{eqnarray}\label{zetazeta-decomp}
  \mathbb{E}\left[ \zeta(p_1)_{11,i_1}^{m} \zeta(p_2)_{11,i_2}^{m} | \mathcal{K}_{i_1}^{m} \right] &=&  \sum_{l_1=i_1}^{i_1+p_1 k_m-1}  \sum_{l_2=i_2}^{i_2+p_2 k_m-1}  \mathbb{E}\left[ \hat{X}_{11,l_2}^{c,m} \hat{X}_{11,l_1}^{c,m} | \mathcal{K}_{i_1}^{m} \right] + \mathbb{E}\left[ \hat{\epsilon}_{11,l_2}^{m} \hat{X}_{11,l_1}^{c,m} | \mathcal{K}_{i_1}^{m} \right] \nonumber\\
  && + 2 \mathbb{E}\left[  \tilde{X}_{1,l_2}^{c,m} \tilde{\epsilon}_{1,l_2}^{m} \hat{X}_{11,l_1}^{c,m} | \mathcal{K}_{i_1}^{m} \right] + \mathbb{E}\left[ \hat{X}_{11,l_2}^{c,m} \hat{\epsilon}_{11,l_1}^{m} | \mathcal{K}_{i_1}^{m} \right] \nonumber \\
  && + \mathbb{E}\left[ \hat{\epsilon}_{11,l_2}^{m} \hat{\epsilon}_{11,l_1}^{m} | \mathcal{K}_{i_1}^{m} \right] + 2 \mathbb{E}\left[  \tilde{X}_{1,l_2}^{c,m} \tilde{\epsilon}_{1,l_2}^{m} \hat{\epsilon}_{11,l_1}^{m} | \mathcal{K}_{i_1}^{m} \right] \nonumber \\
  && + 2 \mathbb{E}\left[ \hat{X}_{11,l_2}^{c,m} \tilde{X}_{1,l_1}^{c,m} \tilde{\epsilon}_{1,l_1}^{m} | \mathcal{K}_{i_1}^{m} \right] + 2 \mathbb{E}\left[ \hat{\epsilon}_{11,l_2}^{m} \tilde{X}_{1,l_1}^{c,m} \tilde{\epsilon}_{1,l_1}^{m} | \mathcal{K}_{i_1}^{m} \right] \nonumber \\
  && + 4 \mathbb{E}\left[  \tilde{X}_{1,i_2}^{c,m} \tilde{\epsilon}_{1,l_2}^{m} \tilde{X}_{1,l_1}^{c,m} \tilde{\epsilon}_{1,l_1}^{m} | \mathcal{K}_{i_1}^{m} \right]
  \text{ a.s.}
\end{eqnarray}
For the first term of the summand on the right hand side of \eqref{zetazeta-decomp}, we have 
\begin{eqnarray*}
  \left|\mathbb{E}\left[ \hat{X}_{11,l_2}^{c,m} \hat{X}_{11,l_1}^{c,m} | \mathcal{K}_{i_1}^{m} \right]\right|  &=& \left|\mathbb{E}\left[ \mathbb{E}\left[ \hat{X}_{11,l_2}^{c,m} | \mathcal{K}_{l_2}^{m} \right] \hat{X}_{11,l_1}^{c,m} | \mathcal{K}_{i_1}^{m} \right]\right| \\
  &\leq& C \Delta_m \mathbb{E}\left[ \left|\hat{X}_{11,l_1}^{c,m}\right|  | \mathcal{K}_{i_1}^{m} \right] \\
  &\leq& C \Delta_m ^{3/2}
  \text{ a.s.}
\end{eqnarray*}
where the equality is due to tower property and the first and second inequalities are due to Lemma A.2 of \citet{jacod2019estimating}.
For the second term of the summand on the right hand side of \eqref{zetazeta-decomp}, we have
\begin{eqnarray*}
  \left|\mathbb{E}\left[ \hat{\epsilon}_{11,l_2}^{m} \hat{X}_{11,l_1}^{c,m} | \mathcal{K}_{i_1}^{m} \right]\right| &=&  \left| \mathbb{E}\left[  \mathbb{E}\left[ \hat{\epsilon}_{11,l_2}^{m} | \mathcal{K}_{l_2}^{m} \right] \hat{X}_{11,l_1}^{c,m} | \mathcal{K}_{i_1}^{m} \right]\right| \\
  &\leq&   \left| \mathbb{E}\left[  C \varPsi_{i}^{m,2} \Delta_m  \left|\hat{X}_{11,l_1}^{c,m}\right|  | \mathcal{K}_{i_1}^{m} \right]\right| \\
  &=&  C \varPsi_{i}^{m,2} \Delta_m   \left| \mathbb{E}\left[  \left|\hat{X}_{11,l_1}^{c,m}\right|  | \mathcal{K}_{i_1}^{m} \right]\right| \\
  &\leq& C \varPsi_{i}^{m,2} \Delta_m ^{3/2}
  \text{ a.s.}
\end{eqnarray*}
where the first and second equalities are due to tower property and the independence of $\varPsi_{i,2}^{m}$ and $\hat{X}_{11,l_1}^{c,m}$, and the first and second inequalities are due to Lemmas A.2 and A.5 of \citet{jacod2019estimating}.
Similarly, we can obtain the following inequalities:
\begin{eqnarray*}
  && \left|\mathbb{E}\left[  \tilde{X}_{1,l_2}^{c,m} \tilde{\epsilon}_{1,l_2}^{m} \hat{X}_{11,l_1}^{c,m} | \mathcal{K}_{i_1}^{m} \right]\right| \leq C \varPsi_{i}^{m,2} \Delta_m ^{3/4 + v/2}, \quad \left|\mathbb{E}\left[ \hat{X}_{11,l_2}^{c,m} \hat{\epsilon}_{11,l_1}^{m} | \mathcal{K}_{i_1}^{m} \right]\right| \leq C \varPsi_{i}^{m,2} \Delta_m ^{3/2},   \nonumber \\
  && \left|\mathbb{E}\left[ \hat{\epsilon}_{11,l_2}^{m} \hat{\epsilon}_{11,l_1}^{m} | \mathcal{K}_{i_1}^{m} \right]\right|  \leq C \varPsi_{i}^{m,1} \Delta_m ^{3/2}, \quad  \left|\mathbb{E}\left[  \tilde{X}_{1,l_2}^{c,m} \tilde{\epsilon}_{1,l_2}^{m} \hat{\epsilon}_{11,l_1}^{m} | \mathcal{K}_{i_1}^{m} \right]\right|  \leq C \varPsi_{i}^{m,1} \Delta_m ^{3/4 + v/2}, \nonumber \\
  && \left|\mathbb{E}\left[ \hat{X}_{11,l_2}^{c,m} \tilde{X}_{1,l_1}^{c,m} \tilde{\epsilon}_{1,l_1}^{m} | \mathcal{K}_{i_1}^{m} \right]\right|  \leq C \varPsi_{i}^{m,1} \Delta_m ^{3/2}, \quad  \left|\mathbb{E}\left[ \hat{\epsilon}_{11,l_2}^{m} \tilde{X}_{1,l_1}^{c,m} \tilde{\epsilon}_{1,l_1}^{m} | \mathcal{K}_{i_1}^{m} \right]\right|  \leq C \varPsi_{i}^{m,1} \Delta_m ^{3/2}, \nonumber \\
  && \left|\mathbb{E}\left[  \tilde{X}_{1,i_2}^{c,m} \tilde{\epsilon}_{1,l_2}^{m} \tilde{X}_{1,l_1}^{c,m} \tilde{\epsilon}_{1,l_1}^{m} | \mathcal{K}_{i_1}^{m} \right]\right|  \leq C \varPsi_{i}^{m,1} \Delta_m ^{3/4 + v/2} \text{ a.s.}
\end{eqnarray*}
Thus, in view of \eqref{zetazeta-decomp}, we have
\begin{equation*}
  \mathbb{E}\left[ \zeta(p_1)_{11,i_1}^{m} \zeta(p_2)_{11,i_2}^{m} | \mathcal{K}_{i_1}^{m} \right] \leq C \varPsi_{i}^{m,1} p_1 p_2 \Delta_m ^{1/2} \text{ a.s.}
\end{equation*}
For the second part of \eqref{zeta-p1p2}, we have
\begin{eqnarray}\label{zetazeta2-decomp}
  && \mathbb{E}\left[ \left( \zeta(p_1)_{11,i_1}^{m} \zeta(p_2)_{11,i_2}^{m} \right)^2  | \mathcal{K}_{i_1}^{m} \right] \nonumber\\
  && \leq C p_1 p_2 k_m^2  \sum_{l_1=i_1}^{i_1+p_1 k_m-1}  \sum_{l_2=i_2}^{i_2+p_2 k_m-1}  \mathbb{E}\left[ \left( \hat{X}_{11,l_2}^{c,m} \hat{X}_{11,l_1}^{c,m} \right)^2  | \mathcal{K}_{i_1}^{m} \right] + \mathbb{E}\left[ \left( \hat{\epsilon}_{11,l_2}^{m} \hat{X}_{11,l_1}^{c,m} \right)^2  | \mathcal{K}_{i_1}^{m} \right] \nonumber\\
  && \qquad \qquad \qquad \qquad + \mathbb{E}\left[  \left( \tilde{X}_{1,l_2}^{c,m} \tilde{\epsilon}_{1,l_2}^{m} \hat{X}_{11,l_1}^{c,m} \right)^2  | \mathcal{K}_{i_1}^{m} \right] + \mathbb{E}\left[ \left( \hat{X}_{11,l_2}^{c,m} \hat{\epsilon}_{11,l_1}^{m} \right)^2  | \mathcal{K}_{i_1}^{m} \right] \nonumber \\
  && \qquad \qquad \qquad \qquad + \mathbb{E}\left[ \left( \hat{\epsilon}_{11,l_2}^{m} \hat{\epsilon}_{11,l_1}^{m} \right)^2  | \mathcal{K}_{i_1}^{m} \right] + \mathbb{E}\left[  \left( \tilde{X}_{1,l_2}^{c,m} \tilde{\epsilon}_{1,l_2}^{m} \hat{\epsilon}_{11,l_1}^{m} \right)^2  | \mathcal{K}_{i_1}^{m} \right] \nonumber \\
  && \qquad \qquad \qquad \qquad + \mathbb{E}\left[ \left( \hat{X}_{11,l_2}^{c,m} \tilde{X}_{1,l_1}^{c,m} \tilde{\epsilon}_{1,l_1}^{m} \right)^2  | \mathcal{K}_{i_1}^{m} \right] + \mathbb{E}\left[ \left( \hat{\epsilon}_{11,l_2}^{m} \tilde{X}_{1,l_1}^{c,m} \tilde{\epsilon}_{1,l_1}^{m} \right)^2  | \mathcal{K}_{i_1}^{m} \right] \nonumber \\
  && \qquad \qquad \qquad \qquad + \mathbb{E}\left[  \left( \tilde{X}_{1,i_2}^{c,m} \tilde{\epsilon}_{1,l_2}^{m} \tilde{X}_{1,l_1}^{c,m} \tilde{\epsilon}_{1,l_1}^{m} \right)^2  | \mathcal{K}_{i_1}^{m} \right] \text{ a.s.} ,
\end{eqnarray}
where the inequality is due to Jensen's inequality.
Similar to the first part of \eqref{zeta-p1p2}, we can obtain the follwing inequalities:
\begin{eqnarray*}
  && \left|\mathbb{E}\left[ \left( \hat{X}_{11,l_2}^{c,m} \hat{X}_{11,l_1}^{c,m} \right)^2  | \mathcal{K}_{i_1}^{m} \right]\right|  \leq C \Delta_m ^{2}, \quad \left|\mathbb{E}\left[ \left( \hat{\epsilon}_{11,l_2}^{m} \hat{X}_{11,l_1}^{c,m} \right)^2  | \mathcal{K}_{i_1}^{m} \right]\right|  \leq C \varPsi_{i}^{m,2} \Delta_m ^{2},   \nonumber \\
  && \left|\mathbb{E}\left[  \left( \tilde{X}_{1,l_2}^{c,m} \tilde{\epsilon}_{1,l_2}^{m} \hat{X}_{11,l_1}^{c,m} \right)^2  | \mathcal{K}_{i_1}^{m} \right]\right| \leq C \varPsi_{i}^{m,2} \Delta_m ^{2}, \quad \left|\mathbb{E}\left[ \left( \hat{X}_{11,l_2}^{c,m} \hat{\epsilon}_{11,l_1}^{m} \right)^2  | \mathcal{K}_{i_1}^{m} \right]\right| \leq C \varPsi_{i}^{m,2} \Delta_m ^{2}  , \nonumber \\
  && \left|\mathbb{E}\left[ \left( \hat{\epsilon}_{11,l_2}^{m} \hat{\epsilon}_{11,l_1}^{m} \right)^2  | \mathcal{K}_{i_1}^{m} \right]\right|  \leq C \varPsi_{i}^{m,1} \Delta_m ^{1+v/4}, \quad  \left|\mathbb{E}\left[  \left( \tilde{X}_{1,l_2}^{c,m} \tilde{\epsilon}_{1,l_2}^{m} \hat{\epsilon}_{11,l_1}^{m} \right)^2  | \mathcal{K}_{i_1}^{m} \right]\right|  \leq C \varPsi_{i}^{m,1} \Delta_m ^{1+v/4}, \nonumber \\
  && \left|\mathbb{E}\left[ \left( \hat{X}_{11,l_2}^{c,m} \tilde{X}_{1,l_1}^{c,m} \tilde{\epsilon}_{1,l_1}^{m} \right)^2  | \mathcal{K}_{i_1}^{m} \right]\right|  \leq C \varPsi_{i}^{m,1} \Delta_m ^{2}, \quad  \left|\mathbb{E}\left[ \left( \hat{\epsilon}_{11,l_2}^{m} \tilde{X}_{1,l_1}^{c,m} \tilde{\epsilon}_{1,l_1}^{m} \right)^2  | \mathcal{K}_{i_1}^{m} \right]\right|  \leq C \varPsi_{i}^{m,1} \Delta_m ^{2}, \nonumber \\
  && \left|\mathbb{E}\left[ \left( \tilde{X}_{1,i_2}^{c,m} \tilde{\epsilon}_{1,l_2}^{m} \tilde{X}_{1,l_1}^{c,m} \tilde{\epsilon}_{1,l_1}^{m} \right)^2   | \mathcal{K}_{i_1}^{m} \right]\right|  \leq C \varPsi_{i}^{m,2} \Delta_m ^{2} \text{ a.s.}
\end{eqnarray*}
Thus, we have $\mathbb{E}\left[ \left( \zeta(p_1)_{1x,i_1}^{m} \right)^2  \left( \zeta(p_2)_{1y,i_2}^{m} \right)^2  | \mathcal{K}_{i_1}^{m} \right] \leq C \varPsi_{i}^{m,1} p_{1}^{2}p_{2}^{2} \Delta_m ^{v/4 - 1} $ a.s.
$\blacksquare$

\begin{lemma}\label{lemma:M}
  Under Assumption 1, we have for any $x \in \left\lbrace 1, 2 \right\rbrace$ and $p \in \mathbb{N} \cap [2,b_m /k_m -2]$,
  \begin{align}\label{M-moment}
    & \left|\mathbb{E}\left[ M(p)_{1x,i}^{m} | \mathcal{K}_{i}^{m}  \right]\right| \leq C \varPsi_{i}^{m,2}  \Delta_m^{1/2}, \quad \left|\mathbb{E}\left[ M'(p)_{1x,i}^{m} | \mathcal{K}_{i}^{m}  \right]\right| \leq C \varPsi_{i}^{m,2}  p^{-1} \Delta_m^{1/2}, \nonumber\\
    & \mathbb{E}\left[ \left| M(p)_{1x,i}^{m} \right|^{2} | \mathcal{K}_{i}^{m}   \right] \leq C \varPsi_{i}^{m,2}  p b_m ^{-1} \Delta_m ^{-1/2}, \quad \mathbb{E}\left[ \left| M(p)_{1x,i}^{m} \right|^{3}   \right] \leq C \varPsi_{i}^{m,2}  p^{3/2} b_m ^{-3/2} \Delta_m ^{-3/4}, \nonumber\\
    & \mathbb{E}\left[ \left| M(p)_{1x,i}^{m} \right|^{4}   \right] \leq C p^2 b_m ^{-2} \Delta_m ^{v/4 - 2} + C p^{3} b_m ^{-3} \Delta_m ^{v/2-7/2}, \nonumber\\
    & \mathbb{E}\left[ \left| M'(p)_{1x,i}^{m} \right|^{2} | \mathcal{K}_{i}^{m}   \right] \leq C \varPsi_{i}^{m,1} p^{-1} b_m^{-1} \Delta_m^{-1/2}, \quad \mathbb{E}\left[ \left| M'(p)_{1x,i}^{m} \right|^{3}   \right] \leq C  p^{-3/2} b_m^{-3/2} \Delta_m^{-3/4}, \nonumber\\
    & \mathbb{E}\left[ \left| M'(p)_{1x,i}^{m} \right|^{4}    \right] \leq  C   \left( p^{-2} b_m ^{-2} \Delta_{m}^{-1} + p^{-1} b_m ^{-3} \Delta_{m} ^{v/2 - 7/2} \right) \text{ a.s.}
  \end{align}
  Furthermore, we have
  \begin{align}\label{barM-moment}
    & \left|\mathbb{E}\left[ \bar{M}(p)_{1x,i}^{m} | \mathcal{K}_{i}^{m} \right]\right| \leq C \varPsi_{i}^{m,2}  \Delta_m ^{1/2}, \quad \left|\mathbb{E}\left[ \bar{M}'(p)_{1x,i}^{m} | \mathcal{K}_{i}^{m} \right]\right| \leq C \varPsi_{i}^{m,2}  p^{-1} \Delta_m ^{1/2}, \nonumber\\
    & \mathbb{E}\left[ \left|\bar{M}(p)_{1x,i}^{m}\right| ^{2} | \mathcal{K}_{i}^{m} \right] \leq C \varPsi_{i}^{m,1}  \Delta_m, \quad \mathbb{E}\left[ \left|\bar{M}'(p)_{1x,i}^{m}\right| ^{2} | \mathcal{K}_{i}^{m} \right] \leq C \varPsi_{i}^{m,1} p^{-2} \Delta_m \text{ a.s.}, \nonumber\\
    & \mathbb{E}\left[ \left|\bar{M}(p)_{1x,i}^{m}\right| ^{w}  \right] \leq C  \Delta_m^{3v/4 - (w-1)/2}, \quad \mathbb{E}\left[ \left|\bar{M}'(p)_{1x,i}^{m}\right| ^{w}  \right] \leq C  p^{-w} \Delta_m^{3v/4 - (w-1)/2}
    ,
  \end{align}
  for any $w \geq 3$.
\end{lemma}
\textbf{Proof of Lemma \ref{lemma:M}.}
We only consider the case $x=1$, since we can similarly show the other case.
For simplicity, we denote $i_{l,j}^{m} = i + (p_m+2) k_m l + j$.
We have
\begin{eqnarray*}
  \left|\mathbb{E}\left[M(p)_{11,i}^{m} | \mathcal{K}_{i}^{m}  \right]\right|  &=&  \left|\frac{1}{(b_m - 2 k_m)\Delta_m k_m \psi_0} \sum_{l=0}^{L(m,p)-1} \mathbb{E}\left[ \mathbb{E}\left[ \zeta(p)^m_{11,i_{l,0}^{m}} | \mathcal{K}_{i_{l,0}^{m}}^{m}  \right] | \mathcal{K}_{i}^{m} \right]\right| \\
  &\leq& C b_m ^{-1} p L(m,p) \left( \frac{1}{L(m,p)} \sum_{l=0}^{L(m,p)-1} \varPsi_{i,l}^{m,2} \right) \\
  &\leq& C \varPsi_{i}^{m,2} \Delta_m^{1/2} \text{ a.s.},
\end{eqnarray*}
where the first inequality is due to the first part of Lemma \ref{lemma:zeta}.
Similarly, we can show the second part of \eqref{M-moment}.
For the third part of \eqref{M-moment}, we have
\begin{eqnarray*}
  \mathbb{E}\left[ \left|M(p)_{11,i}^{m}\right|^{2}  | \mathcal{K}_{i}^{m}  \right] &\leq& C b_m ^{-2} \Delta_m ^{-1} \sum_{l=0}^{L(m,p)-1} \mathbb{E}\left[ \left( \zeta(p)_{11,i_{l,0}^{m}}^{m} \right)^2 | \mathcal{K}_{i}^{m}  \right] \\
  && + C b_m ^{-2} \Delta_m ^{-1} \sum_{l \neq l'}^{L(m,p)-1} \mathbb{E}\left[ \zeta(p)_{11,i_{l,0}^{m}}^{m} \zeta(p)_{11,i_{l',0}^{m}}^{m}  | \mathcal{K}_{i}^{m}  \right] \\
  &\leq& C \varPsi_{i}^{m,1} p b_m ^{-1} \Delta_m ^{-1/2} \text{ a.s.},
\end{eqnarray*}
where the second inequality is due to Lemma \ref{lemma:zeta}.
By Burkholder-Davis-Gundy inequality, we have
\begin{eqnarray*}
  \mathbb{E}\left[ \left| M(p)_{11,i}^{m} \right|^{3}  \right] &\leq& C \mathbb{E}\left[ \left( \sum_{j=0}^{L(m,p) - 1} \left(  \eta(p)_{11,j}^{m,i} - \bar{\eta}(p)_{11,j}^{m,i} \right)^2  \right)^{3/2} \right] \\
  &\leq& C L(m,p)^{1/2} \mathbb{E}\left[ \sum_{j=0}^{L(m,p) - 1} \left|  \eta(p)_{11,j}^{m,i} \right|^3 + \sum_{j=0}^{L(m,p) - 1} \left|  \bar{\eta}(p)_{11,j}^{m,i} \right|^3 \right] \\
  &\leq& C p^{3/2} b_m ^{-3/2} \Delta_m ^{-3/4}
  ,
\end{eqnarray*}
where the second and third inequalities are due to Jensen's inequality and Lemma \ref{lemma:zeta}.
By Burkholder-Davis-Gundy inequality, we have
\begin{eqnarray}\label{M4}
  \mathbb{E}\left[ \left| M(p)_{11,i}^{m} \right|^{4}  \right] &\leq& C \mathbb{E}\left[ \left(  \sum_{j=0}^{L(m,p) - 1} \left(  \eta(p)_{11,j}^{m,i} - \bar{\eta}(p)_{11,j}^{m,i} \right)^2  \right)^2 \right]  \cr
  &\leq&  C \mathbb{E}\left[ \left( \sum_{j=0}^{L(m,p) - 1} \left(  \eta(p)_{11,j}^{m,i} \right) ^2 \right)^2 \right] +  \mathbb{E}\left[\left( \sum_{l=0}^{L(m,p) - 1} \left( \bar{\eta}(p)_{11,j}^{m,i} \right)^2  \right)^2 \right]
  ,
\end{eqnarray}
where the second inequality is due to Jensen's inequality.
For the first term on the right hand side of \eqref{M4}, we have
\begin{eqnarray*}
  \mathbb{E}\left[ \left( \sum_{j=0}^{L(m,p) - 1} \left(  \eta(p)_{11,j}^{m,i} \right) ^2 \right)^2 \right] &=&  \sum_{j=0}^{L(m,p) - 1} \mathbb{E}\left[ \left(  \eta(p)_{11,j}^{m,i} \right) ^4  \right] + \sum_{j \neq j'}^{L(m,p) - 1} \mathbb{E}\left[ \left(  \eta(p)_{11,j}^{m,i} \right)^2 \left( \eta(p)_{11,j'}^{m,i} \right) ^2  \right] \\
  &\leq& C p^2 b_m ^{-2} \Delta_m ^{v/4 - 2} + C p^{3} b_m ^{-3} \Delta_m ^{v/2-7/2}
  ,
\end{eqnarray*}
where the inequality is due to Lemma \ref{lemma:zeta}.
For the first term on the right hand side of \eqref{M4}, we have
\begin{eqnarray*}
  \mathbb{E}\left[\left( \sum_{j=0}^{L(m,p) - 1} \left( \bar{\eta}(p)_{11,j}^{m} \right)^2  \right)^2 \right] &\leq& L(m,p) \sum_{j=0}^{L(m,p) - 1}  \mathbb{E}\left[ \left( \bar{\eta}(p)_{11,j}^{m,i} \right)^4 \right] \\
  &\leq& C p^2 b_m ^{-2} \left( \Delta_m + \Delta_m ^{7v/8 - 3} \right)  \\
  &\leq& C p^2 b_m ^{-2} \Delta_m ^{v/4 - 2} + C p^{3} b_m ^{-3} \Delta_m ^{v/2-7/2}
  ,
\end{eqnarray*}
where the first and second inequalities are due to Jensen's inequality and Lemma \ref{lemma:zeta}, respectively.
Thus, we have $\mathbb{E}\left[ | M(p)_{11,i}^{m} |^{4}  \right] \leq C p^2 b_m ^{-2} \Delta_m ^{v/4 - 2} + C p^{3} b_m ^{-3} \Delta_m ^{v/2-7/2}$.
Similarly, we can show the rest parts of \eqref{M-moment}.

Consider \eqref{barM-moment}.
By Lemma \ref{lemma:zeta}, we have $\left|\mathbb{E}\left[ \bar{M}(p)_{1x,i}^{m} | \mathcal{K}_{i}^{m} \right]\right| \leq C \varPsi_{i}^{m,2}  \Delta_m ^{1/2}$ a.s.
Using Jensen's inequality and Lemma \ref{lemma:zeta}, we have
\begin{eqnarray*}
  \mathbb{E}\left[ \left|\bar{M}(p)_{1x,i}^{m}\right|^{2} | \mathcal{K}_{i}^{m}   \right] &\leq& C  L(m,p) \sum_{j=0}^{L(m,p)-1} b_m ^{-2} \Delta_m ^{-1} \mathbb{E}\left[ \left| \mathbb{E}\left[ \zeta(p)_{1x,i_{l,0}^{m}}^{m} | \mathcal{K}_{i_{l,0}^{m}}^{m}  \right] \right|^{2} \right] \\
  &\leq& C \Delta_m  \text{ a.s.}
\end{eqnarray*}
and
\begin{eqnarray*}
  \mathbb{E}\left[ \left|\bar{M}(p)_{1x,i}^{m}\right|^{w}  \right] &\leq& C  L(m,p)^{w-1} \sum_{l=0}^{L(m,p)-1} b_m ^{-w} \Delta_m ^{-w/2} \mathbb{E}\left[ \left| \mathbb{E}\left[ \zeta(p)_{1x,i_{l,0}^{m}}^{m} | \mathcal{K}_{i_{l,0}^{m}}^{m}  \right] \right|^{w} \right] \\
  &\leq& C \left( \Delta_m ^{w/2} + \Delta_m ^{7v/8 - w/2} \right)  \quad \text{for any} \quad w \geq 3
  .
\end{eqnarray*}
Similarly, we can bound the rest terms of \eqref{barM-moment}.
$\blacksquare$

\begin{lemma}\label{lemma:e}
  Under Assumption 1, we have for any $x,y \in \left\lbrace 1, 2 \right\rbrace$, we have
  \begin{align}\label{e-moment}
    & \left| \mathbb{E}\left[ e_{1x,i}^{m} | \mathcal{K}_{i}^{m}   \right] \right| \leq C \varPsi_{i}^{m,2}  b_m \Delta_m \quad { a.s.}, \quad \mathbb{E}\left[ \left| e_{1x,i}^{m} \right|^{2}  \right] \leq C b_m^{-1} \Delta_m^{-1/2}, \quad \mathbb{E}\left[ \left| e_{1x,i}^{m} \right|^{3}  \right] \leq C b_m \Delta_m, \nonumber\\
    & \mathbb{E}\left[ \left| e_{1x,i}^{m} \right|^{4}  \right] \leq C  b_m ^{-2} \Delta_{m}^{-1}
    .
  \end{align}
  Furthermore, there exists $\upsilon > 0 $ such that
  \begin{equation}\label{e-debiased}
    \left| \mathbb{E}\left[   e_{1x,i}^{m} e_{1y,i}^{m}    - b_m ^{-1} \Delta_{m}^{-1/2}  \Xi(\bSigma_{i}^{m}, \bvartheta_{i}^{m}  )_{x,y} \Big| \mathcal{K}_{i}^{m}  \right] \right| \leq C \varPsi_{i}^{m,1}  \Delta_m ^{\frac{1}{4} + \upsilon} \text{ a.s.}
  \end{equation}
\end{lemma}
\textbf{Proof of Lemma \ref{lemma:e}.}
Since \eqref{e-moment} is trivial consequence of Lemmas \ref{negligible-xi} and \ref{lemma:M} and the fact that $v \geq 4$ in view of \eqref{SpotErrorDecomp}, so we consider \eqref{e-debiased}.

We can decompose $\left (e_{11,i}^{m}\right)^2$ as follows:
\begin{eqnarray}\label{e-decomp}
  \left (e_{11,i}^{m}\right)^2 &=&  \left (\hat{M}(p_m)_{11,i}^{m} + \hat{M}'(p_m)_{11,i}^{m} + \xi_{11,i}^{m,1} + \xi_{11,i}^{m,2}\right )^2 \nonumber\\
  &=& \left( \hat{M}(p_m)_{11,i}^{m} \right) ^2 + 2 \hat{M}(p_m)_{11,i}^{m} \hat{M}'(p_m)_{11,i}^{m} + \left( \hat{M}'(p_m)_{11,i}^{m} \right)  ^2 + \left( \xi_{11,i}^{m,1} \right) ^2 + \left( \xi_{11,i}^{m,2} \right) ^2 \nonumber\\
  && + 2 \xi_{11,i}^{m,2} \left( \hat{M}(p_m)_{11,i}^{m} + \hat{M}'(p_m)_{11,i}^{m}  + \xi_{11,i}^{m,1} \right) \nonumber\\
  && + 2 \xi_{11,i}^{m,1} \left( \hat{M}(p_m)_{11,i}^{m} + \hat{M}'(p_m)_{11,i}^{m} \right)
  ,
\end{eqnarray}
where $p_m$ is a sequence of integers that satisfies $p_m \asymp \Delta_m ^{- \iota }$ and $\iota \in (\frac{3}{2} -2\kappa, (\frac{1}{2} \kappa -\frac{1}{4}) \land (2\kappa -3 \tau - 1))$, $\hat{M}(p_m)_{11,i}^{m} = {M}(p_m)_{11,i}^{m} + \bar{M}(p_m)_{11,i}^{m}$, and $\hat{M}'(p_m)_{11,i}^{m} = {M}'(p_m)_{11,i}^{m} + \bar{M}'(p_m)_{11,i}^{m}$.
Let $\zeta(p_m)_{11,i}^{m,l} = \zeta(p_m)_{11,i+l(p_m+2)k_m}^{m}$.
We have for some $\upsilon > 0$ 
\begin{align}\label{decompM}
  \mathbb{E}\left[\left( \hat{M}(p_m)_{11,i}^{m} \right) ^2 | \mathcal{K}_{i}^{m} \right] =& \mathbb{E}\left[ \left( \frac{1}{(b_m - 2 k_m)\Delta_m k_m \psi_0} \sum_{l=0}^{L(m,p_m)-1} \zeta(p_m)^m_{11,i+l(p_m+2)k_m} \right) ^2 | \mathcal{K}_{i}^{m}\right]  \nonumber\\
  \leq& \left( \frac{1}{(b_m - 2 k_m)\Delta_m k_m \psi_0} \right)^2 \sum_{l=0}^{L(m,p_m)-1} \mathbb{E}\left[\left( \zeta(p_m)^{m,l}_{11,i} \right) ^{2} | \mathcal{K}_{i}^{m} \right] \nonumber \\
  &  + C b_m ^{-2} \Delta_{m}^{-1}  \sum_{l \neq l'}^{L(m,p_m)-1} \mathbb{E}\left[ \zeta(p_m)^{m,l}_{11,i} \zeta(p_m)^{m,l'}_{11,i}  | \mathcal{K}_{i}^{m} \right] \nonumber\\
  \leq& \left( \frac{1}{(b_m - 2 k_m)\Delta_m k_m \psi_0} \right)^2 \sum_{l=0}^{L(m,p_m)-1} \mathbb{E}\left[\left( \zeta(p_m)^{m,l}_{11,i} \right) ^{2} |\mathcal{K}_{i}^{m} \right] \nonumber \\
  & + C \varPsi_{i}^{m,1} \Delta_m ^{1/2}   \nonumber \\
  \leq& \left( \frac{1}{(b_m - 2 k_m)\Delta_m k_m \psi_0} \right)^2 \sum_{l=0}^{L(m,p_m)-1} \mathbb{E}\left[\left( \zeta(p_m)^{m,l}_{11,i} \right) ^{2} | \mathcal{K}_{i}^{m} \right] \nonumber\\
  & + C \varPsi_{i}^{m,1} \Delta_m ^{\frac{1}{4} + \upsilon} \text{ a.s.},
\end{align}
where the second inequality is due to Lemma \ref{lemma:zeta}.
Let $\zeta(2;p_m)^{m,l}_{11,i} = \zeta(2)^m_{11,i+l(p_m+2)k_m + p_mk_m}$.
We have
\begin{align}\label{MM'}
  & \quad \left|\mathbb{E}\left[ \hat{M}(p_m)_{11,i}^{m} \hat{M}'(p_m)_{11,i}^{m} | \mathcal{K}_{i}^{m}  \right]\right|  \nonumber \\
  & \leq C b_m ^{-2} \Delta_m^{-1} \left| \mathbb{E}\left[ \sum_{l=0}^{L(m,p_m)-1} \zeta(p_m)^{m,l}_{11,i} \sum_{l=0}^{L(m,p_m)-1} \zeta(2;p_m)^{m,l}_{11,i} | \mathcal{K}_{i}^{m}  \right] \right| \nonumber \\
  & \quad + C b_m ^{-2} \Delta_m^{-1} \left| \mathbb{E}\left[ \sum_{l=0}^{L(m,p_m)-1} \zeta(p_m)^{m,l}_{11,i} \sum_{l=L(m,p_m)(p_m+2)k_m+1}^{b_m-k_m} \zeta^m_{11,i+l} | \mathcal{K}_{i}^{m}  \right] \right| \nonumber\\
  & \leq C b_m ^{-2} \Delta_m^{-1} \left| \mathbb{E}\left[ \sum_{l=0}^{L(m,p_m)-1} \zeta(p_m)^{m,l}_{11,i} \zeta(2;p_m)^{m,l}_{11,i} | \mathcal{K}_{i}^{m} \right] \right| \nonumber\\
  & \quad + C b_m ^{-2} \Delta_m^{-1} \left| \mathbb{E}\left[ \sum_{l=1}^{L(m,p_m)-1} \zeta(p_m)^{m,l}_{11,i} \zeta(2;p_m)^{m,l-1}_{11,i} | \mathcal{K}_{i}^{m} \right] \right|\nonumber \\
  & \quad + C b_m ^{-2} \Delta_m^{-1} \left| \mathbb{E}\left[ \sum_{l-l' \notin \{0,1\} }^{L(m,p_m)-1} \zeta(p_m)^{m,l}_{11,i} \zeta(2;p_m)^{m,l'}_{11,i} | \mathcal{K}_{i}^{m} \right] \right| \nonumber \\
  & \quad + C b_m ^{-2} \Delta_m^{-1} \left| \mathbb{E}\left[ \sum_{l=0}^{L(m,p_m)-1} \zeta(p_m)^{m,l}_{11,i} \sum_{l=L(m,p_m)(p_m+2)k_m+1}^{b_m-k_m} \zeta^m_{11,i+l} | \mathcal{K}_{i}^{m} \right]  \right| \text{a.s.},
\end{align}
where the second inequality is due to triangular inequality.
For the first term on the right hand side of \eqref{MM'}, we have
\begin{align*}
  & \quad \left| \mathbb{E}\left[ \sum_{l=0}^{L(m,p_m)-1} \zeta(p_m)^{m,l}_{11,i} \zeta(2;p_m)^{m,l}_{11,i} | \mathcal{K}_{i}^{m} \right] \right| \\
  & =  \left| \mathbb{E}\left[ \sum_{l=0}^{L(m,p_m)-1} \left( \zeta(p_m-2)^m_{11,i_{l,0}^{m}} + \zeta(2)^m_{11,i_{l,(p_m-2)k_m}^{m}} \right)   \zeta(2;p_m)^{m,l}_{11,i} | \mathcal{K}_{i}^{m} \right] \right| \\
  & \leq \sum_{l=0}^{L(m,p_m)-1} \mathbb{E}\left[ \zeta(p_m-2)^m_{11,i_{l,0}^{m}} \zeta(2;p_m)^{m,l}_{11,i} | \mathcal{K}_{i}^{m} \right] \\
  & \quad + \mathbb{E}\left[ \left( \zeta(2)^m_{11,i_{l,(p_m-2)k_m}^{m}}  \right)^2 | \mathcal{K}_{i}^{m}  \right]^{1/2} \mathbb{E}\left[ \left( \zeta(2;p_m)^{m,l}_{11,i}  \right)^2 | \mathcal{K}_{i}^{m}  \right]^{1/2} \\
  & \leq C \varPsi_{i}^{m,2} \left( b_m \Delta_m +  p_m^{-1} b_m \Delta_m ^{1/2}  \right)  \text{ a.s.}
  ,
\end{align*}
where the first and second inequalities are due to H\"older's inequality and Lemmas \ref{lemma:zeta}, respectively.
Similarly, the second term on the right hand side of \eqref{MM'} is bounded by
\begin{eqnarray*}
  \left| \mathbb{E}\left[ \sum_{l=0}^{L(m,p_m)-1} \zeta(p_m)^{m,l}_{11,i} \zeta(2;p_m)^{m,l-1}_{11,i} | \mathcal{K}_{i}^{m} \right] \right| \leq  C \varPsi_{i}^{m,1} \left( b_m \Delta_m + p_m^{-1} b_m \Delta_m ^{1/2} \right)  \text{ a.s.},
\end{eqnarray*}
and the third term on the right hand side of \eqref{MM'} is bounded by
\begin{eqnarray*}
  \left| \mathbb{E}\left[ \sum_{l-l' \notin \{0,1\} }^{L(m,p_m)-1} \zeta(p_m)^{m,l}_{11,i} \zeta(2;p_m)^{m,l'}_{11,i} | \mathcal{K}_{i}^{m} \right] \right| &\leq& \sum_{l-l' \notin \{0,1\} }^{L(m,p_m)-1} \left| \mathbb{E}\left[ \zeta(p_m)^{m,l}_{11,i} \zeta(2;p_m)^{m,l'}_{11,i} |\mathcal{K}_{i}^{m} \right] \right| \\
  &\leq& C \varPsi_{i}^{m,1} p_m^{-1} b_m ^{2} \Delta_m ^{3/2} \text{ a.s.},
\end{eqnarray*}
where the second inequality is due to the Lemma \ref{lemma:zeta}.
For the fourth term on the right hand side of \eqref{MM'}, we have
\begin{eqnarray*}
  \left| \mathbb{E}\left[ \sum_{l=0}^{L(m,p_m)-1} \zeta(p_m)^{m,l}_{11,i} \zeta(p_m)^{m}_{*} | \mathcal{K}_{i}^{m}\right] \right| &=& \left| \sum_{l=0}^{L(m,p_m)-1} \mathbb{E}\left[  \zeta(p_m)^{m,l}_{11,i} \zeta(p_m)^{m}_{*} | \mathcal{K}_{i}^{m}\right] \right| \\
  &\leq& C \varPsi_{i}^{m,1} p_m b_m \Delta_m  \text{ a.s.},
\end{eqnarray*}
where $\zeta(p_m)^{m}_{*} =  \sum_{l=L(m,p_m)(p_m+2)k_m}^{b_m-k_m} \zeta^m_{11,i+l}$.
Thus, we have for some $\upsilon > 0$
\begin{equation}\label{EMM'}
  \left|\mathbb{E}\left[ \hat{M}(p_m)_{11,i}^{m} \hat{M}'(p_m)_{11,i}^{m} | \mathcal{K}_{i}^{m} \right]\right|   \leq C \varPsi_{i}^{m,1} \Delta_m ^{\frac{1}{4} + \upsilon} \text{ a.s.}
\end{equation}
By Lemmas \ref{lemma:M}, \ref{negligible-xi}(a) and (b), we have for some $\upsilon > 0$,
\begin{eqnarray}\label{square-negligible}
  && \mathbb{E}\left[ \left( M'(p_m)_{11,i}^{m} \right)^{2} | \mathcal{K}_{i}^{m} \right] \leq C \varPsi_{i}^{m,1} p_m^{-1} b_m^{-1} \Delta_m^{-1/2} \leq C \varPsi_{i}^{m,1} \Delta_m ^{\frac{1}{4} + \upsilon}, \nonumber\\
  && \mathbb{E}\left[ \left( \xi_{11,i}^{m,1} \right)^2 | \mathcal{K}_{i}^{m} \right]  \leq C  b_m \Delta_m \leq C  \Delta_m ^{\frac{1}{4} + \upsilon}, \nonumber\\
  && \mathbb{E}\left[ \left( \xi_{11,i}^{m,2} \right)^2 | \mathcal{K}_{i}^{m} \right]  \leq C \varPsi_{i}^{m,2} (k_m'^{-2(v-1)} + b_m ^{-1} k_m'^3) \leq C \varPsi_{i}^{m,2} \Delta_m ^{\frac{1}{4} + \upsilon} \text{ a.s.},
\end{eqnarray}
respectively.
Similarly, we can obtain that for some $\upsilon > 0$ 
\begin{align}\label{xi2M}
  &\quad \left|\mathbb{E}\left[ \xi_{11,i}^{m,2} \left( \hat{M}(p_m)_{11,i}^{m} + \hat{M}'(p_m)_{11,i}^{m} + \xi_{11,i}^{m,1} \right) | \mathcal{K}_{i}^{m}  \right]\right| \nonumber\\
  &\leq \mathbb{E}\left[ \left( \xi_{11,i}^{m,2} \right)^2 | \mathcal{K}_{i}^{m}  \right]^{1/2} \Biggl( \mathbb{E}\left[ \left( \hat{M}(p_m)_{11,i}^{m} \right)^2 | \mathcal{K}_{i}^{m}   \right]^{1/2} + \mathbb{E}\left[ \left( \hat{M}'(p_m)_{11,i}^{m} \right)^2 | \mathcal{K}_{i}^{m}   \right]^{1/2} \nonumber\\
  &\quad + \mathbb{E}\left[ \left( \xi_{11,i}^{m,1} \right)^2 | \mathcal{K}_{i}^{m}   \right]^{1/2} \Biggl)  \nonumber \\
  &\leq C \varPsi_{i}^{m,1} \Delta_m ^{\frac{1}{4} +\upsilon} \text{ a.s.}
  ,
\end{align}
where the first inequality is due to H\"older's inequality and the second inequality is due to Lemmas \ref{negligible-xi} and \ref{lemma:M}.
Now, consider $\mathbb{E}\left[ \xi_{11,i}^{m,1} \hat{M}(p_m)_{11,i}^{m} | \mathcal{K}_{i}^{m} \right]$.
We have
\begin{eqnarray*}
  \zeta(p_m)_{11,l}^{m,i} = \zeta_1(p_m)_{11,l}^{m,i} + \zeta_2(p_m)_{11,l}^{m,i} + \zeta_3(p_m)_{11,l}^{m,i}
  ,
\end{eqnarray*}
where
\begin{eqnarray*}
  && \zeta_1(p_m)_{11,l}^{m,i} = \sum_{j=0}^{p_m k_m -1} \hat{\epsilon}_{11,i_{l,j}^{m}}, \quad \zeta_2(p_m)_{11,l}^{m,i} = 2 \sum_{j=0}^{p_m k_m -1} \tilde{X}_{1,i_{l,j}^{m}}^{c,m} \tilde{\epsilon}_{1,i_{l,j}^{m}}^{m}, \quad \zeta_3(p_m)_{11,l}^{m,i} = \sum_{j=0}^{p_m k_m -1} \hat{X}_{11,i_{j}^{m,l}}^{c,m}
  .
\end{eqnarray*}
Using (A.25) in \citet{jacod2019estimating}, we can show that
\begin{align}\label{xi1-zeta1}
  \left|\mathbb{E}\left[ \xi_{11,i}^{m,1} \zeta_1(p_m)_{11,l}^{m,i} | \mathcal{K}_{i}^{m} \right]\right| \leq& \sum_{j=0}^{p_m k_m -1} \left|\mathbb{E}\left[ \xi_{11,i}^{m,1} \ \hat{\epsilon}_{11,i_{l,j}^{m}} | \mathcal{K}_{i}^{m}\right]\right| \nonumber\\
  \leq& \sum_{j=0}^{p_m k_m -1} C(\varPsi_{i,j,l}^{m,2} \Delta_m ^{v/2} + \Delta_m ^{3/4}) \mathbb{E}\left[ \left( \xi_{11,i}^{m,1} \right)^2  | \mathcal{F}_{i}^{m} \right]^{1/2} \nonumber\\
  \leq& C \varPsi_{i,l}^{m,2} p_m b_m ^{1/2} \Delta_m ^{3/4} \text{ a.s.}
  ,
\end{align}
where the second inequality is due to Lemma \ref{negligible-xi}(a).
By (A.26) in \citet{jacod2019estimating}, we have
\begin{align}\label{xi1-zeta2}
  \left|\mathbb{E}\left[ \xi_{11,i}^{m,1} \zeta_2(p_m)_{11,l}^{m,i} | \mathcal{K}_{i}^{m}  \right]\right| \leq&  2 \sum_{j=0}^{p_m k_m -1} \left|\mathbb{E}\left[ \xi_{11,i}^{m,1}  \tilde{X}_{1,i_{l,j}^{m}}^{c,m} \tilde{\epsilon}_{1,i_{l,j}^{m}}^{m} | \mathcal{K}_{i}^{m} \right]\right| \nonumber\\
  \leq& C \sum_{j=0}^{p_m k_m -1} \varPsi_{i,j,l}^{m,2} \Delta_m ^{v/2} \mathbb{E}\left[ \left| \xi_{11,i}^{m,1}  \tilde{X}_{1,i_{l,j}^{m}}^{c,m}\right| | \mathcal{K}_{i}^{m} \right] \nonumber\\
  \leq& C \sum_{j=0}^{p_m k_m -1} \varPsi_{i,j,l}^{m,2} \Delta_m ^{v/2} \mathbb{E}\left[ \left| \xi_{11,i}^{m,1} \right|^2 | \mathcal{F}_{i}^{m} \right]^{1/2} \mathbb{E}\left[ \left| \tilde{X}_{1,i_{l,j}^{m}}^{c,m}\right| ^{2} | \mathcal{F}_{i}^{m} \right]^{1/2} \nonumber\\
  \leq& C \varPsi_{i,l}^{m,2} p_m b_m ^{1/2} \Delta_m ^{v/2 + 1/4} \text{ a.s.}
\end{align}
By It\^{o}'s formula, we have $\zeta_3(p_m)_{11,l}^{m,i} = 2 \left( \zeta_{3,1}(p_m)_{11,l}^{m,i} + \zeta_{3,2}(p_m)_{11,l}^{m,i} + \zeta_{3,3}(p_m)_{11,l}^{m,i} \right) $, where
\begin{eqnarray*}
  && \zeta_{3,1}(p_m)_{11,l}^{m,i} = \sum_{j=0}^{p_m k_m -1}  \hat{X}_{11,i_{l,j}^{m}}^{c,m,1}, \quad \zeta_{3,2}(p_m)_{11,l}^{m,i} = \sum_{j=0}^{p_m k_m -1}  \hat{X}_{11,i_{l,j}^{m}}^{c,m,2} ,\quad \zeta_{3,3}(p_m)_{11,l}^{m,i} = \sum_{j=0}^{p_m k_m -1}  \hat{X}_{11,i_{l,j}^{m}}^{c,m,3}, \\
  && \hat{X}_{11,i}^{c,m,1} = \int_{t_i}^{t_{i+k_m-1}} M_{1,u}^{m,i} \mu_{1,u} G_{u}^{m,i} du + \int_{t_i}^{t_{i+k_m-1}} B_{1,u}^{m,i} dB_{1,u}^{m,i} , \quad  \hat{X}_{11,i}^{c,m,2} = \int_{t_i}^{t_{i+k_m-1}} B_{1,u}^{m,i} dM_{1,u}^{m,i} , \\
  && \hat{X}_{11,i}^{c,m,3} = \int_{t_i}^{t_{i+k_m-1}} M_{1,u}^{m,i} dM_{1,u}^{m,i} , \quad  M_{1,u}^{m,i} = \int_{0}^{u} \sigma_{s} G_{s}^{m,i} dB_{s} , \quad B_{1,u}^{m,i} = \int_{0}^{u} \mu_{1,s} G_{s}^{m,i} ds, \\
  && G_{s}^{m,i} = \sum_{j=1}^{k_m - 1} g_{j}^{m} \mathbf{1}_{(t_{i+j-1},t_{i+j}]}(s)
  .
\end{eqnarray*}
By (S.1) in \citet{jacod2019estimating} and Lemma \ref{negligible-xi}(a), we have
\begin{align}\label{xi1-zeta31}
  \left|\mathbb{E}\left[ \xi_{11,i}^{m,1} \zeta_{3,1}(p_m)_{11,l}^{m,i} | \mathcal{K}_{i}^{m} \right]\right| \leq&  \sum_{j=0}^{p_m k_m -1} \left|\mathbb{E}\left[ \xi_{11,i}^{m,1}\hat{X}_{11,i_{l,j}^{m}}^{c,m,1} | \mathcal{F}_{i}^{m} \right]\right| \nonumber\\
  \leq&  \sum_{j=0}^{p_m k_m -1} \mathbb{E}\left[ \left( \xi_{11,i}^{m,1} \right)^2 | \mathcal{F}_{i}^{m}  \right]^{1/2} \mathbb{E}\left[ \left( \hat{X}_{11,i_{l,j}^{m}}^{c,m,1} \right)^2  | \mathcal{F}_{i}^{m} \right]^{1/2} \nonumber\\
  \leq& C p_m b_m ^{1/2} \Delta_m ^{3/4} \text{ a.s.} 
  ,
\end{align}
where the second inequality is due to H\"older's inequality.
Since $\mu_{1,s}$ is bounded and $G_{s}^{m,i}$ is zero for $s \notin (i \Delta_m  ,(i+k_m-1)\Delta_m ]$, we have $B_{1,u}^{m,i} = O_u(\Delta_m ^{1/2})$.
By It\^{o}'s isometry, we have
\begin{eqnarray*}
  \mathbb{E}\left[ \left( \zeta_{3,2}(p_m)_{11,l}^{m,i} \right)^2 | \mathcal{K}_{i}^{m}  \right] &\leq&  p_m k_m \sum_{j=0}^{p_m k_m -1} \mathbb{E}\left[ \left( \int_{t_{i_{l,j}^{m}}}^{t_{i_{l,j+k_m-1}^{m}}} B_{1,u}^{m,i_{l,j}^{m}} dM_{1,u}^{m,i_{l,j}^{m}} \right)^2 | \mathcal{F}_{i}^{m}  \right] \\
  &=& p_m k_m \sum_{j=0}^{p_m k_m -1} \mathbb{E}\left[  \int_{t_{i_{l,j}^{m}}}^{t_{i_{l,j+k_m-1}^{m}}} \left( B_{1,u}^{m,i_{l,j}^{m}} \sigma_{u} G_{u}^{m,i_{l,j}^{m}}  \right)^2  du  | \mathcal{F}_{i}^{m}  \right] \\
  &\leq& C p_m^{2} \Delta_m ^{1/2} \text{ a.s.}
  ,
\end{eqnarray*}
where the first and second inequalities are due to Jensen's inequality and the facts that $B_{1,u}^{m,i} = O_u(\Delta_m ^{1/2})$ and the boundedness of $\sigma_{u}$ and $G_{u}^{m,i_{l,j}^{m}}$, respectively.
Thus, we can show that
\begin{align}\label{xi1-zeta32}
  \left|\mathbb{E}\left[ \xi_{11,i}^{m,1} \zeta_{3,2}(p_m)_{11,l}^{m,i} | \mathcal{K}_{i}^{m} \right]\right| \leq& \mathbb{E}\left[ \left( \xi_{11,i}^{m,1} \right)^2 | \mathcal{F}_{i}^{m}  \right]^{1/2} \mathbb{E}\left[ \left( \zeta_{3,2}(p_m)_{11,l}^{m,i} \right)^2  | \mathcal{F}_{i}^{m} \right]^{1/2} \nonumber\\
  \leq&  C p_m b_m ^{1/2} \Delta_m ^{3/4}\text{ a.s.}
  ,
\end{align}
where the first and second inequalities are due to H\"older's inequality and Lemma \ref{negligible-xi}(a), respectively.

Simple algbra shows that
\begin{eqnarray*}
  \zeta_{3,3}(p_m)_{11,l}^{m,i} &=& \sum_{j=0}^{p_m k_m -1} \int_{i_{j}^{m,l}}^{t_{i_{j+k_m-1}^{m,l}}} M_{1,u}^{c,m,i_{j}^{m,l}} dM_{1,u}^{m,i_{j}^{m,l}} \\
  &=& \sum_{j=0}^{p_m k_m -1} \sum_{r=0}^{k_m -2}  \int_{i_{j+r}^{m,l}}^{t_{i_{j+r+1}^{m,l}}} M_{1,u}^{c,m,i_{j}^{m,l}} dM_{1,u}^{m,i_{j}^{m,l}} \\
  &=& \sum_{j=0}^{p_m k_m -1} \sum_{r=0}^{k_m -2}  \int_{i_{j+r}^{m,l}}^{t_{i_{j+r+1}^{m,l}}} \int_{t_{i_{j}^{m,l}}}^{u} \sigma_{s} G_{s}^{m,i_{j}^{m,l}} dB_{s} \sigma_{u} G_{u}^{m,i_{j}^{m,l}} dB_{u} \\
  &=& \sum_{j=0}^{(p_m + 1)k_m - 3} \sum_{r=0 \lor (j - p_m k_m + 1)}^{(k_m -2) \land j}  \int_{i_{j}^{m,l}}^{t_{i_{j+1}^{m,l}}} \int_{t_{i_{j-r}^{m,l}}}^{u} \sigma_{s} G_{s}^{m,i_{j-r}^{m,l}} dB_{s} \sigma_{u} G_{u}^{m,i_{j-r}^{m,l}} dB_{u} \\
  &=& \zeta_{3,3,1}(p_m)_{11,l}^{m,i} + \zeta_{3,3,2}(p_m)_{11,l}^{m,i}
  ,
\end{eqnarray*}
where
\begin{eqnarray*}
  \zeta_{3,3,1}(p_m)_{11,l}^{m,i} &=& \sum_{j=0}^{(p_m + 1)k_m - 3} \sum_{r=0 \lor (j - p_m k_m + 1)}^{(k_m -2) \land j}  \int_{i_{j}^{m,l}}^{t_{i_{j+1}^{m,l}}} \int_{i_{j}^{m,l}}^{u} \sigma_{11,s} G_{s}^{m,i_{j-r}^{m,l}} dB_{s} \sigma_{11,u} G_{u}^{m,i_{j-r}^{m,l}} dB_{u} ,\\
  \zeta_{3,3,2}(p_m)_{11,l}^{m,i} &=& \sum_{j=0}^{(p_m + 1)k_m - 3 } \sum_{r=0 \lor (j - p_m k_m + 1)}^{(k_m -2) \land j}  \int_{i_{j}^{m,l}}^{t_{i_{j+1}^{m,l}}}  \sigma_{u} G_{u}^{m,i_{j-r}^{m,l}} dB_{u} \int_{t_{i_{j-r}^{m,l}}}^{i_{j}^{m,l}} \sigma_{u} G_{u}^{m,i_{j-r}^{m,l}} dB_{u}
  .
\end{eqnarray*}
We have
\begin{eqnarray*}
  &&\quad \mathbb{E}\left[ \left(  \zeta_{3,3,1}(p_m)_{11,l}^{m,i} \right)^2 | \mathcal{K}_{i}^{m}  \right] \\ 
  &&= \sum_{j=0}^{(p_m + 1)k_m - 3} \mathbb{E}\left[ \left( \sum_{r=0 \lor (j - p_m k_m + 1)}^{(k_m -2) \land j}  \int_{i_{j}^{m,l}}^{t_{i_{j+1}^{m,l}}} \int_{i_{j}^{m,l}}^{u} \sigma_{s} G_{s}^{m,i_{j-r}^{m,l}} dB_{s} \sigma_{u} G_{u}^{m,i_{j-r}^{m,l}} dB_{u} \right)^2 | \mathcal{F}_{i}^{m} \right] \\
  &&\leq C k_m \sum_{j=0}^{(p_m + 1)k_m - 3}  \sum_{r=0 \lor (j - p_m k_m + 1)}^{(k_m -2) \land j} \mathbb{E}\left[ \left( \int_{i_{j}^{m,l}}^{t_{i_{j+1}^{m,l}}} \int_{i_{j}^{m,l}}^{u} \sigma_{s} G_{s}^{m,i_{j-r}^{m,l}} dB_{s} \sigma_{u} G_{u}^{m,i_{j-r}^{m,l}} dB_{u} \right)^2 | \mathcal{F}_{i}^{m} \right] \\
  &&= C k_m \sum_{j=0}^{(p_m + 1)k_m - 3}  \sum_{r=0 \lor (j - p_m k_m + 1)}^{(k_m -2) \land j} \mathbb{E}\left[  \int_{i_{j}^{m,l}}^{t_{i_{j+1}^{m,l}}} \left( \int_{i_{j}^{m,l}}^{u} \sigma_{s} G_{s}^{m,i_{j-r}^{m,l}} dB_{s} \sigma_{u} G_{u}^{m,i_{j-r}^{m,l}} \right)^2 du  | \mathcal{F}_{i}^{m} \right] \\
  &&\leq C k_m \sum_{j=0}^{(p_m + 1)k_m - 3}  \sum_{r=0 \lor (j - p_m k_m + 1)}^{(k_m -2) \land j}   \int_{i_{j}^{m,l}}^{t_{i_{j+1}^{m,l}}} \mathbb{E}\left[ \left( \int_{i_{j}^{m,l}}^{u} \sigma_{s} G_{s}^{m,i_{j-r}^{m,l}} dB_{s}  \right)^2 | \mathcal{F}_{i}^{m} \right] du  \\
  &&= C k_m \sum_{j=0}^{(p_m + 1)k_m - 3}  \sum_{r=0 \lor (j - p_m k_m + 1)}^{(k_m -2) \land j}   \int_{i_{j}^{m,l}}^{t_{i_{j+1}^{m,l}}} \mathbb{E}\left[ \int_{i_{j}^{m,l}}^{u} \left(  \sigma_{s} G_{s}^{m,i_{j-r}^{m,l}}  \right)^2  ds | \mathcal{F}_{i}^{m} \right] du  \\
  &&\leq C p_m \Delta_m ^{1/2}\text{ a.s.}
  ,
\end{eqnarray*}
where the second and third inequalities are due to It\^{o}'s isometry, the first inequality is due to Jensen's inequality, and the second and third inequalities are due to boundedness of $\sigma_{u}$ and $G_{u}^{m,i_{j-r}^{m,l}}$.
Thus, we can show that
\begin{align}\label{xi1-zeta331}
  \left|\mathbb{E}\left[ \xi_{11,i}^{m,1} \zeta_{3,3,1}(p_m)_{11,l}^{m,i} | \mathcal{K}_{i}^{m} \right]\right| \leq& \mathbb{E}\left[ \left( \xi_{11,i}^{m,1} \right)^2 | \mathcal{F}_{i}^{m}  \right]^{1/2} \mathbb{E}\left[ \left( \zeta_{3,3,1}(p_m)_{11,l}^{m,i} \right)^2  | \mathcal{F}_{i}^{m} \right]^{1/2} \nonumber\\
  \leq&  C p_m^{1/2} b_m ^{1/2} \Delta_m ^{3/4}\text{ a.s.}
  ,
\end{align}
where the first and second inequalities are due to H\"older's inequality and Lemma \ref{negligible-xi}(a), respectively.
By It\^{o}'s isometry, H\"older's inequality, and the boundedness of $\sigma_{u}$ and $G_{u}^{m,i}$, we have
\begin{eqnarray}\label{zeta332-2}
  && \mathbb{E}\left[ \left( \zeta_{3,3,2}(p_m)_{11,l}^{m,i} \right)^2 | \mathcal{K}_{i}^{m}  \right] \nonumber\\
  &&= \mathbb{E}\Biggl[ \sum_{j=0}^{(p_m + 1)k_m - 3 } \sum_{r,r' \geq 0 \lor (j - p_m k_m + 1)}^{(k_m -2) \land j}  \left( \int_{i_{j}^{m,l}}^{t_{i_{j+1}^{m,l}}}  \sigma_{u} G_{u}^{m,i_{j-r}^{m,l}} dB_{u} \right)^2  \int_{t_{i_{j-r}^{m,l}}}^{i_{j}^{m,l}} \sigma_{u} G_{u}^{m,i_{j-r}^{m,l}} dB_{u}  \nonumber\\
  &&\quad \times \int_{t_{i_{j-r'}^{m,l}}}^{i_{j}^{m,l}} \sigma_{u} G_{u}^{m,i_{j-r'}^{m,l}} dB_{u} | \mathcal{F}_{i}^{m} \Biggl] \nonumber\\
  &&=  \sum_{j=0}^{(p_m + 1)k_m - 3 } \sum_{r,r' \geq 0 \lor (j - p_m k_m + 1)}^{(k_m -2) \land j} \mathbb{E}\Biggl[ \mathbb{E}\Biggl[  \left( \int_{i_{j}^{m,l}}^{t_{i_{j+1}^{m,l}}}  \sigma_{u} G_{u}^{m,i_{j-r}^{m,l}} dB_{u} \right)^2 | \mathcal{F}_{i_{j}^{m,l}}^{m} \Biggl] \nonumber\\
  &&\quad \times \int_{t_{i_{j-r}^{m,l}}}^{i_{j}^{m,l}} \sigma_{u} G_{u}^{m,i_{j-r}^{m,l}} dB_{u} \int_{t_{i_{j-r'}^{m,l}}}^{i_{j}^{m,l}} \sigma_{u} G_{u}^{m,i_{j-r'}^{m,l}} dB_{u} | \mathcal{F}_{i}^{m} \Biggl] \nonumber\\
  &&\leq C \Delta_m \sum_{j=0}^{(p_m + 1)k_m - 3 } \sum_{r,r' \geq 0 \lor (j - p_m k_m + 1)}^{(k_m -2) \land j} \mathbb{E}\Biggl[ \int_{t_{i_{j-r}^{m,l}}}^{i_{j}^{m,l}} \sigma_{u} G_{u}^{m,i_{j-r}^{m,l}} dB_{u} \int_{t_{i_{j-r'}^{m,l}}}^{i_{j}^{m,l}} \sigma_{u} G_{u}^{m,i_{j-r'}^{m,l}} dB_{u} | \mathcal{F}_{i}^{m} \Biggl] \nonumber\\
  &&\leq C p_m \text{ a.s.}
\end{eqnarray}
Furthermore, we have
\begin{eqnarray}\label{zeta332-1}
  &&\quad \mathbb{E}\left[ \zeta_{3,3,2}(p_m)_{11,l}^{m,i} | \mathcal{K}_{i}^{m} \right] \nonumber\\
  &&=   \sum_{j=0}^{(p_m + 1)k_m - 3 } \sum_{r=0 \lor (j - p_m k_m + 1)}^{(k_m -2) \land j} \mathbb{E}\Bigg[ \mathbb{E}\left[ \int_{i_{j}^{m,l}}^{t_{i_{j+1}^{m,l}}}  \bSigma_{11,u} G_{u}^{m,i_{j-r}^{m,l}} dW_{u} | \mathcal{K}_{i_{j}^{m,l}}^{m} \right] \cr
  &&\qquad\qquad \qquad \qquad \qquad \times \int_{t_{i_{j-r}^{m,l}}}^{i_{j}^{m,l}} \bSigma_{11,u} G_{u}^{m,i_{j-r}^{m,l}} dW_{u} \Bigg | \mathcal{K}_{i}^{m} \Bigg] \nonumber\\
  &&= 0 \text{ a.s.}
\end{eqnarray}
On the other hand, we can rewrite $\xi_{11,i}^{m,1} = \tilde{\xi}_{11,i}^{m,0} + \tilde{\xi}_{11,i}^{m,1}$, where
\begin{eqnarray*}
  && \tilde{\xi}_{11,i}^{m,0} =  \frac{1}{(b_m - 2 k_m) \Delta_m k_m \psi_0} \sum_{l=0}^{b_m- 2 k_m -1}  \breve{C}_{11,i+l}^{m} - \frac{1}{b_m \Delta_m }  \int_{t_i}^{t_{i+b_m}} \bSigma_{11,t} dt ,\\
  && \tilde{\xi}_{11,i}^{m,1} = \frac{1}{b_m \Delta_m }  \int_{t_i}^{t_{i+b_m}} \bSigma_{11,t} dt - \bSigma_{11,i}^{m}
  .
\end{eqnarray*}
Simple algebra shows that
\begin{eqnarray*}\label{tildeC-decomp}
  &&\sum_{l=0}^{b_m- 2 k_m -1}  \breve{C}_{11,i+l}^{m}\cr
   &&= \sum_{l=0}^{b_m- 2 k_m -1} \sum_{j=1}^{k_m-1} (g _{j}^{m})^2 \left( C_{11,i+l+j}^{m} - C_{11,i+l+j-1}^{m} \right) \nonumber\\
  &&= \sum_{j=1}^{k_m-1} (g _{j}^{m})^2 \sum_{l=k_m-1}^{b_m -2 k_m}  \left( C_{11,i+l}^{m} - C_{11,i+l-1}^{m}  \right) + \sum_{j=1}^{k_m -2} \sum_{l=1}^{j} (g _{j}^{m})^2 \left( C_{11,i+l}^{m} - C_{11,i+l-1}^{m} \right) \nonumber\\
  &&\quad  + \sum_{j=2}^{k_m -1} \sum_{l=b_m - 2 k_m - 1 + j}^{b_m- k_m -2} (g _{j}^{m})^2 \left( C_{11,i+l}^{m} - C_{11,i+l-1}^{m} \right)
  .
\end{eqnarray*}
For the second and third terms on the right hand side of \eqref{tildeC-decomp}, by the boundedness of $\bSigma$ and $g$, we have
\begin{eqnarray}\label{tildeC-side}
  &&\left|\sum_{j=1}^{k_m -2} \sum_{l=1}^{j} (g _{j}^{m})^2 \left( C_{11,i+l}^{m} - C_{11,i+l-1}^{m} \right)\right| \leq C \quad \text{and} \quad \nonumber\\
  &&\left|\sum_{j=2}^{k_m -1} \sum_{l=b_m -  2 k_m - 1 + j}^{b_m - k_m -2} (g _{j}^{m})^2 \left( C_{11,i+l}^{m} - C_{11,i+l-1}^{m} \right)\right| \leq C
  .
\end{eqnarray}
For the first term on the right hand side of \eqref{tildeC-decomp}, we have
\begin{eqnarray}\label{tildeC-main}
  \sum_{j=1}^{k_m-1} (g _{j}^{m})^2 \sum_{l=k_m-1}^{b_m -2 k_m}  \left( C_{11,i+l}^{m} - C_{11,i+l-1}^{m}  \right) = \left( k_m \psi_0 + O(1)\right) \int_{t_{i+k_m-2}}^{t_{i+b_m -2 k_m}} \bSigma_{11,t} dt
  ,
\end{eqnarray}
by Riemann integration.
By \eqref{tildeC-side}, \eqref{tildeC-main}, and the boundedness of $\bSigma_{11,t}$, we have
\begin{align}\label{tilde-xi0}
  \left|\tilde{\xi}_{11,i}^{m,0}\right| \leq&   \Biggl| \frac{O(1) \int_{t_{i+k_m-2}}^{t_{i+b_m -2 k_m }} \bSigma_{11,t} dt}{(b_m - 2 k_m) \Delta_m k_m \psi_0} \Biggl| + \Biggl| \frac{2k_m \int_{t_{i+k_m-2}}^{t_{i+b_m -2 k_m}} \bSigma_{11,t} dt}{(b_m - 2 k_m) b_m  \Delta_m} \Biggl| \nonumber\\
  & + \Biggl| \frac{\int_{i}^{t_{i+k_m-2}} \bSigma_{11,t} dt}{b_m \Delta_m } \Biggl| + \Biggl| \frac{\int_{t_{i+b_m -2 k_m }}^{t_{i+b_m }} \bSigma_{11,t} dt}{b_m \Delta_m } \Biggl|  + C b_m^{-1} \Delta_m ^{-1/2} \nonumber\\
  \leq& C b_m^{-1} \Delta_m ^{-1/2}   .
\end{align}
Using It\^{o}'s lemma, we have
\begin{eqnarray*}
  \tilde{\xi}_{11,i}^{m,1} &=&  \frac{1}{b_m \Delta_m } \int_{t_{i}}^{t_{i+b_m }} \bSigma_{11,t} - \bSigma_{11,t_{i}} dt \\
  &=&  - \frac{1}{b_m \Delta_m }  \int_{t_{i}}^{t_{i+b_m }} (t - t_{i+b_m } ) \tilde{\sigma}_{t} d\tilde{B}_{t} \\
  &&- \frac{1}{b_m \Delta_m } \left(  \int_{t_{i}}^{t_{i+b_m }} (t - t_{i+b_m } ) \tilde{\mu}_{t} dt  + \int_{t_{i}}^{t_{i+b_m }} (t - t_{i+b_m } )  \tilde{J}_{1,t}(d{\tilde{\Lambda}}_{1,t}-\tilde{\lambda}_{1,t}dt) \right) \\
  &=& \int_{t_{i}}^{t_{i+b_m }} \frac{t_{i+b_m } - t}{b_m \Delta_m }   \tilde{\sigma}_{t} d\tilde{B}_{t}  + O_{u}(\Delta_m ^{1-\kappa}) + \hat{J}_{m,i} \\
  &=&  \sum_{l=0}^{L(m,p_m)-1} \int_{t_{i_{0}^{m,l}}}^{t_{i_{0}^{m,l+1}}} \frac{t_{i+b_m } - t}{b_m \Delta_m } \tilde{\sigma}_{t} d\tilde{B}_{t} + \int_{t_{i_{0}^{m,L(m,p_m)}}}^{t_{i+b_m }} \frac{t_{i+b_m } - t}{b_m \Delta_m } \tilde{\sigma}_{t} d\tilde{B}_{t} + O_{u}(\Delta_m ^{1-\kappa}) + \hat{J}_{m,i} \\
  &=&  \sum_{l=0}^{L(m,p_m)-1} \int_{t_{i_{0}^{m,l}}}^{t_{i_{0}^{m,l+1}}} \frac{t_{i+b_m } - t}{b_m \Delta_m } \tilde{\sigma}_{t} d\tilde{B}_{t} + O_u(\Delta_m ^{\kappa - 2 \iota}) + O_{u}(\Delta_m ^{1-\kappa}) + \hat{J}_{m,i}
  ,
\end{eqnarray*}
where the third and fifth equalities are due to the boundedness of $\tilde{\mu}$ and $\tilde{\sigma}$, respectively, and $\hat{J}_{m,i} = - \frac{1}{b_m \Delta_m }  \int_{t_{i}}^{t_{i+b_m }} (t - t_{i+b_m } )  \tilde{J}_{1,t} (d{\tilde{\Lambda}}_{1,t}-\tilde{\lambda}_{1,t}dt) $.
Thus, we have
\begin{align}\label{zeta332xi}
  &\mathbb{E}\left[  \zeta_{3,3,2}(p_m)_{11,l}^{m,i} \tilde{\xi}_{11,i}^{m,1} | \mathcal{K}_{i}^{m} \right]  \cr
  &\leq   \sum_{r=0}^{L(m,p_m)-1} \mathbb{E}\left[ \zeta_{3,3,2}(p_m)_{11,l}^{m,i} \int_{t_{i_{0}^{m,r}}}^{t_{i_{0}^{m,r+1}}} \frac{t_{i+b_m } - t}{b_m \Delta_m } \tilde{\sigma}_{t} d\tilde{B}_{t} | \mathcal{F}_{i}^{m} \right] \nonumber\\
  & \quad + C \Delta_m ^{1-\kappa} \mathbb{E}\left[ \left|\zeta_{3,3,2}(p_m)_{11,l}^{m,i}\right| | \mathcal{F}_{i}^{m}  \right] \nonumber\\
  \leq&  \sum_{r > l}^{L(m,p_m)-1} \mathbb{E}\left[ \mathbb{E}\left[ \int_{t_{i_{0}^{m,r}}}^{t_{i_{0}^{m,r+1}}} \frac{t_{i+b_m } - t}{b_m \Delta_m } \tilde{\sigma}_{t} d\tilde{B}_{t} | \mathcal{F}_{i}^{m} \right] \zeta_{3,3,2}(p_m)_{11,l}^{m,i}  | \mathcal{F}_{i_{0}^{m,r}}^{m} \right] \nonumber\\
  & \quad +  \sum_{r < l}^{L(m,p_m)-1} \mathbb{E}\left[ \mathbb{E}\left[ \zeta_{3,3,2}(p_m)_{11,l}^{m,i} | \mathcal{F}_{i_{0}^{m,r}}^{m} \right] \int_{t_{i_{0}^{m,r}}}^{t_{i_{0}^{m,r+1}}} \frac{t_{i+b_m } - t}{b_m \Delta_m } \tilde{\sigma}_{t} d\tilde{B}_{t}  | \mathcal{F}_{i}^{m} \right] \nonumber\\
  &\quad + \mathbb{E}\left[ \zeta_{3,3,2}(p_m)_{11,l}^{m,i} \int_{t_{i_{0}^{m,l}}}^{t_{i_{0}^{m,l+1}}} \frac{t_{i+b_m } - t}{b_m \Delta_m } \tilde{\sigma}_{t} d\tilde{B}_{t} | \mathcal{F}_{i}^{m} \right] + C p_m^{1/2} \Delta_m ^{1-\kappa} \nonumber\\
  &\leq   \mathbb{E}\left[ \zeta_{3,3,2}(p_m)_{11,l}^{m,i} \int_{t_{i_{0}^{m,l}}}^{t_{i_{0}^{m,l+1}}} \frac{t_{i+b_m } - t}{b_m \Delta_m } \tilde{\sigma}_{t} d\tilde{B}_{t} | \mathcal{F}_{i}^{m} \right] + C  \Delta_m ^{1-\kappa - \iota/2} \text{ a.s.},
\end{align}
where the second and third inequalities are due to \eqref{zeta332-2} and \eqref{zeta332-1}, respectively.
We have
\begin{align}\label{zeta332XtildeSigma}
  &\quad \mathbb{E}\left[ \zeta_{3,3,2}(p_m)_{11,l}^{m,i} \int_{t_{i_{0}^{m,l}}}^{t_{i_{0}^{m,l+1}}} \frac{t_{i+b_m } - t}{b_m \Delta_m } \tilde{\sigma}_{t} d\tilde{B}_{t} | \mathcal{F}_{i}^{m} \right] \nonumber\\
  &= \mathbb{E}\Biggl[ \sum_{j=0}^{(p_m + 1)k_m - 3 } \sum_{r=0 \lor (j - p_m k_m + 1)}^{(k_m -2) \land j}  \int_{i_{j}^{m,l}}^{t_{i_{j+1}^{m,l}}}  \sigma_{u} G_{u}^{m,i_{j-r}^{m,l}} dB_{u} \int_{t_{i_{j-r}^{m,l}}}^{i_{j}^{m,l}} \sigma_{u} G_{u}^{m,i_{j-r}^{m,l}} dB_{u} \nonumber\\
  &\qquad \times \sum_{j'=0}^{(p_m+2)k_m - 1}   \int_{t_{i_{j'}^{m,l}}}^{t_{i_{j'+1}^{m,l+1}}} \frac{t_{i+b_m } - t}{b_m \Delta_m } \tilde{\sigma}_{t} d\tilde{B}_{t} | \mathcal{F}_{i}^{m} \Biggl] \nonumber\\
  &= \mathbb{E}\Biggl[ \sum_{j=0}^{(p_m + 1)k_m - 3 } \sum_{r=0 \lor (j - p_m k_m + 1)}^{(k_m -2) \land j}  \mathbb{E}\left[ \int_{i_{j}^{m,l}}^{t_{i_{j+1}^{m,l}}}  \sigma_{u} G_{u}^{m,i_{j-r}^{m,l}} dB_{u} \int_{t_{i_{j}^{m,l}}}^{t_{i_{j+1}^{m,l+1}}} \frac{t_{i+b_m } - t}{b_m \Delta_m } \tilde{\sigma}_{t} d\tilde{B}_{t} | \mathcal{F}_{i_{j}^{m,l}}^{m} \right]  \nonumber\\
  &\qquad \times    \int_{t_{i_{j-r}^{m,l}}}^{i_{j}^{m,l}} \sigma_{u} G_{u}^{m,i_{j-r}^{m,l}} dB_{u} | \mathcal{F}_{i}^{m} \Biggl] \nonumber\\
  &= \mathbb{E}\Biggl[ \sum_{j=0}^{(p_m + 1)k_m - 3 } \sum_{r=0 \lor (j - p_m k_m + 1)}^{(k_m -2) \land j}  \mathbb{E}\left[ \int_{i_{j}^{m,l}}^{t_{i_{j+1}^{m,l}}} \frac{t_{i+b_m } - u}{b_m \Delta_m } G_{u}^{m,i_{j-r}^{m,l}} \sigma_{u}   \tilde{\sigma}_{u} \tilde{\rho}_{u} du | \mathcal{F}_{i_{j}^{m,l}}^{m} \right]  \nonumber\\
  &\qquad \times    \int_{t_{i_{j-r}^{m,l}}}^{i_{j}^{m,l}} \sigma_{u} G_{u}^{m,i_{j-r}^{m,l}} dB_{u} | \mathcal{F}_{i}^{m} \Biggl] \nonumber\\
  &= \mathbb{E}\Biggl[ \sum_{j=0}^{(p_m + 1)k_m - 3 } \sum_{r=0 \lor (j - p_m k_m + 1)}^{(k_m -2) \land j}   \int_{i_{j}^{m,l}}^{t_{i_{j+1}^{m,l}}} \mathbb{E}\left[  F_{u}^{m,i,l,j,r} - F_{t_{i_{j-r}^{m,l}}}^{m,i,l,j,r} | \mathcal{F}_{i_{j}^{m,l}}^{m} \right]  du  \int_{t_{i_{j-r}^{m,l}}}^{i_{j}^{m,l}} \sigma_{u} G_{u}^{m,i_{j-r}^{m,l}} dB_{u} | \mathcal{F}_{i}^{m} \Biggl] \nonumber\\
  &\qquad + \mathbb{E}\Biggl[ \sum_{j=0}^{(p_m + 1)k_m - 3 } \sum_{r=0 \lor (j - p_m k_m + 1)}^{(k_m -2) \land j}  F_{t_{i_{j-r}^{m,l}}}^{m,i,l,j,r} \mathbb{E}\left[ \int_{t_{i_{j-r}^{m,l}}}^{i_{j}^{m,l}} \sigma_{u} G_{u}^{m,i_{j-r}^{m,l}} dB_{u} | \mathcal{F}_{i_{j}^{m,l}}^{m} \right] \Biggl] \nonumber\\
  &=  \sum_{j=0}^{(p_m + 1)k_m - 3 } \sum_{r=0 \lor (j - p_m k_m + 1)}^{(k_m -2) \land j} \mathbb{E}\Biggl[  \int_{i_{j}^{m,l}}^{t_{i_{j+1}^{m,l}}} \mathbb{E}\left[  F_{u}^{m,i,l,j,r} - F_{t_{i_{j}^{m,l}}}^{m,i,l,j,r} | \mathcal{F}_{i_{j}^{m,l}}^{m} \right]  du  \int_{t_{i_{j-r}^{m,l}}}^{i_{j}^{m,l}} \sigma_{u} G_{u}^{m,i_{j-r}^{m,l}} dB_{u} | \mathcal{F}_{i}^{m} \Biggl] \nonumber\\
  &\qquad +  \sum_{j=0}^{(p_m + 1)k_m - 3 } \sum_{r=0 \lor (j - p_m k_m + 1)}^{(k_m -2) \land j} \mathbb{E}\Biggl[   \Delta_m \left( F_{t_{i_{j}^{m,l}}}^{m,i,l,j,r} - F_{t_{i_{j-r}^{m,l}}}^{m,i,l,j,r} \right)    \int_{t_{i_{j-r}^{m,l}}}^{i_{j}^{m,l}} \sigma_{u} G_{u}^{m,i_{j-r}^{m,l}} dB_{u} | \mathcal{F}_{i}^{m} \Biggl] \text{ a.s.}
  ,
\end{align}
where $F_{u}^{m,i,l,j,r} = \frac{t_{i+b_m } - u}{b_m \Delta_m } G_{u}^{m,i_{j-r}^{m,l}} \sigma_{u}   \tilde{\sigma}_{u} \tilde{\rho}_{u}$ and the third equality is due to It\^{o}'s product rule.
For the summand of the first term on the right hand side of \eqref{zeta332XtildeSigma}, we have
\begin{align}\label{zeta332XtildeSigma-first}
  &\quad \mathbb{E}\Biggl[  \int_{i_{j}^{m,l}}^{t_{i_{j+1}^{m,l}}} \mathbb{E}\left[  F_{u}^{m,i,l,j,r} - F_{t_{i_{j}^{m,l}}}^{m,i,l,j,r} | \mathcal{F}_{i_{j}^{m,l}}^{m} \right]  du  \int_{t_{i_{j-r}^{m,l}}}^{i_{j}^{m,l}} \sigma_{u} G_{u}^{m,i_{j-r}^{m,l}} dB_{u} | \mathcal{F}_{i}^{m} \Biggl] \nonumber\\
  &\leq \mathbb{E}\Biggl[  \int_{i_{j}^{m,l}}^{t_{i_{j+1}^{m,l}}} \mathbb{E}\left[ \left(  F_{u}^{m,i,l,j,r} - F_{t_{i_{j}^{m,l}}}^{m,i,l,j,r} \right)^2  | \mathcal{F}_{i_{j}^{m,l}}^{m} \right]^{1/2}  du  \int_{t_{i_{j-r}^{m,l}}}^{i_{j}^{m,l}} \sigma_{u} G_{u}^{m,i_{j-r}^{m,l}} dB_{u} | \mathcal{F}_{i}^{m} \Biggl] \nonumber\\
  &\leq C \Delta_m ^{5/4}\mathbb{E}\left[ \left( \int_{t_{i_{j-r}^{m,l}}}^{i_{j}^{m,l}} \sigma_{u} G_{u}^{m,i_{j-r}^{m,l}} dB_{u} \right)^2 | \mathcal{F}_{i}^{m}   \right]^{1/2} \nonumber\\
  &\leq C \Delta_m ^{3/2}\text{ a.s.}
  ,
\end{align}
where the first and second inequalities are due to H\"older's inequality and Lemma \ref{smooth-process}, respectively, and the third inequality is due to It\^{o}'s isometry and the boundedness of $\sigma_{u}$ and $G_{u}^{m,i}$.
For the summand of the second term on the right hand side of \eqref{zeta332XtildeSigma}, we have
\begin{align}\label{zeta332XtildeSigma-second}
  &\quad \mathbb{E}\Biggl[   \Delta_m \left( F_{t_{i_{j}^{m,l}}}^{m,i,l,j,r} - F_{t_{i_{j-r}^{m,l}}}^{m,i,l,j,r} \right)    \int_{t_{i_{j-r}^{m,l}}}^{i_{j}^{m,l}} \sigma_{u} G_{u}^{m,i_{j-r}^{m,l}} dB_{u} | \mathcal{F}_{i}^{m} \Biggl] \nonumber\\
  &\leq \Delta_m  \mathbb{E}\left[  \left( F_{t_{i_{j}^{m,l}}}^{m,i,l,j,r} - F_{t_{i_{j-r}^{m,l}}}^{m,i,l,j,r} \right)^{2} | \mathcal{F}_{i}^{m}  \right]^{1/2} \mathbb{E}\left[  \left(  \int_{t_{i_{j-r}^{m,l}}}^{i_{j}^{m,l}} \sigma_{u} G_{u}^{m,i_{j-r}^{m,l}} dB_{u}  \right)^2   | \mathcal{F}_{i}^{m} \right]^{1/2} \nonumber\\
  &  \leq C \Delta_m ^{3/2} \text{ a.s.}
  ,
\end{align}
where the first inequality is due to H\"older's inequality, and the second inequality is due to Lemma \ref{smooth-process}, It\^{o}'s isometry, and the boundedness of $\sigma_{u}$ and $G_{u}^{m,i}$.
Then, by \eqref{xi1-zeta331}, \eqref{zeta332-2}, \eqref{tilde-xi0}, \eqref{zeta332xi}, \eqref{zeta332XtildeSigma}, \eqref{zeta332XtildeSigma-first}, and \eqref{zeta332XtildeSigma-second}, we have
\begin{align}\label{xi1-zeta33}
  &\quad \left|\mathbb{E}\left[  \zeta_{3,3}(p_m)_{11,l}^{m,i} \xi_{11,i}^{m,1} | \mathcal{K}_{i}^{m} \right]\right| \nonumber\\
  &\leq \left|\mathbb{E}\left[  \zeta_{3,3,1}(p_m)_{11,l}^{m,i} \xi_{11,i}^{m,1} | \mathcal{K}_{i}^{m} \right]\right| + \left|\mathbb{E}\left[  \zeta_{3,3,2}(p_m)_{11,l}^{m,i} \xi_{11,i}^{m,1} | \mathcal{K}_{i}^{m} \right]\right| \nonumber\\
  &\leq  C p_m^{1/2} b_m ^{1/2} \Delta_m ^{3/4} + \left|\mathbb{E}\left[  \zeta_{3,3,2}(p_m)_{11,l}^{m,i} \tilde{\xi}_{11,i}^{m,0} | \mathcal{K}_{i}^{m} \right]\right| + \left|\mathbb{E}\left[  \zeta_{3,3,2}(p_m)_{11,l}^{m,i} \tilde{\xi}_{11,i}^{m,1} | \mathcal{K}_{i}^{m} \right]\right|  \nonumber\\
  &\leq \mathbb{E}\left[ \zeta_{3,3,2}(p_m)_{11,l}^{m,i} \int_{t_{i_{0}^{m,l}}}^{t_{i_{0}^{m,l+1}}} \frac{t_{i+b_m } - t}{b_m \Delta_m } \tilde{\sigma}_{t} d\tilde{B}_{t} | \mathcal{F}_{i}^{m} \right] + C  \Delta_m ^{1-\kappa - \iota/2} \nonumber\\
  &\quad + \mathbb{E}\left[ \left(  \zeta_{3,3,2}(p_m)_{11,l}^{m,i} \right)^2 | \mathcal{F}_{i}^{m} \right]^{1/2} \mathbb{E}\left[ \left( \tilde{\xi}_{11,i}^{m,0} \right)^2  | \mathcal{K}_{i}^{m} \right]^{1/2}  \nonumber\\
  &\leq C \left( p_m \Delta_m ^{1/2} + \Delta_m ^{1-\kappa - \iota/2} + p_m^{1/2} b_m ^{-1} \Delta_m ^{-1/2} \right) \text{ a.s.}
\end{align}
Using \eqref{xi1-zeta1}, \eqref{xi1-zeta2}, \eqref{xi1-zeta31}, \eqref{xi1-zeta32}, and \eqref{xi1-zeta33}, we conclude that for some $\upsilon > 0$
\begin{align}\label{M-xi1}
  &\quad \left|\mathbb{E}\left[ \hat{M}(p_m)_{11,i}^{m} \xi_{11,i}^{m,1} | \mathcal{K}_{i}^{m} \right]\right| \nonumber\\
  &\leq C b_m^{-1} \Delta_m ^{-1/2} \sum_{l=0}^{L(m,p)-1} \left|\mathbb{E}\left[ \zeta_{1}(p_m)_{11,l}^{m,i} | \mathcal{K}_{i}^{m} \right]\right| + \left|\mathbb{E}\left[ \zeta_{2}(p_m)_{11,l}^{m,i} | \mathcal{K}_{i}^{m} \right]\right| \nonumber \\
  &\quad + \left|\mathbb{E}\left[ \zeta_{3,1}(p_m)_{11,l}^{m,i} | \mathcal{K}_{i}^{m} \right]\right| + \left|\mathbb{E}\left[ \zeta_{3,2}(p_m)_{11,l}^{m,i} | \mathcal{K}_{i}^{m} \right]\right|  + \left|\mathbb{E}\left[ \zeta_{3,3}(p_m)_{11,l}^{m,i} | \mathcal{K}_{i}^{m} \right]\right| \nonumber \\
  &\leq C \left( \varPsi_{i}^{m,2} \Delta_m ^{3/4 - \kappa/2 } + \Delta_m ^{1/2 } + \Delta_m ^{1-\kappa} + \Delta_m ^{\kappa - 1/2 + \iota/2} \right) \nonumber\\
  &\leq C \varPsi_{i}^{m,2} \Delta_m ^{1/4 + \upsilon}  \text{a.s.}
\end{align}
Similarly, we can show that
\begin{equation}\label{hatM-xi1}
  \left|\mathbb{E}\left[ \hat{M}'(p_M)_{11,i}^{m} \xi_{11,i}^{m,1} | \mathcal{K}_{i}^{m} \right]\right| \leq C \varPsi_{i}^{m,2} \Delta_m ^{1/4 + \upsilon} \text{ a.s.}
\end{equation}

Let $\Xi_{11,j}^{m} = \Xi(\bSigma^{m}_{j}, \bvartheta^{m}_{j})_{11}$.
Simple algebra shows that
\begin{eqnarray*}
  && \quad \left( \frac{1}{(b_m - 2 k_m)\Delta_m k_m \psi_0} \right)^2 \sum_{l=0}^{L(m,p_m)-1} \mathbb{E}\left[\left( \zeta(p_m)^{m,l}_{11,i} \right) ^{2} | \mathcal{K}_{i}^{m} \right] - b^{-1} \Delta_m ^{-1/2}  \Xi_{11,i}^{m} \\
  &&= \left( \frac{1}{(b_m - 2 k_m)\Delta_m k_m \psi_0} \right)^2 \left( \mathcal{A}(p_m)_{i}^{m,1} + \mathcal{A}(p_m)_{i}^{m,2} + \mathcal{A}(p_m)_{i}^{m,3} \right) + \mathcal{A}(p_m)_{i}^{m,4}
  ,
\end{eqnarray*}
where
\begin{eqnarray*}
  && \mathcal{A}(p_m)_{i}^{m,1} = \sum_{l=0}^{L(m,p_m)-1} \left( \mathbb{E}\left[ \left (\zeta(p_m)^{m,l}_{11,i}\right)^2 | \mathcal{K}_{i}^{m} \right] - \varXi(p_m)_{11,i+l(p_m+2)k_m}^{m}   \right) , \\
  && \mathcal{A}(p_m)_{i}^{m,2} = \sum_{l=0}^{L(m,p_m)-1} \left( \varXi(p_m)_{11,i+l(p_m+2)k_m}^{m}   -   {C_k}^3 p_m \psi_0^{2} \Xi_{11,i+l(p_m+2)k_m}^{m} \right)  ,  \\
  && \mathcal{A}(p_m)_{i}^{m,3} = {C_k}^3 p_m \psi_0^{2} \sum_{l=0}^{L(m,p_m)-1} \left(  \Xi_{11,i+l(p_m+2)k_m}^{m} - \Xi_{11,i}^{m} \right) , \\
  && \mathcal{A}(p_m)_{i}^{m,4} = \Xi_{11,i}^{m} \left[  \left( \frac{1}{(b_m - 2 k_m)\Delta_m k_m \psi_0} \right)^2 {C_k}^3 p_m \psi_0^{2} \times L(m,p_m) - b_m ^{-1} \Delta_m ^{-1/2}  \right]  .
\end{eqnarray*}
By Lemma \ref{lemma:zeta}, we have
\begin{equation*}
  \left|\mathbb{E}\left[ \mathcal{A}(p_m)_{i}^{m,1} | \mathcal{K}_{i}^{m} \right]\right|  \leq C \varPsi_{i}^{m,2} p_m b_m \Delta_m ^{3/4}  \text{ a.s.}
\end{equation*}
Since $\bSigma$ and $\bvartheta$ are bounded, we have
\begin{align*}
  &\left|\mathbb{E}\left[ \mathcal{A}(p_m)_{i}^{m,2} | \mathcal{K}_{i}^{m} \right]\right| \cr
  &= \left|\sum_{l=0}^{L(m,p_m)-1} \mathbb{E}\left[ \left( \varXi(p_m)_{11,i+l(p_m+2)k_m}^{m}   - {C_k}^3 p_m \psi_0^{2} \Xi_{11,i+l(p_m+2)k_m}^{m} \right) | \mathcal{K}_{i}^{m} \right] \right|  \\
  &= \left|\sum_{l=0}^{L(m,p_m)-1} \mathbb{E}\left[ - 4 \left( \bSigma_{11,i}^{m} \right)^{2}   {C_k}^4 \bar{\Phi}_{00} - 8 \bSigma_{11,i}^{m}  \left(  \vartheta_{1,i}^{m} \right)^2 R {C_k}^2 \bar{\Phi}_{01} - 4 \left( \vartheta_{1,i}^{m} \right)^{4} R^2  \bar{\Phi}_{11} | \mathcal{K}_{i}^{m} \right]\right|    \\
  &\leq C p_m^{-1} b_m k_m^{-1} \text{ a.s.}
\end{align*}
Since $\bSigma$ and $\bvartheta$ are bounded It\^{o} semimartingale, we have
\begin{eqnarray*}
  \left|\mathbb{E}\left[ \mathcal{A}(p_m)_{i}^{m,3} \right]\right| &\leq& {C_k}^3 p \psi_0^{2} \sum_{l=0}^{L(m,p_m)-1} \left| \mathbb{E}\left[ \left(  \Xi_{11,i+l(p+2)k_m}^{m} - \Xi_{11,i}^{m} \right) \right] \right|   \\
  &\leq& C {C_k}^3 p_m \psi_0^{2} L(m,p_m) b_m \Delta_m \\
  &\leq& C b_m ^{2} \Delta_m ^{3/2} \text{ a.s.}
\end{eqnarray*}
Simple algebra shows that
\begin{eqnarray*}
  \left| \mathbb{E}\left[ \mathcal{A}(p_m)_{i}^{m,4} | \mathcal{K}_{i}^{m} \right] \right| &=&  \left| \mathbb{E}\left[ \Xi_{11,i+l(p_m+2)k_m}^{m} \left(  \frac{b_m }{(b_m - 2 k_m)^2}  \frac{p_m}{p_m+2} \Delta_m ^{-1/2}  - b_m ^{-1} \Delta_m ^{-1/2}  \right) | \mathcal{K}_{i}^{m} \right] \right| \\
  &=& \biggl| \mathbb{E} \biggl[ \Xi_{11,i+l(p_m+2)k_m}^{m} \Delta_m ^{-1/2} \\
  && \times \left\lbrace \left( \frac{b_m }{\left( b_m - 2 k_m \right)^2 } - b_m ^{-1} \right) \frac{p_m}{p_m+2} + b_m ^{-1} \left( \frac{p_m}{p_m+2} -1 \right)   \right\rbrace | \mathcal{K}_{i}^{m} \biggl] \biggl| \\
  &=& \left| \mathbb{E}\left[ \Xi_{11,i+l(p_m+2)k_m}^{m} \Delta_m ^{-1/2} \left\lbrace \frac{4 b_m k_m - 4 k_m^2}{\left( b_m - 2 k_m \right)^2 b_m  }  \frac{p_m}{p_m+2} - b_m ^{-1} \frac{2}{p_m+2}    \right\rbrace | \mathcal{K}_{i}^{m} \right] \right| \\
  &\leq& C p_m^{-1} \Delta_m ^{-1/2} b_m ^{-1}  \text{ a.s.}
\end{eqnarray*}
Thus, we have
\begin{eqnarray}\label{ZminusX}
  && \left| \mathbb{E}\left[ \left( \frac{1}{(b_m - 2 k_m)\Delta_m k_m \psi_0} \right)^2 \sum_{l=0}^{L(m,p_m)-1} \mathbb{E}\left[\left( \zeta(p_m)^{m,l}_{11,i} \right) ^{2}\right] - b^{-1} \Delta_m ^{-1/2}  \Xi\left( \bSigma_{11,i}^{m}, \gamma_{11,i}^{m} \right) | \mathcal{K}_{i}^{m} \right] \right| \nonumber\\
  &&\leq C b_m ^{-2} \Delta_m ^{-1} \left( C \varPsi_{i}^{m,2} p_m  b_m\Delta_m ^{3/4} + p_m^{-1} b_m k_m^{-1} + b_m ^2  \Delta_m ^{3/2} \right) + C p_m^{-1} \Delta_m ^{-1/2} b_m ^{-1} \nonumber \\
  &&\leq C \varPsi_{i}^{m,2} p_m^{-1} b_m ^{-1} \Delta_m ^{-1/2} \nonumber \\
  &&\leq C \varPsi_{i}^{m,2} \Delta_m ^{\frac{1}{4} + \upsilon} \text{ a.s.}
  ,
\end{eqnarray}
for some $\upsilon > 0$.
By \eqref{e-decomp}, \eqref{decompM}, \eqref{EMM'}, \eqref{square-negligible}, \eqref{xi2M}, \eqref{M-xi1}, \eqref{hatM-xi1}, and \eqref{ZminusX},  we establish
\begin{equation*}
  \left| \mathbb{E}\left[ \left(  e_{11,i}^{m}  \right)^2   - b_m ^{-1} \Delta_{m}^{-1/2}  \Xi_{11,i}^{m} | \mathcal{K}_{i}^{m} \right] \right| \leq C \varPsi_{i}^{m,2} \Delta_m ^{\frac{1}{4} + \upsilon}    \quad \text{ for some } \quad \upsilon > 0 \text{ a.s.}
\end{equation*}
$\blacksquare$

\subsubsection{Properties of spot volatility: Jump part}

\begin{lemma}\label{discontinuous-error}
  Under the assumptions in Theorem \ref{Theorem-1}, for any $i\geq0$, $1 \leq z \leq 4  $ and sufficiently large $m$, we have almost surely
  \begin{eqnarray*}
    && \max_{1\leq x,y \leq 2}  \left|\hat{\bSigma}_{xy,i}^{m} - \hat{\bSigma}_{xy,i}^{c,m}\right|   \leq C \left( J_i(1) + J_i(2) + J_i(3) \right) , \\
    && \max_{1\leq x,y \leq 2}  \left|\hat{\bvartheta}_{xy,i}^{m} - \hat{\bvartheta}_{xy,i}^{c,m}\right|   \leq C J_i(3) ,\\
    && \max_{1\leq x,y \leq 2} \left|\hat{\bvartheta}_{xy,i}^{m}\right| \leq  C \left( J_i(3) + J_i(4) \right),
  \end{eqnarray*}
  where
  \begin{eqnarray*}
    && \mathbb{E}\left[ \left|J_i(1) \right|^{z}  \right] \leq C_z (b_m \Delta_m )^{-(z-1 )} \Delta_m ^{\varpi_1 z}, \quad \mathbb{E}\left[ \left|J_i(2) \right|^{z}  \right] \leq C_z \Delta_m ^{\frac{1}{2} \left( [v] - z \right) \left( 1- 2 \varpi_2 \right)  }, \\
    && \mathbb{E}\left[ \left|J_i(3) \right|^{z}  \right] \leq C_z l_m \Delta_m ^{1 - \varpi_{2} z }, \quad  \mathbb{E}\left[ \left|J_i(4)\right|^z  \right] \leq C_z
    .
  \end{eqnarray*}
\end{lemma}

\textbf{Proof of Lemma \ref{discontinuous-error}.}
First, we consider $x=1$, $y=1$.
Using the fact that $ \left|\phi_{d}^{m}\right|  < C$ for any $d \in \mathbb{Z}$, we have
\begin{eqnarray*}
&& \frac{1}{\left( b_m - 2 k_m \right)  \Delta _m  k_m \psi_0}| v(Y_1,Y_1,u_{1,m}, u_{1,m}, u_{11,m}, t_i) - v(Y_1^c,Y_1^c,\infty, \infty, \infty, t_i)| \\
& &\leq  C  \left( J_i(1) +J_i(2)  +J_i(3) + J'_i(3) \right) \quad \text{a.s.} ,
\end{eqnarray*} where
\begin{eqnarray*}
&&J_i(1) = b_m ^{-1} \Delta_m ^{-1/2} \sum_{l=0}^{b_m - 2 k_m -1}  \left| \left( \tilde{Y}_{1,i+l}^{m}  \right)^2   \mathbf{1}_{\{|\tilde{Y}_{1,i+l}^{m}|\leq u_{1,m}\}} - \left( \tilde{Y}_{1,i+l}^{c,m}  \right)^2 \mathbf{1}_{\{|\tilde{Y}_{1,i+l}^{c,m}|\leq u_{1,m}\}}\right|, \\
&& J_i(2) =   b_m ^{-1} \Delta_m ^{-1/2} \sum_{l=0}^{b_m - 2 k_m -1} \left|\left( \tilde{Y}_{1,i+l}^{c,m}  \right)^2 \mathbf{1}_{\{|\tilde{Y}_{1,i+l}^{c,m}| > u_{1,m}\}} \right|, \\
&& J_i(3) =  b_m ^{-1} \sum_{l=0}^{b_m - 6 l_m}  \left|\hat{\mathcal{E}}_{Y_1 Y_1,i+l}^{m} \mathbf{1}_{\{ |\hat{\mathcal{E}}_{Y_1 Y_1,i+l}^{m}| \leq u_{11,m}  \}} - \hat{\mathcal{E}}_{Y_1^c Y_1^c,i+l}^{m} \mathbf{1}_{\{ |\hat{\mathcal{E}}_{Y_1^c Y_1^c,i+l}^{m}| \leq u_{11,m}  \}} \right|, \\
&& J'_i(3) =  b_m ^{-1} \sum_{l=0}^{b_m - 6 l_m}  \left| \hat{\mathcal{E}}_{Y_1^c Y_1^c,i+l}^{m} \mathbf{1}_{\{ |\hat{\mathcal{E}}_{Y_1^c Y_1^c,i+l}^{m}| > u_{11,m}  \}} \right| .
\end{eqnarray*}
Consider $J_i(1)$.
We have 
\begin{eqnarray*}
\mathbb{E}\left[\left|\int^{t_{i+ b_m - 2 k_m}}_{t_i} d\Lambda_{1,t}\right|^z\right]
\leq \mathbb{E}\left[\left|{\Lambda}\right|^z\right ] 
\leq Cb_m\Delta_m,
\end{eqnarray*}
where ${\Lambda}$ is a Poisson process with the intensity $C_{\lambda}(b_m - 2 k_m)\Delta_m$ for some $C_{\lambda}>0$.
For each $t_{i}$, the summand of $J_i(1)$ has non-zero value which is equal to or smaller than $4u_{2,m}^2$, only if $\bar{X}^{d,m}_{1,i+l} \neq 0$. 
Thus, we have 
\begin{eqnarray} \label{ji1}
  \mathbb{E}\left[ \left|J_i(1)\right|^{z}  \right]
  &\leq& C   b_m^{-z}\Delta_m^{-z/2}  u_{1,m}^{2z} \mathbb{E} \[ \left| \sum_{l=0}^{ b_m - 2 k_m -1}  \1 _{\{ \tilde{X}^{d,m}_{1,i+l} \neq 0 \}} \right|^z \]   \nonumber  \cr
  &\leq& C   b_m^{-z}\Delta_m^{-z/2}  u_{1,m}^{2z} k_m^z \mathbb{E}\left[ \left| \int^{t_{i+b_m -  k_m}}_{t_i} d\Lambda_{1,t} \right|^z  \right]    \nonumber \\
  &\leq& C (b_m \Delta_m )^{-(z-1 )} \Delta_m ^{\varpi_1 z}.
\end{eqnarray}


Consider $J_i(2)$.
By (A.28) of \citet{jacod2019estimating}, we have for $[v] > 5$ and $1 \leq z \leq 4$,
\begin{eqnarray} \label{ji2}
  \mathbb{E}\left[ \left|J_i(2)\right|^z  \right] &\leq&    b_m ^{-1} \Delta_m ^{-z/2} \sum_{l=0}^{b_m - 2 k_m -1} \mathbb{E}\left[ \left| \tilde{Y}_{1,i+l}^{c,m} \right|^{2[v]} \right]^{z/[v]}  \mathbb{P} \left( |\tilde{Y}_{1,i+l}^{c,m}| > u_{2,m} \right)^{\frac{[v]-z}{[v]} } \nonumber \\
  &\leq&  b_m ^{-1} \Delta_m ^{-z/2} \sum_{l=0}^{b_m - 2  k_m -1} \mathbb{E}\left[ \left| \tilde{Y}_{1,i+l}^{c,m} \right|^{2[v]} \right]^{z/[v]} \mathbb{E}\left[ \left| \tilde{Y}_{1,i+l}^{c,m} \right|^{2[v]} \right]^{\frac{[v]-z}{[v]} } / u_{2,m}^{2 ([v]-z)}  \nonumber \\
  &\leq& C \Delta_m ^{\frac{1}{2} \left( [v] - z \right) \left( 1 - 2 \varpi_{2} \right)  }
  ,
\end{eqnarray}
where the first inequality is due to Jensen's and H\"older's inequalities, the second inequality is due to Markov's inequality.
Similarly, we can show \eqref{ji2} for $[v] = 4$ and $1 \leq z < 4$.
In case of $[v] = 4$ and $z = 4$, we have
\begin{eqnarray} \label{ji2-4}
  \mathbb{E}\left[ \left|J_i(2)\right|^4  \right] &\leq&    b_m ^{-1} \Delta_m ^{-2} \sum_{l=0}^{b_m - 2 k_m -1} \mathbb{E}\left[ \left| \tilde{Y}_{1,i+l}^{c,m} \right|^{8} \right] \nonumber \\
  &\leq& C
  .
\end{eqnarray}

Consider $J_i(3)$.
For each $t_{i}$, the summand of $J_i(3)$ has non-zero value which is equal to or smaller than $2u_{11,m}$, only if $\int_{t_{i+l}}^{t_{i+l+5l_m}} d\Lambda_{1,t} \neq 0$.
Thus, we have 
\begin{eqnarray} \label{ji3}
  \mathbb{E}\left[ \left|J_i(3)\right|^{z}  \right]
  &\leq& C_z   b_m^{-z}  u_{11,m}^{z} \mathbb{E} \[ \left| \sum_{l=0}^{ b_m - 6 l_m }  \1 _{\{ \int_{t_{i+l}}^{t_{i+l+5l_m}} d\Lambda_{1,t} \neq 0 \}} \right|^z \]   \nonumber  \cr
  &\leq& C_z   b_m^{-1}  u_{11,m}^{z} \sum_{l=0}^{ b_m - 6 l_m } \mathbb{E}\left[ \left| \int_{t_{i+l}}^{t_{i+l+5l_m}} d\Lambda_{1,t} \right|^z  \right]    \nonumber \\
  &\leq& C_z l_m \Delta_m ^{1 - \varpi_2 z},
\end{eqnarray}
where the second inequality is due to Jensen's inequality.
In case of $J'_i(3)$, simple algebra shows that
\begin{align}\label{hatMathE}
  \hat{\mathcal{E}}_{Y_1^c Y_1^c,i+l}^{m} =&   \sum_{d=-k'_m}^{k'_m} \phi_{d}^{m} T (0,2 l_m ) _{i+l}^{m,1} T (d,4 l_m ) _{i+l}^{m,1} + \sum_{d=-k'_m}^{k'_m} \phi_{d}^{m} T (0,2 l_m ) _{i+l}^{m,1} T (d,4 l_m ) _{i+l}^{m,2}   \nonumber\\
  &+ \sum_{d=-k'_m}^{k'_m} \phi_{d}^{m} T (0,2 l_m ) _{i+l}^{m,2} T (d,4 l_m ) _{i+l}^{m,1} + \sum_{d=-k'_m}^{k'_m} \phi_{d}^{m} T (0,2 l_m ) _{i+l}^{m,1} T (d,4 l_m ) _{i+l}^{m,1}
  .
\end{align}
By \eqref{theta-moment}, the first, second, and third terms on the right hand side of \eqref{hatMathE} have finite moments of all orders.
For the fourth term on the right hand side of \eqref{hatMathE}, by \citet{shao1995maximal}, we have
\begin{align*}
  &\quad \mathbb{E}\left[ \left|\sum_{d=-k'_m}^{k'_m} \phi_{d}^{m} T (0,2 l_m ) _{i+l}^{m,1} T (d,4 l_m ) _{i+l}^{m,1}\right|^{z}  \right] \\
  &\leq C_z \mathbb{E}\left[ \left|(\vartheta _{1,i+l}^{m})^2 \chi_{1,i+l} \sum_{d=-k'_m}^{k'_m} \phi_{d}^{m}   \chi_{1,i+l+d}\right|^{z}  \right] + C_z \mathbb{E}\left[ \left|(\vartheta _{1,i+l}^{m})^2  \bar{\chi} _{1,i+l+2l_m}^{m} \sum_{d=-k'_m}^{k'_m} \phi_{d}^{m} \chi _{1,i+l+d}^{m}\right|^{z}  \right]  \\
  &\quad + C_z \mathbb{E}\left[ \left|(\vartheta _{1,i+l}^{m})^2 \chi _{1,i+l}^{m} \bar{\chi} _{1,i+l+4l_m}^{m}  \sum_{d=-k'_m}^{k'_m} \phi_{d}^{m}\right|^{z}  \right]   + C_z \mathbb{E}\left[ \left|\bar{\chi} _{1,i+l+2l_m}^{m} \bar{\chi} _{1,i+l+4l_m}^{m} \sum_{d=-k'_m}^{k'_m} \phi_{d}^{m}\right|^{z}  \right] \\
  &\leq C_z k_m^{'z/2}
  .
\end{align*}
Thus, we have
\begin{align}\label{hatMathE-moment}
  \mathbb{E}\left[ \left|\hat{\mathcal{E}}_{Y_1^c Y_1^c,i+l}^{m}\right|^{z}  \right] \leq C_z k_m^{'z/2}
  .
\end{align}
By \eqref{hatMathE-moment}, we have for any $z \geq 1$
\begin{eqnarray}\label{ji_3}
  \mathbb{E}\left[ |J'_{i}(3)|^{z} \right] &\leq& C_z b_m ^{-1} \sum_{l=0}^{b_m - 6 l_m}  \mathbb{E}\left[ \left| \hat{\mathcal{E}}_{Y_1^c Y_1^c,i+l}^{m} \mathbf{1}_{\{ |\hat{\mathcal{E}}_{Y_1^c Y_1^c,i+l}^{m}| > u_{11,m}  \}} \right|^z \right] \nonumber\\
  &\leq& C_z b_m ^{-1} \sum_{l=0}^{b_m - 6 l_m}  \mathbb{E}\left[ \left| \hat{\mathcal{E}}_{Y_1^c Y_1^c,i+l}^{m} \right|^{2z} \right]^{1/2}  \mathbb{P}\left( |\hat{\mathcal{E}}_{Y_1^c Y_1^c,i+l}^{m}|^{(\frac{4}{\tau} + 2 )z} > u_{11,m}^{(\frac{4}{\tau} + 2 )z} \right)^{1/2} \nonumber\\
  &\leq& C_z \Delta_m ^{z} \nonumber\\
  &\leq& C_z l_m \Delta_m ^{1 - \varpi_2 z}
  ,
\end{eqnarray}
where the first and second inequalities are due to H\"older's inequality and Markov's inequality, respectively.
Therefore, from \eqref{ji1}, \eqref{ji2}, \eqref{ji2-4}, \eqref{ji3}, and \eqref{ji_3}, we can show the statement.
Similarly, we can show the statement for the other cases of $x,y$ and $\max_{1\leq x,y \leq 2} \mathbb{E}\left[ \left|\hat{\bvartheta}_{xy,i}^{m} - \hat{\bvartheta}_{xy,i}^{c,m}\right|^{z}  \right]$  $\leq C J_i(3)$.
Furthermore, similar to the proof of \ref{negligible-xi}(b), we can show that
\begin{eqnarray*}
  \left|\hat{\bvartheta}_{xy,i}^{m}\right| &\leq&  \left|\hat{\bvartheta}_{xy,i}^{m} - \hat{\bvartheta}_{xy,i}^{c,m}\right| + \left|\hat{\bvartheta}_{xy,i}^{c,m} - \tilde{\bvartheta}_{xy,i}^{m}  \right|  + \left|\tilde{\bvartheta}_{xy,i}^{m} \right| \\
  &\leq&  C \left( J_i(3) + J_i(4) \right)  
  ,
\end{eqnarray*}
where $\tilde{\bvartheta}_{xy,i}^{m}  = (b_m - 6 k_m)^{-1} \sum_{d=-k'_m}^{k'_m} r_{xy}(|d|) \mathcal{U}_{xy,i}^{m}$.
$\blacksquare$

\subsubsection{A key decomposition}
Let $N_m = \left\lfloor 1/b_m \Delta_m  \right\rfloor$, $\beta_{i b_m }^{c,m}  = \bSigma^{c,m}_{12,i b_m } / \bSigma^{c,m}_{11,i b_m }$,  $\hat{\beta}^{c,m}_{i} = \hat{\bSigma}_{12,ib_m}^{c,m} / \hat{\bSigma}_{11,ib_m}^{c,m,*} $, $\hat{\bSigma}_{11,ib_m}^{c,m,*} = \max (\hat{\bSigma}_{11,ib_m}^{c,m},\delta_m)$, $f(\bc) = (c_{11})^{-1} c_{12}$, $e_{11,i}^{m,*} = \hat{\bSigma}_{11,i}^{c,m,*} - \bSigma_{11,i}^{c,m} $, and $e_{12,i}^{m,*} = e_{12,i}^{m} $.
Simple algebra shows that
\begin{equation*}
  \Delta_m ^{-1/4} \left( RIB - \int_{0}^{1} \beta_{t}^{c} dt \right) = \mathcal{D}_{m,1} + \mathcal{D}_{m,2} + \mathcal{D}_{m,3} + \mathcal{D}_{m,4} + \mathcal{D}_{m,5} 
  ,
\end{equation*}
where
\begin{eqnarray*}
  \mathcal{D}_{m,1} &=&  b_m \Delta_m ^{3/4} \sum_{i=0}^{N_m-1} \left[ \hat{\beta}_{i b_m } - \hat{\beta}_{i b_m }^{c,m} - \left( \hat{B}^{m}_{i} - \hat{B}^{c,m}_{i}  \right)   \right] ,\\
  \mathcal{D}_{m,2} &=& b_m \Delta_m ^{3/4} \sum_{i=0}^{N_m-1} \left[ \hat{\beta}_{i b_m }^{c,m} - \beta_{i b_m }^{c,m} - \sum_{x=1}^{2}  \partial_{1x}f(\bSigma_{ib_m}^{c,m}) e_{1x,i b_m}^{m,*}     - \hat{B}^{c,m}_{i} \right] ,\\
  \mathcal{D}_{m,3} &=& \Delta_m ^{-1/4} \left[ \sum_{i=0}^{N_m - 1} \int_{i b_m \Delta_m }^{(i+1) b_m \Delta_m } \beta_{i b_m }^{c,m} - \beta_{s }^{c} ds + \int_{N_m b_m \Delta_m }^{1}  \beta_{s }^{c}  ds \right] ,\\
  \mathcal{D}_{m,4}(p) &=& b_m \Delta_m ^{3/4} \sum_{i=0}^{N_m-1} \sum_{x=1}^{2}  \partial_{1x}f(\bSigma_{ib_m}^{c,m}) \left[  M'(p)_{1x,i b_m }^{m} + \xi_{1x,i b_m }^{m,1} + \xi_{1x,i b_m }^{m,2} \right] ,\\
  \mathcal{D}_{m,5}(p) &=& b_m \Delta_m ^{3/4} \sum_{i=0}^{N_m-1} \sum_{x=1}^{2}  \partial_{1x}f(\bSigma_{ib_m}^{c,m})  M(p)_{1x,i b_m }^{m} .\\
\end{eqnarray*}
We note that Lemmas \ref{lemma:e} and \ref{discontinuous-error} also hold in view of $\hat{\bSigma}^{*}$ and $\hat{\bSigma}^{c,*}$ due to the fact that
\begin{equation*}
  \left|\hat{\bSigma}^{m,*}_{1x,i} - \hat{\bSigma}^{c,m,*}_{1x,i}\right| \leq \left|\hat{\bSigma}^{m}_{1x,i} - \hat{\bSigma}^{c,m}_{1x,i}\right| \quad \text{and} \quad \left|\hat{\bSigma}^{c,m,*}_{1x,i} - \bSigma^{c,m}_{1x,i}\right| \leq \left|\hat{\bSigma}^{c,m}_{1x,i} - \bSigma^{c,m}_{1x,i}\right|
\end{equation*}
for sufficiently large $m$ and $x \in \left\lbrace 1,2 \right\rbrace$.

\begin{lemma}\label{lemma-D123}
  As $m\rightarrow\infty$, we have $\mathcal{D}_{m,1} \xrightarrow[]{p} 0$, $\mathcal{D}_{m,2} \xrightarrow[]{p} 0$, and $\mathcal{D}_{m,3} \xrightarrow[]{p} 0$.
\end{lemma}
\textbf{Proof of Lemma \ref{lemma-D123}.}
Consider $\mathcal{D}_{m,1}$.
By Taylor's theorem, we have
\begin{eqnarray*}
  \left|f(\hat{\bSigma}_{ib_m}^{m,*}) - f(\hat{\bSigma}_{ib_m}^{c,m,*})\right|  &\leq&  C \left( \left|\hat{\bSigma}_{12,ib_m}^{m}\right| + \left|\hat{\bSigma}_{12,ib_m}^{c,m}\right| \right)  \left|\hat{\bSigma}_{11,ib_m}^{m,*} - \hat{\bSigma}_{11,ib_m}^{c,m,*}\right| \\
  && + C  \left|\hat{\bSigma}_{12,ib_m}^{m} - \hat{\bSigma}_{12,ib_m}^{c,m}\right| \\
  &\leq& C \left( 1 + \left|\hat{\bSigma}_{12,ib_m}^{c,m} - \bSigma_{12,ib_m}^{m}\right| + \left|\hat{\bSigma}_{12,ib_m}^{m} - \hat{\bSigma}_{12,ib_m}^{c,m}\right| \right) \\
  && \times \left|\hat{\bSigma}_{11,ib_m}^{m,*} - \hat{\bSigma}_{11,ib_m}^{c,m,*}\right| ,
\end{eqnarray*}
where the second inequality is due to the triangular inequality and the fact that $\bSigma$ is locally bounded.
By Lemma \ref{discontinuous-error}, we have
\begin{eqnarray}\label{D1-discs1}
  \mathbb{E}\left[ \left|\hat{\bSigma}_{11,ib_m}^{m,*} - \hat{\bSigma}_{11,ib_m}^{c,m,*}\right|  \right] &\leq& \mathbb{E}\left[ \left|J_i(1)\right|  \right] + \mathbb{E}\left[ \left|J_i(2)\right|  \right] + \mathbb{E}\left[ \left|J_i(3)\right|  \right] \nonumber \\
  &\leq& C \Delta_m ^{\frac{1}{4} + \upsilon }
,
\end{eqnarray}
\begin{eqnarray}\label{D1-discs2}
  \mathbb{E}\left[ \left|\hat{\bSigma}_{11,ib_m}^{m,*} - \hat{\bSigma}_{11,ib_m}^{c,m,*}\right|^2  \right] &\leq& \mathbb{E}\left[ \left|J_i(1)\right|^{2}  \right] + \mathbb{E}\left[ \left|J_i(2)\right|^{2}  \right] + \mathbb{E}\left[ \left|J_i(3)\right|^{2}  \right] \nonumber \\
  &\leq& C \Delta_m ^{\frac{1}{4} + \upsilon },
\end{eqnarray}
and
\begin{eqnarray}
  \mathbb{E}\left[ \left|\hat{\bSigma}_{12,ib_m}^{c,m} - \bSigma_{12,ib_m}^{m}\right| \left|\hat{\bSigma}_{11,ib_m}^{*} - \hat{\bSigma}_{11,ib_m}^{c,*}\right| \right] &\leq& \norm{ \hat{\bSigma}_{12,ib_m}^{c} - \bSigma_{12,ib_m} }_{L_2} \norm{ \hat{\bSigma}_{11,ib_m}^{*} - \hat{\bSigma}_{11,ib_m}^{c,*} }_{L_2} \nonumber\\
  &\leq& C b_m ^{-1/2} \Delta_m ^{-1/4} \Delta_m ^{\frac{1}{4} ([v] - 2) (1 -2\varpi_{1})}  \nonumber\\
  &\leq& C \Delta_m ^{\frac{1}{4} + \upsilon }
  ,
\end{eqnarray}
for some $\upsilon > 0$.
Thus, we have
\begin{eqnarray*}
  \mathbb{E}\left[ \left|\hat{\beta}_{i b_m } - \hat{\beta}_{i b_m }^{c,m}\right|  \right] \leq C \Delta_m ^{1/4 + \upsilon}
  ,
\end{eqnarray*}
for some $\upsilon > 0$.
Simple algebra shows that
\begin{eqnarray}\label{B-disc}
  \left|\hat{B}^{m}_{i} - \hat{B}^{c,m}_{i}\right|  &\leq&   C b_m ^{-1} \Delta_m ^{-1/2} \Bigg[ \hat{\bvartheta}_{11,ib_m}^{m} \left| \frac{\hat{\bSigma}_{12,ib_m}}{\left( \hat{\bSigma}_{11,ib_m}^{*} \right)^2 } - \frac{\hat{\bSigma}_{12,ib_m}^{c,*}}{\left( \hat{\bSigma}_{11,ib_m}^{c} \right)^2 }  \right| \nonumber\\
  && + \hat{\bvartheta}_{12,ib_m}^{m} \left| \frac{1}{\left( \hat{\bSigma}_{11,ib_m}^{*} \right) } - \frac{1}{\left( \hat{\bSigma}_{11,ib_m}^{c,*} \right) }  \right| \nonumber\\
  && + \left( \hat{\bvartheta}_{11,ib_m}^{m} \right)^2  \left| \frac{\hat{\bSigma}_{12,ib_m}}{\left( \hat{\bSigma}_{11,ib_m}^{*} \right)^3 } - \frac{\hat{\bSigma}_{12,ib_m}^{c}}{\left( \hat{\bSigma}_{11,ib_m}^{c,*} \right)^3 }  \right| \nonumber\\
  && + \hat{\bvartheta}_{11,ib_m}^{m} \hat{\bvartheta}_{12,ib_m}^{m} \left| \frac{1}{\left( \hat{\bSigma}_{11,ib_m}^{*} \right) } - \frac{1}{\left( \hat{\bSigma}_{11,ib_m}^{c,*} \right) }  \right| \Bigg] , 
\end{eqnarray}
For the third term on the right hand side of \eqref{B-disc}, we have
\begin{align}\label{fourth-term-Bdiff}
  & \quad b_m ^{-1} \Delta_m ^{-1/2} \mathbb{E}\left[ \left|\left( \hat{\bvartheta}_{11,ib_m}^{m} \right)^2  \left| \frac{\hat{\bSigma}_{12,ib_m}}{\left( \hat{\bSigma}_{11,ib_m}^{*}\right)^3 } - \frac{\hat{\bSigma}_{12,ib_m}^{c}}{\left( \hat{\bSigma}_{11,ib_m}^{c,*} \right)^3 }  \right|\right|  \right] \nonumber\\
  & \leq C b_m ^{-1} \Delta_m ^{-1/2} \mathbb{E}\left[ \left( \hat{\bvartheta}_{11,ib_m}^{m} \right)^2 \left\lbrace \left|\hat{\bSigma}_{12,ib_m} - \hat{\bSigma}_{12,ib_m}^{c}\right| + \left( \left|\hat{\bSigma}_{12,ib_m}\right| + \left|\hat{\bSigma}_{12,ib_m}^{c}\right|  \right)  \left|\hat{\bSigma}_{11,ib_m}^{*}- \hat{\bSigma}_{11,ib_m}^{c,*}\right| \right\rbrace \right] \nonumber\\
  & \leq C b_m ^{-1} \Delta_m ^{-1/2} \mathbb{E}\left[ \left( \hat{\bvartheta}_{11,ib_m}^{m} \right)^2 \left( 1 + \left|\hat{\bSigma}_{12,ib_m} - \hat{\bSigma}_{12,ib_m}^{c}\right| + \left|\hat{\bSigma}_{12,ib_m}^{c} - \bSigma_{12,ib_m}\right|  \right)  \left|\hat{\bSigma}_{12,ib_m} - \hat{\bSigma}_{12,ib_m}^{c}\right| \right] \nonumber\\
  & \leq C b_m ^{-1} \Delta_m ^{-1/2} E\Big[ \left( \hat{\bvartheta}_{11,ib_m}^{m} \right)^2  \Big( \left|\hat{\bSigma}_{12,ib_m} - \hat{\bSigma}_{12,ib_m}^{c}\right| + \left|\hat{\bSigma}_{11,ib_m}^{*}- \hat{\bSigma}_{11,ib_m}^{c,*}\right|  \nonumber\\
  & \quad +  \left|\hat{\bSigma}_{12,ib_m} - \hat{\bSigma}_{12,ib_m}^{c}\right|  \left|\hat{\bSigma}_{11,ib_m}^{*}- \hat{\bSigma}_{11,ib_m}^{c,*}\right| +  \left|\hat{\bSigma}_{12,ib_m}^{c} - \bSigma_{12,ib_m}\right|\left|\hat{\bSigma}_{11,ib_m}^{*}- \hat{\bSigma}_{11,ib_m}^{c,*}\right| \Big)    \Big] \nonumber\\
  & \leq C b_m ^{-1} \Delta_m ^{-1/2} \mathbb{E}\left[  \left| J_i(3)^2 + J_i(4)^2\right| \left(\sum_{j=1}^{3} \left|e_{12,ib_m}^{m}J_i(j)\right| +  \left|J_i(j)\right| + \left( J_i(j) \right)^2   \right)      \right] ,
\end{align}
where the first, second, and third inequalities are due to Taylor's theorem, triangular inequality, and Lemma \ref{discontinuous-error}, respectively.
By Lemma \ref{discontinuous-error}, we have
\begin{eqnarray*}
  b_m ^{-1} \Delta_m ^{-1/2} \mathbb{E}\left[ \left|J_i(3)\right|^{4}  \right] &\leq& C b_m ^{-1} \Delta_m ^{-1/2} l_m \Delta_m ^{1  - 4\varpi_2} \\
  &\leq& C \Delta_m ^{1/4 + \upsilon}
  ,
\end{eqnarray*}
for some $\upsilon > 0$.
Furthermore, using Lemma \ref{lemma:e}, we obtain
\begin{eqnarray*}
  b_m ^{-1} \Delta_m ^{-1/2} \mathbb{E}\left[ \left|J_i(3)^{3} e_{12,ib_m}^{m}\right| \right] &\leq&  b_m ^{-1} \Delta_m ^{-1/2} \mathbb{E}\left[ \left|J_i(3)\right|^{4}  \right]^{3/4} \mathbb{E}\left[ \left|e_{12,ib_m}^{m}\right|^{4}  \right]^{1/4} \\
  &\leq& C b_m ^{-\frac{3}{2} } l_m^{\frac{3}{4}} \Delta_m ^{- 3 \varpi_2} \\ 
  &\leq& C \Delta_m ^{1/4 + \upsilon}
  ,
\end{eqnarray*}
for some $\upsilon > 0$.
Similarly, we can bound all other terms on the right hand side of \eqref{fourth-term-Bdiff} by $C \Delta_m ^{1/4 + \upsilon}$ for some $\upsilon > 0$.
Furthermore, we can bound all rest terms on the right hand side of \eqref{B-disc}  by $C \Delta_m ^{1/4 + \upsilon}$ for some $\upsilon > 0$.
Thus, we have
\begin{eqnarray}\label{D1}
  \mathcal{D}_{m,1} \xrightarrow[]{p} 0
  .
\end{eqnarray}

Consider $\mathcal{D}_{m,2}$.
Simple algebra shows that
\begin{equation*}
  \mathcal{D}_{m,2} =  \bar{\mathcal{D}}_{m,1} + \bar{\mathcal{D}}_{m,2} + \bar{\mathcal{D}}_{m,3} + \bar{\mathcal{D}}_{m,4}  ,
\end{equation*}
where
\begin{eqnarray*}
  \bar{\mathcal{D}}_{m,1} &=& b_m \Delta_m ^{3/4} \sum_{i=0}^{N_m-1} \mathcal{Q}_{1,ib_m}^{m}
  ,\quad \bar{\mathcal{D}}_{m,2} = b_m \Delta_m ^{3/4} \sum_{i=0}^{N_m-1} \mathcal{Q}_{2,ib_m}^{m} - \mathbb{E}\left[ \mathcal{Q}_{2,ib_m}^{m} | \mathcal{K}_{ib_m}^{m}  \right], \\
  \bar{\mathcal{D}}_{m,3} &=& b_m \Delta_m ^{3/4} \sum_{i=0}^{N_m-1} \mathbb{E}\left[ \mathcal{Q}_{2,ib_m}^{m} | \mathcal{K}_{ib_m}^{m}  \right] 
  ,\quad \bar{\mathcal{D}}_{m,4} = b_m \Delta_m ^{3/4} \sum_{i=0}^{N_m-1} B_{ib_m}^{c,m} - \hat{B}_{ib_m}^{c,m}, \\
  \mathcal{Q}_{1,ib_m}^{m} &=&  f(\bSigma_{ib_m}^{m}+e_{ib_m}^{m,*}) - f(\bSigma_{ib_m}^{m}) - \sum_{x=1}^{2}  \partial_{1x}f(\bSigma_{ib_m}^{m}) e_{1x,i b_m}^{m,*}\\
  &&  - \sum_{x,y=1}^{2}  \partial_{1x,1y}^{2}f(\bSigma_{ib_m}^{m}) e_{1x,i b_m}^{m,*} e_{1y,i b_m}^{m,*}, \\
  \mathcal{Q}_{2,ib_m}^{m} &=& \sum_{x,y=1}^{2}  \partial_{1x,1y}^{2}f(\bSigma_{ib_m}^{m}) \left[ e_{1x,i b_m}^{m,*} e_{1y,i b_m}^{m,*} - \left( 2 b_m \Delta_m ^{1/2} \right)^{-1} \Xi(\bSigma_{ib_m}^{m},\bvartheta_{i,b_m}^{m})_{x,y} \right].
\end{eqnarray*}
By Taylor's theorem, we have
\begin{eqnarray*}
  \mathbb{E}\left[ \left| \mathcal{Q}_{1,ib_m}^{m} \right| \right] &\leq& C \mathbb{E}\left[ \left( \left| \bSigma_{12,ib_m}^{m} \right| + \left| \hat{\bSigma}_{12,ib_m}^{c,m,*} \right|   \right) \left| e_{11,ib_m}^{m,*} \right|^{3} + \left| e_{12,ib_m}^{m,*} \right|^{3}   \right] \\
  &\leq& C \mathbb{E}\left[ \left| e_{11,ib_m}^{m,*} \right|^{3} + \left| e_{12,ib_m}^{m,*} \right|^{3} + \left| e_{11,ib_m}^{m,*} \right|^{3} \left| e_{12,ib_m}^{m,*} \right|  \right] \\
  &\leq& C \mathbb{E}\left[ \left| e_{11,ib_m}^{m,*} \right|^{3} \right] + \mathbb{E}\left[ \left| e_{12,ib_m}^{m,*} \right|^{3}  \right] + E\left| \left| e_{11,ib_m}^{m,*} \right|^{4} \right]^{3/4} \mathbb{E}\left[ \left| e_{12,ib_m}^{m,*} \right|^{4}  \right]^{1/4} \\
  &\leq& C b_m \Delta_m
  ,
\end{eqnarray*}
Then, we have $\bar{\mathcal{D}}_{m,1} \xrightarrow[]{p} 0$.
By Burkholder-Davis-Gundy inequality, we have
\begin{eqnarray*}
  \mathbb{E}\left[ \left|\bar{\mathcal{D}}_{m,2} \right|^{2}  \right] &\leq& C b_m^{2} \Delta_m ^{3/2} \sum_{i=0}^{N_m-1}  \mathbb{E}\left[ \left( \mathcal{Q}_{2,ib_m}^{m} \right)^2  \right] - \mathbb{E}\left[ \mathbb{E}\left[ \mathcal{Q}_{2,ib_m}^{m} | \mathcal{K}_{ib_m}^{m}  \right]^2 \right] \\
  &\leq& C b_m^2 \Delta_m ^{3/2} \sum_{i=0}^{N_m-1} \left( \mathbb{E}\left[ \left| e_{11,ib_m}^{m,*} \right|^{4} \right] + \mathbb{E}\left[ \left| e_{11,ib_m}^{m,*} \right|^{2} \left| e_{12,ib_m}^{m,*} \right|^{2} \right] +b_m ^{-2} \Delta_m ^{-1} \right)  \\
  &\leq& C b_m^2 \Delta_m ^{3/2} \sum_{i=0}^{N_m-1} \left( \mathbb{E}\left[ \left| e_{11,ib_m}^{m,*} \right|^{4} \right] + \mathbb{E}\left[ \left| e_{11,ib_m}^{m,*} \right|^{4} \right]^{1/2} \mathbb{E}\left[ \left| e_{12,ib_m}^{m,*} \right|^{4} \right]^{1/2} +b_m ^{-2} \Delta_m ^{-1} \right)  \\
  &\leq& C \Delta_m ^{\kappa - \frac{1}{2} }
  ,
\end{eqnarray*}
where the second inequality is due to the fact that $\bSigma$ and $\bvartheta$ are locally bounded and the third and fourth inequalities are due to H\"older's inequality and Lemma \ref{lemma:e}, respectively.
Thus, we have $\bar{\mathcal{D}}_{m,2} \xrightarrow[]{p} 0$.
By Lemma \ref{lemma:e}, we have
\begin{eqnarray*}
  \mathbb{E}\left[ \left|\bar{\mathcal{D}}_{m,3}\right|  \right] &\leq& C b_m \Delta_m ^{3/4} \sum_{i=0}^{N_m-1} \sum_{x=1}^{2} \mathbb{E}\left[ \left| \mathbb{E}\left[ e_{11,i b_m}^{m,*} e_{1x,i b_m}^{m,*} - \left( 2 b_m \Delta_m ^{1/2} \right)^{-1} \Xi(\bSigma_{ib_m}^{c,m},\bvartheta_{i,b_m}^{m})_{1x} | \mathcal{K}_{ib_m}^{m} \right] \right| \right] \\
  &\leq& C \Delta_m ^{\upsilon} \quad \text{for some} \quad \upsilon > 0,
\end{eqnarray*}
where the first inequality is due to the fact that $\bSigma$ is locally bounded.
Therefore, we have $\bar{\mathcal{D}}_{m,3} \xrightarrow[]{p} 0$.
Similar to \eqref{D1}, we can show that $\bar{\mathcal{D}}_{m,4} \xrightarrow[]{p} 0$.
Thus, we have
\begin{equation}\label{D2}
  {\mathcal{D}}_{m,2} \xrightarrow[]{p} 0.
\end{equation}

Consider $\mathcal{D}_{m,3}$.
Since $\beta^c$ is locally bounded, we have
\begin{eqnarray*}
  \mathbb{E}\left[ \left|\Delta_m ^{-1/4} \int_{N_m b_m \Delta_m }^{1}  \beta_{s }^{c}  ds \right|  \right] &\leq& C b_m \Delta_m ^{3/4}
  .
\end{eqnarray*}
Using It\^o's lemma, we have 
\begin{eqnarray*} 
	&&  \Delta_m ^{-1/4} \sum ^{N_m -1}_{i=0}   \int^{(i+1)b_m\Delta_m}_{ib_m\Delta_m} \{ \beta^c_{ib_m\Delta_m} -\beta^c_{s} \} ds \\
	&& = - \Delta_m ^{-1/4}  \sum ^{N_m -1}_{i=0}   \int^{(i+1)b_m\Delta_m}_{ib_m\Delta_m} (t_{(i+1)b_m}-s)d\beta_s^c \\
	&& = - \Delta_m ^{-1/4} \sum ^{N_m -1}_{i=0}   \int^{(i+1)b_m\Delta_m}_{ib_m\Delta_m} (t_{(i+1)b_m}-s) (\mu_{\beta,s}ds+ \sigma_{\beta,s} dW_s).
\end{eqnarray*}
Using It\^o's lemma and It\^o's isometry, we can show
\begin{eqnarray*} 
	&& \mathbb{E}\left[\left( \Delta_m ^{-1/4} \sum ^{N_m -1}_{i=0}  \int^{(i+1)b_m\Delta_m}_{ib_m\Delta_m} (t_{(i+1)b_m}-s) \sigma_{\beta,s} dW_s \right)^2 \right] \\
  && = \mathbb{E}\left[ \Delta_m ^{-1/4} \sum ^{N_m -1}_{i=0}  \int^{(i+1)b_m\Delta_m}_{ib_m\Delta_m} (t_{(i+1)b_m}-s)^2 \sigma_{\beta,s}^2 ds \right] \\
  && \leq C \left( b_m  \Delta_m ^{3/4} \right) ^2
  .
\end{eqnarray*}
Also, we have
\begin{equation*} 
	\mathbb{E}\left[ \Delta_m ^{-1/4} \left|\sum ^{N_m -1}_{i=0}  \int^{(i+1)b_m\Delta_m}_{ib_m\Delta_m} (t_{(i+1)b_m}-s) \mu_{\beta,s}ds\right|  \right] \leq C b_m \Delta_m ^{3/4}
\end{equation*}
Thus, we have
\begin{equation}  \label{D3}
  {\mathcal{D}}_{m,3} \xrightarrow[]{p} 0.
\end{equation}
$\blacksquare$

\begin{lemma}\label{lemma-D4}
  As $m\rightarrow\infty$, we have $\mathbb{E}\left[ \left|{\mathcal{D}}_{m,4}(p)\right|  \right] \leq C/\sqrt{p} $.
\end{lemma}
\textbf{Proof of Lemma \ref{lemma-D4}.}
We have
\begin{equation*}
  \mathcal{D}_{m,4}(p) = \hat{\mathcal{D}}_{m,1}(p) +  \hat{\mathcal{D}}_{m,2}(p)
  ,
\end{equation*}
where
\begin{eqnarray*}
  \hat{\mathcal{D}}_{m,1}(p) &=&  b_m \Delta_m ^{3/4} \sum_{i=0}^{N_m-1} \Biggl( \sum_{x=1}^{2}  \partial_{1x}f(\bSigma_{ib_m}^{m}) \left(  M'(p)_{1x,i b_m }^{m} + \xi_{1x,i b_m }^{m,1} + \xi_{1x,i b_m }^{m,2} \right) \\
  &&- \mathbb{E}\left[ \sum_{x=1}^{2}  \partial_{1x}f(\bSigma_{ib_m}^{m})  \left( M'(p)_{1x,i b_m }^{m} + \xi_{1x,i b_m }^{m,1} + \xi_{1x,i b_m }^{m,2} \right) \Big| \mathcal{K}_{ib_m}^{m}  \right] \Biggl),  \\
  \hat{\mathcal{D}}_{m,2}(p) &=& b_m \Delta_m ^{3/4} \sum_{i=0}^{N_m-1} \mathbb{E}\left[ \sum_{x=1}^{2}  \partial_{1x}f(\bSigma_{ib_m}^{m})  \left( M'(p)_{1x,i b_m }^{m} + \xi_{1x,i b_m }^{m,1} + \xi_{1x,i b_m }^{m,2} \right) \Big| \mathcal{K}_{ib_m}^{m}  \right] 
  .
\end{eqnarray*}
By Burkholder-Davis-Gundy inequality, we have
\begin{eqnarray*}
  \mathbb{E}\left[ \left( \hat{\mathcal{D}}_{m,1}(p) \right)^2  \right] &\leq& C b_m ^2 \Delta_m ^{3/2} \sum_{i=0}^{N_m-1} \sum_{x=1}^{2}  \mathbb{E}\left[ \left(  M'(p)_{1x,i b_m }^{m} \right)^2  + \left( \xi_{1x,i b_m }^{m,1} \right)^2  + \left( \xi_{1x,i b_m }^{m,2} \right)^2  \right] \\
  &\leq& C b_m ^2 \Delta_m ^{3/2} \sum_{i=0}^{N_m-1} \sum_{x=1}^{2} (p^{-1} b_m ^{-1} \Delta_m ^{-1/2} + b_m \Delta_m  + \Delta_m ^{2\tau(v-1)} + \Delta_m ^{\kappa - 3\tau}) \\ 
  &\leq& \frac{C}{p} 
  ,
\end{eqnarray*}
where the second inequality is due to Lemmas \ref{negligible-xi} and \ref{lemma:M}.
By Lemmas \ref{negligible-xi} and \ref{lemma:M}, we have
\begin{eqnarray*}
  \mathbb{E}\left[ \left| \hat{\mathcal{D}}_{m,2}(p) \right|  \right] &\leq& C b_m \Delta_m ^{3/4} \sum_{i=0}^{N_m-1} \sum_{x=1}^{2}  \mathbb{E}\left[ \left| \mathbb{E}\left[  M'(p)_{1x,i b_m }^{m}   +  \xi_{1x,i b_m }^{m,1}  +  \xi_{1x,i b_m }^{m,2} | \mathcal{K}_{ib_m}^{m}  \right] \right| \right] \\
  &\leq& C b_m  \Delta_m ^{3/4} \sum_{i=0}^{N_m-1} \sum_{x=1}^{2} \mathbb{E}\left[ \varPsi_{i,2}^{m} (p^{-1} \Delta_m ^{1/2}  + b_m \Delta_m  + \Delta_m ^{\tau(v-1)} + \Delta_m ^{(v+\frac{1}{2}) \varsigma - \tau} ) \right] \\ 
  &\leq& C ( b_m \Delta_m ^{3/4} + \Delta_m ^{\tau(v-1) - 1/4}) \\
  &\leq& C \Delta_m ^{\upsilon}
  ,
\end{eqnarray*}
for some positive $\upsilon$.
Thus, for sufficiently large $m$, we have
\begin{equation*}
  \mathbb{E}\left[ \left| \mathcal{D}_{m,4}(p) \right|  \right] \leq \mathbb{E}\left[ \left| \hat{\mathcal{D}}_{m,1}(p) \right|  \right] + \mathbb{E}\left[ \left| \hat{\mathcal{D}}_{m,2}(p) \right|  \right] \leq  \mathbb{E}\left[ \left( \hat{\mathcal{D}}_{m,1}(p) \right)^2  \right]^{1/2} + \mathbb{E}\left[ \left| \hat{\mathcal{D}}_{m,1}(p) \right|  \right] \leq \frac{C}{\sqrt{p}} ,
\end{equation*}
where the first and second inequality is due to triangular inequality and H\"older's inequality, respectively.
$\blacksquare$

\begin{lemma}\label{lemma-D5}
  For any fixed $p\geq 2$, the sequence $\mathcal{D}_{m,5}(p)$ of processes converges $\mathcal{F}_{\infty}$-stably in law to the process
  \begin{equation*}
    Z(p) = \int_{0}^{1} \mathcal{R}(p)_s d\tilde{Z}_s
    ,
  \end{equation*}
  where $\tilde{Z}$ is a standard Brownian motion independent of $\mathcal{F}$, $\mathcal{R}(p)_s$ is the square root of
  \begin{eqnarray*}
    && \mathcal{R}(p)_s^2 = \frac{2}{\psi_0^2} \left( \frac{p \Phi_{00} - \bar{\Phi}_{00}}{p+2} \frac{{C_k} q_s^2}{\sigma_s^2} +   \frac{p \Phi_{01} - \bar{\Phi}_{01}}{p+2} \frac{A_{1,s}}{{C_k}}  + \frac{p \Phi_{11} - \bar{\Phi}_{11}}{p+2} \frac{A_{2,s}}{{C_k}^3}    \right) ,
  \end{eqnarray*}
  and
  \begin{eqnarray*}
    A_{1,s} &=&  \frac{\vartheta_{1,s}^2 R_{11} q_s^2 - 2 \beta_{s}^{c} \vartheta_{1,s}\vartheta_{2,s} R_{12} + \vartheta_{2,s}^2 R_{22}}{\sigma_s^2} + \vartheta_{1,s}^2 R_{11} (\beta_{s}^{c})^2 , \\
    A_{2,s} &=&  \frac{\vartheta_{1,s}^{2}}{\sigma_{s}^{4}} \left( 2(\beta_{s}^{c})^2 \vartheta_{1,s}^{2} R_{11}^{2} - 4 \beta_{s}^{c} \vartheta_{1,s} \vartheta_{2,s} R_{11} R_{12} + \vartheta_{2}^{2} (R_{11}R_{22} + R_{12}^2) \right) 
    .
  \end{eqnarray*}
\end{lemma}
\textbf{Proof of Lemma \ref{lemma-D5}.}
Let
\begin{eqnarray*}
  && \tilde{I}_{p,j}^{m} =  \left[  \frac{j}{L(m,p)} \right]    b_m, \quad \hat{I}_{p,j}^{m} = j-1 -  \left[ \frac{j-1}{L(m,p)} \right]   L(m,p), \quad \bar{I}_{p,j}^{m} = \tilde{I}_{p,j}^{m} + (p+2)k_m \hat{I}_{p,j}^{m} , \\
  && I(m,p) = N_m L(m,p), \quad \tilde{\mathcal{H}}(p)_{j}^{m} = \mathcal{K}_{\bar{I}_{p,j}^{m}}^{m} , \quad \tilde{\eta}(p)_{j}^{m} = \sum_{x=1}^{2}  \partial_{1x}f(\bSigma_{\tilde{I}_{p,j}^{m}}^{c,m}) \hat{\eta}(p)_{1x,\hat{I}_{p,j}^{m}}^{m,\tilde{I}_{p,j}^{m}}
  .
\end{eqnarray*}
Then, $\tilde{\eta}(p)_{j}^{m}$ is a martingale difference sequence with respect to a filtration $\tilde{\mathcal{H}}(p)_{j}^{m}$ and we have
\begin{equation*}
  \mathcal{D}_{m,5}(p) = b_m \Delta_m ^{3/4} \sum_{j=1}^{I(m,p)} \tilde{\eta}(p)_{j}^{m}
  .
\end{equation*}
To prove Lemma \ref{lemma-D5}, it suffices to show the following three convergences:
\begin{eqnarray}
  && b_m ^{2} \Delta_m ^{3/2} \sum_{j=1}^{I(m,p)} \mathbb{E}\left[ \left( \tilde{\eta}(p)_{j}^{m} \right)^2 | \tilde{\mathcal{H}}(p)_{j-1}^{m} \right] \xrightarrow[]{p} \int_{0}^{1} \mathcal{R}(p)_{s}^{2} ds , \label{tri-MDS-2} \\
  && b_m ^{4} \Delta_m ^{3} \sum_{j=1}^{I(m,p)} \mathbb{E}\left[ \left( \tilde{\eta}(p)_{j}^{m} \right)^4 | \tilde{\mathcal{H}}(p)_{j-1}^{m} \right] \xrightarrow[]{p} 0 , \label{tri-MDS-4} \\
  && b_m \Delta_m ^{3/4} \sum_{j=1}^{I(m,p)} \mathbb{E}\left[ \tilde{\eta}(p)_{j}^{m} \Delta(V,p)_{j}^{m}   | \tilde{\mathcal{H}}(p)_{j-1}^{m} \right] \xrightarrow[]{p} 0 \quad \text{for any } V \in \mathcal{M}, \label{tri-MDS-ortho}
\end{eqnarray}
where $\mathcal{M} = \mathcal{M}_1 \cup \left\lbrace W \right\rbrace$ and $\mathcal{M}_1$ is the class of all bounded $(\mathcal{F}_t)$-martingales orthogonal to $W$.
Consider \eqref{tri-MDS-2}.
The left hand side of \eqref{tri-MDS-2} is $H(p)_{1}^{m} + H(p)_{2}^{m} - H(p)_{3}^{m}$, where
\begin{eqnarray*}
  H(p)_{1}^{m} &=&  \frac{b_m ^{2} \Delta_m ^{3/2}}{(b_m  - 2 k_m)^2 \Delta_m ^2 k_m^2 \psi_0^2}  \sum_{j=1}^{I(m,p)} \sum_{x,y=1}^{2} \partial_{1x}f(\bSigma_{\tilde{I}_{p,j}^{m}}^{m}) \partial_{1y}f(\bSigma_{\tilde{I}_{p,j}^{m}}^{m})  \\
  && \times \mathbb{E}\left[ \zeta(p)_{1x,\bar{I}_{p,j}^{m}}^{m} \zeta(p)_{1y,\bar{I}_{p,j}^{m}}^{m} - \varXi(p)_{1x,1y,\bar{I}_{p,j}^{m}}^{m} | \tilde{\mathcal{H}}(p)_{j-1}^{m} \right], \\
  H(p)_{2}^{m} &=&  b_m ^{2} \Delta_m ^{3/2} \sum_{j=1}^{I(m,p)} \sum_{x,y=1}^{2} \partial_{1x}f(\bSigma_{\tilde{I}_{p,j}^{m}}^{m}) \partial_{1y}f(\bSigma_{\tilde{I}_{p,j}^{m}}^{m})  \varXi(p)_{1x,1y,\bar{I}_{p,j}^{m}}^{m}, \\
  H(p)_{3}^{m} &=&  b_m ^{2} \Delta_m ^{3/2} \sum_{j=1}^{I(m,p)} \sum_{x,y=1}^{2} \partial_{1x}f(\bSigma_{\tilde{I}_{p,j}^{m}}^{m}) \partial_{1y}f(\bSigma_{\tilde{I}_{p,j}^{m}}^{m}) \bar{\eta}(p)_{1x,\hat{I}_{p,j}^{m}}^{m,\tilde{I}_{p,j}^{m}} \bar{\eta}(p)_{1y,\hat{I}_{p,j}^{m}}^{m,\tilde{I}_{p,j}^{m}}
  .
\end{eqnarray*}
By Lemma \ref{lemma:M} and the fact that $\bSigma$ is locally bounded, we have
\begin{eqnarray}\label{H1}
  \mathbb{E}\left[ \left| H(p)_{1}^{m} \right|  \right] &\leq&  C \Delta_m ^{1/2} \sum_{j=1}^{I(m,p)} \sum_{x,y=1}^{2} \mathbb{E}\left[ \left| \partial_{1x}f(\bSigma_{\tilde{I}_{p,j}^{m}}^{m}) \partial_{1y}f(\bSigma_{\tilde{I}_{p,j}^{m}}^{m}) \varPsi_{\bar{I}_{p,j}^{m},1}^{m} \Delta_m ^{1/4}  \right|  \right] \nonumber\\
  &\leq& C \Delta_m ^{1/4}
  ,
\end{eqnarray}
Using Lemma \ref{lemma:zeta} and the fact that $\bSigma$ is locally bounded, we have
\begin{equation}\label{H3}
  \mathbb{E}\left[ \left| H(p)_{3}^{m} \right| \right] \leq C  b_m ^{2} \Delta_m ^{3/2}  \sum_{j=1}^{I(m,p)} \sum_{x,y=1}^{2} p^2 b_m ^{-2} \leq C p \Delta_m .
\end{equation}
By Riemann integration, we have
\begin{equation}\label{H2}
  H(p)_2^{m} \rightarrow \int_{0}^{1} \mathcal{R}(p)_{s}^{2} ds
  .
\end{equation}
Then, \eqref{tri-MDS-2} follows from \eqref{H1}, \eqref{H3}, and \eqref{H2}.
Using the same arguments as that of proofs of (A.47) and (A.48) in \citet{jacod2019estimating}, we can show \eqref{tri-MDS-4} and \eqref{tri-MDS-ortho}, respectively.
$\blacksquare$

\textbf{Proof of Theorem \ref{Theorem-1}.}
By Lemma \eqref{lemma-D123} and \eqref{lemma-D4}, we have
\begin{equation*}
  \lim_{p\rightarrow\infty} \limsup_{m \rightarrow \infty} \mathbb{P}\left(\left|\Delta_m ^{-1/4} \left( RIB - \int_{0}^{1} \beta_{t}^{c} dt \right) - \mathcal{D}_{m,5}(p) \right| \right) = 0
  ,
\end{equation*}
for all $\varepsilon > 0$.
In addition, with a fixed sample path of Brownian motion $\tilde{W}$ independent of $\mathcal{F}$ we have $\mathcal{R}(p)_s(\omega)^2 \rightarrow \mathcal{R}_s^2$ for all $s$ and $\omega$, and $\mathcal{R}(p)_s^2 \leq C$.
Thus, we have $Z(p) \xrightarrow[]{p} \int_{0}^{1} \mathcal{R}_s d\tilde{Z}_{s}$.
Finally, Lemma \ref{lemma-D5} concludes that
\begin{equation*}
  m ^{1/4}(RIB_{1}-I \beta_1) \rightarrow \int_{0}^{1} \mathcal{R}_s d\tilde{Z}_s \quad \mathcal{F}_{\infty}\text{-stably as }  m \rightarrow \infty
  .
\end{equation*}
$\blacksquare$

\subsection{Proof of Proposition \ref{prop-AsympVar}}

\begin{lemma}\label{lemma:SpotNoise}
  Under Assumptions \ref{assumption-noise} and \ref{assumption-formal}, we have
  \begin{eqnarray*}
    \mathbb{E}\left[ \left| \hat{\bvartheta}_{xy,i}^{m} - \bvartheta_{xy,i}  \right|^{w}  \right] \leq C_w \left( \Delta_m ^{(\frac{1}{4} - \tau) w} + \Delta_m ^{(1-\kappa - 2\tau)w/2} + \Delta_m ^{1 - \varsigma -\varpi_2 w} \right) 
  \end{eqnarray*}
  for any $1 \leq w \leq 4$ and $x,y \in \left\lbrace 1,2 \right\rbrace$.
\end{lemma}
\textbf{Proof of Lemma \ref{lemma:SpotNoise}.}
First, we consider $x=y=1$.
By Lemma \ref{discontinuous-error}, we have
\begin{eqnarray*}
  \mathbb{E}\left[ \left| \hat{\bvartheta}_{11,i}^{m} - \bvartheta_{11,i}  \right|^{w} \right] &\leq& C_w \left( \mathbb{E}\left[ \left|\hat{\bvartheta}_{11,i}^{m} - \hat{\bvartheta}_{11,i}^{c,m}\right|^{w} \right]  + \mathbb{E}\left[  \left|\hat{\bvartheta}_{11,i}^{c,m} - \bvartheta_{11,i}^{m} \right|^{w}   \right] \right) \\
  &\leq& C_w \left( \Delta_m ^{1 - \varsigma -\varpi_2 w} +  \mathbb{E}\left[  \left|\hat{\bvartheta}_{11,i}^{c,m} - \bvartheta_{11,i}^{m} \right|^{w}   \right] \right) 
  ,
\end{eqnarray*}
where the first inequality is due to Jensen's inequality.
Therefore, it is enough to show that
\begin{equation}\label{SpotCtnNoise-spot}
  \mathbb{E}\left[  \left|\hat{\bvartheta}_{11,i}^{c,m} - \bvartheta_{11,i}^{m} \right|^{w}   \right] \leq C_w \left( \Delta_m ^{(\frac{1}{4} - \tau)w} + \Delta_m ^{(1-\kappa - 2\tau)w/2} \right) 
  .
\end{equation}
Simple algebra shows that
\begin{align}\label{SpotNoiseDecomp}
  \hat{\bvartheta}_{11,i}^{c,m} - \bvartheta_{11,i}^{m} =&  \frac{1}{b_m - 6k_m}  \sum_{d = - k'_m}^{k'_m} \sum_{j=i}^{i+b_m - 6 k_m}  \left(  (Y_{1,j}^{c,m} - \bar{Y}_{1,j+2k_m}^{m} ) (Y_{1,j+|d|}^{c,m} - \bar{Y}_{1,j+4k_m}^{m} ) - \epsilon_{1,j}^{m} \epsilon_{1,j+|d|}^{m} \right) \nonumber\\
  &+ \frac{1}{b_m - 6k_m} \sum_{d = - k'_m}^{k'_m}   \sum_{j=i}^{i+ b_m - 6 k_m} ( \vartheta_{1,j}^{m} \vartheta_{1,j+|d|}^{m}  - (\vartheta_{1,i}^{m})^2 ) \chi_{1,j} \chi_{1,j+|d|}   \nonumber\\
  &+ \frac{(\vartheta_{1,j}^{m})^2}{b_m - 6k_m}  \sum_{d = - k'_m}^{k'_m}   \sum_{j=i}^{i+ b_m - 6 k_m} (\chi_{1,j} \chi_{1,j+|d|} - r_{11}(|d|))  \nonumber \\
  &+ \frac{(b_m - 6k_m)\vartheta_{1,j}^{m}}{b_m - 6k_m +1}  \sum_{|d| \geq k'_m + 1} r_{11}(|d|)
  .
\end{align}
For the first term on the right hand side of \eqref{SpotNoiseDecomp}, we have
\begin{align}\label{SpotNoise-D1}
  & \quad \mathbb{E}\left[ \left| \frac{1}{b_m - 6k_m}  \sum_{d = - k'_m}^{k'_m} \sum_{j=i}^{i+b_m - 6 k_m}   (Y_{1,j}^{c,m} - \bar{Y}_{1,j+2k_m}^{m} ) (Y_{1,j+|d|}^{c,m} - \bar{Y}_{1,j+4k_m}^{m} ) - \epsilon_{1,j}^{m} \epsilon_{1,j+|d|}^{m} \right|^{w}  \right] \nonumber\\
  & \leq C_w k'^{w-1}_m b_m ^{-1}   \sum_{d = - k'_m}^{k'_m} \sum_{j=i}^{i+b_m - 6 k_m} \big( \mathbb{E}\left[ \left|(X_{1,j}^{c,m} - \bar{X}_{1,j+2k_m}^{c,m} ) (X_{1,j+|d|}^{c,m} - \bar{X}_{1,j+4k_m}^{c,m} )\right|^{w}   \right]   \nonumber\\
  & \quad + \mathbb{E}\left[ \left|(X_{1,j}^{c,m} - \bar{X}_{1,j+2k_m}^{c,m} ) ( \epsilon_{1,j+|d|}^{m} - \bar{\epsilon}_{1,j+4k_m}^{m} )\right|^{w}   \right] + \mathbb{E}\left[ \left|(X_{1,j+|d|}^{c,m} - \bar{X}_{1,j+4k_m}^{c,m} ) ( \epsilon_{1,j}^{m} - \bar{\epsilon}_{1,j+2k_m}^{m} )\right|^{w}   \right] \nonumber\\
  & \quad + \mathbb{E}\left[ \left|\epsilon_{1,j}^{m} \bar{\epsilon}_{1,j+4k_m}^{m}\right|^{w}  \right] + \mathbb{E}\left[ \left|\epsilon_{1,j+|d|}^{m} \bar{\epsilon}_{1,j+2k_m}^{m}\right|^{w}  \right] + \mathbb{E}\left[ \left|\bar{\epsilon}_{1,j+2k_m}^{m} \bar{\epsilon}_{1,j+4k_m}^{m}\right|^{w}  \right] \big).
\end{align}
By H\"older's inequality and the fact that for any $l \in [1,4k_m]$, we have
\begin{equation*}
  \mathbb{E}\left[ | X_{1,j}^{c,m} - \bar{X}_{1,j+l}^{c,m} |^{w} \right] \leq C_w (k_m \Delta_m )^{w/2} \quad \text{and} \quad \mathbb{E}\left[ \left|\epsilon_{1,j}^{m} - \bar{\epsilon}_{1,j+l}^{m}\right|^{w}  \right] \leq C_{w} 
  ,
\end{equation*}
where the first, second, and third summands on the right hand side of \eqref{SpotNoise-D1} are bounded by $C_w \Delta_m^{w/4}$.
Furthermore, we have
\begin{eqnarray*}
  \mathbb{E}\left[ \left|\bar{\epsilon}_{1,j}^{m}\right|^{w}  \right] &\leq& C_w \left( \mathbb{E}\left[ \left|k_m^{-1} \sum_{j=0}^{k_m - 1} (\vartheta_{1,i+j} - \vartheta_{1,i}) \chi_{1,i+j}  \right|^{w}  \right] + \mathbb{E}\left[ \left|k_m^{-1} \vartheta_{1,i} \sum_{j=0}^{k_m - 1} \chi_{1,i+j}  \right|^{w}  \right] \right) \\
  &\leq& C_w k_m^{-1}   \sum_{j=0}^{k_m - 1} \mathbb{E}\left[ \left| (\vartheta_{1,i+j} - \vartheta_{1,i}) \chi_{1,i+j}  \right|^{w}  \right] + C_w  \mathbb{E}\left[ \left| \vartheta_{1,i} \bar{\chi}_{1,i}^{m}  \right|^{w}  \right]  \\
  &\leq& C_w k_m^{-1} \sum_{j=0}^{k_m - 1} \mathbb{E}\left[ \left| \vartheta_{1,i+j} - \vartheta_{1,i}\right|^{w} \right] \mathbb{E}\left[ \left| \chi_{1,i+j}  \right|^{w}  \right] + C_w  \mathbb{E}\left[ \left| \vartheta_{1,i} \right|^{w} \right] \mathbb{E}\left[ \left| \bar{\chi}_{1,i}^{m}  \right|^{w}  \right] \\
  &\leq& C_w \Delta_m ^{w/4}
  ,
\end{eqnarray*}
where the first and second inequalities are due to Jensen's inequality, and the fourth inequality is due to Lemma \ref{chi-property}(a).
By H\"older's inequality, the fourth, fifth, and sixth summands on the right hand side of \eqref{SpotNoise-D1} are bounded by $C_w \Delta_m^{w/4}$, and thus we have
\begin{align}\label{SpotNoise-D1Done}
  & \mathbb{E}\left[ \left| \frac{1}{b_m - 6k_m}  \sum_{d = - k'_m}^{k'_m} \sum_{j=i}^{i+b_m - 6 k_m}   (Y_{1,j}^{c,m} - \bar{Y}_{1,j+2k_m}^{m} ) (Y_{1,j+|d|}^{c,m} - \bar{Y}_{1,j+4k_m}^{m} ) - \epsilon_{1,j}^{m} \epsilon_{1,j+|d|}^{m} \right|^{w}  \right] \nonumber\\
  & \leq C_w \Delta_m ^{(\frac{1}{4} -\tau)w}
  .
\end{align}
For the second term on the right hand side of \eqref{SpotNoiseDecomp}, we have
\begin{align}\label{SpotNoise-D2Done}
  & \quad \mathbb{E}\left[ \left|\frac{1}{b_m - 6k_m} \sum_{d = - k'_m}^{k'_m}   \sum_{j=i}^{i+ b_m - 6 k_m} ( \vartheta_{1,j}^{m} \vartheta_{1,j+|d|}^{m}  - (\vartheta_{1,i}^{m})^2 ) \chi_{1,j} \chi_{1,j+|d|}\right|^{w}  \right] \nonumber\\
  & \leq C_w k'^{w-1}_{m} b_m ^{-1} \sum_{d = - k'_m}^{k'_m}   \sum_{j=i}^{i+ b_m - 6 k_m} \mathbb{E}\left[ \left|( \vartheta_{1,j}^{m} \vartheta_{1,j+|d|}^{m}  - (\vartheta_{1,i}^{m})^2 ) \chi_{1,j} \chi_{1,j+|d|}\right|^{w}  \right] \nonumber\\
  & \leq C_w k'^{w-1}_{m} b_m ^{-1} \sum_{d = - k'_m}^{k'_m}   \sum_{j=i}^{i+ b_m - 6 k_m} \mathbb{E}\left[ \left|( \vartheta_{1,j}^{m} (\vartheta_{1,j+|d|}^{m} - \vartheta_{1,i}^{m})   - \vartheta_{1,i}^{m}(\vartheta_{1,j}^{m} - \vartheta_{1,i}^{m}) )\right|^{w}  \right] \nonumber\\
  & \leq C_w \Delta_m ^{(1-\kappa - 2\tau)w/2}
  ,
\end{align}
where the first and second inequalities are due to Jensen's inequality and the fact that the process $\chi$ is independent of the $\sigma$-field $\mathcal{F}_{\infty}$ and bounded finite moments of all orders, respectively.
For the third term on the right hand side of \eqref{SpotNoiseDecomp}, we can show that $(\mathcal{X}_{11,(d+1)i+j,d})_{i \in \mathbb{Z}}$ is $v$-polynomially $\rho$-mixing for any $j \in \mathbb{Z}$, where $\mathcal{X}_{11,i,d}$ is defined in Lemma \ref{chi-property}.
Then, by Theorem 1.1 in \citet{shao1995maximal}, we have for any $N > d$
\begin{eqnarray*}
  \mathbb{E}\left[ \left| \sum_{j=1}^{N} \mathcal{X}_{11,i+j,|d|}  \right|^{w}  \right] &=& \mathbb{E}\left[ \left| \sum_{j=0}^{d}  \sum_{l=0}^{[N/(d+1)]} \mathcal{X}_{11,i+(d+1)l+j,|d|} + \sum_{j=1}^{N} \mathcal{X}_{11,i+j,|d|} \right|^{w}  \right] \\
  &\leq& C_w N^{w/2} d^{w/2}
  .
\end{eqnarray*}
Therefore, we have
\begin{align}\label{SpotNoise-D3Done}
  & \quad \mathbb{E}\left[ \left|\frac{(\vartheta_{1,j}^{m})^2}{b_m - 6k_m}  \sum_{d = - k'_m}^{k'_m}   \sum_{j=i}^{i+ b_m - 6 k_m} (\chi_{1,j} \chi_{1,j+|d|} - r_{11}(|d|)) \right|^{w}  \right] \nonumber\\
  & \leq C_w b_m ^{-w} \mathbb{E}\left[ \left| \sum_{d = - k'_m}^{k'_m}   \sum_{j=i}^{i+ b_m - 6 k_m} (\chi_{1,j} \chi_{1,j+|d|} - r_{11}(|d|)) \right|^{w}  \right] \nonumber\\
  & \leq C_w b_m ^{-w} k'^{w-1}_{m}  \sum_{d = - k'_m}^{k'_m}    \mathbb{E}\left[ \left|\sum_{j=i}^{i+ b_m - 6 k_m} (\chi_{1,j} \chi_{1,j+|d|} - r_{11}(|d|)) \right|^{w}  \right] \nonumber\\
  & \leq C_w b_m ^{-w/2} k'^{3w/2}_{m} \nonumber\\
  & \leq C_w \Delta_m ^{(\frac{1}{4} -\tau)w}
  ,
\end{align}
where the first and second inequalities are due to the boundedness of $\vartheta$ and Jensen's inequality, respectively.
For the fourth term on the right hand side of \eqref{SpotNoiseDecomp}, by the fact that $r_{11}(|d|) \leq C (|d|+1)^{-v}$, we have
\begin{equation}\label{SpotNoise-D4Done}
  \mathbb{E}\left[ \left|\frac{(b_m - 6k_m)\vartheta_{1,j}^{m}}{b_m - 6k_m +1}  \sum_{|d| \geq k'_m + 1} r_{11}(|d|)\right|^{w}  \right] \leq C_w \Delta_m ^{(v-1)\tau} \leq C_w \Delta_m ^{1/4}
  .
\end{equation}
Then, \eqref{SpotCtnNoise-spot} follows from \eqref{SpotNoise-D1Done}, \eqref{SpotNoise-D2Done}, \eqref{SpotNoise-D3Done}, and \eqref{SpotNoise-D4Done}.
Similarly, we can show the statement for the other cases of $x$ and $y$.
$\blacksquare$

\textbf{Proof of Proposition \ref{prop-AsympVar}.}
Simple algebra shows that
\begin{eqnarray*}
  \mathcal{R}_{t_i}^{2} &=& \frac{2C_k}{\psi_{0}^{2}} \Biggl[ \Phi_{00}  \left( \frac{{\bSigma}_{22,i}^{m}}{{\bSigma}_{11,i}^{m}} - \frac{({\bSigma}_{12,i}^{m})^2}{({\bSigma}_{11,i}^{m})^2}  \right)     + \frac{\Phi_{01}}{C_k^{2}}  \left( \frac{{\bvartheta}_{22,i}^{m}}{{\bSigma}_{11,i}^{m}} -  \frac{2{\bSigma}_{12,i}^{m} {\bvartheta}_{12,i}^{m}  }{ ( {\bSigma}_{11,i}^{m}  )^2  }  + \frac{{\bSigma}_{22,i}^{m} {\bvartheta}_{11,i}^{m} }{( {\bSigma}_{11,i}^{m}  )^2 }   \right) \\
  &&  + \frac{\Phi_{11}}{C_k^3}  \left( \frac{2  ({\bSigma}_{12,i}^{m} {\bvartheta}_{11,i}^{m} )^2    }{({\bSigma}_{11,i}^{m} )^4} + \frac{{\bvartheta}_{11,i}^{m} {\bvartheta}_{12,i}^{m} }{({\bSigma}_{11,i}^{m} )^2} - 4 \frac{{\bSigma}_{12,i}^{m} {\bvartheta}_{11,i}^{m} {\bvartheta}_{12,i}^{m}  }{({\bSigma}_{11,i}^{m} )^3}  + \frac{({\bvartheta}_{11,i}^{m} )^2}{({\bSigma}_{11,i}^{m} )^2}  \right)     \Biggl]
  .
\end{eqnarray*}
We have
\begin{align}\label{mathR-decomp}
  &\mathbb{E}\left[ \left|\hat{\mathcal{R}}^{2,m}_{i} - \mathcal{R}_{t_i}^{2} \right| \right] \cr
  &\leq C \Biggl[  \mathbb{E}\left[ \left|\frac{\hat{\bSigma}_{22,i}^{m}}{\hat{\bSigma}_{11,i}^{m,*}} - \frac{{\bSigma}_{22,i}^{m}}{{\bSigma}_{11,i}^{m}}\right|  \right]  - \mathbb{E}\left[ \left|\frac{(\hat{\bSigma}_{12,i}^{m})^2}{(\hat{\bSigma}_{11,i}^{m,*})^2} - \frac{({\bSigma}_{12,i}^{m})^2}{({\bSigma}_{11,i}^{m})^2}\right|  \right] +  \mathbb{E}\left[ \left|\frac{\hat{\bvartheta}_{22,i}^{m}}{\hat{\bSigma}_{11,i}^{m,*}} - \frac{{\bvartheta}_{22,i}^{m}}{{\bSigma}_{11,i}^{m,*}}\right|  \right]  \nonumber\\
  & \quad + \mathbb{E}\left[ \left|\frac{\hat{\bSigma}_{12,i}^{m} \hat{\bvartheta}_{12,i}^{m}  }{ ( \hat{\bSigma}_{11,i}^{m,*}  )^2  } - \frac{{\bSigma}_{12,i}^{m} {\bvartheta}_{12,i}^{m}  }{ ( {\bSigma}_{11,i}^{m,*}  )^2  }\right|  \right] + \mathbb{E}\left[ \left|\frac{\hat{\bSigma}_{22,i}^{m} \hat{\bvartheta}_{11,i}^{m} }{( \hat{\bSigma}_{11,i}^{m,*}  )^2 }  - \frac{{\bSigma}_{22,i}^{m} {\bvartheta}_{11,i}^{m} }{( {\bSigma}_{11,i}^{m,*}  )^2 } \right|  \right] \nonumber\\
  & \quad + \mathbb{E}\left[ \left|\frac{  (\hat{\bSigma}_{12,i}^{m} \hat{\bvartheta}_{11,i}^{m} )^2    }{(\hat{\bSigma}_{11,i}^{m,*} )^4} - \frac{ ({\bSigma}_{12,i}^{m} {\bvartheta}_{11,i}^{m} )^2    }{({\bSigma}_{11,i}^{m,*} )^4}\right|  \right]  + \mathbb{E}\left[ \left|\frac{\hat{\bvartheta}_{11,i}^{m} \hat{\bvartheta}_{12,i}^{m} }{(\hat{\bSigma}_{11,i}^{m,*} )^2} - \frac{{\bvartheta}_{11,i}^{m} {\bvartheta}_{12,i}^{m} }{({\bSigma}_{11,i}^{m,*} )^2}\right|  \right]\nonumber\\
  &\quad  + \mathbb{E}\left[ \left|\frac{\hat{\bSigma}_{12,i}^{m} \hat{\bvartheta}_{11,i}^{m} \hat{\bvartheta}_{12,i}^{m}  }{(\hat{\bSigma}_{11,i}^{m,*} )^3} - \frac{{\bSigma}_{12,i}^{m} {\bvartheta}_{11,i}^{m} {\bvartheta}_{12,i}^{m}  }{({\bSigma}_{11,i}^{m,*} )^3}\right|  \right] + \mathbb{E}\left[ \left|\frac{(\hat{\bvartheta}_{11,i}^{m} )^2}{(\hat{\bSigma}_{11,i}^{m,*} )^2} - \frac{({\bvartheta}_{11,i}^{m} )^2}{({\bSigma}_{11,i}^{m,*} )^2}\right|  \right] \Biggl] 
  .
\end{align}
For the sixth term on the right hand side of \eqref{mathR-decomp}, by the boundedness of $1/\hat{\bSigma}_{11,i}^{m,*}$, ${\bSigma}_{12,i}^{m}$, and ${\bvartheta}_{11,i}^{m}$ we have
\begin{eqnarray*}
  &&\quad \mathbb{E}\left[ \left|\frac{  (\hat{\bSigma}_{12,i}^{m} \hat{\bvartheta}_{11,i}^{m} )^2    }{(\hat{\bSigma}_{11,i}^{m,*} )^4} - \frac{ ({\bSigma}_{12,i}^{m} {\bvartheta}_{11,i}^{m} )^2    }{({\bSigma}_{11,i}^{m,*} )^4}\right|  \right] \\
  &&\leq \mathbb{E}\left[ \left|\frac{  (\hat{\bSigma}_{12,i}^{m} \hat{\bvartheta}_{11,i}^{m} )^2 - ({\bSigma}_{12,i}^{m} {\bvartheta}_{11,i}^{m} )^2    }{(\hat{\bSigma}_{11,i}^{m,*} )^4} \right| \right] + \mathbb{E}\left[ \left|  \frac{ ({\bSigma}_{12,i}^{m} {\bvartheta}_{11,i}^{m} )^2    }{({\bSigma}_{11,i}^{m,*} \hat{\bSigma}_{11,i}^{m,*}  )^4 } ((\hat{\bSigma}_{11,i}^{m,*} )^4 - ({\bSigma}_{11,i}^{m,*} )^4)\right|  \right] \\
  &&\leq C \mathbb{E}\left[ \left|  (\hat{\bSigma}_{12,i}^{m} \hat{\bvartheta}_{11,i}^{m} )^2 - ({\bSigma}_{12,i}^{m} {\bvartheta}_{11,i}^{m} )^2     \right| \right] + C  \mathbb{E}\left[ \left|  ((\hat{\bSigma}_{11,i}^{m,*} )^4 - ({\bSigma}_{11,i}^{m,*} )^4)\right|  \right] \\
  &&\leq C \mathbb{E}\left[ \left|  (\hat{\bSigma}_{12,i}^{m})^2 ((\hat{\bvartheta}_{11,i}^{m})^2  - ( {\bvartheta}_{11,i}^{m} )^2) \right|\right] + C\mathbb{E}\left[ \left| ({\bvartheta}_{11,i}^{m} )^2 ((\hat{\bSigma}_{12,i}^{m})^2 - ({\bSigma}_{12,i}^{m}  )^2 )    \right| \right] \\
  &&\quad + C  \mathbb{E}\left[ \left|  ((\hat{\bSigma}_{11,i}^{m,*} )^4 - ({\bSigma}_{11,i}^{m,*} )^4)\right|  \right] \\
  &&\leq C \mathbb{E}\left[ \left|  ((\hat{\bSigma}_{12,i}^{m})^2 - ({\bSigma}_{12,i}^{m})^2) ((\hat{\bvartheta}_{11,i}^{m})^2  - ( {\bvartheta}_{11,i}^{m} )^2) \right|\right] + C\mathbb{E}\left[ \left|  (\hat{\bSigma}_{12,i}^{m})^2 - ({\bSigma}_{12,i}^{m}  )^2     \right| \right] \\
  &&\quad + C  \mathbb{E}\left[ \left|  ((\hat{\bSigma}_{11,i}^{m,*} )^4 - ({\bSigma}_{11,i}^{m,*} )^4)\right|  \right] \\
  &&\leq C \sum_{w_1=1}^{2} \sum_{w_2=0}^{2} \mathbb{E}\left[ \left|\hat{\bSigma}_{12,i}^{m} - {\bSigma}_{12,i}^{m}\right|^{w_1} \left|\hat{\bvartheta}_{11,i}^{m} - {\bvartheta}_{11,i}^{m}\right|^{w_2}   \right] + C \sum_{w=1}^{4} \mathbb{E}\left[ \left|\hat{\bSigma}_{11,i}^{m,*} - {\bSigma}_{11,i}^{m,*}\right|^{w_1} \right] \\
  &&\leq C \Delta_m ^{\upsilon} \quad \text{for some} \quad \upsilon>0  ,
\end{eqnarray*}
where the sixth inequality is due to Lemmas \ref{lemma:e}, \ref{discontinuous-error}, and \ref{lemma:SpotNoise}.
Similarly, we can bound all rest terms on the right hand side of \eqref{mathR-decomp} by $C \Delta_m ^{\upsilon}$ for some $\upsilon>0$.
Thus, we have
\begin{equation*}
  b_m \Delta_m  \sum_{i=0}^{\left[ \frac{1}{b_m\Delta_m} \right]  -1 } (\hat{\mathcal{R}}^{2,m}_{i b_m} - \mathcal{R}^2_{t_{ib_m}} ) \xrightarrow[]{0} 0
  .
\end{equation*}
Thus, using Riemann approximation, we can show that $\hat{S}_m \xrightarrow[]{p} \int_{0}^{1} \mathcal{R}^2_{t_{ib_m}} ds $.
$\blacksquare$

\subsection{Proof of Theorem \ref{Theorem-2}}

For simplicity, we denote derivatives of any given function $f$ at $x_0$ by
\begin{equation*} 
    \frac{\partial f(x_0)}{\partial x}= \left. \dfrac{\partial f\left( x\right) }{\partial x} \right| _{x=x_{0}} ,
\end{equation*}
and define
\begin{eqnarray*}
    &&\hat{L}_{n,m}(\theta)=-\frac{1}{n}\sum^n_{i=1} \{ RIB_i-\hat{h}_i(\theta) \}^2\quad \text{and} \quad \hat{s}_{n,m} (\theta)=\dfrac{\partial \hat{L}_{n,m} (\theta)}{\partial \theta}; \cr
    &&\hat{L}_{n} (\theta)=-\frac{1}{n}\sum^n_{i=1} 
    \left\{ I\beta_i-h_i(\theta) \right \}^2  \quad  \text{and} \quad \hat{s}_{n} (\theta)=\dfrac{\partial \hat{L}_{n} (\theta)}{\partial \theta}; \cr
    &&L_{n} (\theta)=-\frac{1}{n}\sum^n_{i=1} \left[ 
    \left\{ h_i(\theta_0)-h_i(\theta) \right \}^2+D^2_i  
     \right]\quad  \text{and} \quad s_{n} (\theta)=\dfrac{\partial L_{n} (\theta)}{\partial \theta}.
\end{eqnarray*}
Since the dependence of $h_i(\theta)$ on the initial value decays with the order $n^{-1}$, without loss of the generality, we suppose that $h_1(\theta_0)$ is given during the rest of the proofs.

\begin{lemma}\label{Lemma-6}
\begin{enumerate}
Under the assumption of Theorem \ref{Theorem-2}, we have 
\item [(a)] $\mathbb{E}[ I\beta_i ]=\mathbb{E}[h_i(\theta_{ 0})]$, $\sup_{i \in \mathbb{N}} \mathbb{E}[\left|I\beta_i \right|] < \infty$, and $ \sup_{i \in \mathbb{N}}\mathbb{E}[ \sup_{\theta \in \Theta} | h_i(\theta) |] < \infty$ a.s. 
 
\item [(b)]for any $j,k,l \in \{ 1, 2, \ldots, p+ p\vee q + 1 \}$,  
\begin{eqnarray*}
    && \sup_{i \in \mathbb{N}} \mathbb{E}\left[ \sup_{\theta \in \Theta} \left| \frac{\partial {{h}_i(\theta)}}{\partial {\theta_{ j}}} \right| \right]  \leq C, \quad 
    \sup_{i \in \mathbb{N}} \mathbb{E}\left[ \sup_{\theta \in \Theta} \left| \frac{\partial^2 {{h}_i(\theta)}}{\partial {\theta_{j}} {\partial {\theta_{ k}}}}\right|  \right] \leq C, \quad \text{and} \cr
    && \sup_{i \in \mathbb{N}} \mathbb{E}\left[ \sup_{\theta \in \Theta} \left| 
    \frac{\partial^3 {{h}_i(\theta)}}{\partial {\theta_{ j}} {\partial {\theta_{ k}}}{\partial {\theta_{ l}}}} \right|  \right] \leq C, 
\end{eqnarray*}
where $=(\theta_{1},\theta_{ 2}, \ldots, \theta_{ p+ p \vee q +1})=(\omega^g, \gamma_1, \ldots, \gamma_p, \alpha_1^g, \ldots,  \alpha_{p\vee q} ^g)$.
\end{enumerate}
\end{lemma}

\textbf{Proof of Lemma \ref{Lemma-6}.}
Since the proof is similar to the case where $p=q=1$, we show the statements for $p=q=1$. 
$(a)$ By Proposition \ref{Proposition-1}(b), we have 
\begin{equation*}
    \mathbb{E}[I\beta_i]=\mathbb{E}[h_i(\theta_{0})].
\end{equation*}
Using It\^o's isometry and It\^o's lemma, we have
\begin{eqnarray} \label{Equation-A.27}
    \mathbb{E}[D_i^2|\mathcal{F}_{i-1}] 
    &=& \mathbb{E}[4\nu_0^2\alpha_0^{-4} \int^i_{i-1}\{\alpha_0(i-t-\alpha_0^{-1})e^{\alpha_0(i-t)}+1\}^2 Z_t^2dt|\mathcal{F}_{i-1}] \cr
    &=& 4\nu_0^2\alpha_0^{-4} \int^i_{i-1}\{\alpha_0(i-t-\alpha_0^{-1})e^{\alpha_0(i-t)}+1\}^2 (t-i+1)dt \cr
    &\leq&  C \quad  \text{a.s.},
\end{eqnarray}
which also implies $\mathbb{E}\big[ | D_i | \big|\mathcal{F}_{i-1}\big] \leq \mathbb{E}\big[ D_i^2 \big|\mathcal{F}_{i-1}\big]^{1/2} \leq C$ a.s.
Then,  with the iterative relationship in $h_i(\theta_0)$ and $|\alpha^g_0+\gamma_0|<1$, we have
\begin{eqnarray} \label{Equation-A.28}
      \mathbb{E}[\left|h_i(\theta_{0})\right|]
     &\leq& |\omega_0^g|+|\gamma_0|\mathbb{E}[\left|D_{i-1}(\theta_{0})\right|]+|\alpha_0^g+\gamma_0|(\mathbb{E}[\left|h_{i-1}(\theta_{0})\right|]) \nonumber \\
     &\leq& C  |\omega_0^g+\gamma_0|+|\alpha_0^g+\gamma_0|\mathbb{E}[\left|h_{i-1}(\theta_{0})\right|] \nonumber \\
     &\leq& \frac{C\left|\omega_0^g+\gamma_0\right|(1-|\alpha_0^g+\gamma_0|^{i-1})}{1-|\alpha_0^g+\gamma_0|}+\left|\alpha_0^g+\gamma_0 \right|^{i-1}\mathbb{E}[\left|h_1(\theta_{0})\right|] \nonumber \\
     &\leq& 
     \frac{C \left|\omega_0^g+\gamma_0\right|}{1-|\alpha_0^g+\gamma_0|}+  \mathbb{E}[\left|h_1(\theta_{0})\right|] < \infty  \quad  \text{a.s.},
\end{eqnarray} 
for any $i$.
Then, \eqref{Equation-A.27}, \eqref{Equation-A.28}, and Proposition \ref{Proposition-1}(a) derive $ \sup_{i \in \mathbb{N}} \mathbb{E}[\left|I\beta_i\right|] < \infty$. 
Similarly, we can show
\begin{equation*}
    \sup_{i \in \mathbb{N}}\mathbb{E}[ \sup_{\theta \in \Theta} | h_i(\theta) |] < \infty  \quad  \text{a.s.}
\end{equation*}
(b) Consider the first inequality. 
Since $\beta^c_i(\theta)$ is the linear function of $\omega^g$ and $\alpha^g$, we obtain
\begin{equation*}
 \mathbb{E}\left[ \sup_{\theta \in \Theta} \left| \frac{\partial {{h}_i(\theta)}}{\partial {\theta_{ j}}} \right| \right]  \leq C,
\end{equation*}
for $j=1,2$. 
For $j=3$, we have
\begin{eqnarray*}
    \left| \frac{\partial {{h}_i(\theta)}}{\partial {\theta_{j}}}\right|
    &\leq& \Bigg| \sum_{k=1}^{i-2}[k \gamma^{k-1}(\omega^g+\alpha^g 
        I\beta_{i-1-k})+\gamma^k\{ (\varrho_1 - \varrho_2 + 2\varrho_3)\omega_1+(\varrho_2-2\varrho_3) \omega_2\\
        & & -(\varrho_2-2\varrho_3) \nu +2\varrho_3 I\beta_{i-1-k} \}] + (i-1)\gamma^{i-2}\dfrac{\partial {{h}_1(\theta)}}{\partial {\gamma}}
        \Bigg|  \cr
    &\leq&  C \sum_{k=1}^{i-2} k |\gamma|^k \{|\omega^g_u|\vee|\omega^g_l|+|\alpha^g_u|\vee|\alpha^g_l|  | I\beta_{i-1-k}|\} + C. 
\end{eqnarray*}
Then,  from Lemma \ref{Lemma-6}(a) and $|\gamma|<1$, we obtain 
\begin{equation*}
 \sup_{i \in \mathbb{N}} \mathbb{E}\left[ \sup_{\theta \in \Theta}  \left|  \frac{\partial {{h}_i(\theta)}}{\partial {\theta_{j}}}\right| \right] \leq C  \quad  \text{a.s.}
\end{equation*}
Similarly, we can check the boundedness of the second and third derivatives.
\endpf

\begin{lemma} \label{Lemma-7}
Under  the assumption of Theorem \ref{Theorem-2}, we have
\begin{align}
    & \sup_{\theta \in \Theta} \left| \hat{L}_{n,m} (\theta) - \hat{L}_{n} (\theta) \right|= O_{p}\left( m^{-1/4} \sqrt{ \log n} \right), \label{Equation-A.29} \\ 
    & \sup_{\theta \in \Theta} \left| \hat{L}_{n} (\theta) - L_{n} (\theta) \right| = o_{p}(1), \label{Equation-A.30}\\
    & \sup_{\theta \in \Theta} \left| \hat{L}_{n,m} (\theta) - L_{n} (\theta)  \right| =  O_{p}\left( m^{-1/4} \sqrt { \log n }  \right) + o_{p}(1). \label{Equation-A.31}
\end{align}
\end{lemma}

\textbf{Proof of Lemma \ref{Lemma-7}.}
Consider \eqref{Equation-A.29}. We have
\begin{eqnarray*}
    \left| \hat{L}_{n,m} (\theta) - \hat{L}_{n} (\theta) \right| 
    &\leq& \frac{1}{n} \sum^n_{i=1} \left|RIB^2_i- I\beta^2_i \right| + \frac{2}{n} \sum^n_{i=1} \left| RIB_i\{h_i(\theta)-\hat{h}_i(\theta)\} \right| \cr
    & &  +\frac{2}{n} \sum^n_{i=1}  \left|  h_i(\theta)( I\beta_i-RIB_i)\right|+ \frac{1}{n} \sum^n_{i=1} | \hat{h}^2_i(\theta)-h^2_i(\theta)|  \cr
    &=& \left( I\right) +\left( II\right) +\left( III\right) + \left( IV\right).
\end{eqnarray*}
For $\left( I\right)$, we have 
\begin{eqnarray*}
    \frac{1}{n} \sum^n_{i=1} \left| RIB^2_i - I\beta^2_i\right|
    &=& \frac{1}{n} \sum^{n}_{i=1} \Big|  \{RIB_i -I\beta_i \}    \{RIB_i +I\beta_i \} \Big| \\
    &=& O_p(m^{-1/4} \sqrt { \log n } ),
\end{eqnarray*}
where the last inequality is due to Assumption \ref{Assumption-2} (e).
Also, we obtain
\begin{eqnarray} \label{Equation-A.32}
    |\sup_{\theta \in \Theta} \big( \hat{h}_i(\theta)-h_i(\theta) \big)  | 
    &\leq& C \sum^{i-2}_{k=0} |\gamma_u|^k \vee |\gamma_l|^k 
    |  RIB_{i-1-k}-I\beta_{i-1-k} |\cr
    &=& O_p( m^{-1/4} \sqrt { \log n } )  \quad  \text{a.s.}
\end{eqnarray}
For  $\left( II\right)$, by \eqref{Equation-A.32}, we have 
\begin{eqnarray*} 
     \sup_{\theta \in \Theta} \frac{2}{n} \sum^n_{i=1} \left| RIB_i\{h_i(\theta)-\hat{h}_i(\theta)\} \right| 
     &=& O_p(m^{-1/4} \sqrt { \log n }  ).
\end{eqnarray*}
For $(III)$, by Lemma \ref{Lemma-6}(a) and Assumption \ref{Assumption-2}(e), we have
\begin{eqnarray*}
    \sup_{\theta \in \Theta} \frac{2}{n} \sum^n_{i=1}  \left|  h_i(\theta)( I\beta_i-RIB_i)\right| 
    &=& O_p(m^{-1/4} \sqrt { \log n } ).
\end{eqnarray*}
For $(IV)$, by \eqref{Equation-A.32} and Lemma \ref{Lemma-6}(a), we have
\begin{eqnarray*}
    \sup_{\theta \in \Theta} \sum^n_{i=1} \left| \hat{h}^2_i(\theta)-h^2_i(\theta) \right|
    &=& \frac{1}{n} \sum^{n}_{i=1}\sup_{\theta \in \Theta} \Big|  \{\hat{h}_i(\theta)-h_i(\theta) \} \{\hat{h}_i(\theta) + h_i(\theta) \} \Big|  \cr
    &=& O_p(m^{-1/4} \sqrt { \log n }     ).
\end{eqnarray*}
Hence, we have 
\begin{equation*}
    \sup_{\theta \in \Theta} \left| \hat{L}_{n,m} (\theta) - \hat{L}_{n} (\theta) \right|= O_{p}\left( m^{-1/4} \sqrt { \log n }  \right).
\end{equation*}

Consider \eqref{Equation-A.30}. 
We have 
\begin{equation*}
    \hat{L} _{n}(\theta)-L _{n}(\theta)=-\frac{2}{n} \sum_{i=1}^{n} D_i\{h_i(\theta_0)-h_i(\theta) \}.
\end{equation*}
Since $h_i(\theta)$ is adapted to $\mathcal{F}_{i-1}$, $D_i\{h_i(\theta_0)-h_i(\theta) \}$ is also a martingale difference. 
Also, $D_i\{h_i(\theta_0)-h_i(\theta) \}$ is uniform integrable.
Then,  by application of Theorem 2.22 in \citet{hall2014martingale}, we can show 
\begin{equation*}
    \left|\hat{L} _{n}(\theta)-L _{n}(\theta)\right| \rightarrow 0 \quad \text{in probability}.
\end{equation*}
Define 
\begin{equation*}
    G_n(\theta)=\hat{L} _{n}(\theta)-L _{n}(\theta).
\end{equation*}
From Theorem 3 in \citet{andrews1992generic}, the stochastic equicontinuity of $G_n(\theta)$ implies that $G_n(\theta)$ uniformly converges to 0.
Thus, we enough to show $G_n(\theta)$ is stochastic equicontinuous.
By the mean value theorem and Talyor expansion, there exists $\theta^{\ast }$ between $\theta$ and $\theta'$ such that 
\begin{eqnarray*}
    \left| G_n(\theta) -G_n(\theta') \right|
    &=&  \frac{2}{n}\sum_{i=1}^{n} \left| \dfrac{\partial h_i(\theta^{\ast })}{\partial \theta} D_i(\theta-\theta') \right| \cr
    &\leq&  \frac{C}{n}\sum_{i=1}^{n} \left \|  \dfrac{\partial h_i(\theta^{\ast })}{\partial \theta} D_i \right\|_{\max} \|\theta-\theta'\|_{\max} .
\end{eqnarray*}
Similar to  the proofs of Lemma \ref{Lemma-6}(b) and \eqref{Equation-A.27}, we can show that 
\begin{equation*} 
    \mathbb{E}\left[ \left\|  \sup_{\theta^{\ast } \in \Theta} \left| \frac{\partial {h_i(\theta^{\ast })}}{\partial {\theta_{ j}}} D_i \right| \right\|_{\max} \right] \leq C \quad \text{a.s.},
\end{equation*}
which implies that $G_n(\theta)$ is stochastic equicontinuous.

Finally, the triangular inequality concludes \eqref{Equation-A.31}.
\endpf

\begin{proposition} \label{Proposition-2}
Under Assumptions \ref{Assumption-2}  (except for $(n   \log n  ) ^2m^{-1} \rightarrow 0$), there is a unique maximizer of $L_n (\theta)$ and as $m,n \rightarrow \infty$, $\hat{\theta} \rightarrow \theta_0$ in probability.
\end{proposition}

\textbf{Proof of Proposition \ref{Proposition-2}.}
First, we show that there is a unique maximizer of $L_n(\theta)$. 
$L_n(\theta)$ is concave and the solution $\theta$ of $ \partial L_n (\theta) / \partial \theta = 0$ should satisfy $h_i(\theta)=h_i(\theta_0)$ for all $i=1, \ldots , n$. 
Thus, the maximizer $\theta^{\ast}$ should satisfy $h_i(\theta^{\ast })=h_i(\theta_0)$ for all $i=1, \ldots , n$. 
By Proposition \ref{Proposition-1}(a), we have
\begin{eqnarray*}
	h_i (\theta)= \frac{\omega^g}{\varphi_{\theta} (1) } + \frac{\Upsilon_{\theta} (B) }{\varphi_{\theta} (B) } D_{i} \text{ a.s.},
\end{eqnarray*}
where $B$ is the back operator. 
Since $\varphi_{\theta_0} (B)$ and $\Upsilon_{ \theta_0} (B)$  do not have the common root and $D_i$'s are nondegenerating,  to satisfy $h_i(\theta^{\ast })=h_i(\theta_0)$ for all $i=1, \ldots , n$, we should have $\theta^{\ast } = \theta_0$.
 Thus, there is a unique maximizer, $\theta_0$. 
Then, the statement can be shown by Theorem 1 in \citet{xiu2010quasi} with the result of Lemma \ref{Lemma-7}.
\endpf
\\

\textbf{Proof of Theorem \ref{Theorem-2}.}
By the mean value theorem and Taylor expansion, there exists $\theta^{\ast }$ between $\hat{\theta}$ and $\theta_0$ such that
\begin{equation*}
    \hat{s}_{n,m}(\theta_0)-\hat{s}_{n,m}(\hat{\theta})=\hat{s}_{n,m}(\theta_0)=-\triangledown \hat{s}_{n,m}(\theta^{\ast })(\hat{\theta}-\theta_0).
\end{equation*}
Similar to the proofs of Proposition \ref{Proposition-2}, we can show 
\begin{equation*} 
    -\triangledown \hat{s}_{n,m}(\theta^{\ast }) \overset{p}{\rightarrow} -\triangledown s_{n}(\theta_0).
\end{equation*}
Then,  from the concavity of $\hat{L}_{n,m}(\theta)$, the convergence rate of $\left\|\hat{\theta}-\theta_0 \right \|_{\max}$ is the same as that of $ s_{n}(\theta_0)$.
Thus, it is enough to show 
\begin{equation} \label{Equation-A.33}
    \hat{s}_{n,m} (\theta_0)=O_p(m^{-1/4} \sqrt { \log n} )+O_p(n^{-1/2}).
\end{equation}
Similar to the proof of Lemma \ref{Lemma-7}, we can show that
\begin{eqnarray} \label{Equation-A.34}
    \hat{s}_{n,m} (\theta_0)
    &=&s_{n} (\theta_0)+ \frac{2}{n}\sum_{i=1}^{n} \dfrac{\partial h_i(\theta_0)}{\partial \theta} D_i+O_p(m^{-1/4}\sqrt { \log n} ) \cr
    &=&\frac{2}{n}\sum_{i=1}^{n} \dfrac{\partial h_i(\theta_0)}{\partial \theta}  D_i+O_p(m^{-1/4}\sqrt { \log n} ).
\end{eqnarray}
Since $D_i$ is a martingale difference and $h_i(\theta_0)$ is $\mathcal{F}_{i-1}$-adaptive, we have 
\begin{equation*}
    \frac{2}{n}\sum_{i=1}^{n} \dfrac{\partial h_i(\theta_0)}{\partial \theta}  D_i=O_p(n^{-1/2}),
\end{equation*}
which shows \eqref{Equation-A.33} with \eqref{Equation-A.34}.
\endpf 
\\

\subsection{Proof of Theorem \ref{Theorem-3}}
\textbf{Proof of Theorem \ref{Theorem-3}.}
By the mean value theorem and Taylor expansion, we obtain, for some $\theta^{\ast }$ between $\theta_0$ and $\hat{\theta} $,
\begin{eqnarray*}
    -\triangledown \hat{s} _{n,m}(\theta^{\ast })(\hat{\theta} -\theta_0)
    &&= \hat{s} _{n}(\theta_0)+\left\{ \hat{s} _{n,m}(\theta_0)- \hat{s} _{n}(\theta_0)\right\} \cr
    &&= \frac{2}{n} \sum ^n_{i=1} \dfrac{\partial h_i(\theta_0)}{\partial \theta}  D_i+O_p(m^{-1/4}\sqrt { \log n} ),
\end{eqnarray*}
where the last equality is due to \eqref{Equation-A.34}. 
By Assumption \ref{Assumption-2}(b), we can show that $I\beta_i$ and $h_i(\theta)$ can be represented by MA($\infty$) with $D_i$'s. 
Thus,  $I\beta_i$'s and $h_i(\theta)$'s are strictly stationary. 
By the ergodic theorem and the result in the proof of Theorem \ref{Theorem-2}, we have 
\begin{equation*} 
    -\triangledown \hat{s} _{n,m}(\theta^{\ast })  
    \overset{p} \rightarrow 
    2 \mathbb{E} \left[ \left. {\dfrac{\partial h_1(\theta)}{\partial \theta} \dfrac{\partial h_1(\theta)}{\partial \theta^\top} }  \right|_ {\theta=\theta_0} \right],
\end{equation*}
where its asymptotic covariance matrix is positive definite.
The ergodic theorem also provides that
\begin{equation*}
    \sqrt{n}\hat{s}_i (\theta_0)=\sqrt{n}\frac{2}{n} \sum ^n_{i=1} \left\{ \dfrac{\partial h_i(\theta_0)}{\partial \theta}  D_i \right\}
    \overset{d} \rightarrow  N(0,S),
\end{equation*}
where
\begin{eqnarray*}
    &&S =\mathbb{E} \left[ 16 \alpha_{0,1}^{-4}\nu_0^2\int^1_0\{\alpha_{0,1} (1-t-\alpha_{0,1}^{-1})e^{\alpha_{0,1}(1-t)}+1\}^2tdt  
   \left. {\dfrac{\partial h_1(\theta)}{\partial \theta} \dfrac{\partial h_1(\theta)}{\partial \theta^\top} }\right|_{\theta=\theta_0} \right]. \cr
\end{eqnarray*}
Therefore, by Slutsky's theorem, we obtain
\begin{equation*}
    \sqrt{n}(\hat{\theta} -\theta_0) \overset{d} \rightarrow  N(0,V).
\end{equation*}
where
\begin{eqnarray*}
    V
    &=&\left(2 \mathbb{E} \left[ \left. {\dfrac{\partial h_1(\theta)}{\partial \theta} \dfrac{\partial h_1(\theta)}{\partial \theta^\top} }  \right|_ {\theta=\theta_0} \right]\right)^{-1}
    S
    \left(2 \mathbb{E} \left[ \left. {\dfrac{\partial h_1(\theta)}{\partial \theta} \dfrac{\partial h_1(\theta)}{\partial \theta^\top} }  \right|_ {\theta=\theta_0} \right]\right)^{-1} \\
    &=&4 \alpha_{0,1}^{-4}\nu_0^2\int^1_0\{\alpha_{0,1} (1-t-\alpha_{0,1}^{-1})e^{\alpha_{0,1}(1-t)}+1\}^2tdt   \left ( \mathbb{E} \left[  \left. 
    {\dfrac{\partial h_1(\theta)}{\partial \theta} \dfrac{\partial h_1(\theta)}{\partial \theta^\top} }\right|_{\theta=\theta_0}  \right]\right )^{-1}.
\end{eqnarray*}
\endpf

\end{spacing}

\end{document}